\documentclass[prx,twocolumn,english,superscriptaddress,floatfix,longbibliography]{revtex4}

\usepackage{color}
\usepackage{amsmath}
\usepackage{amssymb}
\usepackage{graphicx}
\usepackage{tikz}
\usepackage{multirow}

\usepackage{float}
\usepackage{bbm}
\usepackage{dsfont}

\usepackage{epsfig}
\usepackage{verbatim}
\usepackage{array}
\usepackage{setspace}
\newcommand*{\red}{\textcolor{red}} 

\usepackage[unicode=true,
bookmarks=true,bookmarksnumbered=false,bookmarksopen=false,
breaklinks=false,pdfborder={0 0 1},backref=false,colorlinks=true]
{hyperref}
\hypersetup{
	linkcolor=magenta, urlcolor=blue, citecolor=blue, pdfstartview={FitH}, hyperfootnotes=false, unicode=true}

\makeatletter
\@ifundefined{textcolor}{}
{%
	\definecolor{BLACK}{gray}{0}
	\definecolor{WHITE}{gray}{1}
	\definecolor{RED}{rgb}{1,0,0}
	\definecolor{GREEN}{rgb}{0,1,0}
	\definecolor{BLUE}{rgb}{0,0,1}
	\definecolor{CYAN}{cmyk}{1,0,0,0}
	\definecolor{MAGENTA}{cmyk}{0,1,0,0}
	\definecolor{YELLOW}{cmyk}{0,0,1,0}
}

\usepackage{amsfonts}\usepackage{tabularx}\usepackage{dcolumn}\usepackage{bm}\usepackage{graphicx}\usepackage{epstopdf}

\hypersetup{urlcolor=blue}

\makeatother

\allowdisplaybreaks
\newcolumntype{C}[1]{>{\centering\arraybackslash$}p{#1}<{$}}

\begin{document}
	
	\title{Robust entangling gate for capacitively coupled few-electron singlet-triplet qubits}
	
	\author{Guo Xuan Chan}
	\affiliation{Department of Physics, City University of Hong Kong, Tat Chee Avenue, Kowloon, Hong Kong SAR, China, and City University of Hong Kong Shenzhen Research Institute, Shenzhen, Guangdong 518057, China}
	
	\author{Xin Wang}
	\email{x.wang@cityu.edu.hk}
	\affiliation{Department of Physics, City University of Hong Kong, Tat Chee Avenue, Kowloon, Hong Kong SAR, China, and City University of Hong Kong Shenzhen Research Institute, Shenzhen, Guangdong 518057, China}
	\date{\today}
	
	\begin{abstract}
		The search of a sweet spot, locus in qubit parameters where quantum control is first-order insensitive to noises, is key to achieve high-fidelity quantum gates. Efforts to search for such a sweet spot in conventional double-quantum-dot singlet-triplet qubits where each dot hosts one electron (``two-electron singlet-triplet qubit''), especially for two-qubit operations, have been unsuccessful. Here we consider singlet-triplet qubits allowing each dot to host more than one electron, with a total of four electrons in the double quantum dots (``four-electron singlet-triplet qubit''). We theoretically demonstrate, using configuration-interaction calculations, that sweet spots appear in this coupled qubit system. We further demonstrate that, under realistic charge noise and hyperfine noise, two-qubit operation at the proposed sweet spot could offer gate fidelities ($\sim99\%$) that are higher than conventional two-electron singlet-triplet qubit system ($\sim90\%$). Our results should facilitate realization of high-fidelity two-qubit gates in singlet-triplet qubit systems.
	\end{abstract}
	\maketitle

	\section{Introduction}

	Spin qubits based on quantum dots (QD) are promising candidates for realization of large-scale quantum computation. In semiconductor double-quantum-dot (DQD) devices, previous works have mostly focused on singlet-triplet qubits defined by two-electron spin states  \cite{Petta.05,Shulman.12,Levy.02,Wu.14,Maune.12,Barthel.10,Shi.11,Takeda.20,Cerfontaine.20,Eng.15,Noiri.18,Harvey.17}, or three-electron spin states in the large detuning regime, so called ``hybrid-qubit'' \cite{Koh.12,Shi.14,Shi.12,Thorgrimsson.17,Koh.13}. Although high fidelity single-qubit gates with long coherence time have been achieved in these systems, two-qubit gate is still a key obstacle toward realization of quantum algorithms since the effect of charge noises prevents high-fidelity quantum control \cite{Cao.13,Shinkai.09,Hayashi.03,Petersson.10,Dovzhenko.11,Gorman.05,Shi.13}. 
	
	Two-qubit gates between singlet-triplet qubits can be generally divided into two classes: capacitive gates and exchange gates. Capacitive gates are achieved by introducing charge-dipole to the logical states, usually by modifying the DQD detuning, while suppressing the electron tunneling between DQDs such that the interaction between qubits is solely mediated by the capacitive Coulomb interaction \cite{Shulman.12,Nichol.17,Taylor.05,Nielsen.12,Hiltunen.14,Buterakos.19,Ramon.11,Calderon.15,Wolfe.17,Stepanenko.07,Yang.11,Srinivasa.15,Setser.19,Frees.19,Buterakos.18.2}. Exchange gates, on the other hand, are mediated by the exchange coupling between two neighboring spins between DQDs, which can be manipulated by inter-DQD exchange interaction \cite{Li.12,Klinovaja.12,Mehl.14,Wardrop.14,Buterakos.18,Cerfontaine.20.2,Yang.20}. 
	Practically, capacitive gates are easier to implement \cite{Shulman.12,Nichol.17} as their realization allows for a reasonable inter-DQD distance, which is relatively easy to faricate while the capacitive crosstalk can be suppressed \cite{Buterakos.18.2}. Moreover, leakage into energetically accessible non-logical $S_z=0$ states is forbidden due to the absence of inter-DQD electron tunneling. In contrast, exchange gates require a much more complicated experimental setup, including a non-uniform magnetic field across four dots in a double-DQD device (DDQD) to suppress leakage and pulse design \cite{Buterakos.18,Cerfontaine.20,Wardrop.14,Li.12} to reduce the crosstalk effect by both exchange and capacitive interaction. In this work, we focus on capacitively coupled singlet-triplet qubits. 
	
	Although being free from leakage by inter-DQD tunneling, capacitive gates suffer from charge noises that arise from the coupling between the charge dipoles introduced during manipulation of the DDQD device and nearby charged impurities. Researchers therefore have been actively searching for a sweet spot, where the control fidelity is first-order insensitive to charge noises,  to operate capacitive gates \cite{Yang.11,Wolfe.17,Abadillo.19,Abadillo.21}. In some of these works, a simplified version of Configuration Interaction (CI) method is employed to study possible existence of sweet spots for two-qubit capacitive gates between a pair of singlet-triplet qubits. As an example, Ref.~\onlinecite{Yang.11} has shown that the effective exchange energy of a qubit is insensitive to the corresponding DQD detuning. As another example, Ref.~\onlinecite{Wolfe.17}, by balancing the local exchange energies and capacitive shift by inter-DQD Coulomb interaction, has shown that the effective exchange energies of two qubits can be made simultaneously insensitive to both DQD detunings. However, a more rigorous calculation based on full Configuration Interaction (full CI) shows the sweet spots are absent under the same dot parameters \cite{Chan.21}. 
	
	Other works attempt to establish a sweet spot for single-qubit operation in presence of charge dipoles. Ref.~\onlinecite{Abadillo.19} has proposed that, under a relatively large external magnetic field gradient, sweet spots for a singlet-triplet qubit can be found in \red{the} detuning regime where \red{the} doubly-occupied singlet is energetically accessible to introduce a charge dipole.
	Ref.~\onlinecite{Abadillo.21} has shown that the ``hybrid qubit'' can similarly be operated in the detuning regime where logical states and leakage states are highly mixed, forming charge dipoles. It is however challenging to ramp the system into and out of the proposed sweet spots adiabatically, which may lead to leakage.

	Departing from two- or three-electron systems, some works have shown that single-triplet qubits hosted in multielectron systems offer a higher degree of protection against charge noises due to the screening effect \cite{Barnes.11,Leon.21,Mehl.13}. However, in the regime where charge dipoles are present, sweet spots are not found. Therefore, capacitive gates performed in those multielectron systems are still expected to be sensitive to charge noises.
	
	Recent experimental and theoretical works show that by populating more electrons into QDs, the exchange energy may change non-monotonically due to additional magnetic correlations among the electron spins \cite{Deng.18,Malinowski.18,Martins.17}, suggesting existence of sweet spots. Inspired by these works, we would like to study the effect of multiple electrons on the exchange interaction under multi-qubit situation. In this paper, we show, by projecting full CI results of a few-electron singlet-triplet qubit onto the capacitive Coulomb integrals, effective single-qubit exchange energy sweet spots appear in the coupled singlet-triplet qubits system. Furthermore, these sweet spots lie very closely to the sweet spots for capacitive gates, enabling high fidelity manipulations. We demonstrate that operating at the sweet spots yields entangling gates with high fidelities ($>99.9\%$), even under realistic noise environments. Our results should facilitate realization of high-fidelity two-qubit gates in singlet-triplet qubit systems.
	
	The remainder of the paper is organized as follows. Sec.~\ref{sec:Model} presents the model of the capacitively coupled few-electron singlet-triplet qubits, including the effective Hamiltonian of the two-qubit system (Sec.~\ref{subsec:2QH}), the CI method to obtain the eigenvalues of the logical states as function of control parameters (Sec.~\ref{subsec:CI}), and the analytical interpretation of the qubit parameters based on the extended Hubbard Model (Sec.~\ref{subsec:HubbardModel}). Sec.~\ref{sec:result} shows the results, including exchange energies sweet spots and the corresponding capacitive coupling (Sec.~\ref{subsec:ExSS}), phonon-mediated decoherence effect (Sec.~\ref{subsec:phonon}), and the simulated two-qubit gate fidelities under noisy environment (Sec.~\ref{subsec:Fid}). In the end, we summarize our results in Sec.~\ref{sec:conclusion}.

	\section{Model}\label{sec:Model}
	\subsection{Two-qubit Hamiltonian}\label{subsec:2QH}
	\begin{figure}[t]
		\includegraphics[width=\linewidth]{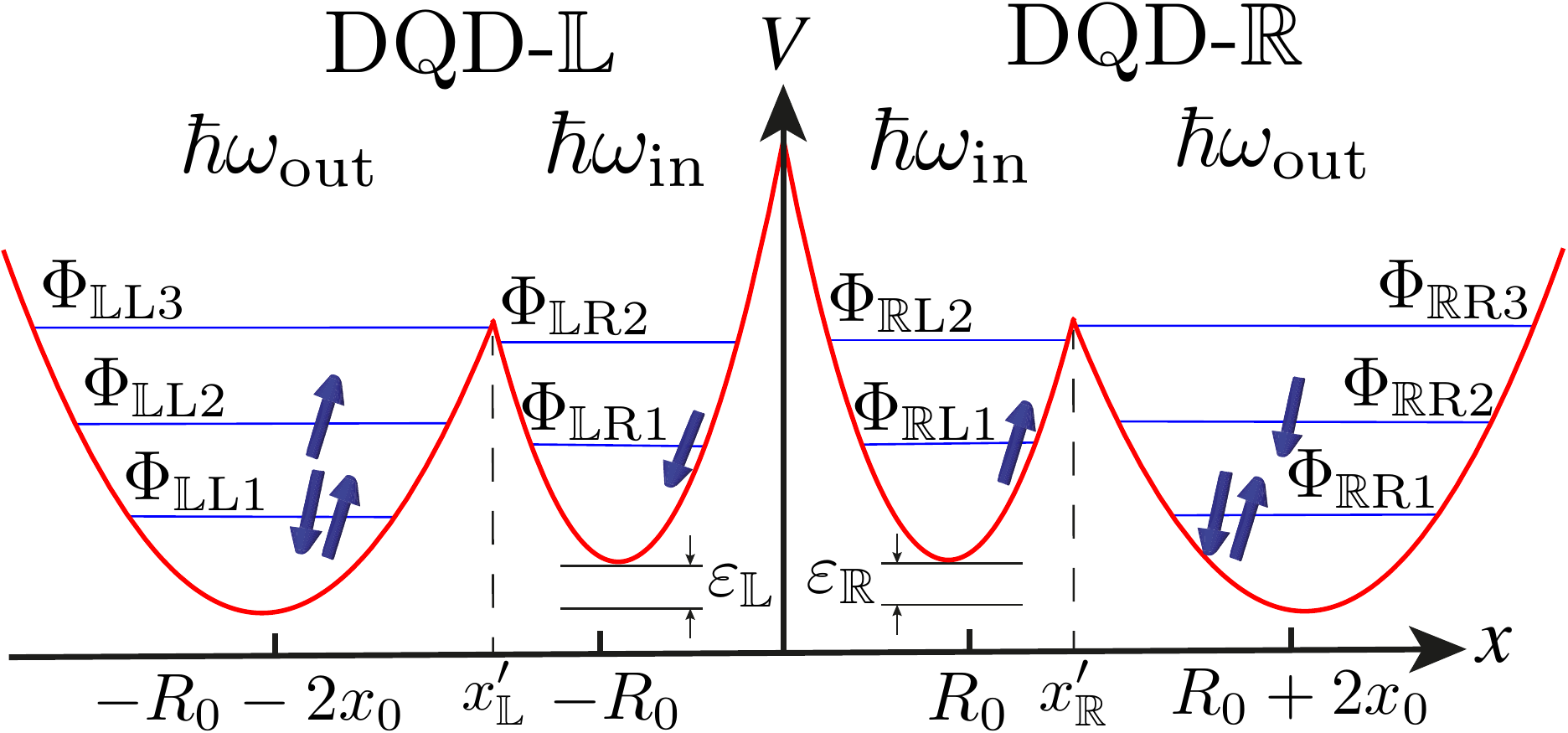}
		\caption{Schematic illustration of the DDQD device, where $\mathbb{L}$ and $\mathbb{R}$ DQD denote left and right DQD respectively, with $x=0$ being the boundary between them. The potential function of each DQD is shown in Eq.~\eqref{eq:V}. The label, $\Phi_j$, denotes the orthonormalized F-D states.}
		\label{fig:DDQDV}
	\end{figure}
	We consider a pair of capacitively coupled singlet-triplet qubits realized in a DDQD device, with each DQD hosting a \emph{four-electron} singlet-triplet qubit, as shown schematically in Fig.~\ref{fig:DDQDV}. The two-qubit Hilbert space is spanned by the products of single-qubit logical states:  the spin-singlet $|S\rangle$, and the unpolarized spin-triplet state $|T\rangle$ ($S_z = 0$). Hence, the two-qubit logical states are $|SS\rangle$, $|ST\rangle$, $|TS\rangle$ and $|TT\rangle$. Under this set of bases, the system Hamiltonian is written as \cite{Calderon.15,Stepanenko.07,Wolfe.17}
	\begin{align}\label{eq:Hint}
		\begin{split}
			&H (\varepsilon_\mathbb{L}, \varepsilon_\mathbb{R},h_\mathbb{L},h_\mathbb{R})
			\\
			&= \sum_{j\in\left\{\mathbb{L},\mathbb{R}\right\}} \Bigg[\left(\frac{J_j(\varepsilon_j)}{2}-\beta_j (\varepsilon_\mathbb{L},\varepsilon_\mathbb{R})\right)\sigma_z^{(j)}
			+\frac{h_j}{2}\sigma_x^{(j)}
			\\
			&\quad+\alpha\left(\varepsilon_\mathbb{L},\varepsilon_\mathbb{R}\right) \sigma_z^{(\mathbb{L})}\sigma_z^{(\mathbb{R})}\Bigg]
			\\
			&=\sum_{j\in\left\{\mathbb{L},\mathbb{R}\right\}} J_j^\text{eff}\left(\varepsilon_{\mathbb{L}},\varepsilon_{\mathbb{R}}\right)\sigma_z^{(j)}+\frac{h_j}{2}\sigma_x^{(j)}
			\\
			&\quad +\alpha\left(\varepsilon_\mathbb{L},\varepsilon_\mathbb{R}\right) \sigma_z^{(\mathbb{L})}\sigma_z^{(\mathbb{R})},
		\end{split}
	\end{align}
	where the index $j$ denotes left $(\mathbb{L})$ and right $(\mathbb{R})$ DQD (denoted by DQD-$\mathbb{L}$ and DQD-$\mathbb{R}$ respectively). In Eq.~\eqref{eq:Hint}, $\vec{\sigma}^{(j)}$ are the Pauli operators of a singlet-triplet qubit in the DQD labeled by $j$ (DQD-$j$), $J_j$ and $h_j$ are the local exchange energy and magnetic field gradient for the same DQD respectively. The inter-DQD Coulomb interaction does two things: it introduces a shift $\beta_j$ in the local exchange energy, and at the same time gives rise to inter-qubit capacitive coupling, $\alpha$, resulting in the effective single-qubit exchange energy, $J_j^\text{eff}$, where $J_j^\text{eff} = J_j/2-\beta_j$.

	\subsection{Configuration Interaction, exchange energies and inter-qubit interaction}\label{subsec:CI}
	We consider an $n$-electron system $H=h_0 + h_C = \sum h_j + \sum{e^2}/4\pi \kappa \left|\mathbf{r}_j-\mathbf{r}_k\right|$
	with the single-particle Hamiltonian $h_0 = \sum h_j = \sum {(-i\hbar \nabla_j+e \mathbf{A})^2}/{2m^*}+V(\mathbf{r})+g^*\mu_B \mathbf{B}\cdot \mathbf{S}$.
	
	The confinement potential of the left $(\mathbb{L})$ and right $(\mathbb{R})$ double-quantum-dot device (DQD) can be modeled as (cf. Fig.~\ref{fig:DDQDV}) \cite{Barrier}: 
	\begin{equation}\label{eq:V}
	\begin{split}
		V_\mathbb{L}(\mathbf{r})&=
		\begin{cases}
			V\left(\mathbf{r}\left|\right.-\mathbf{R}_\text{out},\omega_\text{out}\right) & x < x'_\mathbb{L},\\
			V\left(\mathbf{r}\left|\right.-\mathbf{R}_\text{in},\omega_\text{in}\right) & x'_\mathbb{L}<x<0,
		\end{cases},
	\\
	V_\mathbb{R}(\mathbf{r})&=
	\begin{cases}
		V\left(\mathbf{r}\left|\right.\mathbf{R}_\text{in},\omega_\text{in}\right) & x'_\mathbb{R} > x > 0,\\
		V\left(\mathbf{r}\left|\right.\mathbf{R}_\text{out},\omega_\text{out}\right) & x>x'_\mathbb{R},
	\end{cases},
	\end{split}
	\end{equation}
	where 
	\begin{equation}
		V\left(\mathbf{r}\left|\right.\widetilde{\mathbf{R}},\widetilde{\omega}\right) = \frac{1}{2}m^* \widetilde{\omega}^2\left(\mathbf{r}-\widetilde{\mathbf{R}}\right)^2,
	\end{equation}
	$\mathbf{r} = \left(x,y\right)$ is the vector in $x-y$ plane, $\mathbf{R}_\text{out} = \left(R_0+2x_0,0\right)$ and $\mathbf{R}_\text{in} = \left(R_0,0\right)$
	are the minima of the parabolic wells, $m^*$ the effective mass, and $\widetilde{\omega}$ the confinement strength. The potential cut, $x'_\mathbb{L}$ $(x'_\mathbb{R})$, is determined by locating the value of $x$ at which the potential values of left and right dot in DQD-$\mathbb{L}$ $(\mathbb{R})$ are equal at $y=0$. The parameters used in this work are: $m^*= 0.067m_e$ ($m_e$: electron mass) and $\kappa = 13\varepsilon_0$ ($\varepsilon_0$: vacuum permittivity) in GaAs, $\varepsilon_\mathbb{L}>0$, $\varepsilon_\mathbb{R}>0$, $\hbar \omega_\text{in} = 7.5$ meV, $\hbar \omega_\text{out} = $ 4 meV, $h_j = 0.124$ $\mu$eV $= 2\pi \times 30$ MHz \cite{Shulman.12}, $x_0 = 40$ nm. The qubit parameters in Eq.~\eqref{eq:Hint} are obtained based on the results extracted from CI calculation, for which the details are outlined in the following sections (Sec.~\ref{subsubsec:SQJ} and Sec.~\ref{subsubsec:interQInt}).
	\subsubsection{Single qubit exchange energy}\label{subsubsec:SQJ}
	We consider a singlet-triplet qubit formed by \emph{four} electrons occupying a DQD with one electron occupying the smaller quantum-dot (QD), $\hbar \omega_\text{in}$, while three electrons occupying the larger QD, $\hbar \omega_\text{out}$ \cite{ChanGX.22}. The single qubit exchange energy, $J_j$, of four-electron states in a DQD is obtained using full CI method by taking \cite{Barnes.11}
	\begin{equation}\label{eq:J1Q}
		J_j = E_{|T\rangle}-E_{|S\rangle},
	\end{equation}
	where $E_{|S\rangle}$ and $E_{|T\rangle}$ are the eigenvalues of the lowest singlet and unpolarized triplet states respectively. We retain 6 Fock-Darwin (F-D) states for the QD with a single electron, and 10 F-D states for the QD with three electrons, and our experience with full-CI calculations indicate that this is sufficient for convergence \cite{ChanGX.22}.
The eigenvalues of the system are obtained by diagonalizing the Hamiltonian written in the bases of all possible four-electron Slater determinants for a given number of F-D states.
	
	The eigenstates of the lowest $|S\rangle$ and $|T\rangle$, written in the linear combination of four-electron Slater determinants formed by orthonormalized F-D states, are
	\begin{equation}\label{eq:ST1Q}
		\begin{split}
			|S\rangle &= \sum_{j,m} C_{S_{j,m}}|S(\uparrow_j \downarrow_j)\rangle_m \\
			&\quad+ \sum_{j,k,m} C_{S_{j,k,m}}|S(\uparrow_j \downarrow_k)\rangle_m + \cdots,\\
			|T\rangle &= \sum_{j,k,m} C_{T_{j,k,m}}|T(\uparrow_j \downarrow_k)\rangle_m+ \cdots,
		\end{split}
	\end{equation}
	respectively, where
	\begin{equation}\label{eq:STExplicit}
		\begin{split}
			|S(\uparrow_j \downarrow_j)\rangle_m &= |\uparrow_j \downarrow_j \uparrow_m \downarrow_m\rangle,\\
			|S(\uparrow_j \downarrow_k)\rangle_m &= |\uparrow_j \downarrow_k \uparrow_m \downarrow_m \rangle+|\uparrow_k \downarrow_j \uparrow_m \downarrow_m\rangle,\\
			|T(\uparrow_j \downarrow_k)\rangle_m &= |\uparrow_j \downarrow_k \uparrow_m \downarrow_m \rangle-|\uparrow_k \downarrow_j \uparrow_m \downarrow_m\rangle,
		\end{split}
	\end{equation}
	while the normalization conditions hold ($j,k,m$ are different integers):
	\begin{equation}\label{eq:Slater}
		\begin{split}
			\sum_{S_{j,m}} \left| C_{S_{j,m}}\right|^2+\sum_{S_{j,k,m}} \left|C_{S_{j,k,m}}\right|^2 + \sum \left|\cdots\right|^2 &= 1,\\
			\sum_{T_{j,k,m}} \left|C_{T_{j,k,m}}\right|^2+\sum \left|\cdots\right|^2 & = 1.\\
		\end{split}
	\end{equation}
	In Eq.~\eqref{eq:STExplicit}, $|\uparrow_k \downarrow_l \uparrow_m \downarrow_n \rangle$ is a four-electron Slater determinant with two spin-up electrons occupying the $k$ and $m$ orthonormalized F-D states, denoted as $\Phi_k$ and $\Phi_m$ respectively, while another two spin-down electrons occupying $\Phi_l$ and $\Phi_n$ states, cf. Fig.~\ref{fig:DDQDV}.
	In Eq.~\eqref{eq:ST1Q},  $\left(\cdots\right)$ indicates other possible singlet and triplet configurations which contribute less than those explicitly shown.
	The components of $|S\rangle$ and $|T\rangle$ can be further divided as
	\begin{equation}\label{eq:STDivide}
	\begin{split}
		|S\rangle &= C_{S(1,3)}|S(1,3)\rangle + C_{S(0,4)}|S(0,4)\rangle + \cdots,\\
		|T\rangle &= C_{T(1,3)}|T(1,3)\rangle + C_{T(0,4)} |T(0,4)\rangle + \cdots.\\
	\end{split}
	\end{equation}
	In Eq.~\eqref{eq:STDivide},
		\begin{subequations}\label{eq:STDivideExplicit}
			\begin{align}
		\begin{split}
			|S(1,3)\rangle = \sum_{\substack{j\in\{\mathbb{R}\text{L}1,\mathbb{R}\text{L}2, \cdots,\mathbb{R}\text{L}6\},\\k\in\{\mathbb{R}\text{R}2,\mathbb{R}\text{R}3, \cdots,\mathbb{R}\text{R}10\},\\m\in\{\mathbb{R}\text{R}1,\mathbb{R}\text{R}2, \cdots,\mathbb{R}\text{R}9\},\\m\neq k}} 
			&\frac{C_{S_{j,k,m}}}{\left|C_{S(1,3)}\right|} |S(\uparrow_j \downarrow_k) \rangle_m 
			\\ &+\cdots,
		\end{split}
		\\
		\begin{split}
			|T(1,3)\rangle = \sum_{\substack{j\in\{\mathbb{R}\text{L}1,\mathbb{R}\text{L}2, \cdots,\mathbb{R}\text{L}6\},\\k\in\{\mathbb{R}\text{R}2,\mathbb{R}\text{R}3, \cdots,\mathbb{R}\text{R}10\},\\m\in\{\mathbb{R}\text{R}1,\mathbb{R}\text{R}2, \cdots,\mathbb{R}\text{R}9\},\\m\neq k}} 
			&\frac{C_{T_{j,k,m}}}{\left|C_{T(1,3)}\right|} |T(\uparrow_j \downarrow_k) \rangle_m 
			\\& + \cdots,
		\end{split}\\
		\begin{split}
			|S(0,4)\rangle = \sum_{\substack{\{j,k\}\in\{\mathbb{R}\text{R}2,\mathbb{R}\text{R}3, \cdots,\mathbb{R}\text{R}10\},\\m\in\{\mathbb{R}\text{R}1,\mathbb{R}\text{R}2, \cdots,\mathbb{R}\text{R}9\},\\j = k \text{ or } j \neq k, m\neq j, m\neq k}} 
			&\frac{C_{S_{j,k,m}}}{\left|C_{S(0,4)}\right|} |S(\uparrow_j \downarrow_k) \rangle_m \\
			&\quad+\cdots,
		\end{split}\\
		\begin{split}
			|T(0,4)\rangle = \sum_{\substack{\{j,k\}\in\{\mathbb{R}\text{R}2,\mathbb{R}\text{R}3, \cdots,\mathbb{R}\text{R}10\},\\m\in\{\mathbb{R}\text{R}1,\mathbb{R}\text{R}2, \cdots,\mathbb{R}\text{R}9\},\\j\neq k \neq m}} 
			&\frac{C_{T_{j,k,m}}}{\left|C_{T(0,4)}\right|} |T(\uparrow_j \downarrow_k) \rangle_m \\
			&\quad+\cdots,
		\end{split}
		\end{align}
		\end{subequations}
	where $(n_{\mathbb{R}\text{L}},n_{\mathbb{R}\text{R}})$ indicates the electron occupation of left (L) and right (R) QD in DQD-$\mathbb{R}$ respectively. Physically, $|S(1,3)\rangle$ and $|S(0,4)\rangle$ ($|T(1,3)\rangle$ and $|T(0,4)\rangle$) are singlets (triplets) with dot occupation of $(n_{\mathbb{R}\text{L}},n_{\mathbb{R}\text{R}})=(1,3)$ and $(0,4)$ respectively.
	 Eqs.~\eqref{eq:STDivide} and \eqref{eq:STDivideExplicit} are obtained for DQD-$\mathbb{R}$; similar expressions can be obtained for DQD-$\mathbb{L}$ by making the following replacements: $S(0,4) \rightarrow S(4,0)$, $T(0,4)\rightarrow T(4,0)$, $S(1,3)\rightarrow S(3,1)$, $T(1,3)\rightarrow T(3,1)$, $\mathbb{R}\text{R}\rightarrow \mathbb{L}\text{L}, \mathbb{R}\text{L} \rightarrow \mathbb{L}\text{R}$,  $n_{\mathbb{R}\text{L}}\rightarrow n_{\mathbb{L}\text{R}}$ and $n_{\mathbb{R}\text{R}}\rightarrow n_{\mathbb{L}\text{L}}$.
	\subsubsection{Inter-qubit interaction and capacitive shift}\label{subsubsec:interQInt}
	
	The inter-qubit parameters, $\beta_j$ and $\alpha$, are obtained by projecting the two-qubit logical states onto the inter-DQD Coulomb interaction \cite{Stepanenko.07,Calderon.15,Wolfe.17}, giving
	\begin{subequations}\label{eq:betaAlphaDef}
		\begin{align}
			\beta_\mathbb{L} &= \frac{1}{4}\left(-V_{|SS\rangle}-V_{|ST\rangle}+V_{|TS\rangle}+V_{|TT\rangle}\right),\\
			\beta_\mathbb{R} &= \frac{1}{4}\left(-V_{|SS\rangle}+V_{|ST\rangle}-V_{|TS\rangle}+V_{|TT\rangle}\right),\\
			\alpha &= \frac{1}{4}\left(V_{|SS\rangle}-V_{|ST\rangle}-V_{|TS\rangle}+V_{|TT\rangle}\right),
		\end{align}
	\end{subequations}
	where
	\begin{equation}\label{eq:VExp}
		V_{|jk\rangle} = \langle jk | H_C | jk \rangle,
	\end{equation}
	and $\{j,k\}\in\{S,T\}$. Note that for a \emph{traditional} singlet-triplet qubit formed by two singly-occupied QDs in a DQD, only the singlet states hybridize with each other while the triplet states are typically well-separated in energy and do not hybridize. However, in this case, both $|S\rangle$ and $|T\rangle$ involve hybridization between states of $(n_{\mathbb{R}\mathrm{L}},n_{\mathbb{R}\mathrm{R}})=(1,3)$ $\left[(n_{\mathbb{L}\mathrm{L}},n_{\mathbb{L}\mathrm{R}})=(3,1)\right]$ type and the states of $(0,4)$ $\left[(4,0)\right]$ type.

	\subsection{Extended Hubbard Model and inter-qubit interaction}\label{subsec:HubbardModel}
	It is helpful to obtain the physical descriptions of the qubit parameters in Eq.~\eqref{eq:Hint} based on the extended Hubbard model. We confirm the accuracy of the  model by comparing results derived from it to full CI calculations (cf. Sec.~VI in the Supplemental Material for details \cite{sm}). The explicit form of the effective Hamiltonian given by the extended Hubbard model is presented in Sec.~IV~D in the Supplemental Material \cite{sm}. Here for clarity, we show the effective Hamiltonian with truncated number of bases that is sufficient to understand the qualitative behavior of qubit parameters. Written in the bases of $|S(1,3)\rangle$, $|S(0,4)\rangle$, $|T(1,3)\rangle$, $|T(0,4)\rangle$ (cf.~Eq.~\eqref{eq:STDivideExplicit}), with a global energy shift of $E_\text{shift} = U_{\mathbb{R}\mathrm{L}1,\mathbb{R}\mathrm{R}2}+\varepsilon_\mathbb{R}$, we have
	\begin{equation}\label{eq:Heff}
		\begin{split}
			\widetilde{H}=
			\left(
			\begin{array} {cccc}
				0  & \sqrt{2}  t_{\mathbb{R}\text{L}1,\mathbb{R}\text{R}2}  & 0 & 0 \\
				\sqrt{2}  t_{\mathbb{R}\text{L}1,\mathbb{R}\text{R}2}  & E_S & 0 & 0 \\
				0 & 0 & 0  & -t_{\mathbb{R}\text{L}1,\mathbb{R}\text{R}3} \\
				0 & 0 & -t_{\mathbb{R}\text{L}1,\mathbb{R}\text{R}3} & E_T
			\end{array}
			\right),
		\end{split}
	\end{equation}
	where $t_{\mathbb{R}\mathrm{L}1,\mathbb{R}\mathrm{R}2}$ and $t_{\mathbb{R}\mathrm{L}1,\mathbb{R}\mathrm{R}3}$ are the tunneling energies, $E_S = U_{\mathbb{R}\mathrm{R}2} - (U_{\mathbb{R}\mathrm{L}1,\mathbb{R}\mathrm{R}2}+\varepsilon_{\mathbb{R}})$ and $E_T = U_{\mathbb{R}\mathrm{R}2,\mathbb{R}\mathrm{R}3}+\Delta E-\xi - (U_{\mathbb{R}\mathrm{L}1,\mathbb{R}\mathrm{R}3}+\varepsilon_{\mathbb{R}})$ are the globally shifted eigenvalues of $|S(0,4)\rangle$ and $|T(0,4)\rangle$ respectively. $\Delta E=\varepsilon_{\mathbb{R}\mathrm{R}3}-\varepsilon_{\mathbb{R}\mathrm{R}2}$ is the orbital splitting of two valence orbitals in DQD-$\mathbb{R}$, $\xi$ is the ferromagnetic exchange term \cite{Deng.18,Malinowski.18} (cf.~Sec.~IV B in the Supplemental Material for details \cite{sm}) while $\varepsilon_{\mathbb{R}}$ is the detuning between potential minima in a DQD. The eigenstates of Eq.~\eqref{eq:Heff} are
	\begin{equation}\label{eq:STEigenbasesMain}
		\begin{split}
			|S'(1,3)\rangle &= -\cos \frac{\theta^\mathbb{R}_S}{2}|S(1,3)\rangle + \sin \frac{\theta^\mathbb{R}_S}{2}|S(0,4)\rangle, \\
			|S'(0,4)\rangle &= \sin \frac{\theta^\mathbb{R}_S}{2}|S(1,3)\rangle + \cos \frac{\theta^\mathbb{R}_S}{2}|S(0,4)\rangle, \\
			|T'(1,3)\rangle &= \cos \frac{\theta^\mathbb{R}_T}{2}|T(1,3)\rangle + \sin \frac{\theta^\mathbb{R}_T}{2}|T(0,4)\rangle, \\
			|T'(0,4)\rangle &= -\sin \frac{\theta^\mathbb{R}_T}{2}|T(1,3)\rangle + \cos \frac{\theta^\mathbb{R}_T}{2}|T(0,4)\rangle, \\
		\end{split}
	\end{equation}
	where $\theta^\mathbb{R}_S=\arctan \left(2 \sqrt{2} t_{\mathbb{R}\text{L}1,\mathbb{R}\text{R}2}/E_S\right)$ and $\theta^\mathbb{R}_T=\arctan \left(2 t_{\mathbb{R}\text{L}1,\mathbb{R}\text{R}3}/E_T\right)$ are the admixture angles of $|S\rangle$ and $|T\rangle$ respectively (for more details, see Sec.~V~F in the Supplemental Material \cite{sm}). The logical subspace is defined in the $(n_{\mathbb{R}\mathrm{L}},n_{\mathbb{R}\mathrm{R}})\approx(1,3)$ regime, i.e. $\left|\sin (\theta^\mathbb{R}_S/2)\right|^2 \ll 1$ and $\left|\sin(\theta^\mathbb{R}_T/2)\right|^2 \ll 1$, such that $|S'(1,3)\rangle$ and $|T'(1,3)\rangle$ are the logical states.	Eqs.~\eqref{eq:Heff} and \eqref{eq:STEigenbasesMain} refer to the case for  DQD-$\mathbb{R}$, while the case for DQD-$\mathbb{L}$ can be derived using the replacement rules described at the end of Sec.~\ref{subsubsec:SQJ}. The admixtures can be alternatively perceived as introduced dipoles in the logical eigenstates, $\vert S(1,3)\rangle'$ and $\vert T(1,3)\rangle'$, i.e.
	\begin{equation}\label{eq:dipole}
		d^j_m=\sin^2\left(\frac{\theta^j_m}{2}\right),
	\end{equation} 
where $d^j_m$ is the dipole of the $m$ logical eigenstate in DQD-$j$.
	
	\begin{figure*}[t]
		\includegraphics[width=0.9\linewidth]{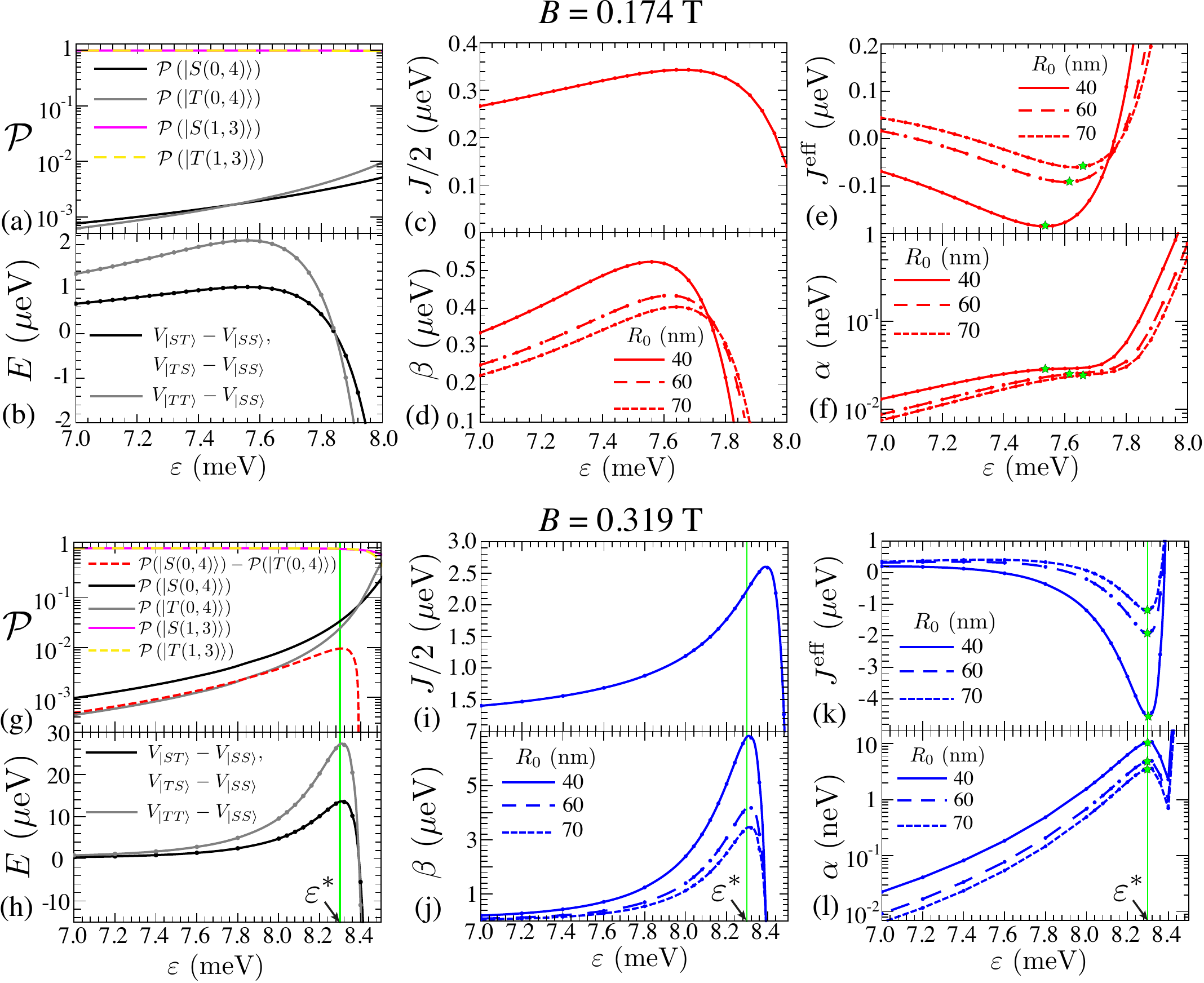}
		\caption{(a, g) The probabilities of various states $\mathcal{P}(|S(0,4)\rangle)=\left|\langle S |S(0,4)\rangle\right|^2$, $\mathcal{P}(|S(1,3)\rangle)=\left|\langle S |S(1,3)\rangle\right|^2$, $\mathcal{P}(|T(0,4)\rangle)=\left|\langle T |T(0,4)\rangle\right|^2$, $\mathcal{P}(|T(1,3)\rangle)=\left|\langle T |T(1,3)\rangle\right|^2$. (b, h) The inter-DQD Coulomb integral, $V_{|jk\rangle}$ (Eq.~\eqref{eq:VExp}). $R_0=40$ nm. (c, i) The local exchange energy, $J=J_\mathbb{L} (\varepsilon)=J_\mathbb{R}(\varepsilon)$. (d, j) The capacitive shift, $\beta=\beta_\mathbb{L} (\varepsilon,\varepsilon)=\beta_\mathbb{R}(\varepsilon,\varepsilon)$. (e, k) Effective exchange energy, $J^\text{eff} = J^\text{eff}_\mathbb{L} = J^\text{eff}_\mathbb{R}$. (f, l) Capacitive coupling, $\alpha(\varepsilon,\varepsilon)$. All the subfigures are plotted as function of the detuning, $\varepsilon$. The green star symbols, green line and the label $\varepsilon^*$ mark the detuning values at which the sweet spots of $J^\text{eff}$ occur. Top panel: (a-f) Results for $B=0.174$ T. Bottom panel: (g-l) Results for $B = 0.319$ T.}
		\label{fig:resultTwoB}
	\end{figure*}

	Considering the case in which DQD-$\mathbb{L}$ and DQD-$\mathbb{R}$ are symmetric, i.e. $\varepsilon=\varepsilon_{\mathbb{L}}=\varepsilon_{\mathbb{R}}$ in Fig.~\ref{fig:DDQDV}, the qubit parameters are
	\begin{subequations}\label{eq:betaAlphaApproxAna}
	\begin{align}
	\begin{split}\label{eq:betaApproxAna}
		\beta_\mathbb{L} (\varepsilon,\varepsilon) &= \beta_\mathbb{R} (\varepsilon,\varepsilon) \\
		&\approx \frac{1}{2}\left[\sin^2\left(\theta_S/2\right)-\sin^2\left(\theta_T/2\right)\right]
		U_{\mathrm{int}}\\
		&=\frac{1}{2}\left(\sqrt{d^\mathbb{L}_S d^\mathbb{R}_S}-\sqrt{d^\mathbb{L}_T d^\mathbb{R}_T}\right)
		U_{\mathrm{int}}
		,
	\end{split}
	\\
	\begin{split}\label{eq:alphaApproxAna}
		\alpha (\varepsilon,\varepsilon)
		&\approx
		\frac{1}{4}\left[\sin ^2\left(\theta_S/2\right)-\sin ^2\left(\theta_T/2\right)\right]^2
		U_{\mathrm{int}}
		\\
		&=\frac{1}{4}\left(\sqrt{d^\mathbb{L}_S d^\mathbb{R}_S}-\sqrt{d^\mathbb{L}_T d^\mathbb{R}_T}\right)^2
		U_{\mathrm{int}}
		,
	\end{split}
	\end{align}
	\end{subequations}
	where the inter-qubit Coulomb interaction is $U_{\mathrm{int}}=
	U_{\mathbb{R}\mathrm{R}2,\mathbb{L}\mathrm{L}2}
	+4U_{\mathbb{R}\mathrm{L}1,\mathbb{L}\mathrm{L}1}
	+U_{\mathbb{R}\mathrm{L}1,\mathbb{L}\mathrm{L}2}
	+U_{\mathbb{R}\mathrm{L}1,\mathbb{L}\mathrm{R}1}$. $U_\text{int}$ consists of the inter-dot Coulomb integral between two inner dots ($U_{\mathbb{R}\mathrm{L}1,\mathbb{L}\mathrm{R}1}$), the Coulomb integral between a left dot in DQD-$\mathbb{R}$ and a left dot in DQD-$\mathbb{L}$ ($U_{\mathbb{R}\mathrm{L}1,\mathbb{L}\mathrm{L}j}$), and the Coulomb integral between two outer dots ($U_{\mathbb{R}\mathrm{R}2,\mathbb{L}\mathrm{L}2}$). In the third line of Eq.~\eqref{eq:betaApproxAna} and the second line of Eq.~\eqref{eq:alphaApproxAna}, we have written the expressions of $\beta_j$ and $\alpha$ in terms of the dipoles, cf.~Eq.~\eqref{eq:dipole}. In Eq.~\eqref{eq:betaAlphaApproxAna}, due to the symmetric dots and detunings, $\theta_S^\mathbb{L}=\theta_S^\mathbb{R}=\theta_S$ and $\theta_T^\mathbb{L}=\theta_T^\mathbb{R}=\theta_T$. Eq.~\eqref{eq:betaAlphaApproxAna} indicates that the main qualitative behavior of $\beta_j$ and $\alpha$ is given by the difference of the hybridization in the singlet states, $\sin(\theta_S/2)$, and the hybridization in the triplet states, $\sin(\theta_T/2)$, i.e.,
	\begin{subequations}\label{eq:diffST}
	\begin{align}
	\begin{split}
		\beta_j \left(\varepsilon,\varepsilon\right)&\propto\sin ^2\left(\theta_S/2\right)-\sin ^2\left(\theta_T/2\right)
	\end{split},
	\\
	\begin{split}
		\alpha \left(\varepsilon,\varepsilon\right)&\propto \left[\sin ^2\left(\theta_S/2\right)-\sin ^2\left(\theta_T/2\right)\right]^2.
	\end{split}
	\end{align}
	\end{subequations}
	Eq.~\eqref{eq:betaAlphaApproxAna} will be used later to interpret the results (for details of derivation, see  Sec.~VII B in the Supplemental Material \cite{sm}). In Eq.~\eqref{eq:betaAlphaApproxAna}, two independent parameters in the parentheses are the detunings in DQD-$\mathbb{L}$ and DQD-$\mathbb{R}$, i.e., $\left(\varepsilon_\mathbb{L},\varepsilon_\mathbb{R}\right)$.

	\section{Results}\label{sec:result}
	
	\subsection{Exchange energy sweet spot and capacitive coupling}\label{subsec:ExSS}

	Figure~\ref{fig:resultTwoB} shows the key results of this paper, showing the composition of the logical states, Coulomb integrals, local exchange energies, capacitive shifts, and most importantly, the effective exchange energies and the capacitive coupling, along with the corresponding sweet spots. The inter-DQD distance, $2R_0$, is determined at the values where the inter-DQD tunneling is absent, see Sec.~II in the Supplemental Material for details \cite{sm}. The left column shows results for the magnetic field $B=0.174$ T while the right column $B=0.319$ T. For simplicity, all results shown in Fig.~\ref{fig:resultTwoB} are obtained with symmetric detuning, i.e. $\varepsilon_{\mathbb{L}} = \varepsilon_{\mathbb{R}} = \varepsilon$. In this case, the local exchange energies, $J_j$, and capacitive shift, $\beta_j$, are also symmetric for DQD-$\mathbb{L}$ and DQD-$\mathbb{R}$, i.e. $J_\mathbb{L} = J_\mathbb{R} = J$, $\beta_\mathbb{L} = \beta_\mathbb{R} = \beta$. The same also applies to the composition of the logical states in DQD-$\mathbb{L}$ and DQD-$\mathbb{R}$, therefore only results for DQD-$\mathbb{R}$ is shown in Fig.~\ref{fig:resultTwoB}(a) and (g). The results are obtained for an \textit{eight}-electron system in a DDQD device with each DQD hosting a \textit{four}-electron singlet-triplet qubit, cf.~Fig.~\ref{fig:DDQDV}. The logical eigenstates of a \textit{four}-electron singlet-triplet qubit are obtained using full CI calculations by retaining 10 F-D states in the few-electron dot (the left and right dot in DQD-$\mathbb{L}$ and DQD-$\mathbb{R}$ respectively) and 6 F-D states in the singly-occupied dot (the right and left dot in DQD-$\mathbb{L}$ and DQD-$\mathbb{R}$ respectively), see Sec.~\ref{subsubsec:SQJ} for details. The two-qubit parameters, i.e., $J^\text{eff}_j$, $\beta_j$ and $\alpha$, are then obtained by projecting the logical eigenstates of two single qubits hosted by two $\textit{four}$-electron systems onto the inter-DQD Coulomb interaction, see Sec.~\ref{subsubsec:interQInt} for details.
	
	We first focus on the case for $B = 0.319$ T (the right column of Fig.~\ref{fig:resultTwoB}). Fig.~\ref{fig:resultTwoB}(i) shows half the local exchange energy, $J/2$, which increases with the detuning when the detuning is small, reaching a maximum value at $\varepsilon = 8.40$ meV, and sharply drops down beyond that point. The point at $\varepsilon = 8.40$ meV is therefore the single-qubit sweet spot. Similar behavior of $J/2$ has been observed in experiments for $(n_\mathbb{RL},n_\mathbb{RR}) = (1,N)$, where $N$ ranges between 50 and 100 \cite{Martins.17,Malinowski.18}. Fig.~\ref{fig:resultTwoB}(j) shows the capacitive shift, $\beta$, also yielding a turning point at $\varepsilon\approx 8.30$meV, for three different values of $R_0$ as indicated. Furthermore, since $V_{|ST\rangle}=V_{|TS\rangle}$ in the symmetric case, $\beta$ is essentially $V_{|TT\rangle}-V_{|SS\rangle}$ (cf.~Eq.~\eqref{eq:betaAlphaDef}). One may readily verify that the $\beta$ shown in Fig.~\ref{fig:resultTwoB}(j) reflects the values of $V_{|TT\rangle}-V_{|SS\rangle}$ shown in Fig.~\ref{fig:resultTwoB}(h) for $R_0=40$ nm.
	
	The behavior of $\beta$ can be understood by looking into the detailed composition of the logical states (cf.~Fig.~\ref{fig:resultTwoB}(g) and Eq.~\eqref{eq:betaApproxAna}). Since $\mathcal{P}(\vert S(0,4)\rangle)=\sin^2 \left(\theta_S/2\right)$ and $\mathcal{P}(\vert T(0,4)\rangle)=\sin^2 \left(\theta_T/2\right)$, as suggested by Eq.~\eqref{eq:betaAlphaApproxAna} and Eq.~\eqref{eq:diffST}, $J^\mathrm{eff}$ and $\alpha$ are proportional to the difference between the hybridization in the singlet states and the hybridization in the triplet states, $\sin^2 \left(\theta_S/2\right)-\sin^2 \left(\theta_T/2\right)$ (cf.~red dashed line in Fig.~\ref{fig:resultTwoB}(g)) \cite{SSConditions}. 
	At small $\varepsilon$ ($\varepsilon < 8.30$meV), $\sin^2 \left(\theta_S/2\right)-\sin^2 \left(\theta_T/2\right)$ yields an increasing positive value, resulting in $\beta$ increasing with $\varepsilon$, as suggested by Eq.~\eqref{eq:betaApproxAna}.
At 8.30meV $<\varepsilon<$ 8.40meV, $\sin^2 \left(\theta_S/2\right)-\sin^2 \left(\theta_T/2\right)$ decreases with $\varepsilon$ but is still positive, giving positive $\beta$ with reduced magnitude when $\varepsilon$ increases. At $\varepsilon>8.40$meV, $\sin^2 \left(\theta_S/2\right)-\sin^2 \left(\theta_T/2\right)<0$, hence $\beta$ switches to negative value. This non-monotonic behavior of $J$ and $\beta$ in combination leads to the emergence of a sweet spot for the effective exchange energy $J^\text{eff} = J/2-\beta$, as indicated in Fig.~\ref{fig:resultTwoB}(k). In addition, the concurrence between the locations of the sweet spots $\varepsilon^*$ (Fig.~\ref{fig:resultTwoB}(k, l)) and the turning point of $\sin^2 \left(\theta_S/2\right)-\sin^2 \left(\theta_T/2\right)$ (red dashed line in Fig.~\ref{fig:resultTwoB}(g)) suggests that the competition between those two hybridizations is the main physical mechanism leading to the existence of the sweet spot, conforming with Eq.~\eqref{eq:diffST}. Since the hybridizations can be alternatively interpreted as the dipoles of the logical eigenstates (Eq.~\eqref{eq:dipole}), the existence of the sweet spots can be attributed to the non-monotonic behavior of the difference of dipoles in the logical singlet and triplet eigenstates, i.e. $\sqrt{d^\mathbb{L}_S d^\mathbb{R}_S}-\sqrt{d^\mathbb{L}_T d^\mathbb{R}_T}$, cf.~Eq.~\eqref{eq:betaAlphaApproxAna}. This also explains the absence of these sweet spots in conventional capacitively coupled two-electron singlet-triplet qubits \cite{Shulman.12,Nichol.17}. In two-electron singlet-triplet qubits, due to the absence of hybridizations between triplet states \cite{Dial.13,Petta.05}, dipole only exists in the logical singlet state, i.e.~$d^j_S>0$ while $d^j_T=0$ (Eq.~\eqref{eq:dipole}). This results in $\beta_j \propto \sqrt{d^\mathbb{L}_S d^\mathbb{R}_S}$ and $\alpha \propto d^\mathbb{L}_S d^\mathbb{R}_S$ (Eq.~\eqref{eq:betaAlphaApproxAna}), conforming with experimental results of capacitively coupled two-electron singlet-triplet qubits, which state that the effective exchange energies, $J^\text{eff}_j=J_j/2-\beta_j$, and capacitive coupling $\alpha$ exhibit monotonic behaviors with respect to the detuning values and $\alpha\propto d^\mathbb{L}_S d^\mathbb{R}_S$ \cite{Shulman.12,Nichol.17}.

Fig.~\ref{fig:resultTwoB}(l) shows the results for $\alpha$ v.s.~$\varepsilon$ for $B=0.319$ T at three different $R_0$ values as indicated. Similar to the above discussions, the behavior of $\alpha$ can be interpreted from Fig.~\ref{fig:resultTwoB}(g) and Eq.~\eqref{eq:alphaApproxAna}. At $\varepsilon<8.40$meV, the qualitative behavior of $\alpha$ is similar to $\beta$, in conformity with Eq.~\eqref{eq:alphaApproxAna}. However, when $\varepsilon>8.40$ meV, as $\sin^2 \left(\theta_S/2\right)-\sin^2 \left(\theta_T/2\right)$ switches from a positive to a negative value, $\alpha$ increases and maintains a positive value due to the $\left[\sin^2 \left(\theta_S/2\right)-\sin^2 \left(\theta_T/2\right)\right]^2$ term in Eq.~\eqref{eq:alphaApproxAna}. In addition, due to the higher exponent of $\sin \theta_{S/T}$ in Eq.~\eqref{eq:alphaApproxAna} as compared to Eq.~\eqref{eq:betaApproxAna}, $\alpha \ll \beta_j$ and is at the scale of neV. The magnitude of $\beta$ and $\alpha$ both show larger values for smaller $R_0$ (cf.~Fig.~\ref{fig:resultTwoB}(j) and (l)), which is evident from Eq.~\eqref{eq:betaAlphaApproxAna}, as smaller $R_0$ yields larger inter-DQD Coulomb integrals.

We proceed to discuss the results for relatively weaker magnetic field, shown in the left column of Fig.~\ref{fig:resultTwoB}. As can be seen by comparing the left and right columns, the results obtained at  $B=0.174$ T are qualitatively similar to those at $B=0.319$ T, so most of the arguments above apply. 

The comparison also shows that $\alpha$ is stronger by more than two orders of magnitude under $B=0.319$ T as compared to $B=0.174$ T. This is due to the fact that the admixture to the states of (0,4) type, making major contribution to $\alpha$ (Eq.~\eqref{eq:alphaApproxAna}), is larger for $B=0.319$ T (compare Fig.~\ref{fig:resultTwoB}(a) and (g)). While this may seemingly suggest that an even larger $\alpha$ could be achieved by further increasing $B$, the situation is unfavorable as sweet spots will enter a regime where $|S(1,3)\rangle$ and $|S(0,4)\rangle$ or $|T(1,3)\rangle$ and $|T(0,4)\rangle$ are highly mixed, and a careful design of adiabatic pulse is required to suppress leakage while ensuring a fast pulse to minimize the charge noise dephasing effect, which is challenging.
	
	As discussed in the previous paragraphs, the existence of the sweet spots for $J^\text{eff}$ and $\alpha$ can be attributed to the competition between the hybridizations in the singlet states and the triplet states, cf.~Eq.~\eqref{eq:diffST}. Since the hybridizations can be equivalently perceived as dipoles in the system (Eq.~\eqref{eq:dipole}), the existence of those sweet spots can be alternatively attributed to the non-monotonic behavior of the difference between the dipoles in the logical singlet and triplet eigenstates, cf.~Eq.~\ref{eq:betaAlphaApproxAna}. The competition is possible only when the energy of $\vert S(0,4)\rangle$ and $\vert T(0,4)\rangle$ are comparable. In addition, the turning point of $\sin^2 \left(\theta_S/2\right)-\sin^2 \left(\theta_T/2\right)$ only exists when the local exchange energy $J_\mathbb{L}$ $\left[J_\mathbb{R}\right]$ is negative in the large detuning regime in which $\left(n_{\mathbb{R}\mathrm{L}},n_{\mathbb{R}\mathrm{R}}\right)=\left(0,4\right)$ $\left[\left(n_{\mathbb{L}\mathrm{L}},n_{\mathbb{L}\mathrm{R}}\right)=\left(4,0\right)\right]$ \cite{ChanGX.22}. Such non-monotonic behaviors of $J_\mathbb{L}$ and $J_\mathbb{R}$ have been demonstrated in experiments \cite{Malinowski.18,Martins.17}. Based on the experimentally measured Hubbard parameters \cite{Malinowski.18}, we confirm that results similar to Fig.~\ref{fig:resultTwoB} (lower panel) can be achieved in a multielectron DDQD device in the detuning regime where $\left(n_{\mathbb{L}\mathrm{L}},n_{\mathbb{L}\mathrm{R}},n_{\mathbb{R}\mathrm{L}},n_{\mathbb{R}\mathrm{R}}\right)=\left(2N+1,1,1,2N+1\right)$ for $N= 28$, see Sec.~VII A in the Supplemental Material \cite{sm} for details.  Moreover, as described above, negative local exchange energies in the fully occupied detuning regimes are part of the key elements to achieve non-monotonic behaviors of $J_\mathbb{L}$ and $J_\mathbb{R}$ \cite{SSConditions}. It has been theoretically demonstrated that negative local exchange energies are also achievable in a QD occupied by 8 and 14 electrons \cite{Deng.18}. Therefore, it is presumed that sweet spots of $J^\text{eff}_j$ and $\alpha$ might also exist in a DDQD device in which the capacitively coupled few-electron singlet-triplet qubits are operated in the detuning regime where $\left(n_{\mathbb{L}\mathrm{L}},n_{\mathbb{L}\mathrm{R}},n_{\mathbb{R}\mathrm{L}},n_{\mathbb{R}\mathrm{R}}\right)=\left(N+1,1,1,N+1\right)$ for $N\in\{7,13\}$. However, confirming those predictions are out of the scope of our work since eight- and fourteen-electron singlet-triplet qubits have not been experimentally demonstrated for the extraction of Hubbard parameters while the computational resources required to numerically simulate them are beyond our capabilities.
	
	\begin{figure*}[t]
		\includegraphics[width=0.8\linewidth]{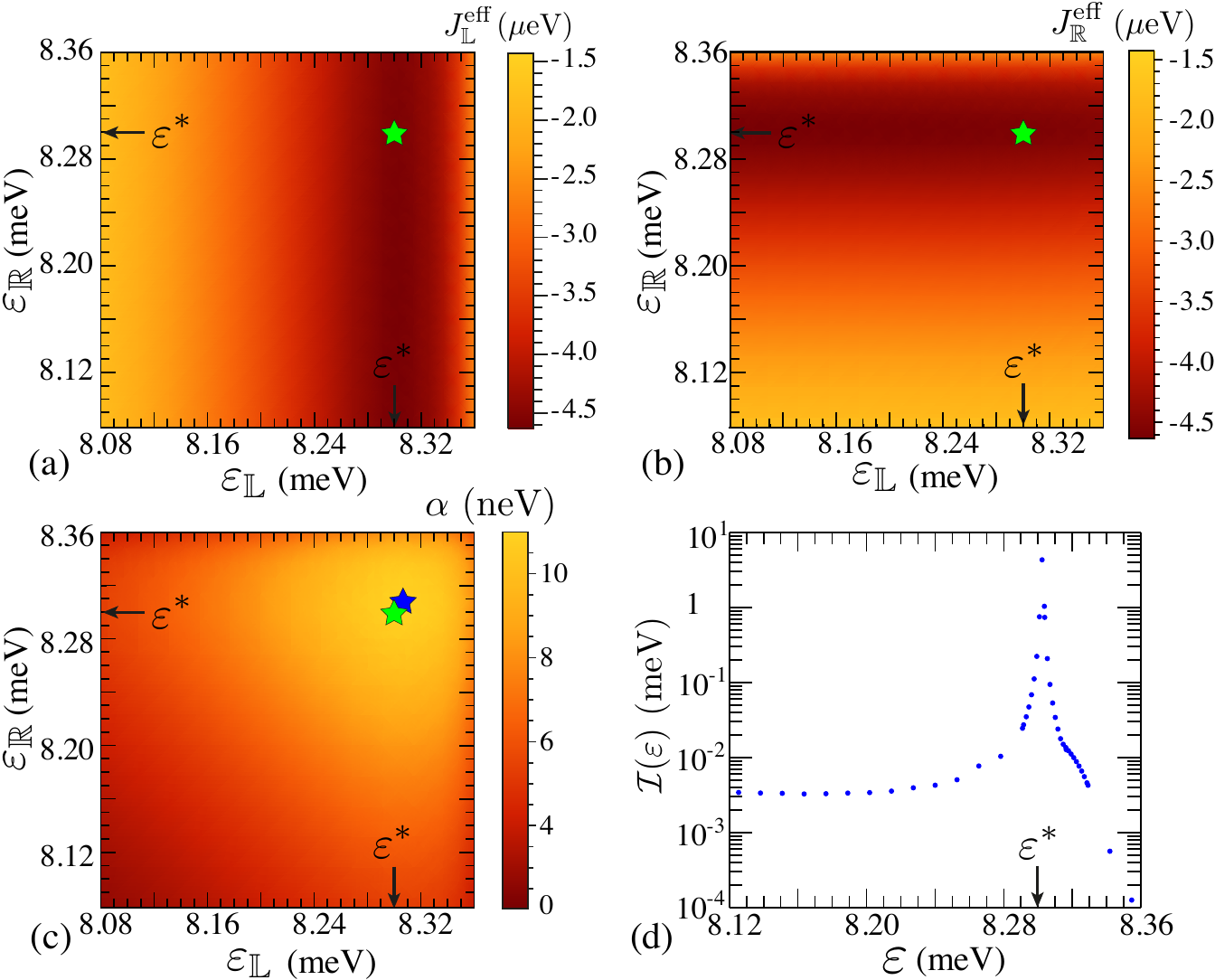}
		\caption{(a) $J_\mathbb{L}^\text{eff}$, (b) $J_\mathbb{R}^\text{eff}$ and (c) $\alpha$ as function of the detuning in DQD-$\mathbb{L}$ ($\varepsilon_{\mathbb{L}}$) and detuning in DQD-$\mathbb{R}$ ($\varepsilon_{\mathbb{R}}$). The green stars mark the $J^{\text{eff}}_j$ sweet spot detunings, $\varepsilon^*$,. The blue star marks the $\alpha$ sweet spot with respect to $\varepsilon_{\mathbb{L}}$ and $\varepsilon_{\mathbb{R}}$, i.e. $\partial \alpha /\partial \varepsilon_{\mathbb{L}} = \partial \alpha /\partial \varepsilon_{\mathbb{R}}=0$. (d) Two-qubit insensitivity, $\mathcal{I}$, as function of the detuning, $\varepsilon = \varepsilon_{\mathbb{L}} = \varepsilon_{\mathbb{R}}$. The parameters are: $B=0.319$ T, $2R_0 = 80$nm.}
		\label{fig:J12AlphaI}
	\end{figure*}
	
	Figure \ref{fig:J12AlphaI}(a), (b) and (c) show the effective exchange energies, $J^\text{eff}_\mathbb{L}$ and $J^\text{eff}_\mathbb{R}$, and the capacitive coupling, $\alpha$, as functions of the individual detunings of both DQDs, $\varepsilon_\mathbb{L}$ and $\varepsilon_\mathbb{R}$, for inter-DQD distance $2R_0 = 80$ nm at $B=0.319$ T. Fig.~\ref{fig:J12AlphaI}(a) indicates that at $\varepsilon_\mathbb{L} = \varepsilon_\mathbb{R} =\varepsilon^* = 8.30$ meV, $J^\text{eff}_\mathbb{L}$ yields a turning point with respect to $\varepsilon_{\mathbb{L}}$, and has a relatively flat dispersion with respect to $\varepsilon_{\mathbb{R}}$. On the other hand, Fig.~\ref{fig:J12AlphaI}(b) shows that at $\varepsilon_\mathbb{L} = \varepsilon_\mathbb{R} =\varepsilon^* = 8.30$ meV, $J^\text{eff}_\mathbb{R}$ yields a turning point with respect to $\varepsilon_{\mathbb{R}}$, and has a relatively flat dispersion with respect to $\varepsilon_{\mathbb{L}}$. Hence, it can be expected that $J^{\text{eff}}_j$ is insensitive to the uncorrelated $\varepsilon_{\mathbb{L}}$ and $\varepsilon_{\mathbb{R}}$ noise (the charge noise environment observed in experiments \cite{Boter.20}), featuring a simultaneous sweet spot with respect to both $\varepsilon_{\mathbb{L}}$ and $\varepsilon_{\mathbb{R}}$. Fig.~\ref{fig:J12AlphaI}(c) shows that there exists a sweet spot for $\alpha$, denoted as a blue star, with respect to both $\varepsilon_{\mathbb{L}}$ and $\varepsilon_{\mathbb{R}}$, i.e. $\partial \alpha /\partial \varepsilon_\mathbb{L} =\partial \alpha /\partial \varepsilon_\mathbb{R} = 0$. The $\alpha$ sweet spot locates at $\varepsilon_{\mathbb{L}}=\varepsilon_{\mathbb{R}} \approx 8.31$meV, which is very close to the $J^\text{eff}_j$ sweet spot at $\varepsilon^* = 8.30$meV.
	
	The effect of charge noise can be roughly captured by evaluating the two-qubit insensitivity, $\mathcal{I}$ \cite{Wolfe.17}, which is defined in analogy to the definition for the single-qubit case \cite{Reed.16} as
	\begin{equation}
		\mathcal{I} \left(\varepsilon_{\mathbb{L}},\varepsilon_{\mathbb{R}}\right) = \frac{\alpha \left(\varepsilon_{\mathbb{L}},\varepsilon_{\mathbb{R}}\right)}{\| \vec{\nabla} H \left(\varepsilon_{\mathbb{L}},\varepsilon_{\mathbb{R}}\right) \|},
	\end{equation}
	where the Frobenius norm, $\| \vec{\nabla} H \|$, is defined as
	\begin{equation}\label{eq:FrobNorm}
		\| \vec{\nabla} H \left(\varepsilon_{\mathbb{L}},\varepsilon_{\mathbb{R}}\right) \| \simeq \sqrt{\sum_{\substack{\{j,k\}\\ \in \{\mathbb{L},\mathbb{R}\}}}\left(\frac{\partial J^\text{eff}_k}{\partial \varepsilon_j}\right)^2+\sum_{j\in \{\mathbb{L},\mathbb{R}\}}\left(\frac{\partial \alpha}{\partial \varepsilon_j}\right)^2},
	\end{equation}
	where we have included the derivatives of $\alpha$ for completeness. To ensure that the two-qubit system is robust against the uncorrelated charge noises in both detunings, $\varepsilon_{\mathbb{L}}$ and $\varepsilon_{\mathbb{R}}$, the derivatives in Eq.~\eqref{eq:FrobNorm} include the off-diagonal elements of the gradient tensor. For example, $\partial J^\text{eff}_\mathbb{L}/\partial \varepsilon_\mathbb{R}$ for $k=\mathbb{L}$ and $j=\mathbb{R}$ indicates the variation of effective exchange energy in DQD-$\mathbb{L}$, $J^\text{eff}_\mathbb{L}$, induced by charge noises in DQD-$\mathbb{R}$, $\varepsilon_{\mathbb{R}}$.
	Fig.~\ref{fig:J12AlphaI}(d) shows $\mathcal{I}$ as a function of symmetric detunings, $\varepsilon = \varepsilon_{\mathbb{L}} = \varepsilon_{\mathbb{R}}$. $\mathcal{I}$ sharply peaks near the $J^\text{eff}_j$ sweet spot at $\varepsilon = \varepsilon^*$, suggesting an optimal operating point for two-qubit gate operation.

	\subsection{Phonon-mediated decoherence effect}\label{subsec:phonon}

	The emergence of an effective exchange sweet spot, $\varepsilon^*$, can be attributed to the competition between the admixtures with $|S(0,4)\rangle$ and $|T(0,4)\rangle$, as discussed in details in Sec.~\ref{subsec:ExSS}. In the context of two-electron singlet-triplet qubit formed by singly-occupied dots in a DQD, the admixture with doubly-occupied singlet was found to be the origin of strong phonon-induced pure-dephasing $\Gamma_\varphi=1/T_\varphi$ \cite{Kornich.14}. In addition, the hyperfine coupling (local magnetic field gradient) and spin-orbit interaction (SOI) are also found to be the sources of phonon-induced relaxation $\Gamma_1=1/T_1$ \cite{Kornich.14}. Hence, the phonon-mediated decoherence effect has to be taken into account to provide a more comprehensive picture on the robustness of two-qubit gate proposed here. However, we have found that the decoherence effect by phonons is largely reduced due to the following factors: (1) Phonon-induced pure dephasing $\Gamma_\varphi$ is suppressed as $\varepsilon^*$ is comparatively far from the transition point to quadruple electron occupation, e.g.~between $|T(1,3)\rangle$ and $|T(0,4)\rangle$, resulting in smaller admixture with states of $(n_{\mathbb{R}\mathrm{L}},n_{\mathbb{R}\mathrm{R}}) = (0,4)$ type. The suppression can also be partially attributed to the smaller inter-dot distance, $2x_0$, adopted in this work. Details on $T_\varphi$ are provided in Sec.~V~F in the Supplemental Material \cite{sm}. 
(2) Because the magnetic field gradient $h$ is much smaller than the local exchange energies at $\varepsilon^*$, resulting in smaller admixture between $|S(1,3)\rangle$ and $|T(1,3)\rangle$, which consequently gives a smaller relaxation rate $\Gamma_1$. (See Sec.~V~G~1 in the Supplemental Material \cite{sm} for details). (3) For a two-electron singlet-triplet qubit, 
$\Gamma_1$ by SOI arises due to the second order coupling between the logical singlet and triplet by SOI integral involving the ground orbital ($\Phi_{\mathbb{R}\mathrm{R}1}$) and first excited orbital ($\Phi_{\mathbb{R}\mathrm{R}2}$) in the same dot. In contrast, for the four-electron singlet-triplet qubit in this work, the ground and first excited valence orbitals are the first ($\Phi_{\mathbb{R}\mathrm{R}2}$) and second ($\Phi_{\mathbb{R}\mathrm{R}3}$) excited orbital respectively, which yields a negligible SOI integral \cite{SOINote}. (See Sec.~V~G~2 in the Supplemental Material \cite{sm} for details). The overall decoherence time are shown in Fig.~S4 in the Supplemental Material \cite{sm}, which is found to be negligible in our system for the dot parameters, gate time and temperature of interest.
	
	\begin{figure}[t]
		\includegraphics[width=\linewidth]{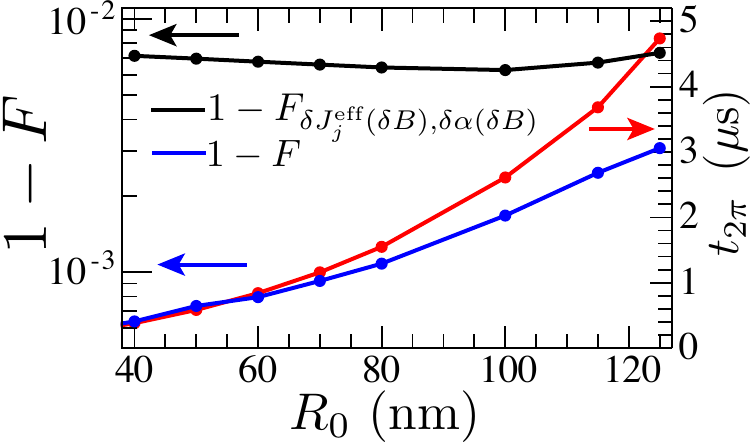}
		\caption{CPHASE gate infidelities ($1-F$) and gate time ($t_{2\pi} = 2\pi \hbar /\alpha$, red circles) as a function of half inter-DQD distance $R_0$. $1-F$ shows the average of 5000 iterations. Blue circles show $1-F$ under charge noises and hyperfine noises while black circles show results with an additional inclusion of magnetic field noises on $J^\text{eff}_j$ and $\alpha$, denoted as $\delta J_j^\mathrm{eff}(\delta B)$ and $\delta \alpha(\delta B)$ respectively. The magnetic field is $B=0.319$ T, cf.~Fig.~\ref{fig:resultTwoB} (lower panel).}
		\label{fig:Fidelity}
	\end{figure}

	\subsection{CPHASE Gate fidelity}\label{subsec:Fid}
	The inter-qubit coupling, $\sigma_z \otimes \sigma_z$, gives rise to a controlled-phase (CPHASE) gate \cite{Nielsen.12}. In the simulations of the CPHASE gate, two noise sources are taken into account, i.e.~magnetic field gradient fluctuations, $\delta h_j$, and charge noise fluctuations, $\delta \varepsilon_j (t)$. $\delta h_j$ can be considered as quasistatic during the gate operation as it is contributed by mostly low-frequency noise with power spectrum $S(\omega) \propto 1/\omega^{2.6}$ \cite{Medford.12,Rudner.11}. We model the hyperfine noise with standard deviation of $\delta h_j = 10$neV, which corresponds to $0.413$ mT (as suggested by the measured hyperfine-noise-limited coherence time, $T_2^*=\hbar \sqrt{2}/\delta h_j=90$ns \cite{Dial.13}). $\delta \varepsilon_j (t)$ consists of both a quasistatic component ($\delta \varepsilon_j^{\text{QS}}$) and a high-frequency part in a $1/f$ form ($\delta \varepsilon_j^{1/f} (t)$). In our simulation, $\delta \varepsilon_j^\text{QS}$ is taken to yield a Gaussian distribution with standard deviation of $8\mu\text{V}\times 1\text{eV}/9.4\text{V}$ \cite{Dial.13}, where $9.4\text{V}/1\text{eV}$ is the lever arm \cite{Dial.13}. $\delta h_j$ and $\delta \varepsilon_j (t)$ are assumed to be independent from each other \cite{Boter.20}. Based on the experimental results, $\delta \varepsilon_j^{1/f} (t)$ yields a power spectrum of $S_\varepsilon(f) = 8\times 10^{-16}\frac{\text{V}^2}{\text{Hz}}(\frac{1 \text{Hz}}{f})^{0.7}$ \cite{Dial.13}. $\delta \varepsilon_j^{1/f} (t)$ is generated using the Fourier transform of a discrete Gaussian white noise sequence, which is then scaled according to $S_\varepsilon(f)$ \cite{Yang.19}, see Sec.~I in the Supplemental Material for details \cite{sm}. To suppress quasistatic noise, we use an echo pulse to perform CPHASE gate \cite{Shulman.12,Nichol.17}. The echo pulse is employed by adding a $\pi$-pulse about the $x$-axis of both qubits, $\exp\left[- i h t_\text{echo}\left(\sigma_x^{(\mathbb{L})} + \sigma_x^{(\mathbb{R})}\right) \right]$, at time $t = t_{2\pi}/2$ for $h \times t_\text{echo} = \pi/2$ and $h_\mathbb{L}=h_\mathbb{R} = h$. We assume that the single qubit gates are errorless in the simulations since experiments have demonstrated fidelities as high as 99\% \cite{Nichol.17}. We note that although the noises in $J^\mathrm{eff}$ can be largely cancelled using the echo pulses, the quasistatic fluctuations on $\alpha$, denoted as $\delta \alpha^\text{QS}$, remains. (see Sec.~IX in the Supplemental Material \cite{sm} for details). However, since the $\alpha$ sweet spot is very close to $\varepsilon^*$, the effect of $\delta \alpha^\text{QS}$ is suppressed as well.
	
	The simulation of the noisy evolution is performed by adding the noisy terms into the control parameters, i.e. $\varepsilon_j(t) \rightarrow \varepsilon_j (t)+\delta \varepsilon_j^\text{QS}+ \delta \varepsilon_j^{1/f}(t)$ and $h_j \rightarrow h_j +\delta h_j$. For each iteration, $\varepsilon_\mathbb{L}^\text{QS}$, $\varepsilon_\mathbb{R}^\text{QS}$, $\delta h_\mathbb{L}$ and $\delta h_\mathbb{R}$ are randomly extracted from a Gaussian distribution with a  standard deviation as explained before, while $\delta \varepsilon_\mathbb{L}^{1/f}(t)$ and $\delta \varepsilon_\mathbb{R}^{1/f}(t)$ are randomly chosen from a sample of generated sequences of size 5000. 
	
	Figure~\ref{fig:Fidelity} shows, for $B=0.319$ T, the CPHASE gate infidelity $1-F$ \cite{FidelityNote} as a function of half inter-DQD distance, $R_0$. We first focus on $1-F$ under the effect of charge noises and hyperfine noises on the qubit parameters, cf.~blue circles in Fig.~\ref{fig:Fidelity}. It can be observed that the gate fidelity is enhanced for smaller $R_0$ due to shorter gate time, $t_{2\pi}$. The shorter gate time is due to the increase of $\alpha$ for a decreasing $R_0$, as discussed in Sec.~\ref{subsec:ExSS}. Since the echo pulses effectively cancel quasistatic noises, $\delta \varepsilon_j^\text{QS}$ and $\delta h_j$, gate infidelity essentially can be attributed to the degree of exposure to the $1/f$ charge noises. When the gate time, $t_{2\pi}$, is longer, the accumulated $1/f$ charge noise dephasing effect is more pronounced, resulting in a lower gate fidelity. Next, we take into account of the magnetic field noises $\delta B$ (hyperfine noises, $\delta h_j$) on $J^\text{eff}$ and $\alpha$. Different values of $J^\text{eff}$ and $\alpha$ for different magnetic field strengths $B$ as shown in Fig.~\ref{fig:resultTwoB} suggests that the variations of $J_j^\text{eff}$ and $\alpha$ induced by $\delta B$, denoted as $\delta J_j^\text{eff}(\delta B)$ and $\delta \alpha(\delta B)$, will contribute to the increase of $1-F$. Fig.~\ref{fig:Fidelity} (black circles) shows the inclusion of $\delta J_j^\text{eff}(\delta B)$ and $\delta \alpha(\delta B)$ results in around an order of increase for the gate infidelities $1-F$. Nevertheless, $1-F$ remains lower than $10^{-2}$. It can be observed that $1-F$ slightly decreases as a function of $R_0$ for smaller $R_0$ ($R_0<100$ nm). This can be attributed to the decrease of $\delta J_j^\text{eff}(\delta B)$ and $\delta \alpha(\delta B)$, for a fixed $\delta B$, as a function of $R_0$ (see Fig.~S8(a, b) in Sec.~X in the Supplemental Material for details \cite{sm}. On the other hand, as a result of longer gate time (longer exposure to environmental noises), $1-F$ slightly increases as a function of $R_0$ for larger $R_0$ ($R_0>100$ nm). The gate fidelities can be further improved by operating the CPHASE gate at a smaller applied magnetic field, e.g. $B=0.290$ T, which results in $1-F\sim10^{-3}$ under the effect of $\delta \varepsilon^\text{QS}_j$, $\delta \varepsilon^{1/f}_j$, $\delta h_j$, $\delta J_j^\text{eff} (\delta B)$, and $\delta \alpha (\delta B)$, see Fig.~S9 in Sec.~X in the Supplemental Material for details \cite{sm}.
	
	\section{Sweet spots of $J^\mathrm{eff}$ and $\alpha$ in other systems}
	In the previous sections, we have presented the results showing, in a GaAs DDQD device, that sweet spots can be found for a pair of capacitively coupled four-electron singlet-triplet qubits and robust CPHASE gates can be performed at those sweet spots. We have also found out that similar results, including the existence of sweet spots of $J^\mathrm{eff}$ and $\alpha$ and largely enhanced $\alpha$ at the sweet spots, can be achieved for silicon and germanium quantum dots. In particular, we have shown that those features can be found in a pair of capacitively coupled four-electron (four-hole) singlet-triplet qubits in a silicon (germanium) DDQD device, see Sec.~XI in the Supplemental Material for details \cite{sm}.
	
	\section{Conclusion and discussion} \label{sec:conclusion}
	We have shown, based on full CI calculations of a few-electron singlet-triplet qubit and projection of the logical eigenstates to inter-DQD Coulomb interaction, that  effective exchange energy sweet spots appear in the coupled few-electron singlet-triplet qubit systems. The sweet spots at the same time are also very close to the capacitive coupling sweet spot. We further show that the sweet spots of the effective exchange energies and capacitive coupling arise due to the competition between the hybridization in the singlet states and the hybridization in the triplet states. The competition is only made possible by a negative local exchange energy in the single-qubit case when all the electrons occupy a QD in a DQD device. These results are in contrast to the two-electron case, in which the local exchange energy is positive in a fully occupied dot \cite{Dial.13,Shulman.12,Nichol.17,Cerfontaine.20}. Therefore, these sweet spots are not found for a pair of capacitively coupled two-electron singlet-triplet qubits \cite{Shulman.12,Nichol.17}.
	By operating CPHASE gates at the effective exchange energy sweet spots, we have demonstrated that gate fidelities above 99\% can be achieved in the presence of charge noises and hyperfine noises. Our results therefore should facilitate realization of high-fidelity two-qubit gates in coupled singlet-triplet qubit systems.
	
	
	\section*{Acknowledgements} We acknowledge support from the Key-Area Research and Development Program of GuangDong Province  (Grant No.~2018B030326001), the National Natural Science Foundation of China (Grant No.~11874312), the Research Grants Council of Hong Kong (Grant Nos.~11303617, 11304018, 11304920), and the Guangdong Innovative and Entrepreneurial Research Team Program (Grant No.~2016ZT06D348). The calculations involved in this work are mostly performed on the Tianhe-2 supercomputer at the National Supercomputer Center in Guangzhou, China.
	
	\bibliographystyle{apsrev4-1}
	%
	
	\setcounter{secnumdepth}{3}  
	\setcounter{equation}{0}
	\setcounter{figure}{0}
	\setcounter{table}{0}
	\renewcommand{\theequation}{S-\arabic{equation}}
	\renewcommand{\thefigure}{S\arabic{figure}}
	\renewcommand{\thetable}{S-\Roman{table}}
	\renewcommand\figurename{Supplementary Figure}
	\renewcommand\tablename{Supplementary Table}
	\newcommand\Scitetwo[2]{[S\citealp{#1}, S\citealp{#2}]}
	\newcommand\citeScite[2]{[\citealp{#1}, S\citealp{#2}]}

	\newcolumntype{M}[1]{>{\centering\arraybackslash}m{#1}}
	\newcolumntype{N}{@{}m{0pt}@{}}
	
	\makeatletter \renewcommand\@biblabel[1]{[S#1]} \makeatother
	
	\newcommand\Scite[1]{[S\citealp{#1}]}
	\makeatletter \renewcommand\@biblabel[1]{[S#1]} \makeatother
	
	\onecolumngrid
	\vspace{1cm}
	
	\newpage
	
	\begin{center}
		{\bf\large Supplementary Material for ``Robust entangling gate for capacitively coupled few-electron singlet-triplet qubits''}
	\end{center}
	\vspace{0.5cm}
	
	\section{$1/f$ charge noise}
	\begin{figure}[t]
		\includegraphics[width=0.7\columnwidth]{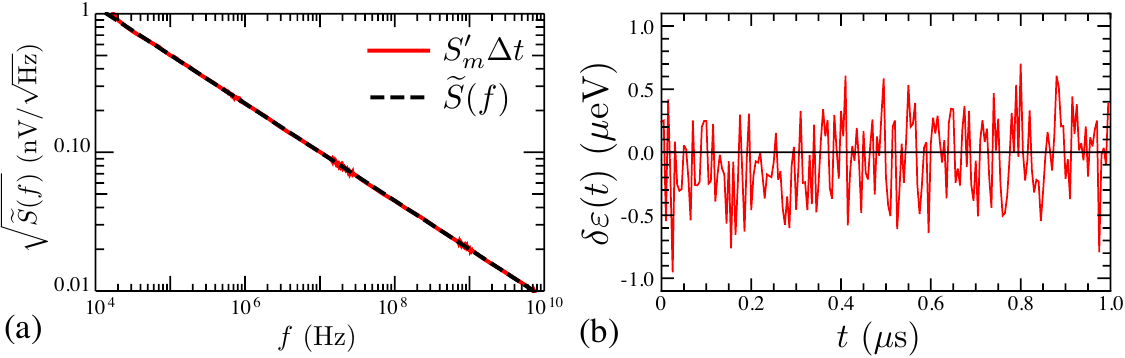}
		\caption{(a) Red solid line shows the PSD of the discrete $1/f$ noise obtained based on Fourier transform discrete Gaussian white noise sequences scaled by Eq.~\eqref{eq:Sf}, averaged for 5000 sequences. Dashed black line shows Eq.~\eqref{eq:Sf} as reference. The noise sequence is generated with total time $t_\text{tot}=100\mu$s, low and high frequency cutoffs are $\omega_l/2\pi=1$ Hz and $\omega_h/2\pi = 10^{11}$ Hz. (b) $\delta \varepsilon (t)$ as function of time, $t$, of a generated $1/f$ noise sequence.}
		\label{fig:PSD}
	\end{figure}
	The $1/f$ charge noise, $\delta \varepsilon^{1/f} (t)$, is generated by scaling a discrete noise sequence in frequency domain, which is obtained by performing Fourier transform on a discrete Gaussian white noise sequence in time domain. The scaling is determined according to the continuous noise spectrum \cite{Yang.19}, 
	\begin{equation}\label{eq:Sf}
		\widetilde{S}(f)=\frac{c_\varepsilon^2}{\text{Hz}}\left|\frac{\text{1 Hz}}{f}\right|^\alpha.
	\end{equation}
	
	The power spectral density (PSD) of the $1/f$ noise, defined as the absolute square of its Fourier transform \cite{Stein.00}, is shown in Fig.~\ref{fig:PSD}(a). The $\alpha$ and $c_\varepsilon^2$ are taken as 0.7 and $8\times10^{-16}\text{V}^2$ \cite{Dial.13}, which gives $\sqrt{\widetilde{S}(f)} = 0.2 \text{nV}/\sqrt{\text{Hz}}$ at $f = 1$ MHz, conforming with the generated power spectrum of $1/f$ noise shown in Fig.~\ref{fig:PSD}(a). The lever arm is taken to be $9.4$ \cite{Dial.13}, such that $\varepsilon=V_\text{applied}/9.4$, where $V_\text{applied}$ is the applied gate voltage and $\varepsilon$ is the detuning. An example of $\delta \varepsilon^{1/f} (t)$ is shown in Fig.~\ref{fig:PSD}(b).
	
	\section{Capacitive limit}
	We assume that $R_0$ in the main text is sufficiently large such that inter-DQD tunneling is suppressed. This can be confirmed by inspecting the tunneling process between the left and right DQD. For a pair of capacitively coupled singlet-triplet qubits in this work, the inter-DQD tunneling can be estimated by considering the tunneling process between two inner dots consisting of singly occupied QDs. Two electrons in the inner dots form singlets, $\left|S(1,1)\right\rangle_\text{in}$ and $\left|S(0,2)\right\rangle_\text{in}$, or triplets, $\left|T(1,1)\right\rangle_\text{in}$ and $\left|T(0,2)\right\rangle_\text{in}$. Explicitly, the two-electron states are
	\begin{equation}
		\begin{split}
			\left|S(1,1)\right\rangle_\text{in} &= \frac{1}{\sqrt{2}} \left(\left|\uparrow_3 \downarrow_2 \right\rangle + \left|\uparrow_2 \downarrow_3 \right\rangle \right),\\
			\left|S(0,2)\right\rangle_\text{in} &= \left|\uparrow_2 \downarrow_2 \right\rangle,\\
			\left|T(1,1)\right\rangle_\text{in} &= \frac{1}{\sqrt{2}} \left(\left|\uparrow_3 \downarrow_2 \right\rangle - \left|\uparrow_2 \downarrow_3 \right\rangle \right),\\
			\left|T(0,2)\right\rangle_\text{in} &= \frac{1}{\sqrt{2}} \left(\left|\uparrow_3 \downarrow_6 \right\rangle - \left|\uparrow_6 \downarrow_3 \right\rangle \right),\\
		\end{split}
	\end{equation}
	where $\left| \uparrow_j \downarrow_k \right\rangle$ is a Slater determinant with a spin-up electron occupying $j$-th orbital ($\Phi_j$) and another spin-down electron occupying $k$-th orbital ($\Phi_k$). One can refer to Fig.~1 in the main text for the indexing of the orbitals. By modeling the potential function of two inner dots in a biquadratic form, the effective Hamiltonians for singlets ($H_S$) and triplets ($H_T$), written in the bases listed above, are
	\begin{equation}
		\begin{split}
			H_S &=\left(\begin{array}{cc}
				U_{2,3}& \sqrt{2}t_{2,3}\\
				\sqrt{2}t_{2,3}& U_2
			\end{array}\right),\\
			H_T &=\left(\begin{array}{cc}
				U_{2,3}& t_{2,7}\\
				t_{2,7}& U_7 + \sqrt{\hbar \omega_\text{in}^2 + \hbar \omega_c^2/4}-\frac{1}{2}\hbar \omega_c
			\end{array}\right),
		\end{split}
	\end{equation} 
	where $t_{j,k}$ is the tunneling energy between $\Phi_j$ and $\Phi_k$, $U_j$ is the on-site Coulomb energy in $\Phi_j$, $U_{j,k}$ is the inter-site Coulomb energy between electrons in $\Phi_j$ and $\Phi_k$, $\hbar \omega_c = e B/m^* c$ while $B$ is the magnetic field, $e$ is the electrical charge of an electron, $m^*$ is the effective mass and $c$ is the speed of light \cite{Barnes.11}. We define leakage as 
	\begin{equation}\label{eq:lea}
		\begin{split}
			\eta_S &= 1- \max\left[\langle S(1,1) | \exp\left(-i H_S t\right)| S(1,1)\rangle\right],\\
			\eta_T &= 1- \max\left[\langle T(1,1) | \exp\left(-i H_T t\right)| T(1,1)\rangle \right],
		\end{split}
	\end{equation}
	where ``max'' computes the maximum value of the inner product of all time $t$. 
	\begin{figure}
		\includegraphics[width=0.4\columnwidth]{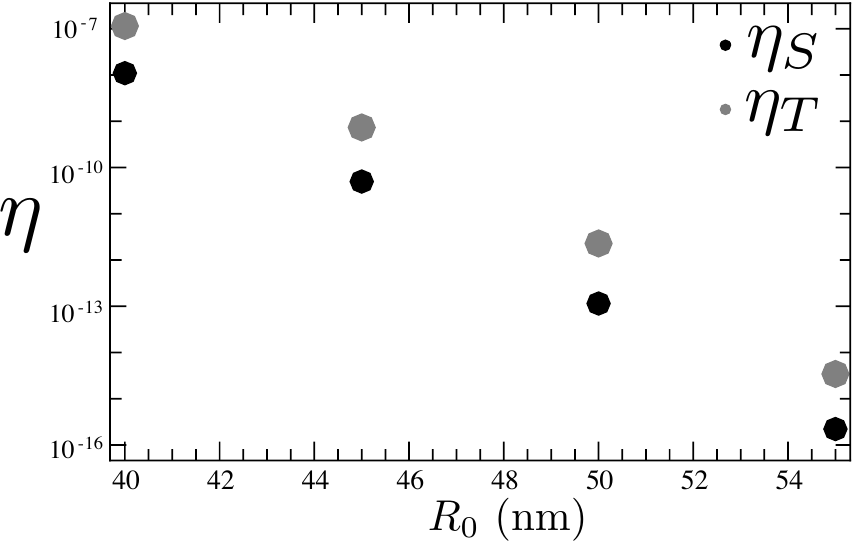}
		\caption{Leakage into singlet and triplet formed in dot $\it 2$ as function of $R_0$, which are denoted as $\eta_S$ and $\eta_T$, respectively. The parameters are: $\hbar \omega_\text{in} = 7.5$meV, $B = 0.319$T.}
		\label{fig:lea}
	\end{figure}
	
	Figure~\ref{fig:lea} shows the leakage, $\eta$, obtained using Eq.~\eqref{eq:lea}. For the values of $R_0$ shown in Fig.~4 in the main text, $\eta$ is negligible. Nevertheless, if $\eta$ is non-negligible, it could always be suppressed by introducing an extra voltage barrier between the inner dots, while ensuring it to be sufficiently sharp such that other aspects of the system remain unchanged.
	
	\section{Electron wavefunctions}
	The electron wavefunctions (F-D states) of a DQD in $x-y$ plane are
	\begin{subequations} \label{eq:FD}
		\begin{align}
			\begin{split}
				\phi_{L1} (x',y')&= \frac{1}{\sqrt{\pi}l_L}e^{-\frac{(x' + x_0)^2+y'^2}{2 l_L^2}}e^{i x_B y'}
			\end{split}\\
			\begin{split}
				\phi_{R1} (x',y')&= \frac{1}{\sqrt{\pi}l_R}e^{-\frac{(x' - x_0)^2+y'^2}{2 l_R^2}}e^{-i x_B y'}
			\end{split}\\
			\begin{split}
				\phi_{R2} (x',y')&= \frac{1}{\sqrt{\pi}l_R}\left[\frac{(x'-x_0)-i y'}{l_R}\right]e^{-\frac{(x' - x_0)^2+y'^2}{2 l_R^2}}e^{-i x_B y'} = \frac{1}{\sqrt{\pi}l_R}\mathcal{F}^{(2)}(x',y') e^{-\frac{(x' - x_0)^2+y'^2}{2 l_R^2}}e^{-i x_B y'}
			\end{split}\\
			\begin{split}
				\phi_{R3} (x',y')&= \frac{1}{\sqrt{\pi}l_R}\left[\frac{(x'-x_0)+i y'}{l_R}\right]e^{-\frac{(x' - x_0)^2+y'^2}{2 l_R^2}}e^{-i x_B y'}= \frac{1}{\sqrt{\pi}l_R}\mathcal{F}^{(3)}(x',y') e^{-\frac{(x' - x_0)^2+y'^2}{2 l_R^2}}e^{-i x_B y'}
			\end{split}
		\end{align}
	\end{subequations}
	where $\eta j$ denotes $\left(j-1\right)\text{-th}$ excited orbital in dot $\eta$ while $j=1$ is the ground orbital. $\eta=L$ and $\eta=R$ denote left and right QD repectively. $x_0$ is the half-interdot distance, $l_\eta = \sqrt{\hbar/(m_\text{eff} \omega_\eta)}$ is the confinement length of the quantum-dot (QD), $x_B = e B x_0 /(2 \hbar c)$ is the out-of-plane magnetic field induced phase, $\mathcal{F}^{(n)}(x,y)$ includes the additional prefactor for $n-1$ excited orbitals when compared to ground orbitals. Note that in contrast to labeling the F-D states as integers in the main text, we use different labeling here to signify the main contribution of the phonon-mediated decoherence, which will be discussed in Sec.~\ref{subsec:HubbardModel}. Also, in the $x-y$ plane, the $x'$ in Eq~\eqref{eq:FD} is aligned at the crystallographic [110] direction, as realized in most experiments \cite{Medford.12,Castin.04,Higginbotham.14,Kornich.14}. 
	
	In the calculation of phonon-mediated decoherence, we use the orthonormal set of F-D states, which are defined as 
	\begin{equation}\label{eq:orthoFD}
		\begin{split}
			\Phi_{L1} (x',y',z)&= \frac{\phi_{L1} (x',y')- g \phi_{R1} (x',y')}{\sqrt{1-2sg+g^2}}\phi_\text{FH}(z),\\
			\Phi_{R1} (x',y',z)&= \frac{\phi_{R1} (x',y')- g \phi_{L1} (x',y')}{\sqrt{1-2sg+g^2}}\phi_\text{FH}(z),\\
			\Phi_{R2} (x',y',z)&= \frac{\phi_{R2} (x',y')- g^{*} \phi_{L1} (x',y')}{\sqrt{1-2s^*g^*+(g^{*})^{2}}}\phi_\text{FH}(z),\\
			\Phi_{R3} (x',y',z)&= \frac{\phi_{R3} (x',y')- g^{**} \phi_{L1} (x',y')}{\sqrt{1-2s^{**}g^{**}+(g^{**})^{2}}}\phi_\text{FH}(z),\\
		\end{split}
	\end{equation}
	where 
	\begin{equation}
		\begin{split}
			\phi_\text{FH}(z)&=\frac{z}{\sqrt{2a_z^3}}e^{-\frac{z}{2a_z}},\\
			s &= \langle \psi_{L1} | \psi_{R1}\rangle,\\
			s^* &= \langle \psi_{L1} | \psi_{R2}\rangle,\\
			s^{**} &= \langle \psi_{L1} | \psi_{R3}\rangle,\\
			g &= \frac{1-\sqrt{1-s^2}}{s},\\
			g^* &= \frac{1-\sqrt{1-(s^{*})^2}}{s^{*}},\\
			g^{**} &= \frac{1-\sqrt{1-(s^{**})^2}}{s^{**}}.\\
		\end{split}
	\end{equation}
	According to Eq.~\eqref{eq:orthoFD}, there exists non-zero overlaps for $\langle \Phi_{R1} | \Phi_{R2}\rangle$, $\langle \Phi_{R1} | \Phi_{R3}\rangle$, $\langle \Phi_{L1} | \Phi_{R2}\rangle$, $\langle \Phi_{L1} | \Phi_{R3}\rangle$, but they are negligibly small \cite{Kornich.14}. Note that the orthogonalization of the F-D states shown in Eq.~\eqref{eq:orthoFD} is adopted only in the calculation of phonon-mediated decoherence \cite{Kornich.14}. For the full CI calculation of four electron states in a DQD device, we use fully orthonormal bases set given by the linear combinations of the electron wavefunctions based on Cholesky decomposition \cite{Barnes.11}.
	
	We denote $x-y$ components of two arbitrary F-D states as
	\begin{equation}
		\begin{split}
			\phi_{\eta j} (x',y')&= \frac{1}{\sqrt{\pi}l_\eta}\mathcal{F}^{(j)}(x',y')e^{-\frac{(\mathbf{r}'_{(2)} - \mathbf{R}_\eta)^2}{2 l_\eta^2}}e^{i h_\eta x_B y'},\\
			\phi_{\widetilde{\eta} j} (x',y')&= \frac{1}{\sqrt{\pi}l_{\widetilde{\eta}}}\mathcal{F}^{(j)}(x',y')e^{-\frac{(\mathbf{r}'_{(2)} - \mathbf{R}_{\widetilde{\eta}})^2}{2 l_{\widetilde{\eta}}^2}}e^{i h_{\widetilde{\eta}} x_B y'},
		\end{split}
	\end{equation}
	where $\{\eta,\widetilde{\eta} \}\in \{L,R\}$, $j \in \{1,2,3\}$, $\mathbf{R}_\eta = (h_\eta x_0,0) = h_\eta \hat{\mathbf{v}}_{[110]}$, $\{h_\eta,h_{\widetilde{\eta}} \} \in \{1,-1\}$ for right and left QD respectively and $\mathbf{r}'_{(2)} = x' \hat{\mathbf{v}}_{[110]} + y' \hat{\mathbf{v}}_{[\bar{1}10]}$ is the position vector in the $x-y$ plane (with the subscript $(2)$ denoting dimension of 2) while $\hat{\mathbf{v}}_{[jkm]} $ is the unit vector in the crystallographic $[jkm]$ direction. The integrand formed by two arbitrary F-D states can be simplified into 
	\begin{equation}\label{eq:2FD}
		\phi_{\eta j}^*  \phi_{\widetilde{\eta} k} = \mathcal{G}_\eta \mathcal{G}_{\widetilde{\eta}} \mathcal{F}^{(j)}(x',y') \mathcal{F}^{(k)}(x',y') K_{\eta j,\widetilde{\eta} k}e^{p_{\eta,\widetilde{\eta}}(\mathbf{r}'_{(2)}-\mathbf{R}_{\eta,\widetilde{\eta}})^2} \mathcal{H}(z)e^{i (h_{\eta}-h_{\widetilde{\eta}}) x_B y'}
	\end{equation}
	where the asterisks, $*$, denotes conjugation, while
	\begin{equation}
		\begin{split}
			K_{\eta j,\widetilde{\eta} k} &= \exp \left[- \frac{\alpha_{\eta} \alpha_{\widetilde{\eta}}}{\alpha_{\eta} +\alpha_{\widetilde{\eta}}} \left(\mathbf{R}_\eta-\mathbf{R}_{\widetilde{\eta}}\right)^2\right],\\
			\mathbf{R}_{\eta,\widetilde{\eta}}&= \frac{\alpha_\eta \mathbf{R}_\eta+\alpha_{\widetilde{\eta}}\mathbf{R}_{\widetilde{\eta}}}{\alpha_\eta +\alpha_{\widetilde{\eta}}},
			\\
			\alpha_\eta &= \frac{1}{2 l_\eta^2},
			\\
			p_{\eta,\widetilde{\eta}} &= \alpha_{\eta}+\alpha_{\widetilde{\eta}},
			\\
			\mathcal{G}_\eta &= \frac{1}{\sqrt{\pi}l_\eta \sqrt{2 a_z^3}}, 
			\\
			\mathcal{H}(z)&= \frac{z^2}{2a_z^3}e^{\frac{z}{a_z}}.\\
		\end{split}
	\end{equation}
	Let $l_{\widetilde{\eta}} = \tilde{c} l_\eta$, we have
	\begin{equation}\label{eq:valueK}
		K_{\eta j,\widetilde{\eta} k} = \exp \left[-\frac{2}{1+\tilde{c}^2}\left(\frac{x_0}{2 l_\eta}\right)^2(h_\eta-h_{\widetilde{\eta}})^2\right].
	\end{equation}
	A typical double-quantum-dot (DQD) usually has $x_0 \geq 2l_\eta$ such that the confinements of electrons in each quantum dot is achieved. Eq.~\eqref{eq:valueK} shows that for an integrand involving two different dots, i.e. $h_\eta \neq h_{\widetilde{\eta}}$, there exist an exponential suppression of the integrand $\phi_{\eta j}^*  \phi_{\widetilde{\eta} k}$.
	\section{Effective Hamiltonian}
	\subsection{Hamiltonian with spin-orbit interaction, phonon bath and electron-phonon interaction, $H$}
	The Hamiltonian of a DQD system is
	\begin{equation}
		\begin{split}\label{eq:fullHam}
			H &= \sum_{j=1}^{4} \left(H_0^{(j)} + H_\text{Z}^{(j)}+H_\text{SOI}^{(j)}+H_\text{hyp}^{(j)} + H_\text{e-ph}^{(j)}\right)	\\
			&\quad +H_C + H_\text{ph}\\
			&=H_\text{e} +H_\text{hyp}+ H_\text{e-ph} +H_\text{ph},
		\end{split}
	\end{equation}
	where 
	\begin{subequations}\label{eq:HamDescrip}
		\begin{align}
			\begin{split}
				H_0^{(j)} &=\frac{\mathbf{p}_j^2}{2m^*}= {(-i\hbar \nabla_j+e \mathbf{A}/c)^2}/(2m^*)+V(\mathbf{r}_j),
			\end{split}
			\\
			\begin{split}
				H_C &= \sum{e^2}/\epsilon\left|\mathbf{r}_j-\mathbf{r}_k\right|,
			\end{split}
			\\
			\begin{split}
				H_\text{Z}^{(j)}&=g^*\mu_B \mathbf{B}\cdot \mathbf{S}_j,
			\end{split}
			\\
			\begin{split}
				H_\text{e}&=\sum_{j=1}^{4} \left(H_0^{(j)} + H_\text{Z}^{(j)}+H_\text{SOI}^{(j)}\right) +H_C,
			\end{split}
			\\
			\begin{split}\label{eq:Hhyp}
				H_\text{hyp}&=\sum_{j=1}^{4} H_\text{hyp}^{(j)} = h \Big( |S(1,3)\rangle \langle T(1,3) | + |S(1,3^*)\rangle \langle T(1,3^*) | + |S(1,3^*)\rangle \langle T(1,3) | + |S(1,3)\rangle \langle T(1,3^*) |\Big)
			\end{split}
			\\
			\begin{split}
				H_\text{e-ph}&= \sum_{j=1}^{4} H_\text{e-ph}^{(j)}
			\end{split}
			\\
			\begin{split}
				H_\text{ph}&= \sum_{\mathbf{q}_m,s} \hbar \omega_s(\mathbf{q}_m) \left[a_s^\dagger(\mathbf{q}_m)a_s(\mathbf{q}_m)+\frac{1}{2}\right],
			\end{split}
		\end{align}
	\end{subequations} 
	where $H_0^{(j)}$ is the single electron Hamiltonian which consists of Kinetic and confinement potential term, $H_C$ is the Coulomb interaction, $H_\text{hyp}$ is the hyperfine coupling, $H_\text{SOI}$ is the spin-orbit interaction (SOI), $H_z$ is the Zeemen energy, $H_\text{ph}$ is the phonon bath and $H_\text{e-ph}$ is the electron phonon interaction. In Eq.~\eqref{eq:fullHam} and \eqref{eq:HamDescrip}, $h$ is the magnetic gradient, $\mathbf{S}$ is the electron spin, $\mu_B$ the Bohr magneton, $g$ the Landé g-factor, $\mathbf{A}$ the vector potential, $m^*$ the effective mass, $V(\mathbf{r})$ the confinement potential, $\epsilon$ the permittivity and $e$ the electron charge. The definition of Slater determinations in Eq.~\eqref{eq:Hhyp} will be given in Eq.~\eqref{eq:4eBases}. $a_s^\dagger(\mathbf{q}_m)$ and $a_s(\mathbf{q}_m)$ is the creation and annihilation operator of the $m$-th phonon of mode $s$ with wave vector, $\mathbf{q}_m$. Diagonalization of $H_\text{e}+H_\text{hyp}$ gives the single-triplet qubit eigenspace while $H_\text{e-ph}$ induces decoherence effect through electron-phonon interaction.
	\subsubsection{Spin-orbit interaction, $H$\textsubscript{\normalfont{SOI}}}
	The SOI term takes the form \cite{Kornich.14,Stano.05,Stano.06,Golovach.08,Raith.12}
	\begin{equation}\label{eq:HamSOI}
		\begin{split}
			H_{\text{SOI}}&=\alpha \left(p_{[100]}\sigma_{[010]}-p_{[010]}\sigma_{[100]}\right)+\beta \left(p_{[010]}\sigma_{[010]}+p_{[100]}\sigma_{[100]}\right)\\
			&=-\left(\alpha p_{[010]}-\beta p_{[100]} \right)  \sigma_{[100]} +\left(\alpha p_{[100]}+\beta p_{[010]}\right)\sigma_{[010]}\\
			&=\widetilde{p}_{[100]}  \sigma_{[100]} +\widetilde{p}_{[010]}\sigma_{[010]},
		\end{split}
	\end{equation}
	where $[jkm]$ denotes the crystallographic direction, $\widetilde{p}_{[100]} = \alpha p_{[010]}-\beta p_{[100]} $, $\widetilde{p}_{[010]} = \alpha p_{[100]}+\beta p_{[010]}$, the Rashba SOI term $\alpha=\hbar /(m^* l_R)$ and the Dresselhaus SOI term$\beta = \hbar /(m^* l_D)$. Note, the readers should distinguish the notations for capacitive shift $\beta$ given in the main text and the Dresselhaus SOI term $\beta$ here in different contexts. In Ref.~\onlinecite{Kornich.14}, the SOI terms are evaluated at their lowest order, which are obtained via a Schrieffer-Wolff transformation,
	\begin{equation}
		\widetilde{H}_\text{SOI} = e^S H e^S \simeq g\mu_B (\mathbf{r}_\text{SOI}\times \mathbf{B})\cdot \boldsymbol{\sigma}-m_\text{eff}\left(\alpha^2+\beta^2\right)+\frac{m_\text{eff}}{\hbar}\left(\beta^2-\alpha^2\right)l_{[001]}\sigma_{[001]},
	\end{equation}
	where the spin vector $\boldsymbol{\sigma} = \{\sigma_{[100]},\sigma_{[010]},\sigma_{[001]}\}$. $\mathbf{r}_\text{SOI}$ and $l_{[001]}$ are defined as
	\begin{equation}
		\begin{split}
			\mathbf{r}_\text{SOI}&=\left(\frac{y'}{l_R}+\frac{x'}{l_D}\right) \mathbf{e}_{[100]} + \left(-\frac{x'}{l_R}-\frac{y'}{l_D}\right) \mathbf{e}_{[010]},\\
			l_{[001]}&=\left(x_{[100]}p_{[010]}-y_{[010]}p_{[100]}\right),
		\end{split}
	\end{equation}
	where $\mathbf{e}_{[jkm]}$ is the unit vector along the direction $[jkm]$ while $x_{[jkm]}, y_{[jkm]}$ and $p_{[jkm]}$ are the length and momentum in the direction of $[jkm]$ respectively.
	
	\subsubsection{Hyperfine interaction, $H$\textsubscript{\normalfont{hyp}}}
	\begin{equation}
		H_\text{hyp} = \frac{h}{4} \left(\mathcal{P}_L - \mathcal{P}_R\right),
	\end{equation}
	where
	\begin{equation}
		\begin{split}
			\mathcal{P}_L&=|\Phi_{L1}\rangle\langle \Phi_{L1}|,\\
			\mathcal{P}_R&=\sum_{j=1}|\Phi_{Rj}\rangle\langle \Phi_{Rj}|,
		\end{split}
	\end{equation}
	while $h$ is the magnetic field gradient between two QDs in a DQD device.
	\subsection{Extended Hubbard Model}\label{subsec:HubbardModel}
	$\sum_{j=1}^4 H_0^{(j)}+H_C$ part of the Hamiltonian (Eq.\eqref{eq:fullHam}) can be fitted into the extended Hubbard model, whose explicit form is
	\begin{equation}\label{eq:HubbardModel2ndQuan}
		\begin{split}
			\sum_{j=1}^4 H_0^{(j)}+H_C &= \sum_{\psi j, \sigma}\varepsilon_{\psi j, \sigma} c^\dagger_{\psi j, \sigma} c_{\psi j, \sigma}+\sum_{\psi j\neq \psi k,\sigma} \left(t_{\psi j,\psi k,\sigma} c^\dagger_{\psi j, \sigma} c_{\psi k, \sigma}\right)
			+\sum_{\psi j} U_{\psi j} n_{\psi j,\downarrow} n_{\psi j,\uparrow} \\
			\quad &+\frac{1}{2}\sum_{\sigma \sigma'}\sum_{\psi j\neq \psi k} U_{\psi j,\psi k} n_{\psi j, \sigma} n_{\psi k, \sigma'}+\frac{1}{2} \sum_{\sigma \sigma'} \sum_{\psi j\neq \psi k} U^e_{\psi j,\psi k} c_{\psi j, \sigma}^\dagger c_{\psi k, \sigma'}^\dagger c_{\psi j, \sigma'} c_{\psi k, \sigma},
		\end{split}
	\end{equation}
	where $\psi j$ and $\psi k$ indicate the orbitals (cf. Eq.~\eqref{eq:FD} and \eqref{eq:orthoFD}), while $\sigma$ and $\sigma'$ are spins. The summations over orbitals $\psi j$ and $\psi k$ are from $R1$ to $R3$ for right dot and $L1$ for left dot, while the spins $(\sigma,\sigma')$ are up and down. $\varepsilon_{\psi j, \sigma}$ denotes the on-site energy at dot $\eta(\psi j)$, where $\eta(\psi j)$ indicates left or right dot ($L$ or $R$) the orbital $\psi j$ occupies. $t_{\psi j,\psi k,\sigma}$ denotes the tunneling between $\psi j$ and $\psi k$, while $U_{\psi j}$ denotes the on-site Coulomb interaction in $\psi j$, $U_{\psi j,\psi k}$ and $U^e_{\psi j,\psi k}$ the direct and exchange Coulomb interaction between $\psi j$ and $\psi k$ respectively. These parameters are calculated from,
	\begin{equation}\label{eq:HubbardPara}
		\begin{split}
			U_{\psi j} &= \int{\Psi^*_{\psi j} (\mathbf{r}_1)\Psi^*_{\psi j} (\mathbf{r}_2) C(\mathbf{r}_1,\mathbf{r}_2) \Psi_{\psi j} (\mathbf{r}_1)\Psi_{\psi j} (\mathbf{r}_2) \text{d}\mathbf{r}^2},\\
			U_{\psi j,\psi k} &= \int{\Psi^*_{\psi j} (\mathbf{r}_1)\Psi^*_{\psi k} (\mathbf{r}_2) C(\mathbf{r}_1,\mathbf{r}_2) \Psi_{\psi j} (\mathbf{r}_1)\Psi_{\psi k} (\mathbf{r}_2) \text{d}\mathbf{r}^2},\\
			U^e_{\psi j,\psi k} &= \int{\Psi^*_{\psi j} (\mathbf{r}_1)\Psi^*_{\psi k} (\mathbf{r}_2) C(\mathbf{r}_1,\mathbf{r}_2) \Psi_{\psi k} (\mathbf{r}_1)\Psi_{\psi j} (\mathbf{r}_2) \text{d}\mathbf{r}^2},\\
			t_{\psi j,\psi k}&=  \int{\Psi^*_{\psi j} (\mathbf{r}) \left[\frac{\hbar^2}{2m^*}\nabla^2+V(\mathbf{r}) \right]\Psi_{\psi k} (\mathbf{r}) \text{d}\mathbf{r}},\\
			\varepsilon_{\psi j} &=  \int{\Psi^*_{\psi j} (\mathbf{r}) \left[\frac{\hbar^2}{2m^*}\nabla^2+V(\mathbf{r}) \right]\Psi_{\psi j} (\mathbf{r}) \text{d}\mathbf{r}},\\
			& C(\mathbf{r}_1,\mathbf{r}_2)=\frac{e^2}{\kappa|\mathbf{r}_1-\mathbf{r_2}|},
		\end{split}
	\end{equation}
	where $\psi k \neq \psi j$. The onsite Coulomb exchange term, $U^e_{\Psi_{\psi j},\Psi_{\psi k}}$ can be rewritten as
	\begin{equation}\label{eq:exchangeTerm}
		\begin{split}
			& \sum_{\sigma \sigma'} \sum_{\psi j\neq \psi k} U^e_{\psi j,\psi k} c_{\psi j, \sigma}^\dagger c_{\psi k, \sigma'}^\dagger c_{\psi j, \sigma'} c_{\psi k, \sigma}
			= - 2 \xi \left(\mathbf{S}_{\psi j}\cdot \mathbf{S}_{\psi k}+\frac{1}{4} n_{\psi j} n_{\Psi k}\right),
		\end{split}
	\end{equation}
	where $\xi$ is the ferromagnetic exchange term.
	Note that in Eq.~\eqref{eq:HubbardPara}, the electron wavefunctions are denoted as $\Psi_{\psi j}$, which are the linear combinations of $\Phi_{\psi j}$ shown in Eq.~\eqref{eq:orthoFD}. The explicit numeric form of $\Psi_{\psi j}$, which can be obtained from full-CI results, is not important as we are only interested in the fitted Hubbard parameters (Eq.~\eqref{eq:HubbardPara}). In regard to the notations of F-D states shown in Eq.~\eqref{eq:FD} and \eqref{eq:orthoFD}, the correspondence between the notations in the main text and Supplemental Material here is: $L1 \rightarrow 2$, $R1 \rightarrow 1$, $R2 \rightarrow 5$, $R3 \rightarrow 9$ for $\mathbb{R}$ DQD while $L1 \rightarrow 3$, $R1 \rightarrow 4$, $R2 \rightarrow 8$, $R3 \rightarrow 12$ for DQD-$\mathbb{L}$.
	
	\subsection{Two electrons occupying a DQD device}\label{subsec:twoEs}
	In Ref.~\onlinecite{Kornich.14}, phonon-mediated decoherence is discussed for two regimes: (1) the regime of ``large'' detuning where the decoherence time, $T_2$, is dominated by one-phonon relaxation, $\Gamma_1^{\text{1p}}$, and two-phonon pure-dephasing, $\Gamma_\varphi^{2p}$ and (2) the regime of ``small'' detuning where $T_2$ is dominated by one-phonon relaxation, $T_1^{1p}=1/\Gamma_1^{\text{1p}}$, such that $T_2 \approx T_1^{1p}$. The effective Hamiltonian of both regions are shown in Ref.~\onlinecite{Kornich.14}. 
	We denote the two-electron Slater determinants as
	\begin{equation}\label{eq:2esState}
		\begin{split}
			|S(\uparrow_{\psi j} \downarrow_{\psi j})\rangle_{2e}&= |\uparrow_{\psi j} \downarrow_{\psi j} \rangle,\\
			|S(\uparrow_{\psi j} \downarrow_{\psi k})\rangle_{2e}&= \left(|\uparrow_{\psi j} \downarrow_{\psi k} \rangle + |\uparrow_{\psi k} \downarrow_{\psi j} \rangle\right)/\sqrt{2},\\
			|T(\uparrow_{\psi j} \downarrow_{\psi k})\rangle_{2e}&= \left(|\uparrow_{\psi j} \downarrow_{\psi k} \rangle - |\uparrow_{\psi k} \downarrow_{\psi j} \rangle\right)/\sqrt{2},\\
			|T_+(\uparrow_{\psi j} \uparrow_{\psi k})\rangle_{2e}&= |\uparrow_{\psi j} \uparrow \rangle,\\
			|T_-(\downarrow_{\psi k} \downarrow_{\psi k})\rangle_{2e}&= |\downarrow_{\psi j} \downarrow_{\psi k} \rangle,
		\end{split}
	\end{equation}
	where $\psi k\neq \psi j$. $|\uparrow_{\psi j} \downarrow_{\psi k} \rangle$  denotes a two-electron Slater determinant with an electron with spin up occupying orbital $\psi j$ while another electron with spin down occupying orbital $\psi k$ while the subscript, $2e$, indicates a two-electron Slater determinant. The relevant two-electron Slater determinants are \cite{Kornich.14}
	\begin{equation}\label{eq:2eBases}
		\begin{split}
			|S(1,1)\rangle &= |S(\uparrow_{L1}\downarrow_{R1})\rangle_{2e},\\
			|S(0,2)\rangle&= |S(\uparrow_{R1}\downarrow_{R1})\rangle_{2e},\\
			|S(2,0)\rangle &= |S(\uparrow_{L1}\downarrow_{L1})\rangle_{2e},\\
			|T(1,1)\rangle &= |T(\uparrow_{L1}\downarrow_{R1})\rangle_{2e},\\
			|S(1,1^\star)\rangle &= |S(\uparrow_{L1}\downarrow_{R2})\rangle_{2e},\\
			|T(1,1^\star)\rangle &= |T(\uparrow_{L1}\downarrow_{R2})\rangle_{2e},\\
			|T_+(1,1)\rangle &= |T(\uparrow_{L1}\uparrow_{R1})\rangle_{2e},\\
			|T_-(1,1)\rangle &= |T(\downarrow_{L1}\downarrow_{R1})\rangle_{2e},\\
			|T_+(1,1^\star)\rangle &= |T(\uparrow_{L1}\uparrow_{R2})\rangle_{2e},\\
			|T_-(1,1^\star)\rangle &= |T(\downarrow_{L1}\downarrow_{R2})\rangle_{2e},\\
		\end{split}
	\end{equation} 
	where $|S(n_L,n_R)\rangle$ and $|T(n_L,n_R)\rangle$ indicate a singlet and triplet, respectively, with $n_L$ and $n_R$ electrons occupying left and right dot respectively. The star symbol, $\star$, indicates an electron occupying the first excited orbital from the ground configuration.
	Following the notations in Ref.~\onlinecite{Kornich.14}, the SOI terms are
	\begin{equation}
		\begin{split}
			\Omega &= E_\text{Z} \left(\langle \Phi_{L1} | R_\text{SOI}| \Phi_{L1} \rangle - \langle \Phi_{R1} | R_\text{SOI}| \Phi_{R1} \rangle\right)\\
			\Omega_1 &= E_\text{Z} \langle \Phi_{L1} | R_\text{SOI}| \Phi_{L2} \rangle \\
			\Omega_2 &= E_\text{Z} \left(\langle \Phi_{L2} | R_\text{SOI}| \Phi_{L2} \rangle  - \langle \Phi_{R1} | R_\text{SOI}| \Phi_{R1} \rangle\right)\\
			\Omega_3 &= E_\text{Z} \left(\langle \Phi_{L2} | R_\text{SOI}| \Phi_{L2} \rangle  + \langle \Phi_{R1} | R_\text{SOI}| \Phi_{R1} \rangle\right),\\
		\end{split}
	\end{equation}
	where $R_\text{SOI}=(\mathbf{r}_\text{SOI}\times \mathbf{e}_\mathbf{B})_{[001]}$ while $\mathbf{e}_\mathbf{B}$ denotes the unit vector of spin quantization axis by magnetic field, $\mathbf{B}$.
	\subsection{Four electrons occupying a DQD device}\label{eq:4esHam}
	Here, we outline the effective Hamiltonian, $H_\text{e}$, of four-electron system considered in this work. In contrast to Ref.~\onlinecite{Kornich.14}, the magnetic field is aligned in the direction of $\mathbf{e}_{[001]}$. In this case, since the spin quantization axis lies in $\mathbf{e}_{[001]}$, the relation of spin operator and Pauli matrices in Eq.~\eqref{eq:HamSOI} are:
	\begin{equation}
		\begin{split}
			\sigma_{[100]}&=\sigma_x, \\
			\sigma_{[010]}&=\sigma_y, \\
		\end{split}
	\end{equation}
	where $\sigma_x=\left(\sum_{\psi j, \psi k} c_{\psi j,\uparrow}^\dagger c_{\psi k,\downarrow}+c_{\psi j,\downarrow}^\dagger c_{\psi k,\uparrow}\right)$ and $\sigma_y=i \left(\sum_{\psi j, \psi k} c_{\psi j,\downarrow}^\dagger c_{\psi k,\uparrow}- c_{\psi j,\uparrow}^\dagger c_{\psi k,\downarrow}\right)$.
	We denote the four-electron Slater determinants as
	\begin{equation}
		\begin{split}
			|S(\uparrow_{\psi j} \downarrow_{\psi j})\rangle&= |\uparrow_{\psi j} \downarrow_{\psi j} \uparrow_{R1} \downarrow_{R1}\rangle\\
			|S(\uparrow_{\psi j} \downarrow_{\psi k})\rangle&= \Big(|\uparrow_{\psi j} \downarrow_{\psi k} \uparrow_{R1} \downarrow_{R1} \rangle + |\uparrow_k \downarrow_j \uparrow_{R1} \downarrow_{R1} \rangle\Big)/\sqrt{2},\\
			|T(\uparrow_{\psi j} \downarrow_{\psi k})\rangle&= \Big(|\uparrow_{\psi j} \downarrow_{\psi k} \uparrow_{R1} \downarrow_{R1} \rangle - |\uparrow_k \downarrow_j \uparrow_{R1} \downarrow_{R1} \rangle\Big)/\sqrt{2},\\
			|T_+(\uparrow_{\psi j} \uparrow_{\psi k})\rangle&= |\uparrow_{\psi j}\uparrow_{\psi k} \uparrow_{R1} \downarrow_{R1} \rangle,\\
			|T_-(\downarrow_{\psi j} \downarrow_{\psi k})\rangle&= |\downarrow_{\psi j} \downarrow_{\psi k} \uparrow_{R1} \downarrow_{R1}\rangle.
		\end{split}
	\end{equation}
	where $\psi k\neq \psi j$. The relevant four-electron Slater determinants which give the lowest energy subspace are
	\begin{equation}\label{eq:4eBases}
		\begin{split}
			|S(1,3)\rangle&=|S(\uparrow_{L1}\downarrow_{R2})\rangle,\\
			|S(0,4)\rangle&=|S(\uparrow_{R2}\downarrow_{R2})\rangle,\\
			|S(1,3^\star)\rangle&=|S(\uparrow_{L1}\downarrow_{R3})\rangle,\\
			|T(1,3)\rangle&=|T(\uparrow_{L1}\downarrow_{R2})\rangle,\\
			|T(0,4)\rangle&=|T(\uparrow_{R2}\downarrow_{R3})\rangle,\\
			|T(1,3^\star)\rangle&=|T(\uparrow_{L1}\downarrow_{R3})\rangle,\\
			|T_+(1,3)\rangle&=|T(\uparrow_{L1}\uparrow_{R2})\rangle,\\
			|T_-(1,3)\rangle&=|T(\downarrow_{L1}\downarrow_{R2})\rangle,\\
			|T_+(0,4)\rangle&=|T(\uparrow_{R2}\uparrow_{R3})\rangle,\\
			|T_-(0,4)\rangle&=|T(\downarrow_{R2}\downarrow_{R3})\rangle,\\
			|T_+(1,3^\star)\rangle&=|T(\uparrow_{L1}\uparrow_{R3})\rangle,\\
			|T_-(1,3^\star)\rangle&=|T(\downarrow_{L1}\downarrow_{R3})\rangle.
		\end{split}
	\end{equation}
	Written in the bases in Eq.~\eqref{eq:4eBases}, the effective Hamiltonian is
	
	\begin{equation}
		\rotatebox{270}{$
			H_\text{e}=
			\left(
			\begin{array}{cccccccccccc}
				0 & 
				t_{L1,R2} &
				0 & 
				\frac{h}{2} & 
				0 & 
				\frac{h}{2} & 
				-\frac{\widetilde{\Omega}^-}{\sqrt{2}} &
				\frac{\widetilde{\Omega} ^+}{\sqrt{2}} & 
				\frac{\widetilde{\Omega} _{R3,L1}^-}{\sqrt{2}} &
				-\frac{\widetilde{\Omega} _{R3,L1}^+}{\sqrt{2}} &
				\frac{\widetilde{\Omega}_{1^*}^{-}}{\sqrt{2}} &
				-\frac{\widetilde{\Omega} _{1^*}^+}{\sqrt{2}} 
				\\
				t_{L1,R2} & 
				E_S & 
				0 & 
				0 & 
				0 & 
				0 & 
				-\widetilde{\Omega}_{L1,R2}^- & 
				\widetilde{\Omega}_{L1,R2}^+ & 
				\widetilde{\Omega}_{1^*}^- & 
				-\widetilde{\Omega} _{1^*}^+ &
				0 &
				0
				\\
				0 & 
				0 & 
				E_{S^\star} & 
				\frac{h}{2} & 
				0 & 
				0 & 
				\frac{\widetilde{\Omega}_1^-}{\sqrt{2}} & 
				-\frac{\widetilde{\Omega} _1^+}{\sqrt{2}} & 
				-\frac{\widetilde{\Omega}_{R2,L1}^-}{\sqrt{2}} & 
				\frac{\widetilde{\Omega}_{R2,L1}^+}{\sqrt{2}} &
				-\frac{\widetilde{\Omega} _2^-}{\sqrt{2}} & 
				\widetilde{\Omega}_2^+ 
				\\
				\frac{h}{2} & 
				0 & 
				\frac{h}{2} & J^{(1,3)} & 
				t_{L1,R3} & 
				0 & 
				0 & 
				0 & 
				-\frac{\widetilde{\Omega} _{R3,L1}^-}{\sqrt{2}} & 
				-\frac{\widetilde{\Omega}_{R3,L1}^+}{\sqrt{2}} &
				\frac{\widetilde{\Omega}_{1^*}^-}{\sqrt{2}} & 
				\frac{\widetilde{\Omega} _{1^*}^+}{\sqrt{2}}
				\\
				0 & 
				0 & 
				0 & 
				t_{L1,R3} & 
				E_T & 
				t_{L1,R2} & 
				\frac{\widetilde{\Omega}_{L1,R3}^+}{\sqrt{2}} & 
				-\frac{\widetilde{\Omega}_{L1,R3}^+}{\sqrt{2}} &
				\frac{\widetilde{\Omega}_4^-}{\sqrt{2}} & 
				\frac{\widetilde{\Omega} _4^+}{\sqrt{2}} &
				\frac{\widetilde{\Omega}_{L1,R2}^-}{\sqrt{2}} & 
				\frac{\widetilde{\Omega}_{L1,R2}^+}{\sqrt{2}}
				\\
				\frac{h}{2} & 
				0 & 
				0 & 
				0 & 
				t_{L1,R2} & 
				E_{T^\star}+J^{(1,3^\star)} & 
				\frac{\widetilde{\Omega}_1^-}{\sqrt{2}} & 
				\frac{\widetilde{\Omega}_1^+}{\sqrt{2}} & 
				\frac{\widetilde{\Omega}_{R2,L1}^-}{\sqrt{2}} & 
				\frac{\widetilde{\Omega}_{R2,L1}^+}{\sqrt{2}} &
				\frac{\widetilde{\Omega} _3^-}{\sqrt{2}} &
				\frac{\widetilde{\Omega} _3^+}{\sqrt{2}} 
				\\
				-\frac{\widetilde{\Omega}^+}{\sqrt{2}} & 
				-\widetilde{\Omega} _{R2,L1}^+ & 
				\frac{\widetilde{\Omega} _{1^*}^+}{\sqrt{2}} & 
				0 & 
				-\frac{\widetilde{\Omega}_{R3,L1}^+}{\sqrt{2}} & 
				\frac{\widetilde{\Omega} _{1^*}^+}{\sqrt{2}} & 
				-E_\text{Z} & 
				0 & 
				0 & 
				0 &
				0 & 
				0
				\\
				\frac{\widetilde{\Omega}^-}{\sqrt{2}} & 
				\widetilde{\Omega}_{R2,L1}^- & 
				-\frac{\widetilde{\Omega}_{1^*}^-}{\sqrt{2}} & 
				0 & 
				-\frac{\widetilde{\Omega}_{R3,L1}^-}{\sqrt{2}} & 
				\frac{\widetilde{\Omega}_{1^*}^-}{\sqrt{2}} & 
				0 & 
				E_\text{Z} & 
				0 & 
				0 &
				0 & 
				0  
				\\
				\frac{\widetilde{\Omega}_{L1,R3}^+}{\sqrt{2}} & 
				\widetilde{\Omega} _1^+ & 
				-\frac{\widetilde{\Omega}_{L1,R2}^+}{\sqrt{2}} &
				-\frac{\widetilde{\Omega}_{L1,R3}^+}{\sqrt{2}} & 
				\frac{\widetilde{\Omega} _4^+}{\sqrt{2}} & 
				\frac{\widetilde{\Omega}_{L1,R2}^+}{\sqrt{2}} & 
				0 & 
				\frac{\widetilde{\Omega} _4^+}{\sqrt{2}} & 
				E_T -E_\text{Z} & 
				0 &
				0 & 
				0 
				\\
				\frac{\widetilde{\Omega}_{L1,R3}^+}{\sqrt{2}} & 
				-\widetilde{\Omega}_1^- & 
				\frac{\widetilde{\Omega}_{L1,R2}^-}{\sqrt{2}} &
				\frac{\widetilde{\Omega}_{L1,R3}^+}{\sqrt{2}} & 
				\frac{\widetilde{\Omega}_4^-}{\sqrt{2}} & 
				\frac{\widetilde{\Omega}_{L1,R2}^-}{\sqrt{2}} & 
				0 & 
				0 & 
				0 & 
				E_T+E_\text{Z} &
				0 & 
				0 
				\\
				\frac{\widetilde{\Omega} _1^+}{\sqrt{2}} & 
				0 & 
				-\frac{\widetilde{\Omega}_2^+}{\sqrt{2}} & 
				\frac{\widetilde{\Omega} _1^+}{\sqrt{2}} & 
				\frac{\widetilde{\Omega}_{R2,L1}^+}{\sqrt{2}} & 
				\frac{\widetilde{\Omega} _3^+}{\sqrt{2}} & 
				0 & 
				0 & 
				0 & 
				0 &
				E_{T^\star}-E_\text{Z} & 
				0  
				\\
				-\frac{\widetilde{\Omega}_1^-}{\sqrt{2}} & 
				0 & 
				\frac{\widetilde{\Omega} _2^-}{\sqrt{2}} & 
				\frac{\widetilde{\Omega}_1^-}{\sqrt{2}} & 
				\frac{\widetilde{\Omega}_{R2,L1}^-}{\sqrt{2}} & 
				\frac{\widetilde{\Omega} _3^-}{\sqrt{2}} & 
				0 & 
				0 & 
				0 & 
				0 &
				0 & 
				E_{T^\star} +E_\text{Z}
			\end{array}
			\right)
			\label{eq:Ham4es}
			$}
	\end{equation}
	where $\widetilde{\Omega}^\pm = \widetilde{\Omega}^x \pm i \widetilde{\Omega} ^y $, $\widetilde{\Omega}_j^\pm = \widetilde{\Omega}_j^x \pm i \Omega_j^y$, $\widetilde{\Omega}^\pm_{\psi j,\psi k}= \widetilde{\Omega}_{\psi j,\psi k}^x \pm i \widetilde{\Omega}_{\psi j,\psi k}^y$ and
	\begin{equation}
		\begin{split}
			\widetilde{\Omega}^v &= \langle \Phi_{L1} | \widetilde{p}_{[jkm]} |\Phi_{L1}\rangle - \langle \Phi_{R1} | \widetilde{p}_{[jkm]} |\Phi_{R1}\rangle,\\
			\widetilde{\Omega}_1^v &= \langle \Phi_{R2} | \widetilde{p}_{[jkm]} |\Phi_{R3}\rangle,\\
			\widetilde{\Omega}_{1^*}^v &= \langle \Phi_{R3} | \widetilde{p}_{[jkm]} |\Phi_{R2}\rangle,\\
			\widetilde{\Omega}_2^v &= \langle \Phi_{L1} | \widetilde{p}_{[jkm]} |\Phi_{L1}\rangle - \langle \Phi_{R2} | \widetilde{p}_{[jkm]} |\Phi_{R2}\rangle,\\
			\widetilde{\Omega}_4^v &= \langle \Phi_{R1} | \widetilde{p}_{[jkm]} |\Phi_{R1}\rangle + \langle \Phi_{R3} | \widetilde{p}_{[jkm]} |\Phi_{R3}\rangle,\\
			\widetilde{\Omega}_{\psi j,\psi k}^v &= \langle \Phi_{\psi j} | \widetilde{p}_{[jkm]} |\Phi_{\psi k}\rangle,\\
			[jkm] &= \begin{cases} [100]  & v=x\\ [010]  & v=y \end{cases}.
		\end{split}
	\end{equation}
	In Eq.~\eqref{eq:Ham4es}, $E_S=U_{R2}-E_\text{shift}$, $E_{S^\star}=U_{L1,R3}+\Delta E-E_\text{shift}$, $E_T= U_{R2,R3}+\Delta E-\xi - E_\text{shift}$, $E_{T^\star}=U_{L1,R3}+\Delta E-E_\text{shift}$, $E_\text{shift} = U_{L1,R2}+2\Delta$, $\Delta E=\varepsilon_{R3}-\varepsilon_{R2}$ is the orbital splitting, $E_\text{Z}=g\mu_B |\mathbf{B}|$, $\mathbf{B}=|\mathbf{B}|\mathbf{e}_{[001]}=B\mathbf{e}_{[001]}$ and $2\Delta =\varepsilon=\varepsilon_{L1}-\varepsilon_{R1}$ is the detuning, while $J^{(1,3)}=\langle T(1,3)| H_e | T(1,3)\rangle-\langle S(1,3)| H | S(1,3)\rangle$ and $J^{(1,3^\star)}=\langle T(1,3^\star)| H_e | T(1,3^\star)\rangle-\langle S(1,3^\star)| H | S(1,3^\star)\rangle$ are the exchange splittings in the limit of $\Delta = 0$. Note that the diagonal terms in Eq.~\eqref{eq:Ham4es} are obtained by considering only the valence orbitals in right QD, i.e. $\psi j \in \{R2, R3, \cdots\}$. The effect of electrons occupying the core orbitals has been encoded in the Hubbard parameters in Eq.~\eqref{eq:Ham4es}, which is confirmed by the correspondence between full CI results and Hubbard model, see Sec.~\ref{sec:CoressHubbardCI}.
	\section{Phonon-mediated decoherence}\label{sec:phononDecoherence}
	\subsection{Electron-phonon coupling}
	$H_\text{e-ph}$ can be split into two parts \cite{Kornich.14}, i.e.
	\begin{equation}
		\begin{split}
			H_\text{e-ph}&=H_\text{dp}+H_\text{pe},\\
			&=\sum_{\mathbf{q}_m,s} C_{s,\text{cpl}} (\mathbf{q}_m) \left[a_s(\mathbf{q}_m)\mp_s a_s^\dagger(-\mathbf{q}_m)\right]e^{i \mathbf{q}_m \cdot \mathbf{r}}
		\end{split}
	\end{equation}
	where $e^{i \mathbf{q}_m \cdot \mathbf{r}}$ is the electron density operator while $\text{cpl} = \text{dp}$ and $\text{cpl} = \text{pe}$ stands for deformation potential and piezoelectric coupling respectively. The label $s\in \{l,t_1,t_2\}$ stands for the longitudinal $(l)$ and two transverse ($t_1$, $t_2$) phonon modes such that $\mp_l = -1$ while $\mp_{t_1,t_2}=1$. $C_{s,\text{cpl}}$ is the prefactor of the electron-phonon interaction terms, where
	\begin{equation}\label{eq:prefactor}
		\begin{split}
			C_{l,\text{dp}} (\mathbf{q}_m)&=  i \Xi  \sqrt{\frac{\hbar}{2\rho V \omega_l(\mathbf{q}_m)}} \left|\mathbf{q}_m\right|,\\
			C_{s,\text{pe}} (\mathbf{q}_m) &= \frac{e h_{14}}{\epsilon_0\epsilon_r} f_{\mathbf{q}_m,s}\sqrt{\frac{\hbar}{2\rho V \omega_s (\mathbf{q}_m)}}.
		\end{split}
	\end{equation}
	In Eq.~\eqref{eq:prefactor}, $\omega_s(\mathbf{q}_m) = |\mathbf{q}_m|v_s $ with $\nu_s$ being the speed of phonon mode $s$, while
	\begin{equation}
		\begin{split}
			f_{\mathbf{q}_m,l} &= 3 \cos(\theta_{\mathbf{q}_m})\sin^2 \theta_{\mathbf{q}_m} \sin(2 \phi_{\mathbf{q}_m}),\\
			f_{\mathbf{q}_m,t_1}&=-\sin(2\theta_{\mathbf{q}_m})\cos(2\phi_{\mathbf{q}_m}),\\
			f_{\mathbf{q}_m,t_2}&=-(3\sin^2\theta_{\mathbf{q}_m}-2)\sin\theta_{\mathbf{q}_m}\sin(2\phi_{\mathbf{q}_m}).
		\end{split}
	\end{equation}
	The $f_{\mathbf{q}_m,s}$ is defined when $\mathbf{q}_m$ are aligned in the main crystallographic directions, i.e.
	\begin{equation}\label{eq:qDef}
		\begin{split}
			\mathbf{q}_m &= q_x^{(m)} \hat{\mathbf{v}}_{[100]} + q_y^{(m)} \hat{\mathbf{v}}_{[010]} + q_z^{(m)} \hat{\mathbf{v}}_{[001]} \\
			&= q_m \sin \theta_{\mathbf{q}_m} \cos \phi_{\mathbf{q}_m} \hat{\mathbf{v}}_{[100]} + q_m \sin \theta_{\mathbf{q}_m} \sin \phi_{\mathbf{q}_m} \hat{\mathbf{v}}_{[010]} + q_m \cos \theta_{\mathbf{q}_m} \hat{\mathbf{v}}_{[001]},
		\end{split}
	\end{equation} 
	where $0\leq \theta_{\mathbf{q}_m}\leq \pi$ and $0\leq \phi_{\mathbf{q}_m} \leq 2\pi$ are the polar and azimuthal angle respectively.. We denote the inner products between the electron wavefunctions and electron-phonon interaction term of $m$-th phonon as
	\begin{subequations}\label{eq:Ppsi12Notation}
		\begin{align}
			\begin{split}
				P_{\psi 1,\psi 2} \left(\mathbf{q}_m\right)
				&=\langle \Phi_{\psi 1} | H_\text{e-ph} | \Phi_{\psi 2} \rangle\\
				&=\sum_{s,\text{cpl}} \sum_{\mathbf{q}_m} C_{s,\text{cpl}} (\mathbf{q}_m) P_{\psi1,\psi2}^{s,\text{cpl}}(\mathbf{q}_m)\\
				&=\sum_{s,\text{cpl}} \sum_{\mathbf{q}_m} C_{s,\text{cpl}} (\mathbf{q}_m) \langle \Phi_{\psi1} (\mathbf{r})| \left[a_s(\mathbf{q}_m)\mp_s a_s^\dagger(-\mathbf{q}_m)\right]e^{i \mathbf{q}_m \cdot \mathbf{r}}|\Phi_{\psi2} (\mathbf{r})\rangle,
			\end{split}\\
			\begin{split}
				P_{\psi 1,\psi 2} \left(\mathbf{q}_m, \tau\right)
				&=\langle \Phi_{\psi 1} | H_\text{e-ph}(\tau) | \Phi_{\psi 2} \rangle\\
				&=\sum_{s,\text{cpl}} \sum_{\mathbf{q}_m} C_{s,\text{cpl}} (\mathbf{q}_m) P_{\psi1,\psi2}^{s,\text{cpl}}(\mathbf{q}_m,\tau)\\
				&=\sum_{s,\text{cpl}} \sum_{\mathbf{q}_m} C_{s,\text{cpl}} (\mathbf{q}_m) \langle \Phi_{\psi1} (\mathbf{r})| \left[a_s(\mathbf{q}_m)e^{-i \omega_s \tau}\mp_s a_s^\dagger(-\mathbf{q}_m)e^{i \omega_s \tau}\right]e^{i \mathbf{q}_m \cdot \mathbf{r}}|\Phi_{\psi2} (\mathbf{r})\rangle,
			\end{split}
		\end{align}
	\end{subequations}
	where $\psi j \in \{L1,R1,R2,R3\}$ and $\tau$ is the time.
	\subsection{Bloch-Redfield theory}
	A singlet-triplet qubit can be formed by diagonalizing $H_\text{e}+H_\text{hyp}$ and extract the lowest singlet and triplet eigenstate, denoted as $|\widetilde{S}\rangle$ and $|\widetilde{T}\rangle$ respectively. We denote $H_\text{e}+H_\text{hyp}$ and $H_\text{e-ph}$ written in the singlet and triplet bases as $H_\text{q}$ and $H_\text{q-ph}$ respectively, i.e.
	\begin{equation}
		\begin{split}
			\mathcal{P}_{\widetilde{S}\widetilde{T}} U^\dagger \left(H_\text{e}+H_\text{hyp}\right) U \mathcal{P}_{\widetilde{S}\widetilde{T}} = H_\text{q}, \\
			\mathcal{P}_{\widetilde{S}\widetilde{T}} U^\dagger H_\text{e-ph} U \mathcal{P}_{\widetilde{S}\widetilde{T}} = H_\text{q-ph}, \\
		\end{split}
	\end{equation}
	where $U$ is the unitary matrix that diagonalize $H_\text{e}+H_\text{hyp}$ while $\mathcal{P}_{\widetilde{S}\widetilde{T}}$ is the projection to the lowest singlet and triplet eigenspace. $H_\text{q}$ is diagonal in the eigenbases of $H_\text{e}+H_\text{hyp}$ while $H_\text{q-ph}$ is not.
	Reformulating $H_\text{q}$ and $H_\text{q-ph}$ in the bases of Pauli matrices, we have
	\begin{equation}
		\begin{split}
			H_\text{q} = \mathbf{B}_\text{eff} \cdot \boldsymbol{\sigma}',\\
			H_\text{q-ph}(\tau) = \delta \mathbf{B}(\tau) \cdot \boldsymbol{\sigma}',\\
		\end{split}
	\end{equation}
	where $\boldsymbol{\sigma}'=\{\sigma'_x,\sigma'_y,\sigma'_z\}$ is the vector of spin-half Pauli matrices for the singlet-triplet qubit and $\tau$ is the time, such that
	\begin{equation}
		\begin{split}
			H_\text{q-ph}(\tau) = e^{i H_\text{ph} \tau/\hbar} H_\text{q-ph} e^{-i H_\text{ph}\tau/\hbar}.\\
		\end{split}
	\end{equation}
	Note that $\mathbf{B}_\text{eff}$ and $\delta \mathbf{B}$ yield the unit of energy. Similarly, 
	\begin{equation}
		\begin{split}
			H_\text{e-ph}(\tau) = e^{i H_\text{ph} \tau/\hbar} H_\text{e-ph} e^{-i H_\text{ph}\tau/\hbar}.\\
		\end{split}
	\end{equation}
	Defining the spectral functions
	\begin{equation}\label{eq:Jjk}
		\begin{split}
			J_{jk}(\omega_s) &= \frac{1}{2\hbar^2}\int_{0}^{\infty} e^{-i \omega_s \tau} \langle \delta B_j(0)\delta B_k(\tau) \rangle_\beta d\tau\\,
			J^+_{jj} (\omega_s) &= \frac{1}{2\hbar^2}\int_{-\infty}^{\infty} \cos (\omega_s \tau) \langle \delta B_j(0)\delta B_j(\tau) \rangle_\beta d\tau\\,
		\end{split}
	\end{equation}
	where the angle bracket with subscript $\beta$, $\langle \cdot \rangle_\beta$, is the thermal average, $j,k\in\{x,y,z\}$, and $\omega_s$ is the angular frequency of phonon with mode $s$ such that $\omega_s = q_s \nu_s$. Following \cite{Kornich.14}, the relaxation time, $T_1$, pure dephasing time, $T_\varphi$, and the decoherence time, $T_2$, are defined as 
	\begin{equation}
		\begin{split}
			\frac{1}{T_1} &= \Gamma_1 =J^+_{xx} (\omega_z)+J^+_{yy} (\omega_z),\\
			\frac{1}{T_\varphi} &= \Gamma_\varphi = J^+_{zz} (0),\\
			\frac{1}{T_2}&=\frac{1}{2T_1}+\frac{1}{T_\varphi},
		\end{split}
	\end{equation}
	where $\hbar \omega_z$ is the energy splitting between singlet and triplet state.
	\subsection{Inner product of electron wavefunctions and electron density operator}
	\label{subsec:innerproductDoublet}
	Here, we provide a general expression of the inner product, $\langle \Phi_{\eta j} | \exp\left( i \mathbf{q}_m \cdot \mathbf{r} \right) | \Phi_{\widetilde{\eta} k }\rangle$. The position vector in three dimension is defined as $\mathbf{r} = x' \hat{\mathbf{v}}_{[110]} + y' \hat{\mathbf{v}}_{[\bar{1}10]} + z \hat{\mathbf{v}}_{[001]} = x \hat{\mathbf{v}}_{[100]} + y \hat{\mathbf{v}}_{[100]} + z \hat{\mathbf{v}}_{[001]}$ and $\mathbf{r} = \mathbf{r}'_{(2)}+z \hat{\mathbf{v}}_{[001]}$, where $\hat{\mathbf{v}}_{[jkm]}$ denotes the unit vector in the crystallographic $[jkm]$ direction such that
	\begin{equation}
		\begin{split}
			\hat{\mathbf{v}}_{[110]}&=\frac{1}{\sqrt{2}}\left(\hat{\mathbf{v}}_{[100]}+\hat{\mathbf{v}}_{[010]}\right),\\
			\hat{\mathbf{v}}_{[\bar{1}10]}&=\frac{1}{\sqrt{2}}\left(-\hat{\mathbf{v}}_{[100]}+\hat{\mathbf{v}}_{[010]}\right)\\
			\mathbf{r} &= \frac{1}{\sqrt{2}}(x'-y') \hat{\mathbf{v}}_{[100]} + \frac{1}{\sqrt{2}} (x'+y') \hat{\mathbf{v}}_{[100]}.
		\end{split}
	\end{equation}
	The expression of $\mathbf{q}_m$ is defined in the main crystallographic direction, cf. Eq.~\eqref{eq:qDef}, such that
	\begin{equation}
		\begin{split}
			\mathbf{q}_m \cdot \mathbf{r} &= q_x^{(m)} \frac{1}{\sqrt{2}}(x'-y') + q_y^{(m)} \frac{1}{\sqrt{2}}(x'+y')\\
			&= \frac{1}{\sqrt{2}}\left(q_x^{(m)}+q_y^{(m)}\right)x' + \frac{1}{\sqrt{2}}\left(-q_x^{(m)}+q_y^{(m)}\right)y'
		\end{split}
	\end{equation}
	Adopting Eq.~\eqref{eq:2FD}, the expression of electron density operator yields
	\begin{equation}\label{eq:doublets}
		\begin{split}
			\langle \Phi_{\eta j} | \exp\left( i \mathbf{q}_m \cdot \mathbf{r} \right) | \Phi_{\widetilde{\eta} k }\rangle & =\mathcal{G}_\eta \mathcal{G}_{\widetilde{\eta}} \int  dx' \int dy' \int dz' \mathcal{F}^{(j)}\mathcal{F}^{(k)} e^{p_{\eta,\widetilde{\eta}}(\mathbf{r}'_{(2)}-\mathbf{R}_{\eta,\widetilde{\eta}})^2} \exp\left( i \mathbf{q}_m \cdot \mathbf{r} \right)\mathcal{H}(z),
		\end{split}
	\end{equation}
	where we have dropped the extra terms arisen due to the orthogonalization process for analytical purposes. We keep only the terms which give $\eta = \widetilde{\eta}$, as the contribution from $\eta \neq \widetilde{\eta}$ is exponentially small for typical DQD devices, cf. Eq.~\eqref{eq:valueK}. Any extra term dropped in the analytical description of the phonon-induced decoherence process is included when the problem is evaluated numerically. From Eq.~\eqref{eq:doublets}, we have
	\begin{equation}\label{eq:doubletIntegral}
		\begin{split}
			&\langle \Phi_{\eta j} | \exp\left( i \mathbf{q}_m \cdot \mathbf{r} \right) | \Phi_{\eta k }\rangle  =\mathcal{G}_\eta^2\int  dx' \int dy'  \mathcal{F}^{(j)}\mathcal{F}^{(k)} e^{2\alpha_\eta \left[(x'-h_\eta)^2+y'^2\right]}e^{ \frac{i}{\sqrt{2}}\left(q_x^{(m)}+q_y^{(m)}\right)x'} e^{ \frac{i}{\sqrt{2}}\left(-q_x^{(m)}+q_y^{(m)}\right)x'}\\
			&\quad\quad\quad\quad\quad\quad\quad\quad\quad\quad\quad \times \int e^{i q_z^{(m)}z}dz'\mathcal{H}(z)\\
			&=\mathcal{G}_\eta^2 \mathcal{M}_{\eta}^{j,k}\left(q_x^{(m)},q_y^{(m)}\right)\frac{\pi}{\alpha_{\eta}}\frac{1}{\left(\frac{1}{a_z}-i q_m \cos \theta_{\mathbf{q}_m}\right)^3}\exp\left[- \frac{q_m^2 \sin^2 \theta_{\mathbf{q}_m}}{8 \alpha_\eta}\right]\exp\left[-\frac{i h_\eta x_0 q_m \sin \theta_{\mathbf{q}_m}\left(\cos \phi_{\mathbf{q}_m} + \sin \phi_{\mathbf{q}_m}\right)}{\sqrt{2}}\right]
		\end{split}
	\end{equation}
	where $\alpha_\eta = 1/(2l_\eta^2)$ and $q_{x,y,z}^{(m)}$ in the exponents have been replaced with variables in the spherical coordinate. $\mathcal{M}_\eta^{j,k}$ in Eq.~\eqref{eq:doubletIntegral} is the prefactor which is not a unity when $\mathcal{F}^{(j)}\mathcal{F}^{(k)}\neq 1$, for example
	\begin{equation}
		\mathcal{F}^{(j)}\mathcal{F}^{(k)} = x' \xrightarrow{\int\int\int dx' dy' dz'} \mathcal{M}^{j,k}_\eta = -\frac{i q_m \sin \theta_{\mathbf{q}_m}\left(\cos\phi_{\mathbf{q}_m}+\sin\phi_{\mathbf{q}_m}\right)}{4\sqrt{2}\alpha_{\eta}}+i\alpha_\eta h_\eta
	\end{equation}
	In the calculation of the phonon decoherence rate, the eigenvalues and eigenstates of electron Hamiltonian, $H_\text{e}+H_\text{hyp}$, are obtained based on the fitted Hubbard parameters, e.g. $U_{\psi j}$, $U_{\psi j,\psi k }$, $t_{\psi j,\psi k}$ etc.~, which are extracted from full CI calculation. As a result, the real eigenstates are written as a linear combination of F-D states. For simplicity, we use raw orthonormal F-D states as the electron wavefunctions for the calculation of phonon decoherence to capture the main picture. A more rigorous method would be taking the electron wavefunctions of the eigenstates from full CI results, which is out of the scope of this work and is left for future studies.
	
	\newpage
	
	\subsection{Calculation of $J^+_{jj}(\omega)$ for two-phonon process}\label{subsec:JjjTwoPhonon}
	An arbitrary $\langle \delta B_j(0)\delta B_j(\tau)\rangle$ for two-phonon process can be expanded in the form of
	
	\begin{equation}
		\begin{split}
			&\frac{1}{2\hbar^2}\int_{-\infty}^{\infty} d\tau \left(e^{i \omega_s \tau}+e^{-i \omega_s \tau}\right)\langle\delta B_j(0)\delta B_j(\tau)\rangle_\beta \\
			&=\frac{1}{2\hbar^2} \int_{-\infty}^{\infty}d\tau \left(e^{i \omega_s \tau}+e^{-i \omega_s \tau}\right) \sum\limits_{\substack{\psi1,\psi2,\psi3,\psi4, \\ \psi5,\psi6,\psi7,\psi8}}  \gamma^{\substack{\psi1,\psi2,\psi3,\psi4}}_j \gamma^{\substack{\psi5,\psi6, \psi7,\psi8}}_j \\
			&\qquad\Big\langle \langle \Phi_{\psi1} | H_\text{e-ph}(0) | \Phi_{\psi2}\rangle \langle \Phi_{\psi3} | H_\text{e-ph}(0) | \Phi_{\psi4}\rangle \langle \Phi_{\psi5} | H_\text{e-ph}(\tau) | \Phi_{\psi6}\rangle
			\Phi_{\psi7} | H_\text{e-ph}(\tau) | \Phi_{\psi8} \rangle\Big\rangle_\beta\\
			&=\sum\limits_{\substack{\psi1,\psi2,\psi3,\psi4, \\ \psi5,\psi6,\psi7,\psi8}} \gamma^\psi \mathcal{I}^\psi
		\end{split}
	\end{equation}
	where $\gamma^\psi = \gamma^{\psi1,\psi2,\psi3,\psi4}_j \gamma^{\psi5,\psi6,\psi7,\psi8}_j$ and $\gamma^{\psi1,\psi2,\psi3,\psi4}_j$, $\gamma^{\psi5,\psi6,\psi7,\psi8}_j$ are the coefficients whose explicit values depend on the electronic Hamiltonian, $H_\text{e}+H_\text{hyp}$, while $\psi$ denotes the orbitals involved, e.g. $\psi = (L1,L1,R1,R2,R3,L1,R2,R3)$ indicates $\psi 1 = L1$, $\psi 2=L1$, $\psi 3=R1$, $\psi 4=R2$, $\psi 5=R3$, $\psi 6=L1$, $\psi 7=R2$ and $\psi 8=R3$. The larger outer angle bracket, $\langle \cdots \rangle_\beta$, denotes the thermal average, while the inner angle brackets, $\langle \Phi_{\psi j} | H_\text{e-ph} | \Phi_{\psi k}\rangle$, indicate the inner product of electron wavefunctions with electron density operator, $e^{i \mathbf{q}\cdot \mathbf{r}}$. From there, we have
	\begin{equation}\label{eq:allInt}
		\begin{split}
			&(2\hbar^2 )\mathcal{I}^\psi\\
			&=\int_{-\infty}^{\infty}d\tau\left(e^{i \omega_s \tau}+e^{-i \omega_s \tau}\right)\Big\langle \langle \Phi_{\psi1} | H_\text{e-ph}(0) | \Phi_{\psi2}\rangle \langle \Phi_{\psi3} | H_\text{e-ph}(0) | \Phi_{\psi4}\rangle \langle \Phi_{\psi5} | H_\text{e-ph}(\tau) | \Phi_{\psi6}\rangle \langle \Phi_{\psi7} | H_\text{e-ph}(\tau) | \Phi_{\psi8} \rangle\Big\rangle_\beta\\
			&=\sum_{s,\text{cpl}}\int_{-\infty}^{\infty}d\tau\left(e^{i \omega_s \tau}+e^{-i \omega_s \tau}\right)\sum_{\mathbf{q}_1,\mathbf{q}_2,\mathbf{q}_3,\mathbf{q}_4} C_{s,\text{cpl}} (\mathbf{q}_1)C_{s,\text{cpl}} (\mathbf{q}_2)C_{s,\text{cpl}} (\mathbf{q}_3)C_{s,\text{cpl}} (\mathbf{q}_4) \\
			&\quad\Big\langle\langle\Phi_{\psi1} (\mathbf{r})| \left[ a_s(\mathbf{q}_1)\mp_s a_s^\dagger(-\mathbf{q}_1)\right]e^{i \mathbf{q}_1 \cdot \mathbf{r}}|\Phi_{\psi2} (\mathbf{r})\rangle
			\langle \Phi_{\psi3} (\mathbf{r})| \left[a_s(\mathbf{q}_2)\mp_s a_s^\dagger(-\mathbf{q}_2)\right]e^{i \mathbf{q}_2 \cdot \mathbf{r}}|\Phi_{\psi4} (\mathbf{r})\rangle\\
			&\quad\langle \Phi_{\psi5} (\mathbf{r})| \left[a_s(\mathbf{q}_3)e^{-i \omega_s(\mathbf{q}_3)\tau}\mp_s a_s^\dagger(-\mathbf{q}_3)e^{i \omega_s(\mathbf{q}_3)\tau}\right]e^{i \mathbf{q}_3 \cdot \mathbf{r}}|\Phi_{\psi6} (\mathbf{r})\rangle\\
			&\quad\langle \Phi_{\psi7} (\mathbf{r})| \left[a_s(\mathbf{q}_4)e^{-i \omega_s(\mathbf{q}_4)\tau}\mp_s a_s^\dagger(-\mathbf{q}_4)e^{i \omega_s(\mathbf{q}_4)\tau}\right]e^{i \mathbf{q}_4 \cdot \mathbf{r}}|\Phi_{\psi8} (\mathbf{r})\rangle\Big\rangle_\beta	\\
			&=\sum_{s,\text{cpl}}\sum_{\mathbf{q}_1,\mathbf{q}_2,\mathbf{q}_3,\mathbf{q}_4} C_{s,\text{cpl}} (\mathbf{q}_1)C_{s,\text{cpl}} (\mathbf{q}_2)C_{s,\text{cpl}} (\mathbf{q}_3)C_{s,\text{cpl}} (\mathbf{q}_4) \int_{-\infty}^{\infty}d\tau\left(e^{i \omega_s \tau}+e^{-i \omega_s \tau}\right)\\
			&\quad\langle\Phi_{\psi1} (\mathbf{r})| e^{i \mathbf{q}_1 \cdot \mathbf{r}}|\Phi_{\psi2} (\mathbf{r})\rangle
			\langle \Phi_{\psi3} (\mathbf{r})| e^{i \mathbf{q}_2 \cdot \mathbf{r}}|\Phi_{\psi4} (\mathbf{r})\rangle
			\langle \Phi_{\psi5} (\mathbf{r})| e^{i \mathbf{q}_3 \cdot \mathbf{r}}|\Phi_{\psi6} (\mathbf{r})\rangle
			\langle \Phi_{\psi7} (\mathbf{r})| e^{i \mathbf{q}_4 \cdot \mathbf{r}}|\Phi_{\psi8} (\mathbf{r})\rangle \\
			&\quad\Big\langle\left[ a_s(\mathbf{q}_1)\mp_s a_s^\dagger(-\mathbf{q}_1)\right]\left[a_s(\mathbf{q}_2)\mp_s a_s^\dagger(-\mathbf{q}_2)\right]\left[a_s(\mathbf{q}_3)e^{-i \omega_s(\mathbf{q}_3)\tau}\mp_s a_s^\dagger(-\mathbf{q}_3)e^{i \omega_s(\mathbf{q}_3)\tau}\right]\\
			&\qquad\times\left[a_s(\mathbf{q}_4)e^{-i \omega_s(\mathbf{q}_4)\tau}\mp_s a_s^\dagger(-\mathbf{q}_4)e^{i 	\omega_s(\mathbf{q}_4)\tau}\right]\Big\rangle_\beta\\
			&=\sum_{s,\text{cpl}}\sum_{\mathbf{q}_1,\mathbf{q}_2,\mathbf{q}_3,\mathbf{q}_4} C_{s,\text{cpl}} (\mathbf{q}_1)C_{s,\text{cpl}} (\mathbf{q}_2)C_{s,\text{cpl}} (\mathbf{q}_3)C_{s,\text{cpl}} (\mathbf{q}_4) \int_{-\infty}^{\infty}d\tau\left(e^{i \omega_s \tau}+e^{-i \omega_s \tau}\right)\mathcal{N}(\mathbf{q}_1,\mathbf{q}_2,\mathbf{q}_3,\mathbf{q}_4)\\
			&\quad\Big\langle\left[ a_s(\mathbf{q}_1)\mp_s a_s^\dagger(-\mathbf{q}_1)\right]\left[a_s(\mathbf{q}_2)\mp_s a_s^\dagger(-\mathbf{q}_2)\right]\left[a_s(\mathbf{q}_3)e^{-i \omega_s(\mathbf{q}_3)\tau}\mp_s a_s^\dagger(-\mathbf{q}_3)e^{i \omega_s(\mathbf{q}_3)\tau}\right]\\
			&\qquad \times\left[a_s(\mathbf{q}_4)e^{-i \omega_s(\mathbf{q}_4)\tau}\mp_s a_s^\dagger(-\mathbf{q}_4)e^{i \omega_s(\mathbf{q}_4)\tau}\right]\Big\rangle_\beta\\
			&=(2\hbar^2)\sum_{s,\text{cpl}} \mathcal{I}_{s,\text{cpl}}^\psi(q_3,\phi_{\mathbf{q}_3},\phi_{\mathbf{q}_4},\theta_{\mathbf{q}_3},\theta_{\mathbf{q}_4})\\
		\end{split}
	\end{equation}
	where we have dropped the cross-terms between different phonon modes and coupling mechanisms. The former is valid as different phonon modes are decoupled while the latter is justified as it is found that the contribution of deformation potential dominates the decoherence effect in GaAs, by orders of magnitude compared to piezoelectric coupling. In Eq.~\eqref{eq:allInt}, we denote the energy of electron-phonon interaction of each mode, $s$, and coupling mechanism, cpl, whose unit is $(\text{electron volt}^2 \times \text{second})$, as 
	\begin{equation}
		\begin{split}
			&(2\hbar^2)\mathcal{I}^\psi_{s,\text{cpl}}(q_3,\phi_{\mathbf{q}_3},\phi_{\mathbf{q}_4},\theta_{\mathbf{q}_3},\theta_{\mathbf{q}_4})\\
			&=\int_{-\infty}^{\infty}d\tau\left(e^{i \omega_s \tau}+e^{-i \omega_s \tau}\right)\sum_{\mathbf{q}_1,\mathbf{q}_2,\mathbf{q}_3,\mathbf{q}_4} C_{s,\text{cpl}}  (\mathbf{q}_1)C_{s,\text{cpl}} (\mathbf{q}_2)C_{s,\text{cpl}} (\mathbf{q}_3)C_{s,\text{cpl}} (\mathbf{q}_4) \mathcal{N}(\mathbf{q}_1,\mathbf{q}_2,\mathbf{q}_3,\mathbf{q}_4) \\
			&\quad\times\Big\langle\langle\Phi_{\psi1} (\mathbf{r})| \left[ a_s(\mathbf{q}_1)\mp_s a_s^\dagger(-\mathbf{q}_1)\right]e^{i \mathbf{q}_1 \cdot \mathbf{r}}|\Phi_{\psi2} (\mathbf{r})\rangle
			\langle \Phi_{\psi3} (\mathbf{r})| \left[a_s(\mathbf{q}_2)\mp_s a_s^\dagger(-\mathbf{q}_2)\right]e^{i \mathbf{q}_2 \cdot \mathbf{r}}|\Phi_{\psi4} (\mathbf{r})\rangle \\
			&\qquad\times\left[a_s(\mathbf{q}_4)e^{-i \omega_s(\mathbf{q}_4)\tau}\mp_s a_s^\dagger(-\mathbf{q}_4)e^{i \omega_s(\mathbf{q}_4)\tau}\right]\Big\rangle_\beta
		\end{split}
	\end{equation}
	and the integral of electron wavefunctions as 
	\begin{equation}
		\mathcal{N}(\mathbf{q}_1,\mathbf{q}_2,\mathbf{q}_3,\mathbf{q}_4)=\langle\Phi_{\psi1} (\mathbf{r})| e^{i \mathbf{q}_1 \cdot \mathbf{r}}|\Phi_{\psi2} (\mathbf{r})\rangle
		\langle \Phi_{\psi3} (\mathbf{r})| e^{i \mathbf{q}_2 \cdot \mathbf{r}}|\Phi_{\psi4} (\mathbf{r})\rangle
		\langle \Phi_{\psi5} (\mathbf{r})| e^{i \mathbf{q}_3 \cdot \mathbf{r}}|\Phi_{\psi6} (\mathbf{r})\rangle
		\langle \Phi_{\psi7} (\mathbf{r})| e^{i \mathbf{q}_4 \cdot \mathbf{r}}|\Phi_{\psi8} (\mathbf{r})\rangle.
	\end{equation}
	Note that the number of independent variables of the whole integrand, $\mathcal{I}_{s,\text{cpl}}$ is less than those involved in the rest of the expression in Eq.~\eqref{eq:allInt}. This is the result of algebraic manipulation as shown in the following discussion. 
	
	Using Wick's theorem \cite{Castin.04}, the thermal average term in Eq.~\eqref{eq:allInt} $\left(\int d\tau \mathcal{N}  [\exp (i \omega \tau)+\exp (-i \omega \tau)]\langle \cdots \rangle_\beta\right)$, and the identity
	\begin{equation}
		\int_{\infty}^{\infty} e^{i (q-q_0) \nu_s \tau} d\tau= \frac{2\pi}{\nu_s} \delta(q-q_0),
	\end{equation}
	we have
	\begin{subequations}\label{eq:thermalAverage4phonons}
		\begin{align}
			\begin{split}\label{eq:Itilde}
				&(2\hbar^2)\mathcal{\widetilde{I}}^\psi_{s,\text{cpl}}(q_3,\phi_{\mathbf{q}_3},\phi_{\mathbf{q}_4},\theta_{\mathbf{q}_3},\theta_{\mathbf{q}_4})
			\end{split}\\
			\begin{split}\label{eq:IdivC}
				&=\frac{\mathcal{I}^\psi_{s,\text{cpl}}(q_3,\phi_{\mathbf{q}_3},\phi_{\mathbf{q}_4},\theta_{\mathbf{q}_3},\theta_{\mathbf{q}_4})}{C_{s,\text{cpl}} (\mathbf{q}_1)C_{s,\text{cpl}} (\mathbf{q}_2)C_{s,\text{cpl}} (\mathbf{q}_3)C_{s,\text{cpl}} (\mathbf{q}_4)}
			\end{split}\\
			\begin{split}
				&=\int_{-\infty}^{\infty}d\tau\left(e^{i \omega_s \tau}+e^{-i \omega_s \tau}\right)
				\sum_{\mathbf{q}_1,\mathbf{q}_2,\mathbf{q}_3,\mathbf{q}_4}
				\mathcal{N}(\mathbf{q}_1,\mathbf{q}_2,\mathbf{q}_3,\mathbf{q}_4)
				\Bigg\{\langle a_s^\dagger (-\mathbf{q}_1) a_s(\mathbf{q}_3)\rangle_\beta \langle a_s^\dagger (-\mathbf{q}_2) a_s(\mathbf{q}_4)\rangle_\beta e^{-i (q_3+q_4)\nu_s \tau} \\
				&+\langle a_s(\mathbf{q}_1) a_s^\dagger(-\mathbf{q}_3)\rangle_\beta \langle a_s^\dagger (-\mathbf{q}_2) a_s(\mathbf{q}_4)\rangle_\beta e^{i (q_3-q_4)\nu_s \tau} \\
				&+\langle a_s^\dagger (-\mathbf{q}_1) a_s(\mathbf{q}_3)\rangle_\beta \langle a_s(\mathbf{q}_2) a_s^\dagger(-\mathbf{q}_4)\rangle_\beta e^{i (-q_3+q_4)\nu_s \tau} +\langle a_s(\mathbf{q}_1) a_s^\dagger(-\mathbf{q}_3)\rangle_\beta \langle a_s(\mathbf{q}_2) a_s^\dagger(-\mathbf{q}_4)\rangle_\beta e^{i (q_3+q_4)\nu_s \tau} \\
				&+\langle a_s^\dagger (-\mathbf{q}_1) a_s(\mathbf{q}_4)\rangle_\beta \langle a_s^\dagger (-\mathbf{q}_2) a_s(\mathbf{q}_3)\rangle_\beta e^{-i (q_3+q_4)\nu_s \tau}  +\langle a_s^\dagger (-\mathbf{q}_1) a_s(\mathbf{q}_4)\rangle_\beta \langle a_s(\mathbf{q}_2) a_s^\dagger(-\mathbf{q}_3)\rangle_\beta e^{i (q_3-q_4)\nu_s \tau} \\
				&+\langle a_s(\mathbf{q}_1) a_s^\dagger(-\mathbf{q}_4)\rangle_\beta \langle a_s^\dagger (-\mathbf{q}_2) a_s(\mathbf{q}_3)\rangle_\beta e^{i (-q_3+q_4)\nu_s \tau} +\langle a_s(\mathbf{q}_1) a_s^\dagger(-\mathbf{q}_4)\rangle_\beta \langle a_s(\mathbf{q}_2) a_s^\dagger (-\mathbf{q}_3)\rangle_\beta e^{i (q_3+q_4)\nu_s \tau} \Bigg\}
			\end{split}\\
			\begin{split}\label{eq:q4toq2}
				&=\int_{-\infty}^{\infty}d\tau\left(e^{i \omega_s \tau}+e^{-i \omega_s \tau}\right) \sum_{\mathbf{q}_3,\mathbf{q}_4} 
				\Bigg\{ \mathcal{N}(\mathbf{q}_1,\mathbf{q}_2,\mathbf{q}_3,\mathbf{q}_4)\Big|_{\substack{\mathbf{q}_1=-\mathbf{q}_3,\\ \mathbf{q}_2=-\mathbf{q}_4}}\left[\langle a_s^\dagger (\mathbf{q}_3) a_s(\mathbf{q}_3)\rangle_\beta \langle a_s^\dagger (\mathbf{q}_4) a_s(\mathbf{q}_4)\rangle_\beta e^{-i (q_3+q_4)\nu_s \tau}\right. \\
				&+\langle a_s(-\mathbf{q}_3) a_s^\dagger(-\mathbf{q}_3)\rangle_\beta \langle a_s^\dagger (\mathbf{q}_4) a_s(\mathbf{q}_4)\rangle_\beta e^{i (q_3-q_4)\nu_s \tau} +\langle a_s^\dagger (\mathbf{q}_3) a_s(\mathbf{q}_3)\rangle_\beta \langle a_s(-\mathbf{q}_4) a_s^\dagger(-\mathbf{q}_4)\rangle_\beta e^{i (-q_3+q_4)\nu_s \tau} \\
				&+\left.\langle a_s(-\mathbf{q}_3) a_s^\dagger(-\mathbf{q}_3)\rangle_\beta \langle a_s(-\mathbf{q}_4) a_s^\dagger(-\mathbf{q}_4)\rangle_\beta e^{i (q_3+q_4)\nu_s \tau}\right] \\
				&+\mathcal{N}(\mathbf{q}_1,\mathbf{q}_2,\mathbf{q}_3,\mathbf{q}_4)\Big|_{\substack{\mathbf{q}_1=-\mathbf{q}_4,\\ \mathbf{q}_2=-\mathbf{q}_3}}\left[\langle a_s^\dagger (\mathbf{q}_4) a_s(\mathbf{q}_4)\rangle_\beta \langle a_s^\dagger (\mathbf{q}_3) a_s(\mathbf{q}_3)\rangle_\beta e^{-i (q_3+q_4)\nu_s \tau}  \right. \\
				&+\langle a_s^\dagger (\mathbf{q}_4) a_s(\mathbf{q}_4)\rangle_\beta \langle a_s(-\mathbf{q}_3) a_s^\dagger(-\mathbf{q}_3)\rangle_\beta e^{i (q_3-q_4)\nu_s \tau} +\langle a_s(-\mathbf{q}_4) a_s^\dagger(-\mathbf{q}_4)\rangle_\beta \langle a_s^\dagger (\mathbf{q}_3) a_s(\mathbf{q}_3)\rangle_\beta e^{i (-q_3+q_4)\nu_s \tau} +\\
				&\left.\langle a_s(-\mathbf{q}_4) a_s^\dagger(-\mathbf{q}_4)\rangle_\beta \langle a_s(-\mathbf{q}_3) a_s^\dagger (-\mathbf{q}_3)\rangle_\beta e^{i (q_3+q_4)\nu_s \tau}\right]\Bigg\}
			\end{split}\\
			\begin{split}\label{eq:nB}
				&=\sum_{\mathbf{q}_3,\mathbf{q}_4}\int_{-\infty}^{\infty}d\tau\left(e^{i \omega_s \tau}+e^{-i \omega_s \tau}\right)\Bigg\{\mathcal{N}(\mathbf{q}_1,\mathbf{q}_2,\mathbf{q}_3,\mathbf{q}_4)\Big|_{\substack{\mathbf{q}_1=-\mathbf{q}_3,\\\mathbf{q}_2=-\mathbf{q}_4}}\left[n_s (q_3) n_s (q_4)  e^{-i (q_3+q_4)\nu_s \tau} \right. \\
				&+\left[n_s (q_3)+1\right] n_s (q_4) e^{i (q_3-q_4)\nu_s \tau} +\left. n_s (q_3) \left[n_s (q_4)+1\right] e^{i (-q_3+q_4)\nu_s \tau} +\left[n_s (q_3)+1\right] \left[n_s (q_4)+1\right] e^{i (q_3+q_4)\nu_s \tau}\right] \\
				&+\mathcal{N}(\mathbf{q}_1,\mathbf{q}_2,\mathbf{q}_3,\mathbf{q}_4)\Big|_{\substack{\mathbf{q}_1=-\mathbf{q}_4,\\\mathbf{q}_2=-\mathbf{q}_3}}\left[n_s (q_4) n_s (q_3) e^{-i (q_3+q_4)\nu_s \tau}  +\left[n_s (q_4)+1\right] n_s (q_3) e^{i (q_3-q_4)\nu_s \tau}\right. \\
				&+\left. n_s (q_4) \left[n_s (q_3)+1\right] e^{i (-q_3+q_4)\nu_s \tau} +\left[n_s (q_4)+1\right] \left[n_s (q_3)+1\right] e^{i (q_3+q_4)\nu_s \tau}\right]\Bigg\}
			\end{split}\\
			\begin{split}
				&=\sum_{\mathbf{q}_3,\mathbf{q}_4}\int_{-\infty}^{\infty}d\tau\Bigg\{\mathcal{N}(\mathbf{q}_1,\mathbf{q}_2,\mathbf{q}_3,\mathbf{q}_4)\Big|_{\substack{\mathbf{q}_1=-\mathbf{q}_3,\\ \mathbf{q}_2=-\mathbf{q}_4}}\left[n_s (q_3) n_s (q_4)  \left(e^{-i (q_3+q_4+q_s)\nu_s \tau}+e^{-i (q_3+q_4-q_s)\nu_s \tau}\right) \right.\\
				&\qquad\quad+\left[n_s (q_3)+1\right] n_s (q_4) \left(e^{i (q_3-q_4+q_s)\nu_s \tau}+e^{i (q_3-q_4-q_s)\nu_s \tau}\right) \\
				&\qquad\quad+ n_s (q_3) \left[n_s (q_4)+1\right] \left(e^{i (-q_3+q_4+q_s)\nu_s \tau} + e^{i (-q_3+q_4-q_s)\nu_s \tau}  \right)\\
				&\qquad\quad+\left.\left[n_s (q_3)+1\right] \left[n_s (q_4)+1\right] \left(e^{i (q_3+q_4+q_s)\nu_s \tau}+e^{i (q_3+q_4-q_s)\nu_s \tau}\right)\right] \\
				&\qquad\quad+\left[\mathcal{N}(\mathbf{q}_1,\mathbf{q}_2,\mathbf{q}_3,\mathbf{q}_4)\Big|_{\substack{\mathbf{q}_1=-\mathbf{q}_4,\\ \mathbf{q}_2=-\mathbf{q}_3}}n_s (q_4) n_s (q_3) \left(e^{-i (q_3+q_4+q_s)\nu_s \tau}+e^{-i (q_3+q_4-q_s)\nu_s \tau} \right) \right.\\
				&\qquad\quad+\left[n_s (q_4)+1\right] n_s (q_3) \left(e^{i (q_3-q_4+q_s)\nu_s \tau}+e^{i (q_3-q_4-q_s)\nu_s \tau}\right) \\
				&\qquad\quad+ n_s (q_4) \left[n_s (q_3)+1\right] \left(e^{i (-q_3+q_4+q_s)\nu_s \tau}+e^{i (-q_3+q_4-q_s)\nu_s \tau} \right) \\
				&\qquad\quad+\left.\left[n_s (q_4)+1\right] \left[n_s (q_3)+1\right] \left(e^{i (q_3+q_4+q_s)\nu_s \tau}+e^{i (q_3+q_4-q_s)\nu_s \tau}\right)\right]_{\substack{\mathbf{q}_1=-\mathbf{q}_4,\\ \mathbf{q}_2=-\mathbf{q}_3}}\Bigg\}
			\end{split}\\
			\begin{split}
				&=\frac{2\pi}{\nu_s}\int_{0}^{\pi} \sin \theta_{\mathbf{q}_3} d\theta_{\mathbf{q}_3} \int_{0}^{2\pi} d\phi_{\mathbf{q}_3} \int_{0}^{\pi} \sin \theta_{\mathbf{q}_4} d\theta_{\mathbf{q}_4}\int_{0}^{2\pi} d \phi_{\mathbf{q}_4} \int_{0}^{\infty} q_4^2 dq_4 \int_{0}^{\infty} q_3^2 dq_3\\
				&\qquad\quad\Bigg\{\mathcal{N}(\mathbf{q}_1,\mathbf{q}_2,\mathbf{q}_3,\mathbf{q}_4)\Big|_{\substack{\mathbf{q}_1=-\mathbf{q}_3,\\ \mathbf{q}_2=-\mathbf{q}_4}}\left[n_s (q_3) n_s (q_4)  \left(\delta(q_3+q_4+q_s)+\delta (q_3+q_4-q_s)\right) \right.\\
				&\qquad\quad+\left[n_s (q_3)+1\right] n_s (q_4) \left(\delta (q_3-q_4+q_s)+\delta(q_3-q_4-q_s)\right) \\
				&\qquad\quad+ n_s (q_3) \left[n_s (q_4)+1\right] \left(\delta(-q_3+q_4+q_s) + \delta(-q_3+q_4-q_s) \right)\\
				&\qquad\quad+\left.\left[n_s (q_3)+1\right] \left[n_s (q_4)+1\right] \left(\delta(q_3+q_4+q_s)+\delta(q_3+q_4-q_s)\right)\right]\\
				&\qquad\quad+\mathcal{N}(\mathbf{q}_1,\mathbf{q}_2,\mathbf{q}_3,\mathbf{q}_4)\Big|_{\substack{\mathbf{q}_1=-\mathbf{q}_4,\\ \mathbf{q}_2=-\mathbf{q}_3}}\left[n_s (q_4) n_s (q_3) \left(\delta(q_3+q_4+q_s)+\delta (q_3+q_4-q_s) \right) \right.\\
				&\qquad\quad+\left[n_s (q_4)+1\right] n_s (q_3) \left(\delta (q_3-q_4+q_s)+\delta(q_3-q_4-q_s)\right) \\
				&\qquad\quad+ n_s (q_4) \left[n_s (q_3)+1\right] \left(\delta(-q_3+q_4+q_s)+\delta(-q_3+q_4-q_s)\right) \\
				&\qquad\quad+\left.\left[n_s (q_4)+1\right] \left[n_s (q_3)+1\right] \left(\delta(q_3+q_4+q_s)+\delta(q_3+q_4-q_s)\right)\right]\Bigg\}
			\end{split}\\
			\begin{split}
				&=\frac{2\pi}{\nu_s}\int_{0}^{\pi} \sin \theta_{\mathbf{q}_3} d\theta_{\mathbf{q}_3} \int_{0}^{2\pi} d\phi_{\mathbf{q}_3} \int_{0}^{\pi} \sin \theta_{\mathbf{q}_4} d\theta_{\mathbf{q}_4}\int_{0}^{2\pi} d \phi_{\mathbf{q}_4} \\
				&\Bigg\{\int_{0}^{q_s}dq_3q_3^2q_4^2\left. \mathcal{N}(\mathbf{q}_1,\mathbf{q}_2,\mathbf{q}_3,\mathbf{q}_4) \right|_{\substack{\mathbf{q}_1=-\mathbf{q}_3,\\ \mathbf{q}_2=-\mathbf{q}_4,\\q_4=q_s-q_3}}n_s (q_3) n_s (q_s-q_3)   \\
				&+\int_{0}^{\infty}dq_3q_3^2q_4^2\left. \mathcal{N}(\mathbf{q}_1,\mathbf{q}_2,\mathbf{q}_3,\mathbf{q}_4) \right|_{\substack{\mathbf{q}_1=-\mathbf{q}_3,\\ \mathbf{q}_2=-\mathbf{q}_4,\\q_4=q_3+q_s}}\left[n_s (q_3)+1\right] n_s (q_3+q_s)\\
				&+\int_{q_s}^{\infty}dq_3q_3^2q_4^2\left. \mathcal{N}(\mathbf{q}_1,\mathbf{q}_2,\mathbf{q}_3,\mathbf{q}_4) \right|_{\substack{\mathbf{q}_1=-\mathbf{q}_3,\\ \mathbf{q}_2=-\mathbf{q}_4,\\q_4=q_3-q_s}}\left[n_s (q_3)+1\right] n_s (q_3-q_s)  \\
				&+\int_{q_s}^{\infty}dq_3q_3^2q_4^2\left. \mathcal{N}(\mathbf{q}_1,\mathbf{q}_2,\mathbf{q}_3,\mathbf{q}_4) \right|_{\substack{\mathbf{q}_1=-\mathbf{q}_3,\\ \mathbf{q}_2=-\mathbf{q}_4,\\q_4=q_3-q_s}} n_s (q_3) \left[n_s (q_3-q_s)+1\right]\\
				&+\int_{0}^{\infty}dq_3q_3^2q_4^2\left. \mathcal{N}(\mathbf{q}_1,\mathbf{q}_2,\mathbf{q}_3,\mathbf{q}_4) \right|_{\substack{\mathbf{q}_1=-\mathbf{q}_3,\\ \mathbf{q}_2=-\mathbf{q}_4,\\q_4=q_3+q_s}} n_s (q_3) \left[n_s (q_3+q_s)+1\right] \\
				&+\int_{0}^{q_s}dq_3q_3^2q_4^2\left. \mathcal{N}(\mathbf{q}_1,\mathbf{q}_2,\mathbf{q}_3,\mathbf{q}_4) \right|_{\substack{\mathbf{q}_1=-\mathbf{q}_3,\\ \mathbf{q}_2=-\mathbf{q}_4,\\q_4=q_s-q_3}}\left[n_s (q_3)+1\right] \left[n_s (q_s-q_3)+1\right] \\
				&+\int_0^{q_s}dq_3q_3^2q_4^2\left. \mathcal{N}(\mathbf{q}_1,\mathbf{q}_2,\mathbf{q}_3,\mathbf{q}_4) \right|_{\substack{\mathbf{q}_1=-\mathbf{q}_4,\\ \mathbf{q}_2=-\mathbf{q}_3,\\q_4=q_s-q_3}}n_s (q_s-q_3) n_s (q_3) \\
				&+\int_{0}^{\infty}dq_3q_3^2q_4^2\left. \mathcal{N}(\mathbf{q}_1,\mathbf{q}_2,\mathbf{q}_3,\mathbf{q}_4) \right|_{\substack{\mathbf{q}_1=-\mathbf{q}_4,\\ \mathbf{q}_2=-\mathbf{q}_3,\\q_4=q_3+q_s}}\left[n_s (q_3+q_s)+1\right] n_s (q_3)\\
				&+\int_{q_s}^{\infty}dq_3q_3^2q_4^2\left. \mathcal{N}(\mathbf{q}_1,\mathbf{q}_2,\mathbf{q}_3,\mathbf{q}_4) \right|_{\substack{\mathbf{q}_1=-\mathbf{q}_4,\\ \mathbf{q}_2=-\mathbf{q}_3,\\q_4=q_3-q_s}}\left[n_s (q_3-q_s)+1\right] n_s (q_3)  \\
				&+ \int_{q_s}^{\infty}dq_3q_3^2q_4^2\left. \mathcal{N}(\mathbf{q}_1,\mathbf{q}_2,\mathbf{q}_3,\mathbf{q}_4) \right|_{\substack{\mathbf{q}_1=-\mathbf{q}_4,\\ \mathbf{q}_2=-\mathbf{q}_3,\\q_4=q_3-q_s}}n_s (q_3-q_s) \left[n_s (q_3)+1\right]\\
				&+ \int_{0}^{\infty}dq_3q_3^2q_4^2\left. \mathcal{N}(\mathbf{q}_1,\mathbf{q}_2,\mathbf{q}_3,\mathbf{q}_4) \right|_{\substack{\mathbf{q}_1=-\mathbf{q}_4,\\ \mathbf{q}_2=-\mathbf{q}_3,\\q_4=q_3+q_s}}n_s (q_3+q_s) \left[n_s (q_3)+1\right]  \\
				&+\int_{0}^{q_s}dq_3q_3^2q_4^2\left. \mathcal{N}(\mathbf{q}_1,\mathbf{q}_2,\mathbf{q}_3,\mathbf{q}_4) \right|_{\substack{\mathbf{q}_1=-\mathbf{q}_4,\\ \mathbf{q}_2=-\mathbf{q}_3,\\q_4=q_s-q_3}}\left[n_s (q_s-q_3)+1\right] \left[n_s (q_3)+1\right] \Bigg\}
			\end{split}\\
			\begin{split}\label{eq:final}
				&=\frac{2\pi}{\nu_s}\int_{0}^{\pi} \sin \theta_{\mathbf{q}_3} d\theta_{\mathbf{q}_3} \int_{0}^{2\pi} d\phi_{\mathbf{q}_3} \int_{0}^{\pi} \sin \theta_{\mathbf{q}_4} d\theta_{\mathbf{q}_4}\int_{0}^{2\pi} d \phi_{\mathbf{q}_4} \\
				&\Bigg\{\int_{0}^{q_s}dq_3
				q_3^2\left(\left.q_4^2\mathcal{N}(\mathbf{q}_1,\mathbf{q}_2,\mathbf{q}_3,\mathbf{q}_4)\right|_{\substack{\mathbf{q}_1=-\mathbf{q}_3,\\ \mathbf{q}_2=-\mathbf{q}_4,\\q_4=q_s-q_3}}+\left.q_4^2\mathcal{N}(\mathbf{q}_1,\mathbf{q}_2,\mathbf{q}_3,\mathbf{q}_4)\right|_{\substack{\mathbf{q}_1=-\mathbf{q}_4,\\ \mathbf{q}_2=-\mathbf{q}_3,\\q_4=q_s-q_3}}\right)\\
				&\qquad\times\left(n_s (q_3) n_s (q_s-q_3)   +\left[n_s (q_3)+1\right] \left[n_s (q_s-q_3)+1\right]\right)\\
				&+\int_{0}^{\infty}dq_3 q_3^2\left(\left.q_4^2\mathcal{N}(\mathbf{q}_1,\mathbf{q}_2,\mathbf{q}_3,\mathbf{q}_4)\right|_{\substack{\mathbf{q}_1=-\mathbf{q}_3,\\ \mathbf{q}_2=-\mathbf{q}_4,\\q_4=q_3+q_s}}+\left.q_4^2\mathcal{N}(\mathbf{q}_1,\mathbf{q}_2,\mathbf{q}_3,\mathbf{q}_4)\right|_{\substack{\mathbf{q}_1=-\mathbf{q}_4,\\ \mathbf{q}_2=-\mathbf{q}_3,\\q_4=q_3+q_s}}\right)\\
				&\qquad\times\left(\left[n_s (q_3)+1\right] n_s (q_3+q_s)+n_s (q_3) \left[n_s (q_3+q_s)+1\right]\right)\\
				&+\int_{q_s}^{\infty}dq_3q_3^2\left(\left.q_4^2\mathcal{N}(\mathbf{q}_1,\mathbf{q}_2,\mathbf{q}_3,\mathbf{q}_4)\right|_{\substack{\mathbf{q}_1=-\mathbf{q}_3,\\ \mathbf{q}_2=-\mathbf{q}_4,\\q_4=q_3-q_s}}+q_4^2\mathcal{N}(\mathbf{q}_1,\mathbf{q}_2,\mathbf{q}_3,\mathbf{q}_4)\Big|_{\substack{\mathbf{q}_1=-\mathbf{q}_4,\\ \mathbf{q}_2=-\mathbf{q}_3,\\q_4=q_3-q_s}}\right)\\
				&\qquad\times\left(\left[n_s (q_3)+1\right] n_s (q_3-q_s)+n_s (q_3) \left[n_s (q_3-q_s)+1\right]\right) \Bigg\}
			\end{split}
		\end{align}
	\end{subequations}
	where $\omega_s({\mathbf{q}_m}) = q_m  \nu_s= |\pm \mathbf{q}_m| \nu_s$ and the subscript at the straight line ($\mathcal{A}|_{\text{rule}}$) denotes the replacement ``rule'' to be applied to the variables in the expression $\mathcal{A}$. 
	Eq.~\eqref{eq:q4toq2} and \eqref{eq:nB} are obtained by recognizing that the term $\langle a_s^\dagger (-\mathbf{q}_m)a_s(\mathbf{q}_n)\rangle_\beta$ is non-zero only when $-\mathbf{q}_m = \mathbf{q}_n$, such that 
	\begin{equation}
		\langle a_s^\dagger (-\mathbf{q}_m)a_s(-\mathbf{q}_m)\rangle_\beta = n_s(q_m) = \frac{1}{e^{\hbar q_m \nu_s/(k_B T)}-1},
	\end{equation}
	where $n_s(q_m)$ is the Bose-Einstein distribution, $k_B$ is the Boltzmann constant, $T$ is the temperature. The subscripts $\mathbf{q}_m=-\mathbf{q}_{n\neq m}$ in Eq.~\eqref{eq:q4toq2} and \eqref{eq:nB} are retained as it will affect the pairing of electron orbitals in the inner product $\prod_{j=1}^{j=4}\langle \Phi_{\psi(2j-1)}(\mathbf{r})|\exp(i \mathbf{q}\cdot \mathbf{r})|\Phi_{\psi(2j)}\mathbf{r}\rangle$. 
	
	Taking into account of the algebraic manipulations in Eq.~\eqref{eq:thermalAverage4phonons}, the inner product part in Eq.~\eqref{eq:allInt} leads to
	\begin{subequations}
		\begin{align}\label{eq:electronWFIntegral}
			\begin{split}
				&\left.\mathcal{N}(\mathbf{q}_1,\mathbf{q}_2,\mathbf{q}_3,\mathbf{q}_4)\right|_{\substack{\mathbf{q}_1=-\mathbf{q}_3,\mathbf{q}_2=-\mathbf{q}_4,q_4=q_3+q_s}}
			\end{split}\\
			\begin{split}
				&=\langle\Phi_{\psi1} (\mathbf{r})| e^{i -\mathbf{q}_3 \cdot \mathbf{r}}|\Phi_{\psi2} (\mathbf{r})\rangle
				\langle \Phi_{\psi3} (\mathbf{r})| e^{i -\mathbf{q}_4 \cdot \mathbf{r}}|\Phi_{\psi4} (\mathbf{r})\rangle
				\langle \Phi_{\psi5} (\mathbf{r})| e^{i \mathbf{q}_3 \cdot \mathbf{r}}|\Phi_{\psi6} (\mathbf{r})\rangle
				\langle \Phi_{\psi7} (\mathbf{r})| e^{i \mathbf{q}_4 \cdot \mathbf{r}}|\Phi_{\psi8} (\mathbf{r})\rangle
				\Big|_{q_4=q_3+q_s}
			\end{split}\\
			\begin{split}
				&=
				\langle\Phi_{\psi1} (\mathbf{r})| e^{-i \mathbf{q}_3\cdot \mathbf{r}}|\Phi_{\psi2} (\mathbf{r})\rangle
				\langle \Phi_{\psi5} (\mathbf{r})| e^{i \mathbf{q}_3 \cdot \mathbf{r}}|\Phi_{\psi6} (\mathbf{r})\rangle
				\langle \Phi_{\psi3} (\mathbf{r})| e^{-i \mathbf{q}_4 \cdot \mathbf{r}}|\Phi_{\psi4} (\mathbf{r})\rangle
				\langle \Phi_{\psi7} (\mathbf{r})| e^{i \mathbf{q}_4 \cdot \mathbf{r}}|\Phi_{\psi8} (\mathbf{r})\rangle
				\Big|_{q_4=q_3+q_s}
			\end{split}\\
			\begin{split}\label{eq:octetInt2}
				&=\left(\prod_{j\in \{1,3,5,7\}} \frac{\pi \mathcal{G}_{\eta(\psi j)}^2}{\alpha_{\eta(\psi j)}}\right) \Bigg\{\mathcal{M}_{\eta(\psi 1)}^{\psi 1,\psi 2}\left(-q_x^{(3)},-q_y^{(3)}\right)  \mathcal{M}_{\eta(\psi 5)}^{\psi 5,\psi 6}\left(q_x^{(3)},q_y^{(3)}\right) \frac{1}{\left[\left(\frac{1}{a_z}\right)^2+\left( q_3 \cos \theta_{\mathbf{q}_3}\right)^2\right]^3} \\
				&\qquad\times\exp\left[- q_3^2 \sin^2 \theta_{\mathbf{q}_3}\left(\frac{1}{8 \alpha_{\eta(\psi1)}}+\frac{1}{8 \alpha_{\eta(\psi5)}}\right)\right] \exp\left[\frac{i \left(h_{\eta(\psi 1)}-h_{\eta(\psi 5)}\right) x_0 q_3 \sin \theta_{\mathbf{q}_3}\left(\cos \phi_{\mathbf{q}_3} + \sin \phi_{\mathbf{q}_3}\right)}{\sqrt{2}}\right] \Bigg\}\\
				&\qquad\Bigg\{\mathcal{M}_{\eta(\psi 3)}^{\psi 3,\psi 4}\left(-q_x^{(4)},-q_y^{(4)}\right) \mathcal{M}_{\eta(\psi 7)}^{\psi 7,\psi 8}\left(q_x^{(4)},q_y^{(4)}\right) \frac{1}{\left[\left(\frac{1}{a_z}\right)^2+\left( q_4 \cos \theta_{\mathbf{q}_4}\right)^2\right]^3} \\
				&\qquad \times \exp\left[- q_4^2 \sin^2 \theta_{\mathbf{q}_4}\left(\frac{1}{8 \alpha_{\eta(\psi3)}}+\frac{1}{8 \alpha_{\eta(\psi7)}}\right)\right] 
				\\
				&\qquad\times \exp\left[\frac{i \left(h_{\eta(\psi 3)}-h_{\eta(\psi 7)}\right) x_0 q_4 \sin \theta_{\mathbf{q}_4}\left(\cos \phi_{\mathbf{q}_4} + \sin \phi_{\mathbf{q}_4}\right)}{\sqrt{2}}\right]\Bigg\}
				\bigg|_{q_4=q_3+q_s}
			\end{split}\\
			\begin{split}\label{eq:octetInt21}
				&=\left(\prod_{j\in \{1,3,5,7\}} \frac{\pi \mathcal{G}_{\eta(\psi j)}^2}{\alpha_{\eta(\psi j)}}\right) \mathcal{M}_{\eta(\psi 1)}^{\psi 1,\psi 2}\left(-q_x^{(3)},-q_y^{(3)}\right)  \mathcal{M}_{\eta(\psi 5)}^{\psi 5,\psi 6}\left(q_x^{(3)},q_y^{(3)}\right)\mathcal{M}_{\eta(\psi 3)}^{\psi 3,\psi 4}\left(-q_x^{(4)},-q_y^{(4)}\right) \\
				&\qquad\times\mathcal{M}_{\eta(\psi 7)}^{\psi 7,\psi 8}\left(q_x^{(4)},q_y^{(4)}\right)\frac{1}{\left[\left(\frac{1}{a_z}\right)^2+\left( q_3 \cos \theta_{\mathbf{q}_3}\right)^2\right]^3}\frac{1}{\left[\left(\frac{1}{a_z}\right)^2+\left( q_4 \cos \theta_{\mathbf{q}_4}\right)^2\right]^3} \\
				&\qquad\times\exp\left[- q_3^2 \sin^2 \theta_{\mathbf{q}_3}\left(\frac{1}{8 \alpha_{\eta(\psi1)}}+\frac{1}{8 \alpha_{\eta(\psi5)}}\right)\right]  \exp\left[- q_4^2 \sin^2 \theta_{\mathbf{q}_4}\left(\frac{1}{8 \alpha_{\eta(\psi3)}}+\frac{1}{8 \alpha_{\eta(\psi7)}}\right)\right]\\
				&\qquad\times\exp\left[\frac{i \left(h_{\eta(\psi 1)}-h_{\eta(\psi 5)}\right) x_0 q_3 \sin \theta_{\mathbf{q}_3}\left(\cos 	\phi_{\mathbf{q}_3} + \sin \phi_{\mathbf{q}_3}\right)}{\sqrt{2}}\right] 
				\\
				&\qquad\times\exp\left[\frac{i \left(h_{\eta(\psi 3)}-h_{\eta(\psi 7)}\right) x_0 q_4 \sin \theta_{\mathbf{q}_4}\left(\cos \phi_{\mathbf{q}_4} + \sin \phi_{\mathbf{q}_4}\right)}{\sqrt{2}}\right]\bigg|_{q_4=q_3+q_s}
			\end{split}\\
			\begin{split}
				&=\widetilde{\mathcal{M}}^\psi(\mathbf{q}_3,\mathbf{q}_4)
				\mathcal{Q}_z(\mathbf{q}_3,\mathbf{q}_4)\exp\left[- q_3^2 \sin^2 \theta_{\mathbf{q}_3}\left(\frac{1}{8 \alpha_{\eta(\psi1)}}+\frac{1}{8 \alpha_{\eta(\psi5)}}\right)\right]  \exp\left[- q_4^2 \sin^2 \theta_{\mathbf{q}_4}\left(\frac{1}{8 \alpha_{\eta(\psi3)}}+\frac{1}{8 \alpha_{\eta(\psi7)}}\right)\right]\\
				&\qquad\times\exp\left[i \left(h_{\eta(\psi 1)}-h_{\eta(\psi 5)}\right) x_0 q_4 \sin \theta_{\mathbf{q}_3}\cos (\phi_{\mathbf{q}_3}-\pi/4) \right]
				\\
				&\qquad\times \exp\left[i \left(h_{\eta(\psi 3)}-h_{\eta(\psi 7)}\right) x_0 q_4 \sin \theta_{\mathbf{q}_4}\cos (\phi_{\mathbf{q}_4}-\pi/4) \right]
				\bigg|_{q_4=q_3+q_s},
			\end{split}
		\end{align}
	\end{subequations}
	where
	\begin{subequations}
		\begin{align}
			\begin{split}
				\widetilde{\mathcal{M}}^\psi(\mathbf{q}_3,\mathbf{q}_4)&=\left(\prod_{j\in \{1,3,5,7\}} \frac{\pi \mathcal{G}_{\eta(\psi j)}^2}{\alpha_{\eta(\psi j)}}\right) 
				\mathcal{M}_{\eta(\psi 1)}^{\psi 1,\psi 2}\left(-q_x^{(3)},-q_y^{(3)}\right)  \mathcal{M}_{\eta(\psi 5)}^{\psi 5,\psi 6}\left(q_x^{(3)},q_y^{(3)}\right)\mathcal{M}_{\eta(\psi 3)}^{\psi 3,\psi 4}\left(-q_x^{(4)},-q_y^{(4)}\right) \\
				&\qquad\times\mathcal{M}_{\eta(\psi 7)}^{\psi 7,\psi 8}\left(q_x^{(4)},q_y^{(4)}\right),\\
			\end{split}\\
			\begin{split}
				&\mathcal{Q}_z(\mathbf{q}_3,\mathbf{q}_4)=\frac{1}{\left[\left(\frac{1}{a_z}\right)^2+\left( q_3 \cos \theta_{\mathbf{q}_3}\right)^2\right]^3}\frac{1}{\left[\left(\frac{1}{a_z}\right)^2+\left( q_4 \cos \theta_{\mathbf{q}_4}\right)^2\right]^3},
			\end{split}
		\end{align}
	\end{subequations}
	while $\eta (\psi j)$ indicates the electron wavefunction $\psi j $ is located in dot $\eta\in \{L,R\}$. In the derivation of Eq.~\eqref{eq:electronWFIntegral}, we have assumed the electron wavefunctions in each doublet, $\langle \Phi_{\psi j} (\mathbf{r})| e^{i \mathbf{q}\cdot \mathbf{r}} | \Phi_{\psi k} (\mathbf{r}) \rangle $, belong to the same dot, i.e. $\eta (\psi j) = \eta (\psi k)$, such that $K_{\psi j,\psi k} =K_{\eta j,\widetilde{\eta} k} = 1$. As discussed before, we can safely neglect terms with $\eta (\psi j) \neq \eta (\psi k)$, cf. Eq.~\eqref{eq:valueK} and Sec.~\ref{subsec:innerproductDoublet}. The same derivation procedure can be applied to other replacement rules, e.g. $\mathbf{q}_1 = -\mathbf{q}_4$, $\mathbf{q}_2=-\mathbf{q}_3$, $q_4 = q_3 \pm q_s$ and $q_4 = q_s-q_3$, cf.~Eq.~\eqref{eq:thermalAverage4phonons}. 
	
	When all the electron wavefunctions are ground orbitals, $\psi j = L1$ or $\psi j = R1$ $\forall j$, such that $\mathcal{M}^{\psi j,\psi k}_{\eta(\psi j)} = 1$ $ \forall j,k$, we have 
	\begin{equation}\label{eq:groundOctets}
		\begin{split}
			&\int_{0}^{2\pi} d \phi_{\mathbf{q}_3}\int_{0}^{2\pi} d \phi_{\mathbf{q}_4}
			\left.\mathcal{N}(\mathbf{q}_1,\mathbf{q}_2,\mathbf{q}_3,\mathbf{q}_4)\right|_{\substack{\mathbf{q}_1=-\mathbf{q}_3,\\ \mathbf{q}_2=-\mathbf{q}_4,\\q_4=q_3+q_s}}\\
			&=\left(\prod_{j\in \{1,3,5,7\}} \frac{\pi \mathcal{G}_{\eta(\psi j)}^2}{\alpha_{\eta(\psi j)}}\right) \mathcal{Q}_z(\mathbf{q}_3,\mathbf{q}_4)
			\exp\left[- q_3^2 \sin^2 \theta_{\mathbf{q}_3}\left(\frac{1}{8 \alpha_{\eta(\psi1)}}+\frac{1}{8 \alpha_{\eta(\psi5)}}\right)\right]  \\
			&\qquad\times \exp\left[- q_4^2 \sin^2 \theta_{\mathbf{q}_4}\left(\frac{1}{8 \alpha_{\eta(\psi3)}}+\frac{1}{8 \alpha_{\eta(\psi7)}}\right)\right] 
			\\
			&\qquad \times \int_{0}^{2\pi} d \phi_{\mathbf{q}_3}\exp\left[i \left(h_{\eta(\psi 1)}-h_{\eta(\psi 5)}\right) x_0 q_4 \sin \theta_{\mathbf{q}_3}\cos (\phi_{\mathbf{q}_3}-\pi/4) \right]\\
			&\qquad \times
			\int_{0}^{2\pi} d \phi_{\mathbf{q}_4}\exp\left[i \left(h_{\eta(\psi 3)}-h_{\eta(\psi 7)}\right) x_0 q_4 \sin \theta_{\mathbf{q}_4}\cos (\phi_{\mathbf{q}_4}-\pi/4) \right]
			\Big|_{q_4=q_3+q_s}\\
			&=4\pi^2\left(\prod_{j\in \{1,3,5,7\}} \frac{\pi \mathcal{G}_{\eta(\psi j)}^2}{\alpha_{\eta(\psi j)}}\right) \mathcal{Q}_z(\mathbf{q}_3,\mathbf{q}_4)
			\exp\left[- q_3^2 \sin^2 \theta_{\mathbf{q}_3}\left(\frac{1}{8 \alpha_{\eta(\psi1)}}+\frac{1}{8 \alpha_{\eta(\psi5)}}\right)\right] 
			\\
			&\qquad\times \exp\left[- q_4^2 \sin^2 \theta_{\mathbf{q}_4}\left(\frac{1}{8 \alpha_{\eta(\psi3)}}+\frac{1}{8 \alpha_{\eta(\psi7)}}\right)\right]\\
			&\qquad \times
			\text{BesselJ}\left[0,(\Delta h_{15}) x_0 q_4 \sin \theta_{\mathbf{q}_3}\right]\text{BesselJ}\left[0,(\Delta h_{37}) x_0 q_4 \sin \theta_{\mathbf{q}_4}\right]
			\Big|_{q_4=q_3+q_s},
		\end{split}
	\end{equation}
	where $\Delta h_{jk}= \left|h_{\eta(\psi j)}-h_{\eta(\psi k)}\right|$ and $\text{BesselJ}[a,v]$ is the Bessel function of first kind for $a$-th integer order. If not all the electron wavefunctions are ground orbitals, we have
	\begin{equation}\label{eq:notAllgroundOctets}
		\begin{split}
			&\int_{0}^{2\pi} d \phi_{\mathbf{q}_3}\int_{0}^{2\pi} d \phi_{\mathbf{q}_4}
			\widetilde{\mathcal{M}}^\psi(\mathbf{q}_3,\mathbf{q}_4)\mathcal{N}(\mathbf{q}_1,\mathbf{q}_2,\mathbf{q}_3,\mathbf{q}_4)
			\Big|_{\substack{\mathbf{q}_1=-\mathbf{q}_3,\\ \mathbf{q}_2=-\mathbf{q}_4,\\q_4=q_3+q_s}}\\
			&=\left(\prod_{j\in \{1,3,5,7\}} \frac{\pi \mathcal{G}_{\eta(\psi j)}^2}{\alpha_{\eta(\psi j)}}\right)
			\mathcal{Q}_z(\mathbf{q}_3,\mathbf{q}_4)
			\exp\left[- q_3^2 \sin^2 \theta_{\mathbf{q}_3}\left(\frac{1}{8 \alpha_{\eta(\psi1)}}+\frac{1}{8 \alpha_{\eta(\psi5)}}\right)\right]  \exp\left[- q_4^2 \sin^2 \theta_{\mathbf{q}_4}\left(\frac{1}{8 \alpha_{\eta(\psi3)}}+\frac{1}{8 \alpha_{\eta(\psi7)}}\right)\right]\\
			&\times\int_{0}^{2\pi} d \phi_{\mathbf{q}_3}\mathcal{M}_{\eta(\psi 1)}^{\psi 1,\psi 2}\left(-q_x^{(3)},-q_y^{(3)}\right)  \mathcal{M}_{\eta(\psi 5)}^{\psi 5,\psi 6}\left(q_x^{(3)},q_y^{(3)}\right)\exp\left[i \left(h_{\eta(\psi 1)}-h_{\eta(\psi 5)}\right) x_0 q_4 \sin \theta_{\mathbf{q}_3}\cos (\phi_{\mathbf{q}_3}-\pi/4) \right]\\
			&\times
			\int_{0}^{2\pi} d \phi_{\mathbf{q}_4}\mathcal{M}_{\eta(\psi 3)}^{\psi 3,\psi 4}\left(-q_x^{(4)},-q_y^{(4)}\right) \mathcal{M}_{\eta(\psi 7)}^{\psi 7,\psi 8}\left(q_x^{(4)},q_y^{(4)}\right)\exp\left[i \left(h_{\eta(\psi 3)}-h_{\eta(\psi 7)}\right) x_0 q_4 \sin \theta_{\mathbf{q}_4}\cos (\phi_{\mathbf{q}_4}-\pi/4) \right]
			\Big|_{q_4=q_3+q_s}\\
			&=4\pi^2\left(\prod_{j\in \{1,3,5,7\}} \frac{\pi \mathcal{G}_{\eta(\psi j)}^2}{\alpha_{\eta(\psi j)}}\right) \mathcal{Q}_z(\mathbf{q}_3,\mathbf{q}_4)
			\exp\left[- q_3^2 \sin^2 \theta_{\mathbf{q}_3}\left(\frac{1}{8 \alpha_{\eta(\psi1)}}+\frac{1}{8 \alpha_{\eta(\psi5)}}\right)\right] \\
			&
			\quad \exp\left[- q_4^2 \sin^2 \theta_{\mathbf{q}_4}\left(\frac{1}{8 \alpha_{\eta(\psi3)}}+\frac{1}{8 \alpha_{\eta(\psi7)}}\right)\right] \Big(\sum_k c_{k(\mathcal{X},\mathcal{X'})}^{\psi}\mathcal{X}(l_\eta,x_0,q_3) \mathcal{X}'(l_\eta,x_0,q_4)\\
			&\quad\times
			\text{BesselJ}\left[a_{k(\mathcal{X},\mathcal{X'})},(\Delta h_{15}) x_0 q_3 \sin \theta_{\mathbf{q}_3}\right] \text{BesselJ}\left[b_{k(\mathcal{X},\mathcal{X'})}, (\Delta h_{37}) x_0 q_4 \sin \theta_{\mathbf{q}_4}\right]\Big)
			\Big|_{q_4=q_3+q_s},
		\end{split}
	\end{equation}
	where $c_{k(\mathcal{X},\mathcal{X'})}^{\psi}$ are constant values giving the linear combination of Bessel functions, $\sum_k c_{k(\mathcal{X},\mathcal{X'})}^{\psi}$, which are specific to the orbitals involved, $\psi$. $\mathcal{X}(l_\eta,x_0,q_m)$ is a function with variables $l_\eta$, $x_0$ and $q_m$, as shown below. For example, if $\psi = (L1,L1,L1,L1,L1,L1,R3,R3)$ such that $\psi 1 = \psi 2 = \psi 3 = \psi 4 =L1$, $\psi 5=\psi 6=R1$ while $\psi 7 = \psi 8=R3$, we have
	\begin{equation}
		\begin{split}
			&\int_{0}^{2\pi} d \phi_{\mathbf{q}_3}\int_{0}^{2\pi} d \phi_{\mathbf{q}_4}
			\widetilde{\mathcal{M}}^{\psi}(\mathbf{q}_3,\mathbf{q}_4)\mathcal{N}(\mathbf{q}_1,\mathbf{q}_2,\mathbf{q}_3,\mathbf{q}_4)
			\Big|_{\substack{\mathbf{q}_1=-\mathbf{q}_3,\\ \mathbf{q}_2=-\mathbf{q}_4,\\q_4=q_3+q_s}}\\
			&=4\pi^2\left(\prod_{j\in \{1,3,5,7\}} \frac{\pi \mathcal{G}_{\eta(\psi j)}^2}{\alpha_{\eta(\psi j)}}\right) \mathcal{Q}_z(\mathbf{q}_3,\mathbf{q}_4)
			\exp\left[- \frac{\left(l_L^2+l_R^2\right)q_3^2 \sin^2 \theta_{\mathbf{q}_3}}{4}\right]  \exp\left[- \frac{\left(l_L^2+l_R^2\right)q_4^2 \sin^2 \theta_{\mathbf{q}_4}}{4}\right]\\
			&\quad
			\left( 1-\frac{l_R^2 q_4^2 \sin^2 \theta_{\mathbf{q}_4}}{4} \right) \text{BesselJ}\left[0,2 x_0 q_3 \sin \theta_{\mathbf{q}_3}\right]\text{BesselJ}\left[0,2 x_0 q_4 \sin \theta_{\mathbf{q}_4}\right]
			\Big|_{q_4=q_3+q_s}
		\end{split}
	\end{equation}
	\subsubsection{Pure-dephasing rate, $\Gamma_\varphi=J_{zz}^+(0)/(2\hbar^2)$}\label{subsubsec:pureDephase}
	The pure-dephasing rate is evaluated with $q_s=0$. From Eq.~\eqref{eq:thermalAverage4phonons}, \eqref{eq:electronWFIntegral} and \eqref{eq:notAllgroundOctets}, we have
	\begin{equation}
		\begin{split}
			&(2\hbar^2)\mathcal{I}^{\psi}_{s,\text{cpl}}(q_3,\phi_{\mathbf{q}_3},\phi_{\mathbf{q}_4},\theta_{\mathbf{q}_3},\theta_{\mathbf{q}_4})\\
			&=C_{s,\text{cpl}} (\mathbf{q}_1)C_{s,\text{cpl}} (\mathbf{q}_2)C_{s,\text{cpl}} (\mathbf{q}_3)C_{s,\text{cpl}} (\mathbf{q}_4)\frac{8\pi}{\nu_s}\int_{0}^{\pi} \sin \theta_{\mathbf{q}_3} d\theta_{\mathbf{q}_3} \int_{0}^{2\pi} d\phi_{\mathbf{q}_3} \int_{0}^{\pi} \sin \theta_{\mathbf{q}_4} d\theta_{\mathbf{q}_4}\int_{0}^{2\pi} d \phi_{\mathbf{q}_4}\int_{0}^{\infty}dq_3 q_3^4\\
			&\quad\times n_s (q_3)\left[n_s (q_3)+1\right] \left(
			\left.\mathcal{N}(\mathbf{q}_1,\mathbf{q}_2,\mathbf{q}_3,\mathbf{q}_4)\right|_{\substack{\mathbf{q}_1=-\mathbf{q}_3,\\ \mathbf{q}_2=-\mathbf{q}_4,\\q_4=q_3}}
			+\left.\mathcal{N}(\mathbf{q}_1,\mathbf{q}_2,\mathbf{q}_3,\mathbf{q}_4)\right|_{\substack{\mathbf{q}_1=-\mathbf{q}_4,\\ \mathbf{q}_2=-\mathbf{q}_3,\\q_4=q_3}}
			\right)\\
			&=C_{s,\text{cpl}} (\mathbf{q}_1)C_{s,\text{cpl}} (\mathbf{q}_2)C_{s,\text{cpl}} (\mathbf{q}_3)C_{s,\text{cpl}} (\mathbf{q}_4)\frac{32\pi^3}{\nu_s}\int_{0}^{\pi} \sin \theta_{\mathbf{q}_3} d\theta_{\mathbf{q}_3} \int_{0}^{\pi} \sin \theta_{\mathbf{q}_4} d\theta_{\mathbf{q}_4}\int_{0}^{\infty}dq_3 q_3^4\\
			&\quad \times n_s (q_3)\left[n_s (q_3)+1\right] \left(\prod_{j\in \{1,3,5,7\}} \frac{\pi \mathcal{G}_{\eta(\psi j)}^2}{\alpha_{\eta(\psi j)}}\right) \mathcal{Q}_z(\mathbf{q}_3,\mathbf{q}_4)\exp\left[- \frac{q_3^2 \sin^2 \theta_{\mathbf{q}_3}}{4 \alpha_\eta}\right]  \exp\left[- \frac{q_4^2 \sin^2 \theta_{\mathbf{q}_4}}{4 \alpha_\eta}\right]\\
			&\quad\times\Bigg[\Big(\sum_j c_{j(\mathcal{X},\mathcal{X'})}^{\psi} \mathcal{X}(l_\eta,x_0,q_3) \mathcal{X}'(l_\eta,x_0,q_3)
			\text{BesselJ}\left[a_{j(\mathcal{X},\mathcal{X'})},(\Delta h_{15}) x_0 q_3 \sin \theta_{\mathbf{q}_3}\right]
			\\
			&\qquad\qquad\text{BesselJ}\left[b_{j(\mathcal{X},\mathcal{X'})},(\Delta h_{37}) x_0 q_3 \sin \theta_{\mathbf{q}_3}\right]\Big)\\
			&\quad\,\,+\Big(\sum_k c_{k(\mathcal{X},\mathcal{X'})}^{\psi} \mathcal{X}(l_\eta,x_0,q_3) \mathcal{X}'(l_\eta,x_0,q_3)
			\text{BesselJ}\left[a_{k(\mathcal{X},\mathcal{X'})},(\Delta h_{17}) x_0 q_3 \sin \theta_{\mathbf{q}_3}\right]
			\\
			&\qquad\qquad\text{BesselJ}\left[b_{k(\mathcal{X},\mathcal{X'})},(\Delta h_{35}) x_0 q_3 \sin \theta_{\mathbf{q}_3}\right]\Big)\Bigg].
		\end{split}
	\end{equation}
	If all the electron wavefunctions involved are ground orbitals in the same dot, i.e. $\phi j = L1$ or $\phi j = R1$ $\forall j$, we have 
	\begin{equation}
		\begin{split}
			&(2\hbar^2)\mathcal{I}^{\psi}_{s,\text{cpl}}(q_3,\phi_{\mathbf{q}_3},\phi_{\mathbf{q}_4},\theta_{\mathbf{q}_3},\theta_{\mathbf{q}_4})\\
			&=C_{s,\text{cpl}} (\mathbf{q}_1)C_{s,\text{cpl}} (\mathbf{q}_2)C_{s,\text{cpl}} (\mathbf{q}_3)C_{s,\text{cpl}} (\mathbf{q}_4)\frac{32\pi^3}{\nu_s}\int_{0}^{\pi} \sin \theta_{\mathbf{q}_3} d\theta_{\mathbf{q}_3} \int_{0}^{\pi} \sin \theta_{\mathbf{q}_4} d\theta_{\mathbf{q}_4}\int_{0}^{\infty}dq_3 q_3^4\\
			&\quad \times n_s (q_3)\left[n_s (q_3)+1\right] \left(\prod_{j\in \{1,3,5,7\}} \frac{\pi \mathcal{G}_{\eta(\psi j)}^2}{\alpha_{\eta(\psi j)}}\right) \mathcal{Q}_z(\mathbf{q}_3,\mathbf{q}_4)\exp\left[- \frac{q_3^2 \sin^2 \theta_{\mathbf{q}_3}}{4 \alpha_\eta}\right]  \exp\left[- \frac{q_4^2 \sin^2 \theta_{\mathbf{q}_4}}{4 \alpha_\eta}\right],
		\end{split}
	\end{equation}
	as $\Delta h_{jk}=0$, $\forall j,k$.
	$\mathcal{I}^{\psi}_{s,\text{cpl}}(q_3,\phi_{\mathbf{q}_3},\phi_{\mathbf{q}_4},\theta_{\mathbf{q}_3},\theta_{\mathbf{q}_4})$ generally can be divided into three classes: (1) $\Delta h_{jk} = 0$ $\forall j,k$ and (2) two of $\Delta h_{jk} \neq 0$ and (3) four of $\Delta h_{jk} \neq 0$. We denote the number of terms $\Delta h_{jk} \neq 0$ to be $\mathcal{D}$ and the values of $\mathcal{I}^{\psi}_{s,\text{cpl}}$ follows the rules:
	\begin{equation}\label{eq:BesselSmall}
		\mathcal{I}^{\psi}_{s,\text{cpl}} (\mathcal{D} = 0) > \mathcal{I}^{\psi}_{s,\text{cpl}} (\mathcal{D} = 2)> \mathcal{I}^{\psi}_{s,\text{cpl}} (\mathcal{D} = 4),
	\end{equation}
	as $\mathcal{D}$ directly related to the number of BesselJ functions included in the integral, $\mathcal{I}$. The effect of BesselJ functions is adding oscillations as function $q_3$ into the integrand, resulting in smaller $\mathcal{I}$.
	
	\subsection{Calculation of $J^+_{jj}(\omega)$ for one-phonon process}\label{subsec:JjjOnePhonon}
	Using the similar notations in Sec.~\ref{subsec:JjjTwoPhonon}, an arbitrary $\langle \delta B_j(0)\delta B_j(\tau)\rangle$ for one-phonon process can be expanded in the form of
	
	\begin{equation}
		\begin{split}
			&\frac{1}{2\hbar^2}\int_{-\infty}^{\infty} d\tau \left(e^{i \omega_z \tau}+e^{-i \omega_z \tau}\right)\langle\delta B_j(0)\delta B_j(\tau)\rangle_\beta \\
			&= \frac{1}{2\hbar^2}\int_{-\infty}^{\infty}d\tau \left(e^{i \omega_z \tau}+e^{-i \omega_z \tau}\right) \sum\limits_{\substack{\psi1,\psi2,\psi3,\psi4}}  \varrho^{\psi1,\psi2}_j \varrho^{\psi3,\psi4}_j \Big\langle \langle \Phi_{\psi1} | H_\text{e-ph}(0) | \Phi_{\psi2}\rangle \langle \Phi_{\psi3} | H_\text{e-ph}(\tau) | \Phi_{\psi4}\rangle\Big\rangle_\beta\\
			&=\sum\limits_{\psi1,\psi2,\psi3,\psi4} \varrho^\psi \mathcal{I}^{\text{(1p)}\psi}
		\end{split}
	\end{equation}
	where $\varrho^\psi = \gamma^{\psi1,\psi2}_j \varrho^{\psi3,\psi4}_j$ and $\varrho^{\psi3,\psi4}_j$, $\varrho^{\psi3,\psi4}_j$ are the coefficients whose explicit values depend on the electronic Hamiltonian, $H_\text{e}+H_\text{hyp}$, while $\psi$ denote the orbitals involved. From there, we have
	\begin{equation}\label{eq:allIntOnephonon}
		\begin{split}
			&(2\hbar^2)\mathcal{I}^{(\text{1p})\psi}\\
			&=\int_{-\infty}^{\infty}d\tau\left(e^{i \omega_z \tau}+e^{-i \omega_z \tau}\right)\Big\langle \langle \Phi_{\psi1} | H_\text{e-ph}(0) | \Phi_{\psi2}\rangle \langle \Phi_{\psi3} | H_\text{e-ph}(\tau) | \Phi_{\psi4}\rangle \Big\rangle_\beta\\
			&=\sum_{s,\text{cpl}}\int_{-\infty}^{\infty}d\tau\left(e^{i \omega_z \tau}+e^{-i \omega_z \tau}\right)\sum_{\mathbf{q}_1,\mathbf{q}_2} C_{s,\text{cpl}} (\mathbf{q}_1)C_{s,\text{cpl}} (\mathbf{q}_2) \\
			&\quad\Bigg\langle\left\langle\Phi_{\psi1} (\mathbf{r})| \left[ a_s(\mathbf{q}_1)\mp_s a_s^\dagger(-\mathbf{q}_1)\right]e^{i \mathbf{q}_1 \cdot \mathbf{r}}|\Phi_{\psi2} (\mathbf{r})\right\rangle \left\langle \Phi_{\psi3} (\mathbf{r})| \left[a_s(\mathbf{q}_2)e^{-i \omega_s(\mathbf{q}_2)\tau}\mp_s a_s^\dagger(-\mathbf{q}_2)e^{i \omega_s(\mathbf{q}_2)\tau}\right]e^{i \mathbf{q}_2 \cdot \mathbf{r}}|\Phi_{\psi4} (\mathbf{r})\right\rangle\Bigg\rangle_\beta	\\
			&=\sum_{s,\text{cpl}}\sum_{\mathbf{q}_1,\mathbf{q}_2} C_{s,\text{cpl}} (\mathbf{q}_1)C_{s,\text{cpl}} (\mathbf{q}_2) \int_{-\infty}^{\infty}d\tau\left(e^{i \omega_z \tau}+e^{-i \omega_z \tau}\right)\\
			&\quad\left\langle\Phi_{\psi1} (\mathbf{r})| e^{i \mathbf{q}_1 \cdot \mathbf{r}}|\Phi_{\psi2} (\mathbf{r})\right\rangle
			\left\langle \Phi_{\psi3} (\mathbf{r})| e^{i \mathbf{q}_2 \cdot \mathbf{r}}|\Phi_{\psi4} (\mathbf{r})\right\rangle \Bigg\langle\left[ a_s(\mathbf{q}_1)\mp_s a_s^\dagger(-\mathbf{q}_1)\right]
			\left[a_s(\mathbf{q}_2)e^{-i \omega_s(\mathbf{q}_2)\tau}\mp_s a_s^\dagger(-\mathbf{q}_2)e^{i \omega_s(\mathbf{q}_2)\tau}\right]\Bigg\rangle_\beta\\
			&=\sum_{s,\text{cpl}}\sum_{\mathbf{q}_1,\mathbf{q}_2} C_{s,\text{cpl}} (\mathbf{q}_1)C_{s,\text{cpl}} (\mathbf{q}_2)\int_{-\infty}^{\infty}d\tau\left(e^{i \omega_z \tau}+e^{-i \omega_z \tau}\right)\mathcal{N}^{(\text{1p})}(\mathbf{q}_1,\mathbf{q}_2)\\
			&\quad\Big\langle\left[ a_s(\mathbf{q}_1)\mp_s a_s^\dagger(-\mathbf{q}_1)\right]\left[a_s(\mathbf{q}_2)e^{-i \omega_s(\mathbf{q}_2)\tau}\mp_s a_s^\dagger(-\mathbf{q}_2)e^{i \omega_s(\mathbf{q}_2)\tau}\right]\Big\rangle_\beta\\
			&=(2\hbar^2)\sum_{s,\text{cpl}} \mathcal{I}_{s,\text{cpl}}^{(\text{1p})\psi}(\phi_{\mathbf{q}_2},\theta_{\mathbf{q}_2})\\
		\end{split}
	\end{equation}
	In Eq.~\eqref{eq:allIntOnephonon}, we denote the energy of electron-phonon interaction of each mode, $s$, and coupling mechanism, cpl, whose unit is $(\text{electron volt}^2 \times \text{second})$, as 
	\begin{equation}
		\begin{split}
			&(2\hbar^2)\mathcal{I}^{(\text{1p})\psi}_{s,\text{cpl}}(\phi_{\mathbf{q}_2},\theta_{\mathbf{q}_2})
			=\sum_{\mathbf{q}_1,\mathbf{q}_2} C_{s,\text{cpl}} (\mathbf{q}_1)C_{s,\text{cpl}} (\mathbf{q}_2)\int_{-\infty}^{\infty}d\tau\left(e^{i \omega_z \tau}+e^{-i \omega_z \tau}\right)\mathcal{N}^{(\text{1p})}(\mathbf{q}_1,\mathbf{q}_2)\\
			&\quad\times\Big\langle\left[ a_s(\mathbf{q}_1)\mp_s a_s^\dagger(-\mathbf{q}_1)\right]\left[a_s(\mathbf{q}_2)e^{-i \omega_s(\mathbf{q}_2)\tau}\mp_s a_s^\dagger(-\mathbf{q}_2)e^{i \omega_s(\mathbf{q}_2)\tau}\right]\Big\rangle_\beta
		\end{split}
	\end{equation}
	and the integral of electron wavefunctions as 
	\begin{equation}
		\mathcal{N}^{\text{(1p)}}(\mathbf{q}_1,\mathbf{q}_2)=\langle\Phi_{\psi1} (\mathbf{r})| e^{i \mathbf{q}_1 \cdot \mathbf{r}}|\Phi_{\psi2} (\mathbf{r})\rangle
		\langle \Phi_{\psi3} (\mathbf{r})| e^{i \mathbf{q}_2 \cdot \mathbf{r}}|\Phi_{\psi4} (\mathbf{r})\rangle.
	\end{equation}
	From Eq.~\eqref{eq:allIntOnephonon}, keeping the non-zero terms after thermal average, we have
	\begin{subequations}\label{eq:thermalAverage2phonons}
		\begin{align}
			\begin{split}\label{eq:ItildeOnePhonon}
				&(2\hbar^2)\mathcal{\widetilde{I}}^{\text{(1p)}\psi}_{s,\text{cpl}}(\phi_{\mathbf{q}_2},\theta_{\mathbf{q}_2})
			\end{split}\\
			\begin{split}\label{eq:IdivCOnePhonon}
				&=\frac{\mathcal{I}^{\text{(1p)}\psi}_{s,\text{cpl}}(\phi_{\mathbf{q}_2},\theta_{\mathbf{q}_2})}{C_{s,\text{cpl}} (\mathbf{q}_1)C_{s,\text{cpl}} (\mathbf{q}_2)}
			\end{split}\\
			\begin{split}
				&=\mp_s\int_{-\infty}^{\infty}d\tau\left(e^{i \omega_z \tau}+e^{-i \omega_z \tau}\right)
				\sum_{\mathbf{q}_1,\mathbf{q}_2}
				\mathcal{N}^{\text{(1p)}}(\mathbf{q}_1,\mathbf{q}_2)
				\left(\langle a_s^\dagger (-\mathbf{q}_1) a_s(\mathbf{q}_2)\rangle_\beta e^{-i q_2\nu_s \tau} +\langle a_s(\mathbf{q}_1) a_s^\dagger(-\mathbf{q}_2)\rangle_\beta e^{i q_2 \nu_s \tau} \right)
			\end{split}\\
			\begin{split}\label{eq:q4toq21p}
				&=\int_{-\infty}^{\infty}d\tau\left(e^{i \omega_z \tau}+e^{-i \omega_z \tau}\right) \sum_{\mathbf{q}_2} 
				\mathcal{N}(\mathbf{q}_1,\mathbf{q}_2)\Big|_{\mathbf{q}_1=-\mathbf{q}_2} \left(\langle a_s^\dagger (\mathbf{q}_2) a_s(\mathbf{q}_2)\rangle_\beta e^{-i q_2 \nu_s \tau} +\langle a_s(-\mathbf{q}_2) a_s^\dagger(-\mathbf{q}_2)\rangle_\beta e^{i q_2 \nu_s \tau} \right)
			\end{split}\\
			\begin{split}\label{eq:nB1p}
				&=\sum_{\mathbf{q}_2}\int_{-\infty}^{\infty}d\tau\left(e^{i \omega_z \tau}+e^{-i \omega_z \tau}\right)\mathcal{N}(\mathbf{q}_1,\mathbf{q}_2)\Big|_{\mathbf{q}_1=-\mathbf{q}_2}
				\left( n_s (q_2) e^{-i q_2 \nu_s \tau} + \left[n_s (q_2)+1\right] e^{i q_2 \nu_s \tau} \right)
			\end{split}\\
			\begin{split}
				&=\sum_{\mathbf{q}_2}\int_{-\infty}^{\infty}d\tau\mathcal{N}(\mathbf{q}_1,\mathbf{q}_2)\Big|_{\mathbf{q}_1=-\mathbf{q}_2}
				\Big[ n_s (q_2) \left(e^{-i \left(q-q_2\right) \nu_s \tau}+e^{-i \left(-q-q_2\right) \nu_s \tau}\right) \\
				&\quad+ \left[n_s (q_2)+1\right] \left(e^{-i \left(q+q_2\right) \nu_s \tau}+e^{-i \left(-q+q_2\right) \nu_s \tau}\right) \Big]\
			\end{split}\\
			\begin{split}
				&=\frac{2\pi}{\nu_s}
				\int_{0}^{\infty} q_2^2 dq_2 
				\int_{0}^{\pi} d\theta_{\mathbf{q}_2}
				\int_{0}^{2\pi} d\phi_{\mathbf{q}_2}
				\mathcal{N}(\mathbf{q}_1,\mathbf{q}_2)\Big|_{\mathbf{q}_1=-\mathbf{q}_2}
				\Big( n_s (q_2) \left[\delta \left(q-q_2\right) +\delta \left(-q-q_2\right)\right]\\
				&\quad+ \left[n_s (q_2)+1\right] \left[\delta\left(q+q_2\right) + \delta \left(-q+q_2\right)\right] \Big)
			\end{split}\\
			\begin{split}
				&=\frac{2\pi}{\nu_s} q^2 
				\Big( 2 n_s (q)+ 1 \Big)
				\int_{0}^{\pi} d\theta_{\mathbf{q}_2}
				\int_{0}^{2\pi} d\phi_{\mathbf{q}_2}
				\mathcal{N}(\mathbf{q}_1,\mathbf{q}_2)\Big|_{\mathbf{q}_1=-\mathbf{q}_2, |\mathbf{q}_2| = q}
			\end{split}\\
		\end{align}
	\end{subequations}
	where $q = \omega_z/\nu_s$ and the subscript at the straight line, $\mathcal{A}|_{\text{rule}}$, denotes the replacement ``rule'' to be applied to the variables in the expression $\mathcal{A}$. 
	
	Taking into account of the algebraic manipulations in Eq.~\eqref{eq:thermalAverage2phonons}, the inner product part in Eq.~\eqref{eq:allIntOnephonon} leads to
	\begin{subequations}
		\begin{align}\label{eq:electronWFIntegral1p}
			\begin{split}
				&\left.\mathcal{N}(\mathbf{q}_1,\mathbf{q}_2)\right|_{\substack{\mathbf{q}_1=-\mathbf{q}_2,q_2=q}}
			\end{split}\\
			\begin{split}
				&=\langle\Phi_{\psi1} (\mathbf{r})| e^{i -\mathbf{q}_2 \cdot \mathbf{r}}|\Phi_{\psi2} (\mathbf{r})\rangle
				\langle \Phi_{\psi3} (\mathbf{r})| e^{i \mathbf{q}_2 \cdot \mathbf{r}}|\Phi_{\psi4} (\mathbf{r})\rangle
				\Big|_{q_4=q_3+q_s}
			\end{split}\\
			\begin{split}\label{eq:quartetInt}
				&=\left(\prod_{j\in \{1,3\}} \frac{\pi \mathcal{G}_{\eta(\psi j)}^2}{\alpha_{\eta(\psi j)}}\right) \Bigg\{\mathcal{M}_{\eta(\psi 1)}^{\psi 1,\psi 2}\left(-q_x,-q_y\right) \mathcal{M}_{\eta(\psi 3)}^{\psi 3,\psi 4}\left(q_x,q_y\right) \frac{1}{\left[\left(\frac{1}{a_z}\right)^2+\left( q \cos \theta_{\mathbf{q}_2}\right)^2\right]^3} \\
				&\qquad\times\exp\left[- q^2 \sin^2 \theta_{\mathbf{q}_2}\left(\frac{1}{8 \alpha_{\eta(\psi1)}}+\frac{1}{8 \alpha_{\eta(\psi3)}}\right)\right] \exp\left[\frac{i \left(h_{\eta(\psi 1)}-h_{\eta(\psi 3)}\right) x_0 q \sin \theta_{\mathbf{q}_2}\left(\cos \phi_{\mathbf{q}_2} + \sin \phi_{\mathbf{q}_2}\right)}{\sqrt{2}}\right] \Bigg\}\\
			\end{split}\\
			\begin{split}
				&=\widetilde{\mathcal{M}}^\psi(\mathbf{q})
				\mathcal{Q}_z(\mathbf{q})\exp\left[- q^2 \sin^2 \theta_{\mathbf{q}_2}\left(\frac{1}{8 \alpha_{\eta(\psi1)}}+\frac{1}{8 \alpha_{\eta(\psi3)}}\right)\right] \exp\left[\frac{i \Delta h_{13} x_0 q \sin \theta_{\mathbf{q}_2}\left(\cos \phi_{\mathbf{q}_2} + \sin \phi_{\mathbf{q}_2}\right)}{\sqrt{2}}\right],
			\end{split}
		\end{align}
	\end{subequations}
	where
	\begin{subequations}
		\begin{align}
			\begin{split}
				\widetilde{\mathcal{M}}^\psi(\mathbf{q})&=\left(\prod_{j\in \{1,3\}} \frac{\pi \mathcal{G}_{\eta(\psi j)}^2}{\alpha_{\eta(\psi j)}}\right) 
				\mathcal{M}_{\eta(\psi 1)}^{\psi 1,\psi 2}\left(-q_x,-q_y\right)  \mathcal{M}_{\eta(\psi 3)}^{\psi 3,\psi 4}\left(q_x,q_y\right),\\
			\end{split}\\
			\begin{split}
				&\mathcal{Q}_z(\mathbf{q})=\frac{1}{\left[\left(\frac{1}{a_z}\right)^2+\left( q_3 \cos \theta_{\mathbf{q}}\right)^2\right]^3},
			\end{split}
		\end{align}
	\end{subequations}
	In the derivation of Eq.~\eqref{eq:electronWFIntegral}, we have assumed the electron wavefunctions in each doublet, $\langle \Phi_{\psi j} (\mathbf{r})| e^{i \mathbf{q}\cdot \mathbf{r}} | \Phi_{\psi k} (\mathbf{r}) \rangle $, belong to the same dot, i.e. $\eta (\psi j) = \eta (\psi k)$, such that $K_{\psi j,\psi k} =K_{\eta j,\widetilde{\eta} k} = 1$, as discussed, cf. Eq.~\eqref{eq:valueK} and Sec.~\ref{subsec:innerproductDoublet}. 
	
	When all the electron wavefunctions are ground orbitals, $\psi j = L1$ or $\psi j = R1$ $\forall j$, such that $\mathcal{M}^{\psi j,\psi k}_{\eta(\psi j)} = 1$ $ \forall j,k$, we have 
	\begin{equation}\label{eq:groundQuartets}
		\begin{split}
			&\int_{0}^{2\pi} d \phi_{\mathbf{q}_2}
			\mathcal{N}(\mathbf{q}_1,\mathbf{q}_2)\big|_{\mathbf{q}_1=-\mathbf{q}_2,q_2=q}\\
			&=\widetilde{\mathcal{M}}^\psi(\mathbf{q})
			\mathcal{Q}_z(\mathbf{q})
			\exp\left[- q^2 \sin^2 \theta_{\mathbf{q}_2}\left(\frac{1}{8 \alpha_{\eta(\psi1)}}+\frac{1}{8 \alpha_{\eta(\psi3)}}\right)\right] 
			\exp\left[\frac{i \Delta h_{13} x_0 q \sin \theta_{\mathbf{q}_2}\left(\cos \phi_{\mathbf{q}_2} + \sin \phi_{\mathbf{q}_2}\right)}{\sqrt{2}}\right]\\
			&=2\pi\left(\prod_{j\in \{1,3\}} \frac{\pi \mathcal{G}_{\eta(\psi j)}^2}{\alpha_{\eta(\psi j)}}\right) \mathcal{Q}_z(\mathbf{q})
			\exp\left[- q^2 \sin^2 \theta_{\mathbf{q}_2}\left(\frac{1}{8 \alpha_{\eta(\psi1)}}+\frac{1}{8 \alpha_{\eta(\psi3)}}\right)\right]
			\text{BesselJ}\left[0,(\Delta h_{13}) x_0 q \sin \theta_{\mathbf{q}_2}\right],
		\end{split}
	\end{equation}
	On the other hand, if not all the electron wavefunctions are ground orbitals, we have
	\begin{equation}\label{eq:notAllgroundQuartets}
		\begin{split}
			&\int_{0}^{2\pi} d \phi_{\mathbf{q}_2}
			\widetilde{\mathcal{M}}^\psi(\mathbf{q})\mathcal{N}(\mathbf{q}_1,\mathbf{q}_2)
			\Big|_{\mathbf{q}_1=-\mathbf{q}_2,q_2=q}\\
			&=\left(\prod_{j\in \{1,3\}} \frac{\pi \mathcal{G}_{\eta(\psi j)}^2}{\alpha_{\eta(\psi j)}}\right)
			\mathcal{Q}_z(\mathbf{q})
			\exp\left[- q^2 \sin^2 \theta_{\mathbf{q}_2}\left(\frac{1}{8 \alpha_{\eta(\psi1)}}+\frac{1}{8 \alpha_{\eta(\psi3)}}\right)\right] \\
			&\quad\times\int_{0}^{2\pi} d \phi_{\mathbf{q}_2}\mathcal{M}_{\eta(\psi 1)}^{\psi 1,\psi 2}\left(-q_x,-q_y\right)  \mathcal{M}_{\eta(\psi 3)}^{\psi 3,\psi 4}\left(q_x,q_y\right)\exp\left[i \Delta h_{13} x_0 q_4 \sin \theta_{\mathbf{q}_2}\cos (\phi_{\mathbf{q}_2}-\pi/4) \right]\\
			&=2\pi\left(\prod_{j\in \{1,3\}} \frac{\pi \mathcal{G}_{\eta(\psi j)}^2}{\alpha_{\eta(\psi j)}}\right) \mathcal{Q}_z(\mathbf{q})
			\exp\left[- q^2 \sin^2 \theta_{\mathbf{q}_2}\left(\frac{1}{8 \alpha_{\eta(\psi1)}}+\frac{1}{8 \alpha_{\eta(\psi3)}}\right)\right]  \\
			&
			\quad \Big(\sum_k c_{k(\mathcal{X})}^{\psi}\mathcal{X}(l_\eta,x_0,q) 
			\text{BesselJ}\left[a_{k(\mathcal{X})},(\Delta h_{13}) x_0 q \sin \theta_{\mathbf{q}_2}\right]\Big)
			,
		\end{split}
	\end{equation}
	where $c_{k(\mathcal{X})}^{\psi}$ are constant values giving the linear combination of Bessel functions, $\sum_k c_{k(\mathcal{X})}^{\psi}$, which are specific to the orbitals involved, $\psi$. $\mathcal{X}(l_\eta,x_0,q)$ is a function with variables $l_\eta$, $x_0$ and $q$, as shown below. For example, if $\psi = (L1,L1,R3,R3)$ such that $\psi 1 = \psi 2 =L1$ and $\psi 3 = \psi 4=R3$, we have
	\begin{equation}
		\begin{split}
			&\int_{0}^{2\pi} d \phi_{\mathbf{q}_2}
			\widetilde{\mathcal{M}}^\psi(\mathbf{q})\mathcal{N}(\mathbf{q}_1,\mathbf{q}_2)
			\Big|_{\mathbf{q}_1=-\mathbf{q}_2,q_2=q}\\
			&=2\pi\left(\prod_{j\in \{1,3\}} \frac{\pi \mathcal{G}_{\eta(\psi j)}^2}{\alpha_{\eta(\psi j)}}\right) \mathcal{Q}_z(\mathbf{q})
			\exp\left[- \frac{\left(l_L^2+l_R^2\right)q^2 \sin^2 \theta_{\mathbf{q}_2}}{4}\right] 
			\left( 1-\frac{l_R^2 q^2 \sin^2 \theta_{\mathbf{q}_2}}{4} \right) \text{BesselJ}\left[0,2 x_0 q \sin \theta_{\mathbf{q}_2}\right]
		\end{split}
	\end{equation}
	
	\subsection{Phonon pure-dephasing, $T_\varphi$} \label{subsec:pureDep}
	
	In the context of capacitively coupled four-electron singlet-triplet qubtis, around the sweet spot, the admixture to quadruple occupation of outer QDs, though not dominating, is not negligible. This corresponds to the ``large'' detuning regime discussed in Ref. \onlinecite{Kornich.14}. The phonon induced pure-dephasing can be understood by focusing on the effective Hamiltonian that gives the admixture between $|S(1,3)\rangle = |S(\uparrow_{L1}\downarrow_{R2})\rangle$, $ |T(1,3)\rangle=|T(\uparrow_{L1}\downarrow_{R2})\rangle$ and $|S(0,4)\rangle=|S(\uparrow_{R2}\downarrow_{R2})\rangle$, $ |T(0,4^*)\rangle=|T(\uparrow_{R2}\downarrow_{R3})\rangle$ \cite{Kornich.14}, where the asterisk indicates an excited valence orbital. Written in the bases of $|S(1,3)\rangle$, $|S(0,4)\rangle$, $|T(1,3)\rangle$ and $|T(0,4^*)\rangle$, which is equivalent to applying a projector $\mathcal{P}^{(S1)}$ onto the electron Hamiltonian, $H_\text{e}$, we have
	\begin{equation}
		\begin{split}
			H_\text{e}^{(S1)}&= \mathcal{P}^{(S1)} H_\text{e} \mathcal{P}^{(S1)}\\
			&=\left(
			\begin{array} {cccc}
				U_{ L1,R2 }+2\Delta  & \sqrt{2}  t_{ L1,R2 }   & 0 & 0 \\
				\sqrt{2} t_{ L1,R2 }   & U_{ R2 } & 0 & 0 \\
				0 & 0 & J^{ (1,3) }+U_{ L1,R2 }+2\Delta  & -t_{ L1,R3 } \\
				0 & 0 & -t_{ L1,R3 } & U_{ R2,R3 }+\Delta E-\xi  
			\end{array}
			\right),
		\end{split}
	\end{equation}
	where the hyperfine coupling term is dropped as it does not affect the main discussion here while the projector
	\begin{equation}
		\mathcal{P}^{(S1)} = |S(1,3)\rangle\langle S(1,3)| + |S(0,4)\rangle\langle S(0,4)| + |T(1,3)\rangle\langle T(1,3) | + |T(0,4^*)\rangle\langle T(0,4^*)\rangle|.
	\end{equation}
	To simplify the discussion, an energy shift is applied on $\widetilde{H}$ with respect to the energy of $|S(1,3)\rangle$,
	
	\begin{equation}
		\begin{split}
			H_\text{e}^{(S1)}=
			\left(
			\begin{array} {cccc}
				0  & \sqrt{2}  t_{L1,R2}  & 0 & 0 \\
				\sqrt{2}  t_{ L1,R2 }  & E_S & 0 & 0 \\
				0 & 0 & 0  & -t_{ L1,R3 } \\
				0 & 0 & -t_{ L1,R3 } & E_T
			\end{array}
			\right)
			+ E_\text{shift},
		\end{split}
	\end{equation}
	where $E_\text{shift} = U_{L1,R2}+2\Delta, E_S = U_{R2} - (U_{L1,R2}+2\Delta), E_T = U_{R2,R3}+\Delta E-\xi - (U_{L1,R2}+2\Delta)$. We drop the term $J^{(1,3)}$ for $E_T$ due to its smallness compared to other terms. Note, for negative exchange in the (0,4) regime, $E_T < E_S$. Also, we are focusing in the detuning regime where the admixture with states of (0,4) type starts to come into play with the dominant part of the eigenstates in (1,3) regime, hence $E_S>0$ and $E_T>0$. In the same bases, the electron-phonon interaction yields
	
	\begin{equation}
		\begin{split}
			&H_\text{e-ph}^{(S1)}\\
			&=\mathcal{P}^{(S1)} H_\text{e-ph} \mathcal{P}^{(S1)}\\
			&=\left(
			\begin{array}{cccc}
				2 P_{ R1,R1 }+P_{ L1,L1 }+P_{ R2,R2 } & \sqrt { 2 } P_{ R2,L1 } & 0 & 0 \\
				\sqrt { 2 } P_{ R2,L1 }^{ \dagger  } & 2 P_{ R1,R1 }+2P_{ R2,R2 } & 0 & 0 \\
				0 & 0 & 2 P_{ R1,R1 }+P_{ L1,L1 }+P_{ R2,R2 } & -P_{ R3,L1 } \\
				0 & 0 & -P_{ R3,L1 } & 2 P_{ R1,R1 }+P_{ R2,R2 }+P_{ R3,R3 } 
			\end{array}
			\right),
		\end{split}
	\end{equation}
	where $P_{j,k} = \langle \Phi_{\psi j} | H_\text{e-ph} | \Phi_{\psi k} \rangle$ for $\{\psi j,\psi k\}\in\{L1,R1,R2,R3\}$. Here the notation of $m$-th phonon is dropped as only one phonon is involved. On the other hand, when multi-phonon process is considered, the notation $m$ has to be imposed to explicitly recognize the couplings between different phonons, as shown in the later part of this section.
	
	By diagonalizing each block individually, ${H}_\text{e}^{(S1)}$ can be rewritten in the eigenbases $\{|S'(1,3)\rangle$, $|S'(0,4)\rangle$, $|T'(1,3)\rangle$,  $|T'(0,4^*)\rangle\}$ as
	\begin{equation}\label{eq:STEigenH}
		\begin{split}
			H'^{(S1)}_\text{e}&=\widetilde{U}^\dagger H_\text{e}^{(S1)} \widetilde{U} \\
			&=E_S/2 \left(1-\sqrt{1+8 t^{'2}_{L1,R2}}\right) |S'(1,3)\rangle\langle S'(1,3)|+ E_S/2 \left(1+\sqrt{1+8 t^{'2}_{L1,R2}}\right) |S'(0,4)\rangle\langle S'(0,4)|\\ 
			&\quad+E_T/2 \left(1-\sqrt{1+4 t^{'2}_{L1,R3}}\right) |T'(1,3)\rangle\langle T'(1,3)|+E_T/2 \left(1+\sqrt{1+4 t^{'2}_{L1,R3}}\right)|T'(0,4)\rangle\langle T'(0,4)|\\
			&=E_{|S'(1,3)\rangle} |S'(1,3)\rangle\langle S'(1,3)|+E_{|S'(0,4)\rangle} |S'(0,4)\rangle\langle S'(0,4)|+ E_{|T'(1,3)\rangle} |T'(1,3)\rangle\langle T'(1,3)|\\
			&\quad+E_{|T'(0,4)\rangle} |T'(0,4)\rangle\langle T'(0,4)|,
		\end{split}
	\end{equation}
	where $t'_{L1,R2}=t_{L1,R2}/E_S$ and $t'_{L1,R3}=t_{L1,R3}/E_S$.
	
	The eigenbases are
	\begin{equation}\label{eq:STEigenbases}
		\begin{split}
			|S'(1,3)\rangle &= -\cos \frac{\theta_S}{2}|S(1,3)\rangle + \sin \frac{\theta_S}{2}|S(0,4)\rangle, \\
			|S'(0,4)\rangle &= \sin \frac{\theta_S}{2}|S(1,3)\rangle + \cos \frac{\theta_S}{2}|S(0,4)\rangle, \\
			|T'(1,3)\rangle &= \cos \frac{\theta_T}{2}|T(1,3)\rangle + \sin \frac{\theta_T}{2}|T(0,4^*)\rangle, \\
			|T'(0,4^*)\rangle &= -\sin \frac{\theta_T}{2}|T(1,3)\rangle + \cos \frac{\theta_T}{2}|T(0,4^*)\rangle, \\
		\end{split}
	\end{equation}
	where 
	\begin{equation}\label{eq:thetaExp}
		\begin{split}
			\tan \theta_S&=\frac{2 \sqrt{2} t_{L1,R2}}{E_S}\\
			\tan \theta_T&=\frac{2 t_{L1,R3}}{E_T}.
		\end{split}
	\end{equation}
	Under the same eigenbases, the electron-phonon interaction yields
	\begin{subequations}
		\begin{align}
			H'^{({S1})}_\text{e-ph}=\widetilde{U}^\dagger H^{(S1)}_\text{e-ph} \widetilde{U} = H'^{(S1)}_{S,\text{e-ph}} \oplus H'^{(S1)}_{T,\text{e-ph}},
		\end{align}
		\begin{align}
			H'^{(S1)}_{S,\text{e-ph}}=
			\left(
			\begin{array}{cc}
				2(P_{ R1,R1 }+P_{ R2,R2 })+\cos ^{ 2 } \frac { \theta _{ S } }{ 2 } \left( P_{ L1,L1 }-P_{ R2,R2 } \right)  & -\frac { \sin  \theta _{ S } }{ 2 } \left( P_{ L1,L1 }-P_{ R2,R2 } \right) \\
				-\frac { \sin  \theta _{ S } }{ 2 } \left( P_{ L1,L1 }-P_{ R2,R2 } \right)  & 2(P_{ R1,R1 }+P_{ R2,R2 })+\cos ^{ 2 } \frac { \theta _{ S } }{ 2 } \left( P_{ L1,L1 }-P_{ R2,R2 } \right)  
			\end{array}
			\right),
		\end{align}
		\begin{align}
			H'^{(S1)}_{T,\text{e-ph}}=
			\left(
			\begin{array}{c}
				2P_{ R1,R1 }+\sum_{j= 2,3}
				P_{ Rj,Rj }+\cos ^{ 2 } \frac { \theta _{ T } }{ 2 } \left( P_{ L1,L1 }-P_{ R3,R3 } \right)  \\
				-\frac { \sin  \theta _{ T } }{ 2 } \left( P_{ L1,L1 }-P_{ R3,R3 } \right)  
			\end{array}
			\right.
			\\
			\left.
			\begin{array}{c}\notag
				2P_{ R1,R1 }+\sum_{j= 2,3} -\frac 	{ \sin  \theta _{ T } }{ 2 } \left( P_{ L1,L1 }-P_{ R3,R3 } \right)  \\
				2P_{ R1,R1 }+\sum_{j= 2,3}
				P_{ Rj,Rj }+\cos ^{ 2 } 	\frac { \theta _{ T } }{ 2 } \left( P_{ L1,L1 }-P_{ R3,R3 } \right)
			\end{array}
			\right).
		\end{align}
	\end{subequations}
	
	\begin{subequations}
		\begin{align}
			H'^{(S1)}_\text{e-ph}=\widetilde{U}^\dagger H^{(S1)}_\text{e-ph} \widetilde{U} = H'^{(S1)}_{S,\text{e-ph}} \oplus H'^{(S1)}_{T,\text{e-ph}},
		\end{align}
		\begin{align}
			H'^{(S1)}_{S,\text{e-ph}}=
			\left(
			\begin{array}{cc}
				2(P_{ R1,R1 }+P_{ R2,R2 })+\cos ^{ 2 } \frac { \theta _{ S } }{ 2 } \widetilde{P}  & -\frac { \sin  \theta _{ S } }{ 2 } \widetilde{P} \\
				-\frac { \sin  \theta _{ S } }{ 2 } \widetilde{P}  & 2(P_{ R1,R1 }+P_{ R2,R2 })+\cos ^{ 2 } \frac { \theta _{ S } }{ 2 } \widetilde{P}
			\end{array}
			\right),
		\end{align}
		\begin{align}
			H'^{(S1)}_{T,\text{e-ph}}=
			\left(
			\begin{array}{cc}
				2P_{ R1,R1 }+\sum_{j= 2,3}
				P_{ Rj,Rj }+\cos ^{ 2 } \frac { \theta _{ T } }{ 2 } \overline{P}  & -\frac 	{ \sin  \theta _{ T } }{ 2 } \overline{P}  \\
				-\frac { \sin  \theta _{ T } }{ 2 } \overline{P}  & 2P_{ R1,R1 }+\sum_{j= 2,3}
				P_{ Rj,Rj }+\cos ^{ 2 } 	\frac { \theta _{ T } }{ 2 } \overline{P}
			\end{array}
			\right),
		\end{align}
	\end{subequations}
	where $\widetilde{P}=P_{ L1,L1 }-P_{ R2,R2 }$ and $\overline{P} = P_{ L1,L1 }-P_{ R3,R3 }$ while $S$ and $T$ denote singlet and triplet subspace respectively. Performing Schrieffer-Wolff transformation on  $H'^{(S1)}_\text{e-ph}$ up to second-order in $P_{j,k}/(E_m - E_n)$, where $E_m$ and $E_n$ are the energy values of the eigenbases, we have
	\begin{subequations}
		\begin{align}
			\begin{split}
				H''^{(S1)}_{\text{e-ph}}&=H''^{(S1)}_{S,\text{e-ph}}\oplus H''^{(S1)}_{T,\text{e-ph}},
			\end{split}\\
			\begin{split}
				\langle S'(1,3) | H''^{(S1)}_{\text{e-ph}} | S'(1,3)\rangle &\approx \sum_{\mathbf{q}_1}\bigg(2\left[P_{ R1,R1 }(\mathbf{q}_1)+P_{ R2,R2 }(\mathbf{q}_1)\right]+\cos ^{ 2 } \frac { \theta _{ S } }{ 2 } \widetilde{P}(\mathbf{q}_1)\\
				&\quad+\frac{1}{4\Delta E_{S'}} \sin^2 \theta_{S} \left[P_{L1,L1}(\mathbf{q}_1)-P_{R2,R2}(\mathbf{q}_1)\right]\left[P_{L1,L1}(\mathbf{q}_2)-P_{R2,R2}(\mathbf{q}_2)\right]\bigg),
			\end{split}\\
			\begin{split}
				\langle T'(1,3) | H''^{(S1)}_{\text{e-ph}} | T'(1,3)\rangle &\approx \sum_{\mathbf{q}_1} \bigg(2P_{ R1,R1 }(\mathbf{q}_1)+\sum_{j= 2,3}
				P_{ Rj,Rj }(\mathbf{q}_1)+\cos ^{ 2 } \frac { \theta _{ T } }{ 2 } \overline{P}(\mathbf{q}_1) \\
				&\quad+\frac{1}{4\Delta E_{T'}} \sin^2 \theta_{T} \left[P_{L1,L1}(\mathbf{q}_1)-P_{R3,R3}(\mathbf{q}_1)\right]\left[P_{L1,L1}(\mathbf{q}_2)-P_{R3,R3}(\mathbf{q}_2)\right]\bigg),
			\end{split}
		\end{align}
	\end{subequations}
	where the terms with $\mathbf{q}_2$ involves two-phonon process and $\Delta E_{S'} = E_{|S'(1,3)\rangle}-E_{|S'(0,4)\rangle}<0$ and $\Delta E_{T'} = E_{|T'(1,3)\rangle}-E_{|T'(0,4)\rangle}<0$. Projecting $H''^{(S1)}_\text{e-ph}$ into $\boldsymbol{\sigma}'$, we have
	\begin{equation}\label{eq:deltaB}
		\begin{split}
			\delta B_z&=\langle T'(1,3) | H''^{(S1)}_{\text{e-ph}} | T'(1,3)\rangle - \langle S'(1,3) | H''_{\text{e-ph}} | S'(1,3)\rangle\\
			&= \sum_{\mathbf{q}_1} \left[\cos ^{ 2 } \frac { \theta _{ S } }{ 2 } \widetilde{P}(\mathbf{q}_1) - \cos ^{ 2 } \frac { \theta _{ T } }{ 2 } \overline{P}(\mathbf{q}_1)\right]\\
			&+\sum_{\mathbf{q}_1,\mathbf{q}_2}\Bigg[\frac{1}{4}\left(\frac{1}{\Delta E_{T'}}\sin^2 \theta_T\right)\left[P_{L1,L1}(\mathbf{q}_1)-P_{R3,R3}(\mathbf{q}_1)\right]\left[P_{L1,L1}(\mathbf{q}_2)-P_{R3,R3}(\mathbf{q}_2)\right] \\
			&-\frac{1}{4}\left(\frac{1}{\Delta E_{S'}}\sin^2 \theta_S\right)\left[P_{L1,L1}(\mathbf{q}_1)-P_{R2,R2}(\mathbf{q}_1)\right]\left[P_{L1,L1}(\mathbf{q}_2)-P_{R2,R2}(\mathbf{q}_2)\right]\Bigg]
		\end{split}
	\end{equation}
	
	For conventional singlet-triplet qubit hosted in a DQD with two electrons, only the admixture to doubly occupied singlet state is considered \cite{Kornich.14}. On the other hand, for four-electron DQD device, admixture with $|S(0,4)\rangle$ and $|T(0,4)\rangle$ are comparable, as an extra $\sqrt{2}$ in the definition of $\theta_S$ to compensate for larger $|\Delta E_{S'}|$, cf. Eq.~\eqref{eq:thetaExp}. To understand the phonon induced pure-dephasing due to the admixture with $|S(0,4)\rangle$ and $|T(0,4)\rangle$, we consider only one of them to facilitate discussion, which the main qualitative result can be derived when both admixtures are taken into account. Here, we take $\theta_S\neq0$ while $\theta_T = 0$. 
	From Eq.~\eqref{eq:deltaB}, the two-phonon induced pure-dephasing rate, $\Gamma_\varphi=1/T_\varphi$ is
	\begin{subequations}\label{eq:ratePureDep2Phonon}
		\begin{align}
			\begin{split}
				&\hbar^2 \Gamma_\varphi^\text{2p} = 2\hbar^2 \left(J^+_{zz} (0)\right)^\text{2p}
			\end{split}\\
			\begin{split}
				&= \int_{-\infty}^{\infty} \cos (\omega \tau) \langle \delta B_z(0)\delta B_z(\tau) \rangle_\beta d\tau \big|_{\omega\rightarrow 0}
			\end{split}\\
			\begin{split}
				&= \frac{1}{16\Delta E_{S'}^2}\sin^4 \theta_S \int_{-\infty}^{\infty} d\tau\cos (\omega \tau) \Big\langle 	\left[P_{L1,L1}(\mathbf{q}_1)-P_{R2,R2}(\mathbf{q}_1)\right]\left[P_{L1,L1}(\mathbf{q}_2)-P_{R2,R2}(\mathbf{q}_2)\right]\\
				&\quad\times \left[P_{L1,L1}(\mathbf{q}_3)-P_{R2,R2}(\mathbf{q}_3)\right]\left[P_{L1,L1}(\mathbf{q}_4)-P_{R2,R2}(\mathbf{q}_4)\right] \Big\rangle_\beta \Bigg|_{\omega\rightarrow 0}
			\end{split}\\
			\begin{split}
				&=\frac{1}{16\Delta E_{S'}}\sin^4 \theta_S \sum_{s,\text{cpl}}  \sum_{\mathbf{q}_1,\mathbf{q}_2,\mathbf{q}_3,\mathbf{q}_4}\int_{-\infty}^{\infty} d\tau\cos (\omega_s \tau)  C_{s,\text{cpl}} (\mathbf{q}_1) C_{s,\text{cpl}} (\mathbf{q}_2) C_{s,\text{cpl}} (\mathbf{q}_3) C_{s,\text{cpl}} (\mathbf{q}_4)\\ &\quad \times \Bigg( \left\langle P_{R2,R2}^{s,\text{cpl}}(\mathbf{q}_1) P_{R2,R2}^{s,\text{cpl}}(\mathbf{q}_2) P_{R2,R2}^{s,\text{cpl}}(\mathbf{q}_3) P_{R2,R2}^{s,\text{cpl}}(\mathbf{q}_4)- P_{R2,R2}^{s,\text{cpl}}(\mathbf{q}_1) P_{R2,R2}^{s,\text{cpl}}(\mathbf{q}_2) P_{R2,R2}^{s,\text{cpl}}(\mathbf{q}_3) P_{L1,L1}^{s,\text{cpl}}(\mathbf{q}_4) \right\rangle_\beta\\
				&\quad +\left\langle P_{L1,L1}^{s,\text{cpl}}(\mathbf{q}_1) P_{L1,L1}^{s,\text{cpl}}(\mathbf{q}_2) P_{L1,L1}^{s,\text{cpl}}(\mathbf{q}_3) P_{L1,L1}^{s,\text{cpl}}(\mathbf{q}_4)- P_{L1,L1}^{s,\text{cpl}}(\mathbf{q}_1) P_{L1,L1}^{s,\text{cpl}}(\mathbf{q}_2) P_{L1,L1}^{s,\text{cpl}}(\mathbf{q}_3) P_{R2,R2}^{s,\text{cpl}}(\mathbf{q}_4) \right\rangle_\beta\\
				&\quad +\left\langle P_{R2,R2}^{s,\text{cpl}}(\mathbf{q}_1) P_{L1,L1}^{s,\text{cpl}}(\mathbf{q}_2) P_{R2,R2}^{s,\text{cpl}}(\mathbf{q}_3) P_{L1,L1}^{s,\text{cpl}}(\mathbf{q}_4)- P_{R2,R2}^{s,\text{cpl}}(\mathbf{q}_1) P_{L1,L1}^{s,\text{cpl}}(\mathbf{q}_2) P_{R2,R2}^{s,\text{cpl}}(\mathbf{q}_3) P_{R2,R2}^{s,\text{cpl}}(\mathbf{q}_4) \right\rangle_\beta\\
				&\quad +\left\langle P_{L1,L1}^{s,\text{cpl}}(\mathbf{q}_1) P_{R2,R2}^{s,\text{cpl}}(\mathbf{q}_2) P_{L1,L1}^{s,\text{cpl}}(\mathbf{q}_3) P_{R2,R2}^{s,\text{cpl}}(\mathbf{q}_4)- P_{L1,L1}^{s,\text{cpl}}(\mathbf{q}_1) P_{R2,R2}^{s,\text{cpl}}(\mathbf{q}_2) P_{R2,R2}^{s,\text{cpl}}(\mathbf{q}_3) P_{R2,R2}^{s,\text{cpl}}(\mathbf{q}_4) \right\rangle_\beta\\
				&\quad +\left\langle P_{R2,R2}^{s,\text{cpl}}(\mathbf{q}_1) P_{L1,L1}^{s,\text{cpl}}(\mathbf{q}_2) P_{L1,L1}^{s,\text{cpl}}(\mathbf{q}_3) P_{R2,R2}^{s,\text{cpl}}(\mathbf{q}_4)- P_{R2,R2}^{s,\text{cpl}}(\mathbf{q}_1) P_{L1,L1}^{s,\text{cpl}}(\mathbf{q}_2) P_{L1,L1}^{s,\text{cpl}}(\mathbf{q}_3) P_{L1,L1}^{s,\text{cpl}}(\mathbf{q}_4) \right\rangle_\beta\\
				&\quad +\left\langle P_{L1,L1}^{s,\text{cpl}}(\mathbf{q}_1) P_{R2,R2}^{s,\text{cpl}}(\mathbf{q}_2) P_{R2,R2}^{s,\text{cpl}}(\mathbf{q}_3) P_{L1,L1}^{s,\text{cpl}}(\mathbf{q}_4)- P_{L1,L1}^{s,\text{cpl}}(\mathbf{q}_1) P_{R2,R2}^{s,\text{cpl}}(\mathbf{q}_2) P_{L1,L1}^{s,\text{cpl}}(\mathbf{q}_3) P_{L1,L1}^{s,\text{cpl}}(\mathbf{q}_4) \right\rangle_\beta\\
				&\quad +\left\langle P_{R2,R2}^{s,\text{cpl}}(\mathbf{q}_1) P_{R2,R2}^{s,\text{cpl}}(\mathbf{q}_2) P_{L1,L1}^{s,\text{cpl}}(\mathbf{q}_3) P_{L1,L1}^{s,\text{cpl}}(\mathbf{q}_4)- P_{R2,R2}^{s,\text{cpl}}(\mathbf{q}_1) P_{R2,R2}^{s,\text{cpl}}(\mathbf{q}_2) P_{L1,L1}^{s,\text{cpl}}(\mathbf{q}_3) P_{R2,R2}^{s,\text{cpl}}(\mathbf{q}_4) \right\rangle_\beta\\
				&\quad +\left\langle P_{L1,L1}^{s,\text{cpl}}(\mathbf{q}_1) P_{L1,L1}^{s,\text{cpl}}(\mathbf{q}_2) P_{R2,R2}^{s,\text{cpl}}(\mathbf{q}_3) P_{R2,R2}^{s,\text{cpl}}(\mathbf{q}_4)- P_{L1,L1}^{s,\text{cpl}}(\mathbf{q}_1) P_{L1,L1}^{s,\text{cpl}}(\mathbf{q}_2) P_{R2,R2}^{s,\text{cpl}}(\mathbf{q}_3) P_{L1,L1}^{s,\text{cpl}}(\mathbf{q}_4) \right\rangle_\beta\Bigg)\Bigg|_{\omega_s\rightarrow 0},
			\end{split}\\
			\begin{split}\label{eq:expand}
				&=\frac{1}{16\Delta E_{S'}}\sin^4 \theta_S   \sum_{\mathbf{q}_1,\mathbf{q}_2,\mathbf{q}_3,\mathbf{q}_4}\int_{-\infty}^{\infty} d\tau\cos (\omega_s \tau)  \\ &\quad \times \Bigg( \left\langle P_{R2,R2}(\mathbf{q}_1) P_{R2,R2}(\mathbf{q}_2) P_{R2,R2}(\mathbf{q}_3) P_{R2,R2}(\mathbf{q}_4)- P_{R2,R2}(\mathbf{q}_1) P_{R2,R2}(\mathbf{q}_2) P_{R2,R2}(\mathbf{q}_3) P_{L1,L1}(\mathbf{q}_4) \right\rangle_\beta\\
				&\quad +\left\langle P_{L1,L1}(\mathbf{q}_1) P_{L1,L1}(\mathbf{q}_2) P_{L1,L1}(\mathbf{q}_3) P_{L1,L1}(\mathbf{q}_4)- P_{L1,L1}(\mathbf{q}_1) P_{L1,L1}(\mathbf{q}_2) P_{L1,L1}(\mathbf{q}_3) P_{R2,R2}(\mathbf{q}_4) \right\rangle_\beta\\
				&\quad +\left\langle P_{R2,R2}(\mathbf{q}_1) P_{L1,L1}(\mathbf{q}_2) P_{R2,R2}(\mathbf{q}_3) P_{L1,L1}(\mathbf{q}_4)- P_{R2,R2}(\mathbf{q}_1) P_{L1,L1}(\mathbf{q}_2) P_{R2,R2}(\mathbf{q}_3) P_{R2,R2}(\mathbf{q}_4) \right\rangle_\beta\\
				&\quad +\left\langle P_{L1,L1}(\mathbf{q}_1) P_{R2,R2}(\mathbf{q}_2) P_{L1,L1}(\mathbf{q}_3) P_{R2,R2}(\mathbf{q}_4)- P_{L1,L1}(\mathbf{q}_1) P_{R2,R2}(\mathbf{q}_2) P_{R2,R2}(\mathbf{q}_3) P_{R2,R2}(\mathbf{q}_4) \right\rangle_\beta\\
				&\quad +\left\langle P_{R2,R2}(\mathbf{q}_1) P_{L1,L1}(\mathbf{q}_2) P_{L1,L1}(\mathbf{q}_3) P_{R2,R2}(\mathbf{q}_4)- P_{R2,R2}(\mathbf{q}_1) P_{L1,L1}(\mathbf{q}_2) P_{L1,L1}(\mathbf{q}_3) P_{L1,L1}(\mathbf{q}_4) \right\rangle_\beta\\
				&\quad +\left\langle P_{L1,L1}(\mathbf{q}_1) P_{R2,R2}(\mathbf{q}_2) P_{R2,R2}(\mathbf{q}_3) P_{L1,L1}(\mathbf{q}_4)- P_{L1,L1}(\mathbf{q}_1) P_{R2,R2}(\mathbf{q}_2) P_{L1,L1}(\mathbf{q}_3) P_{L1,L1}(\mathbf{q}_4) \right\rangle_\beta\\
				&\quad +\left\langle P_{R2,R2}(\mathbf{q}_1) P_{R2,R2}(\mathbf{q}_2) P_{L1,L1}(\mathbf{q}_3) P_{L1,L1}(\mathbf{q}_4)- P_{R2,R2}(\mathbf{q}_1) P_{R2,R2}(\mathbf{q}_2) P_{L1,L1}(\mathbf{q}_3) P_{R2,R2}(\mathbf{q}_4) \right\rangle_\beta\\
				&\quad +\left\langle P_{L1,L1}(\mathbf{q}_1) P_{L1,L1}(\mathbf{q}_2) P_{R2,R2}(\mathbf{q}_3) P_{R2,R2}(\mathbf{q}_4)- P_{L1,L1}(\mathbf{q}_1) P_{L1,L1}(\mathbf{q}_2) P_{R2,R2}(\mathbf{q}_3) P_{L1,L1}(\mathbf{q}_4) \right\rangle_\beta\Bigg)\Bigg|_{\omega_s\rightarrow 0},\\
			\end{split}
		\end{align}
	\end{subequations}
	where we have dropped the variable of time for simplicity, $\tau$, which should be considered implicitly such that $P_{\psi j,\psi k}(\mathbf{q}_m)=P_{\psi j,\psi k}(\mathbf{q}_m,\tau)$ while $\tau = 0$ for $m=1,2$ and $\tau \neq 0$ for $m=3,4$. The superscript ``2p'' in $\Gamma_\varphi^\text{2p}$ denotes two-phonon process. Following the notations in Sec.~\ref{subsec:JjjTwoPhonon}, Eq.~\eqref{eq:Ppsi12Notation} and Eq.~\eqref{eq:Itilde}-\eqref{eq:IdivC}, we have
	\begin{subequations}
		\begin{align}
			\begin{split}
				\gamma^\psi &= \frac{1}{16\Delta E_{S'}}\sin^4 \theta_S,
			\end{split}\\
			\begin{split}
				\widetilde{\mathcal{I}}^{\psi = \substack{\psi1,\psi2,\psi3,\psi4, \psi5,\psi6,\psi7,\psi8}}_{s,\text{cpl}} &= \frac{1}{2\hbar^2}P^{s,\text{cpl}}_{\psi 1, \psi 2}(\mathbf{q}_1)P^{s,\text{cpl}}_{\psi 3, \psi 4}(\mathbf{q}_2)P^{s,\text{cpl}}_{\psi 5, \psi 6}(\mathbf{q}_3)P^{s,\text{cpl}}_{\psi 7, \psi 8}(\mathbf{q}_4),
			\end{split}\\
			\begin{split}
				\mathcal{I}^{\psi}_{s,\text{cpl}} &=\frac{1}{2\hbar^2}\sum_{\mathbf{q}_1,\mathbf{q}_2,\mathbf{q}_3,\mathbf{q}_4}\int_{-\infty}^{\infty} d\tau\cos (\omega_s \tau)  C_{s,\text{cpl}}(\mathbf{q}_1)C_{s,\text{cpl}}(\mathbf{q}_2)C_{s,\text{cpl}}(\mathbf{q}_3)C_{s,\text{cpl}}(\mathbf{q}_4)\widetilde{\mathcal{I}}^{\psi}_{s,\text{cpl}}.
			\end{split}\\
			\begin{split}
				\mathcal{I}^{\psi} &=\sum_{s,\text{cpl}} \mathcal{I}^{\psi}_{s,\text{cpl}},
			\end{split}
		\end{align}
	\end{subequations}
	where $\mathcal{I}^\psi \approx \mathcal{I}^\psi_{l,\text{dp}}$ for GaAs QD device.
	For the case in which two electrons occupying a DQD device \cite{Kornich.14}, $\Gamma_\varphi^\text{2p}$ has the same form as Eq.~\eqref{eq:ratePureDep2Phonon} with all the subscripts $R2$ replaced by $R1$ and the modified Hubbard parameters defined as 
	\begin{equation}\label{eq:2esCase}
		\begin{split}
			\tan \theta_S &= 2\sqrt{2}t_{L1,R1}/E_S,\\
			\Delta E_{S'} &= E_{|S'(1,1)\rangle} - E_{|S'(0,2)\rangle},\\
			E_{|S'(1,1)\rangle} &= E_S /2 \left(1-\sqrt{1+8 t'^2_{L1,R1}}\right),\\
			E_{|S'(0,2)\rangle} &= E_S /2 \left(1+\sqrt{1+8 t'^2_{L1,R1}}\right),\\
			t'^2_{L1,R1} &= t_{L1,R1}/E_S, \\
			E_S&=U_{R1}-(U_{L1,R1}+2\Delta).
		\end{split}
	\end{equation}
	
	Here, we show explicitly some of $\mathcal{I}^\psi$ terms presented in Eq.~\eqref{eq:expand}:
	\begin{subequations}\label{eq:IpsiExplicit}
		\begin{align}
			\begin{split}\label{eq:allGround}
				&\mathcal{I}_{l,\text{dp}}^{L1,L1,L1,L1,L1,L1,L1,L1}\\
				&=\frac{32\pi^2}{2\hbar^2\nu_s}\left(\prod_{j\in \{1,3,5,7\}} \frac{\pi \mathcal{G}_{\eta(\psi j)}^2}{\alpha_{\eta(\psi j)}}\right) 
				\int_{0}^{\infty} dq_3 
				\int_{0}^{\pi} \sin \theta_{\mathbf{q}_3} d\theta_{\mathbf{q}_3} 
				\int_{0}^{\pi} \sin \theta_{\mathbf{q}_4} d\theta_{\mathbf{q}_4} 
				\Bigg\{ n_s (q_3)\left[n_s (q_3)+1\right]\mathcal{Q}_z(\mathbf{q}_3,\mathbf{q}_3)C^4_{s,\text{cpl}} (|\mathbf{q}_3|)\\
				&\quad\times\exp\left[-\frac{l_L^2  q_3^2 \sin^2 \theta_{\mathbf{q}_3}}{2}\right] 
				\exp\left[-\frac{l_L^2  q_3^2 \sin^2 \theta_{\mathbf{q}_4}}{2}\right]\Bigg\},
			\end{split}\\
			\begin{split}\label{eq:oneGroundDiffDot}
				&\mathcal{I}_{l,\text{dp}}^{L1,L1,L1,L1,L1,L1,R1,R1}\\
				&=\frac{32\pi^2}{2\hbar^2\nu_s}\left(\prod_{j\in \{1,3,5,7\}} \frac{\pi \mathcal{G}_{\eta(\psi j)}^2}{\alpha_{\eta(\psi j)}}\right) 
				\int_{0}^{\infty} dq_3 
				\int_{0}^{\pi} \sin \theta_{\mathbf{q}_3} d\theta_{\mathbf{q}_3} 
				\int_{0}^{\pi} \sin \theta_{\mathbf{q}_4} d\theta_{\mathbf{q}_4} 
				\Big\{ n_s (q_3)\left[n_s (q_3)+1\right]\mathcal{Q}_z(\mathbf{q}_3,\mathbf{q}_3)C^4_{s,\text{cpl}} (|\mathbf{q}_3|)\\
				&\quad\times\exp\left[-\frac{l_L^2  q_3^2 \sin^2 \theta_{\mathbf{q}_3}}{2}\right] 
				\exp\left[-\frac{(l_L^2+l_R^2)  q_3^2 \sin^2 \theta_{\mathbf{q}_4}}{4}\right]  
				\text{BesselJ}\left[0,2 x_0 q_3 \sin \theta_{\mathbf{q}_4}\right]\Big\},
			\end{split}\\
			\begin{split}\label{eq:oneGroundDiffDotDiffOrbital}
				&\mathcal{I}_{l,\text{dp}}^{L1,L1,L1,L1,L1,L1,R2,R2}\\
				&=\frac{32\pi^2}{2\hbar^2\nu_s}\left(\prod_{j\in \{1,3,5,7\}} \frac{\pi \mathcal{G}_{\eta(\psi j)}^2}{\alpha_{\eta(\psi j)}}\right) \int_{0}^{\infty} dq_3 
				\int_{0}^{\pi} \sin \theta_{\mathbf{q}_3} d\theta_{\mathbf{q}_3} 
				\int_{0}^{\pi} \sin \theta_{\mathbf{q}_4} d\theta_{\mathbf{q}_4} \Big\{ n_s (q_3)\left[n_s (q_3)+1\right]\mathcal{Q}_z(\mathbf{q}_3,\mathbf{q}_3)C^4_{s,\text{cpl}} (|\mathbf{q}_3|)\\
				&\quad\times\exp\left[-\frac{l_L^2  q_3^2 \sin^2 \theta_{\mathbf{q}_3}}{2}\right] 
				\exp\left[-\frac{(l_L^2+l_R^2)  q_3^2 \sin^2 \theta_{\mathbf{q}_4}}{4}\right]  
				\left( 1-\frac{l_R^2q_3^2 \sin^2 \theta_{\mathbf{q}_4}}{4} \right) \text{BesselJ}\left[0,2 x_0 q_3 \sin \theta_{\mathbf{q}_4}\right]\Big\}.
			\end{split}
		\end{align}
	\end{subequations}
	Eq.~\eqref{eq:IpsiExplicit} shows $\mathcal{I}_{l,\text{dp}}$ for six of the inner products, $\langle \Phi_{\psi j} | H_\text{dp}| \Phi_{\psi k}\rangle$, are all consist of the ground orbitals in the same dot L (L1) while the last inner product consists of the ground orbital in dot L (L1), Eq.~\eqref{eq:allGround}, ground orbital in dot R (R1), Eq.~\eqref{eq:oneGroundDiffDot} and first excited orbital in dot R (R2), Eq.~\eqref{eq:oneGroundDiffDotDiffOrbital}. Eq.~\eqref{eq:oneGroundDiffDot} shows that an introduction of ground orbital in different QD gives rise to a Bessel function as $\Delta h_{7,8}\neq 0$, cf. Eq.~\eqref{eq:groundOctets}, while Eq.~\eqref{eq:oneGroundDiffDotDiffOrbital} shows that an introduction of first excited orbital in different QD gives rise to a Bessel function and a prefactor $1-(l_R^2q_3^2\sin^2\theta_{\mathbf{q}_4})/4$. We denote the integrand of the integral, $\mathcal{I}_{s,\text{cpl}}^\psi$, as $\widehat{\mathcal{I}}_{s,\text{cpl}}^\psi (q_3,\theta_{\mathbf{q}_3},\theta_{\mathbf{q}_4})$, such that $\mathcal{I}_{s,\text{cpl}}^\psi = \int \int\int dq_3 d\theta_{\mathbf{q}_3}\theta_{\mathbf{q}_4} \widehat{\mathcal{I}}_{s,\text{cpl}}^\psi (q_3,\theta_{\mathbf{q}_3},\theta_{\mathbf{q}_4})$.
	
	\begin{table}[t]
		\begin{tabular}{p{0.5cm}|p{1.8cm}|p{4.4cm}|p{2.2cm}|p{0.4cm}|p{4.4cm}|p{2.2cm}|p{0.4cm}|}
			\cline{2-8} 
			& &\multicolumn{3}{c|}{} & \multicolumn{3}{c|}{} \\[-10pt]
			& \hfil \multirow{3}{*}{$\gamma^\psi$ (eV$^{-2}$)}&\multicolumn{3}{c|}{ 1$^{\text{st}}$ $\mathcal{I}^\psi$ in $\langle \cdots \rangle_\beta$}  & \multicolumn{3}{c|}{ 2$^{\text{nd}}$ $\mathcal{I}^\psi$ in $\langle \cdots \rangle_\beta$} \\
			\cline{3-8}
			& &\hfil  & \hfil  & \hfil  & \hfil   & \hfil & \\[-10pt]
			& &\hfil \multirow{2}{*}{$\mathcal{I}^\psi$} & \hfil numeric &\hfil \multirow{2}{*}{$\mathcal{D}$} & \hfil \multirow{2}{*}{$\mathcal{I}^\psi$} &  \hfil numeric & \hfil \multirow{2}{*}{$\mathcal{D}$} \\
			& & & \hfil ($10^{-3}$eV$^2$Hz)  &  &  &  \hfil ($10^{-3}$eV$^2$Hz) & \\
			\cline{2-8}
			& &\hfil  & \hfil  & \hfil   & \hfil   & \hfil   & \\[-10pt]
			&\hfil \multirow{11}{*}{$1.846\times10^5$} &\hfil $ P_{R1,R1} P_{R1,R1} P_{R1,R1} P_{R1,R1}$ & \hfil 2.585&\hfil 0 &  \hfil $P_{R1,R1}P_{R1,R1}P_{R1,R1}P_{L1,L1}$ & \hfil  2.018 & \hfil 2\\
			\cline{3-8}
			&&\hfil  & \hfil  & \hfil   & \hfil & \hfil & \\[-10pt]
			&&\hfil $ P_{L1,L1} P_{L1,L1} P_{L1,L1} P_{L1,L1}$ & \hfil 2.585 & \hfil 0 & \hfil  $ P_{L1,L1} P_{L1,L1} P_{L1,L1} P_{R1,R1}$ & \hfil 2.018 &\hfil 2\\
			\cline{3-8}
			&&\hfil  & \hfil  & \hfil   & \hfil & \hfil & \\[-10pt]
			&&\hfil  $ P_{R1,R1} P_{L1,L1} P_{R1,R1} P_{L1,L1}$ & \hfil 2.166 & \hfil 2 & \hfil  $ P_{R1,R1} P_{L1,L1} P_{R1,R1} P_{R1,R1}$ &\hfil 2.018 &\hfil 2 \\
			\cline{3-8}
			&&\hfil  & \hfil  & \hfil   & \hfil & \hfil & \\[-10pt]
			&&\hfil $ P_{L1,L1} P_{R1,R1} P_{L1,L1} P_{R1,R1}$ & \hfil 2.166 & \hfil 2  & \hfil $ P_{L1,L1} P_{R1,R1} P_{R1,R1} P_{R1,R1}$  & \hfil 2.018 & \hfil 2 \\ 
			\cline{3-8}
			&&\hfil  & \hfil  & \hfil   & \hfil & \hfil & \\[-10pt]
			&&\hfil  $ P_{R1,R1} P_{L1,L1} P_{L1,L1} P_{R1,R1}$ & \hfil 2.166 &\hfil 2 &\hfil $ P_{R1,R1} P_{L1,L1} P_{L1,L1} P_{L1,L1}$  & \hfil  2.018 & \hfil 2  \\
			\cline{3-8}
			&&\hfil  & \hfil  & \hfil   & \hfil & \hfil & \\[-10pt]
			&&\hfil $ P_{L1,L1} P_{R1,R1} P_{R1,R1} P_{L1,L1}$ & \hfil 2.166 & \hfil 4 & \hfil  $ P_{L1,L1} P_{R1,R1} P_{L1,L1} P_{L1,L1}$ & \hfil 2.018 & \hfil 2 \\
			\cline{3-8}
			&&\hfil  & \hfil  & \hfil   & \hfil & \hfil &\\[-10pt]
			&&\hfil $ P_{R1,R1} P_{R1,R1} P_{L1,L1} P_{L1,L1}$ & \hfil 1.746 & \hfil 4 & \hfil  $ P_{R1,R1} P_{R1,R1} P_{L1,L1} P_{R1,R1}$ & \hfil 2.018& \hfil 2\\
			\cline{3-8}
			&&\hfil  & \hfil  & \hfil   & \hfil & \hfil & \\[-10pt]
			(a)&&\hfil $ P_{L1,L1} P_{L1,L1} P_{R1,R1} P_{R1,R1}$  & \hfil 1.746 & \hfil 4 & \hfil  $ P_{L1,L1} P_{L1,L1} P_{R1,R1} P_{L1,L1}$ & \hfil 2.018 & \hfil 2\\
			\cline{2-8}
			\multicolumn{8}{c}{} \\[5pt]
			\cline{2-8} 
			&&\multicolumn{3}{c|}{} & \multicolumn{3}{c|}{} \\[-10pt]
			&\hfil \multirow{3}{*}{$\gamma^\psi$ (eV$^{-2}$)}&\multicolumn{3}{c|}{ 1$^{\text{st}}$ $\mathcal{I}^\psi$ in $\langle \cdots \rangle_\beta$}  & \multicolumn{3}{c|}{ 2$^{\text{nd}}$ $\mathcal{I}^\psi$ in $\langle \cdots \rangle_\beta$} \\
			\cline{3-8}
			&&\hfil  & \hfil  & \hfil  & \hfil   & \hfil & \\[-10pt]
			&&\hfil \multirow{2}{*}{$\mathcal{I}^\psi$} & \hfil numeric &\hfil \multirow{2}{*}{$\mathcal{D}$} & \hfil \multirow{2}{*}{$\mathcal{I}^\psi$} &  \hfil numeric & \hfil \multirow{2}{*}{$\mathcal{D}$} \\
			&& & \hfil ($10^{-3}$eV$^2$Hz)  &  &  &  \hfil ($10^{-3}$eV$^2$Hz) & \\
			\cline{2-8}
			&&\hfil  & \hfil  & \hfil   & \hfil   & \hfil   & \\[-10pt]
			&\hfil \multirow{11}{*}{$1.230\times10^3$} &\hfil $ P_{R2,R2} P_{R2,R2} P_{R2,R2} P_{R2,R2}$ & \hfil 2.764&\hfil 0 &  \hfil $P_{R2,R2}P_{R2,R2}P_{R2,R2}P_{L1,L1}$ & \hfil  2.729 & \hfil 2\\
			\cline{3-8}
			&&\hfil  & \hfil  & \hfil   & \hfil & \hfil & \\[-10pt]
			&&\hfil $ P_{L1,L1} P_{L1,L1} P_{L1,L1} P_{L1,L1}$ & \hfil 2.774 & \hfil 0 & \hfil  $ P_{L1,L1} P_{L1,L1} P_{L1,L1} P_{R2,R2}$ & \hfil 2.733 &\hfil 2\\
			\cline{3-8}
			&&\hfil  & \hfil  & \hfil   & \hfil & \hfil & \\[-10pt]
			&&\hfil  $ P_{R2,R2} P_{L1,L1} P_{R2,R2} P_{L1,L1}$ & \hfil 2.732 & \hfil 2 & \hfil  $ P_{R2,R2} P_{L1,L1} P_{R2,R2} P_{R2,R2}$ &\hfil 2.729 &\hfil 2 \\
			\cline{3-8}
			&&\hfil  & \hfil  & \hfil   & \hfil & \hfil & \\[-10pt]
			&&\hfil $ P_{L1,L1} P_{R2,R2} P_{L1,L1} P_{R2,R2}$ & \hfil 2.732 & \hfil 2  & \hfil $ P_{L1,L1} P_{R2,R2} P_{R2,R2} P_{R2,R2}$  & \hfil 2.729 & \hfil 2 \\ 
			\cline{3-8}
			&&\hfil  & \hfil  & \hfil   & \hfil & \hfil & \\[-10pt]
			&&\hfil  $ P_{R2,R2} P_{L1,L1} P_{L1,L1} P_{R2,R2}$ & \hfil 2.732 &\hfil 2 &\hfil $ P_{R2,R2} P_{L1,L1} P_{L1,L1} P_{L1,L1}$  & \hfil  2.733 & \hfil 2  \\
			\cline{3-8}
			&&\hfil  & \hfil  & \hfil   & \hfil & \hfil & \\[-10pt]
			&&\hfil $ P_{L1,L1} P_{R2,R2} P_{R2,R2} P_{L1,L1}$ & \hfil 2.732 & \hfil 4 & \hfil  $ P_{L1,L1} P_{R2,R2} P_{L1,L1} P_{L1,L1}$ & \hfil 2.733 & \hfil 2 \\
			\cline{3-8}
			&&\hfil  & \hfil  & \hfil   & \hfil & \hfil &\\[-10pt]
			&&\hfil $ P_{R2,R2} P_{R2,R2} P_{L1,L1} P_{L1,L1}$ & \hfil 2.695 & \hfil 4 & \hfil  $ P_{R2,R2} P_{R2,R2} P_{L1,L1} P_{R2,R2}$ & \hfil 2.729& \hfil 2\\
			\cline{3-8}
			&&\hfil  & \hfil  & \hfil   & \hfil & \hfil & \\[-10pt]
			(b)&&\hfil $ P_{L1,L1} P_{L1,L1} P_{R2,R2} P_{R2,R2}$  & \hfil 2.695 & \hfil 4 & \hfil  $ P_{L1,L1} P_{L1,L1} P_{R2,R2} P_{L1,L1}$ & \hfil 2.733 & \hfil 2\\
			\cline{2-8}
		\end{tabular}
		\caption{(a) Two-phonon pure dephasing rate, $\Gamma_\varphi^{\text{2p}}$, of two-electrons occupying a DQD device. The parameters are \cite{Kornich.14}: $\hbar \omega_L = \hbar \omega_R = 0.124$meV, $x_0=200$nm, $B_{||}=0.7$ T, $B_{\perp}=0$, $T = 50$mK, $l_\text{D} = 1\mu$m, $l_\text{R}=2\mu$m, $h=0.14\mu$eV, $a_z=2$nm,$U_{R1}=1$meV, $t_{L1,R1}=7.25\mu$eV, $2\Delta\sim 0.9$meV such that the exchange splitting $\hbar \omega_z =1.43\mu$eV. (b) Two-phonon pure dephasing rate, $\Gamma_\varphi^{\text{2p}}$, of four-electrons occupying a DQD device. The parameters are : $\hbar \omega_L =7.5$meV, $\hbar \omega_R = 4$meV, $x_0=40$nm, $B_{||}=0$ T, $B_{\perp}=0.319$T, $T = 50$mK, $l_\text{D} = 1.26\mu$m, $l_\text{R}=1.72\mu$m, $h=0.14\mu$eV, $a_z=2$nm, $U_{R2}=50.349$meV, $U_{L1,R2}=41.790$meV, $U_{R2,R3}=50.209$meV, $\Delta E = 0.564$meV, $\xi=0.485$meV, $t_{L1,R2}=27.700\mu$eV, $t_{L1,R3}=12.000\mu$eV, $\Delta= 4.154$meV such that the exchange splitting $\hbar \omega_z =4.414\mu$eV. Note that the tunneling values here is slightly small than that shown in Sec.~\ref{sec:CoressHubbardCI}. This is due to the fact when $2\Delta < \varepsilon^*$, the tunneling value has to be modified to give a more accurate description of the exchange energy using Hubbard model. This issue can be attributed to the smallness of the exchange energy compared to the raw eigenvalues of the four-electron bases, which the former is in the scale of $1\mu$eV, while the later  are in the scale of $10$meV.}
		\label{tab:2esDQDPuredephase}
	\end{table}
	
	Ref.~\onlinecite{Kornich.14} shows that phonon-mediated pure-dephasing rate, $\Gamma_\varphi$, is stronger than relaxation rate, $\Gamma_1$ for $T<1$K. In particular, at $T=50$mK, which is within the temperature range realized for GaAs QD device \cite{Dial.13,Nichol.17}, $\Gamma_\varphi \approx 30 \Gamma_1$. Since one-phonon process does not contribute to the pure-dephasing effect, $\Gamma_\varphi$ solely consist of two-phonon process, i.e. $\Gamma_\varphi= \Gamma_\varphi^\text{2p}$, where higher number of phonon process is not taken into account \cite{Kornich.14}. Table.~\ref{tab:2esDQDPuredephase}(a) shows the two-phonon pure-dephasing rate, $\Gamma_\varphi^\text{2p}$, when the parameters in Ref.~\onlinecite{Kornich.14} is employed, while Table.~\ref{tab:2esDQDPuredephase}(b) shows $\Gamma_\varphi^\text{2p}$ for the parameters of the DQD devices we propose to host a robust capacitive entangling gate. To obtain an insight of $\Gamma_\varphi^\text{2p}$, the main contributing terms are listed out in Table.~\ref{tab:2esDQDPuredephase}(a) and \ref{tab:2esDQDPuredephase}(b) based on Eq.~\eqref{eq:expand}. Note that the terms $P_{\psi j, \psi k}$ in Eq.~\eqref{eq:expand} have been grouped such that each thermal average term, $\langle \cdots \rangle_\beta$, encloses two $\prod_{j\in \{1,3,5,7\}} P_{\psi (2j-1) ,\psi (2j)}$ terms, which are inspected side-by-side in Table.~\ref{tab:2esDQDPuredephase}(a) and \ref{tab:2esDQDPuredephase}(b). 
	
	Table.~\ref{tab:2esDQDPuredephase}(a) shows that the strong pure-dephasing can be attributed to the difference of 1$^\text{st}$ and 2$^\text{nd}$ $\mathcal{I}^\psi$ term. Although 7-th and 8-th row shows a negative contribution to the $\Gamma_\varphi^\text{2p}$, however, the amount of suppression is compensated by other terms, e.g. the positive contribution to $\Gamma_\varphi^\text{2p}$ by 3-rd to 6-th row, resulting in considerable value of $\Gamma_\varphi^\text{2p}$. On the other hand, Table.~\ref{tab:2esDQDPuredephase}(b) shows that the difference between 1$^\text{st}$ and 2$^\text{nd}$ $\mathcal{I}^\psi$ term is largely suppressed, expecting a weak two-phonon pure-dephasing effect. 
	
	\begin{figure}[t]
		\includegraphics[width=0.7\columnwidth]{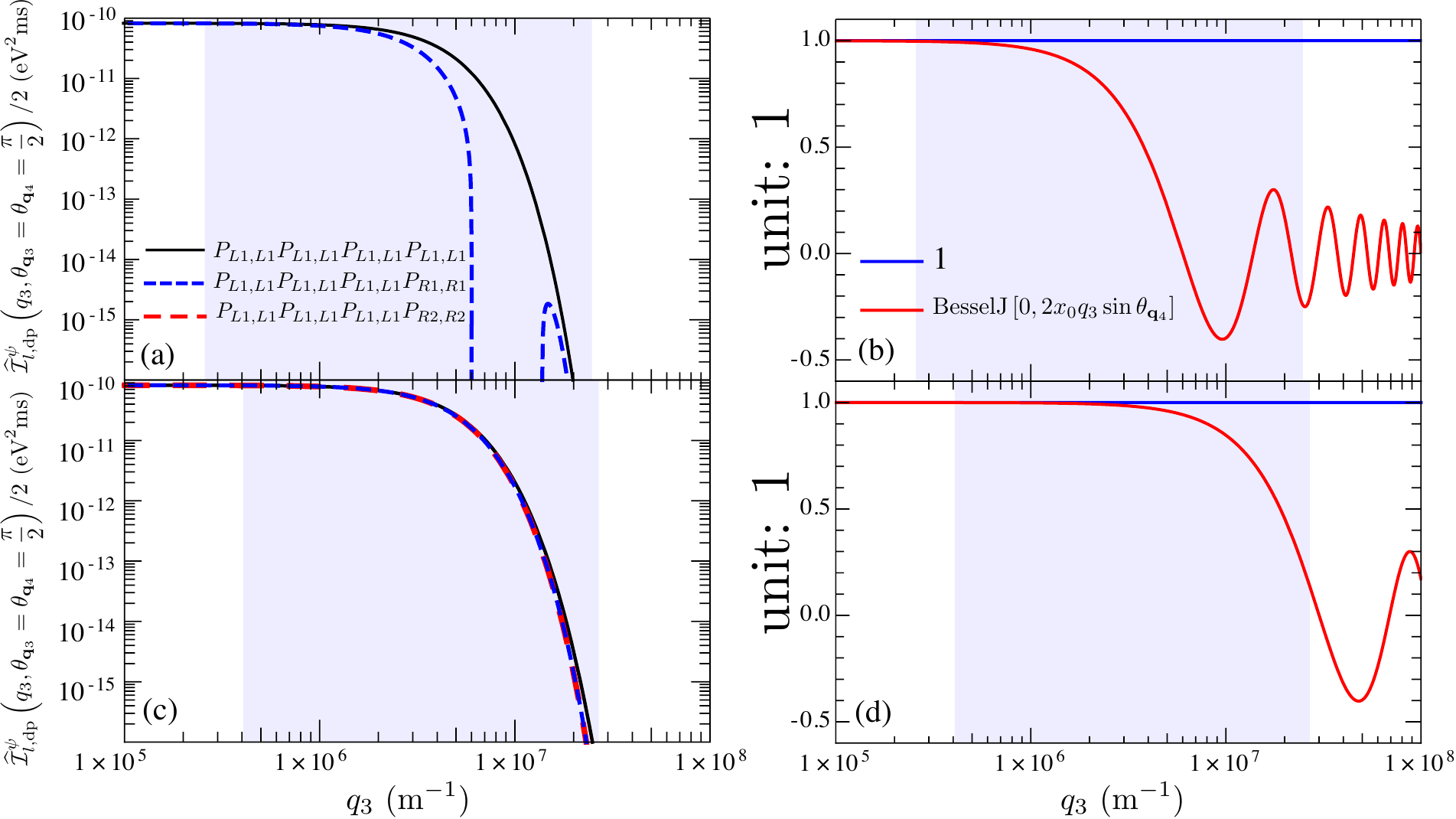}
		\caption{(a) and (c) The integrand $\widehat{\mathcal{I}}_{l,dp}^\psi/2$ as function of $q_3$ at $\theta_{\mathbf{q}_3} = \theta_{\mathbf{q}_4} = \pi/2$. (b) and (d) The plot of Bessel function in Eq.~\eqref{eq:oneGroundDiffDot} and Eq.~\eqref{eq:oneGroundDiffDotDiffOrbital} (red solid curve) as function of $q_3$. (a) and (b) show the results for parameters \cite{Kornich.14}: $\hbar \omega_L = \hbar \omega_R = 0.124$meV, $x_0=200$nm, $T = 50$mK. (c) and (d) show the results for parameters: $\hbar \omega_L = 7.5$meV $\hbar \omega_R = 4$meV, $x_0=40$nm, $T = 50$mK. The light blue region denotes the range of $q_3$ which gives rise to at least $90\%$ of the integral at $\theta_{\mathbf{q}_3} = \theta_{\mathbf{q}_4} = \pi/2$.}
		\label{fig:compare2es4esCase}
	\end{figure}
	
	Figure~\ref{fig:compare2es4esCase}(a) and (c) shows the integrand, $\widehat{\mathcal{I}}_{l,dp}$, by deformation potential (dp), of two set of parameters shown in Table.~\ref{tab:2esDQDPuredephase}(a) and Table.~\ref{tab:2esDQDPuredephase}(b), respectively. $\widehat{\mathcal{I}}_{l,dp}$ is plotted as function of $q_3$ at $\theta_{\mathbf{q}_3}=\theta_{\mathbf{q}_4}=\pi/2$, the point where $\widehat{\mathcal{I}}_{l,dp}$ yields that largest value in the integration domain. For the parameters set in Table.~\ref{tab:2esDQDPuredephase}(a) \cite{Kornich.14}, Fig.~\ref{fig:compare2es4esCase}(a) shows that an additional Bessel function prefactor suppresses the integrand at large $q_3$, confirming with the claim in Eq.~\eqref{eq:BesselSmall} (black solid versus dash blue lines). The suppression by Bessel function can be understood from Fig.~\ref{fig:compare2es4esCase}(b). In the range of $q_3$ which gives rise to most of the integral, cf. light blue region in Fig.~\ref{fig:compare2es4esCase}(b), Bessel function introduces oscillation to the integrand, $\widehat{\mathcal{I}}_{l,dp}$, with suppressed absolute value as function of $q_3$. On the other hand, for the parameters set in Table.~\ref{tab:2esDQDPuredephase}(b), Fig.~\ref{fig:compare2es4esCase} (c) shows negligible modification by additional Bessel function in the $\widehat{\mathcal{I}}_{l,dp}$. From Fig.~\ref{fig:compare2es4esCase} (d), it can be observed that Bessel function starts to oscillate at larger $q_3$, which can be attributed to smaller inter-dot distance, $x_0 = 40$nm for Table.~\ref{tab:2esDQDPuredephase}(b) while $x_0 = 200$nm for Table.~\ref{tab:2esDQDPuredephase}(a) (cf. Eq.~\eqref{eq:IpsiExplicit}). Furthermore, in the former case, the integrand yields larger value as function of $q_3$ (cf. Eq.~\eqref{eq:IpsiExplicit}) due to smaller QD length (stronger confinement). The above mentioned cases give rise to the $\widehat{\mathcal{I}}_{l,dp}$ being insensitive to the added parity (left or right dot), as seen in Fig.~\ref{fig:compare2es4esCase}(c) that all curves are close to each other. In addition, $\gamma^\psi$ in the four-electron case (cf.~Table.~\ref{tab:2esDQDPuredephase}(b)) is more than an order smaller than $\gamma^\psi$ in the two-electron case (cf.~Table.~\ref{tab:2esDQDPuredephase}(a)) as $\Delta E_{S'}\approx251\mu$eV for the former while $\Delta E_{S'}\approx 50\mu$eV, resulting in smaller $\theta_S$ for the former case.
	
	\subsection{Phonon relaxation (One-phonon process), $\Gamma_1^\mathrm{1p}$}\label{subsec:1pRel}
	One-phonon relaxation, $\Gamma_1^\text{1p}$, arises due to the admixture between singlet and triplet states, which can be attributed to two mechanisms: (1) First-order (direct) admixture by magnetic gradient, $h$, and (2) Second-order (indirect) admixture mediated by spin-orbit interaction (SOI) with polarized triplet states, $|T_\pm(1,3)\rangle$ and $|T_\pm(1,3^*)\rangle$, where an asterisk, $*$ indicates an electron occupying firstly excited orbital from ground configuration. The details of the former are discussed in Sec.~\ref{subsubsec:dBFstOrdRel} while the latter in Sec.~\ref{subsubsec:SOIFstOrdRel}. 
	\subsubsection{$\Gamma_1^\mathrm{1p}$ by magnetic gradient, $h$}\label{subsubsec:dBFstOrdRel}
	$H_\text{e}+H_\text{hyp}$, written in the singlet, $|S'(1,3)\rangle$, and triplet, $|T'(1,3)\rangle$ eigenbases (cf. Eq.~\eqref{eq:STEigenH}-\eqref{eq:STEigenbases}), is
	\begin{equation}\label{eq:effHyp}
		\begin{split}
			H_\text{e}^{(S2)} + H_\text{hyp}^{(S2)}=
			&=\mathcal{P}^{(S2)} \widetilde{U}^\dagger \mathcal{P}^{(S1)} \left(H_\text{e}+H_\text{hyp}\right) \mathcal{P}^{(S1)} \widetilde{U} \mathcal{P}^{(S2)} \\
			&=\left(\begin{array}{cc}
				E_{|S'(1,3)\rangle} & -h \cos \frac{\theta_S}{2}\cos \frac{\theta_T}{2}   \\
				-h \cos \frac{\theta_S}{2}\cos \frac{\theta_T}{2} & E_{|T'(1,3)\rangle}
			\end{array}\right),
		\end{split}
	\end{equation}
	where
	\begin{equation}
		\mathcal{P}^{(S2)} = |S'(1,3)\rangle \langle S'(1,3) | + |T'(1,3)\rangle \langle T'(1,3) |,
	\end{equation}
	while $\widetilde{U}$ diagonalize the Hamiltonian $\mathcal{P}^{(S1)} \left(H_\text{e}+H_\text{hyp}\right) \mathcal{P}^{(S1)}$. 
	The eigenvectors of Eq.~\eqref{eq:effHyp} are
	\begin{equation}\label{eq:eigenvectorsdB}
		\begin{split}
			|S'(1,3)\rangle^{(S2)} = \cos \frac{\zeta}{2} |S'(1,3)\rangle+\sin \frac{\zeta}{2} |T'(1,3)\rangle,\\
			|T'(1,3)\rangle^{(S2)} = -\sin \frac{\zeta}{2} |S'(1,3)\rangle+\cos \frac{\zeta}{2} |T'(1,3)\rangle,
		\end{split}
	\end{equation}
	where 
	\begin{equation}
		\tan \zeta =  \frac{2h \cos (\theta_S/2) \cos (\theta_T/2) }{E_{|T'(1,3)\rangle}-E_{|S'(1,3)\rangle}}
	\end{equation}
	while $E_{|T'(1,3)\rangle}-E_{|S'(1,3)\rangle}>0$ for the parameter of concern. Written in the eigenbases of $S2$ subspace, $\delta B_x (\tau)$ (Eq.~\eqref{eq:Jjk}) is
	\begin{subequations}
		\begin{align}
			\begin{split}
				\delta B_x (\tau)&=\left(|S'(1,3)\rangle^{(S2)}\right)^\dagger H_\text{e-ph} |T'(1,3)\rangle^{(S2)}
			\end{split}\\
			\begin{split}
				&=\frac{\sin \zeta}{4} \Big[-\left(\cos \theta_S-1\right)\left(P_{L1,L1}-P_{R2,R2}\right)+\left(\cos \theta_T -1 \right)\left(P_{L1,L1}-P_{R3,R3}\right)+\sqrt{2}\sin \theta_S \left(P_{L1,R2}+P_{R2,L1}\right)\\
				&\quad -\sin \theta_T \left(P_{L1,R3}+P_{R3,L1}\right)\Big]
			\end{split}\\
			\begin{split}\label{eq:finalFstOrddB}
				&\approx\frac{\sin \zeta}{4} \Big[-\left(\cos \theta_S-1\right)\left(P_{L1,L1}-P_{R2,R2}\right)+\left(\cos \theta_T -1 \right)\left(P_{L1,L1}-P_{R3,R3}\right)\Big],
			\end{split}
		\end{align}
	\end{subequations}
	where the term $P_{\psi j,\psi k}$ for $\eta(\psi j) \neq \eta{(\psi k)}$ are dropped, as discussed in Sec.~\ref{subsec:innerproductDoublet} (cf. Eq.~\eqref{eq:valueK}). $\delta B_y (\tau)=0$ as the imaginary part is absent in this case. From Eq.~\eqref{eq:finalFstOrddB}, dropping cross terms between $\theta_S$ and $\theta_T$ for analytical purposes, we have
	
	\begin{subequations}
		\begin{align}
			\begin{split}
				&\hbar^2 \Gamma_1^\text{1p} = 2\hbar^2 \left(J^+_{xx} (\omega_z)+J^+_{yy} (\omega_z)\right)^\text{1p}
			\end{split}\\
			\begin{split}
				&= \int_{-\infty}^{\infty} \cos (\omega \tau) \langle \delta B_x(0)\delta B_x(\tau) \rangle_\beta d\tau \big|_{\omega\rightarrow \omega_z}
			\end{split}\\
			\begin{split}
				&= \frac{\sin^2 \zeta}{16} \int_{-\infty}^{\infty} d\tau\cos (\omega_z \tau) \\
				&\times\Bigg\langle \left(\cos \theta_S-1\right)^2\bigg[P_{L1,L1}(\mathbf{q}_1)P_{L1,L1}(\mathbf{q}_2)+P_{R2,R2}(\mathbf{q}_1)P_{R2,R2}(\mathbf{q}_2)-P_{L1,L1}(\mathbf{q}_1)P_{R2,R2}(\mathbf{q}_2)-P_{R2,R2}(\mathbf{q}_1)P_{L1,L1}(\mathbf{q}_2)\bigg]\\
				&+\left(\cos \theta_T -1 \right)^2\bigg[P_{L1,L1}(\mathbf{q}_1)P_{L1,L1}(\mathbf{q}_2)+P_{R3,R3}(\mathbf{q}_1)P_{R3,R3}(\mathbf{q}_2)-P_{L1,L1}(\mathbf{q}_1)P_{R3,R3}(\mathbf{q}_2)-P_{R3,R3}(\mathbf{q}_1)P_{L1,L1}(\mathbf{q}_2)\bigg]
				\Bigg\rangle_\beta 
			\end{split}\\
			\begin{split}
				&= \frac{\sin^2 \zeta}{16} \int_{-\infty}^{\infty} d\tau\cos (\omega_z \tau) \\
				&\times \Bigg[\left(\cos \theta_S-1\right)^2\bigg(\left\langle P_{L1,L1}(\mathbf{q}_1)P_{L1,L1}(\mathbf{q}_2)-P_{L1,L1}(\mathbf{q}_1)P_{R2,R2}(\mathbf{q}_2)\right\rangle_\beta
				\\
				&\qquad\qquad\qquad\qquad+\left\langle P_{R2,R2}(\mathbf{q}_1)P_{R2,R2}(\mathbf{q}_2)-P_{R2,R2}(\mathbf{q}_1)P_{L1,L1}(\mathbf{q}_2)\right\rangle_\beta\bigg)\\
				&\quad+\left(\cos \theta_T -1 \right)^2\bigg(\left\langle P_{L1,L1}(\mathbf{q}_1)P_{L1,L1}(\mathbf{q}_2)-P_{L1,L1}(\mathbf{q}_1)P_{R3,R3}(\mathbf{q}_2)\right\rangle_\beta
				\\
				&\qquad\qquad\qquad\qquad+\left\langle P_{R3,R3}(\mathbf{q}_1)P_{R3,R3}(\mathbf{q}_2)-P_{R3,R3}(\mathbf{q}_1)P_{L1,L1}(\mathbf{q}_2)\right\rangle_\beta\bigg)\Bigg]		
			\end{split}
		\end{align}
	\end{subequations}
	where $\hbar \omega_z=J$ is the exchange energy while the superscript ``1p'' denotes one-phonon process. To facilitate discussion, considering only the admixtur e among singlet states, i.e. $\theta_T = 0$ while $\theta_S > 0$, we have
	\begin{equation}\label{eq:Rel1p}
		\begin{split}
			&\hbar^2 \Gamma_1^\text{1p} = 2\hbar^2 \left(J^+_{xx} (\omega_z)+J^+_{yy} (\omega_z)\right)^\text{1p}\\
			&= \frac{\sin^2 \zeta}{16} \int_{-\infty}^{\infty} d\tau\cos (\omega_z \tau) \\
			&\times \left(\cos \theta_S-1\right)^2\bigg(\left\langle P_{L1,L1}(\mathbf{q}_1)P_{L1,L1}(\mathbf{q}_2)-P_{L1,L1}(\mathbf{q}_1)P_{R2,R2}(\mathbf{q}_2)\right\rangle_\beta+\left\langle P_{R2,R2}(\mathbf{q}_1)P_{R2,R2}(\mathbf{q}_2)-P_{R2,R2}(\mathbf{q}_1)P_{L1,L1}(\mathbf{q}_2)\right\rangle_\beta\bigg).
		\end{split}
	\end{equation}
	For the case in which two electrons occupying a DQD device \cite{Kornich.14}, $\Gamma_1^\text{1p}$ has the same form as Eq.~\eqref{eq:Rel1p} with all the subscripts $R2$ are replaced by $R1$ and the Hubbard parameters are modified as shown in Eq.~\eqref{eq:2esCase}. Also,
	\begin{equation}
		\begin{split}
			\theta_T &= 0,
			\\
			|T(1,1)\rangle &=|T(\uparrow_{L1}\downarrow_{R1})\rangle,
			\\
			\tan \zeta &= \frac{2 h \cos (\theta_S/2)}{E_{|T(1,1)\rangle}-E_{|S'(1,1)\rangle}},
			\\
			E_{|T(1,1)\rangle}&=U_{L1,R1}+2\Delta.
		\end{split}
	\end{equation}
	
	Following the notations in Sec.~\ref{subsec:JjjOnePhonon}, Eq.~\eqref{eq:Ppsi12Notation} and Eq.~\eqref{eq:ItildeOnePhonon}-\eqref{eq:IdivCOnePhonon}, we have
	\begin{subequations}
		\begin{align}
			\begin{split}
				\varrho^\psi &= \frac{1}{16}\sin^2 \zeta,
			\end{split}
			\\
			\begin{split}
				\tan \zeta &=  \frac{2h \cos (\theta_S/2) \cos (\theta_T/2) }{E_{|T(1,1)\rangle}-E_{|S'(1,1)\rangle}},
			\end{split}
			\\
			\begin{split}
				\widetilde{\mathcal{I}}^{\text{(1p)}\psi = \psi1,\psi2,\psi3,\psi4}_{s,\text{cpl}} &= \frac{1}{2\hbar^2}P^{s,\text{cpl}}_{\psi 1, \psi 2}(\mathbf{q}_1)P^{s,\text{cpl}}_{\psi 3, \psi 4}(\mathbf{q}_2),
			\end{split}\\
			\begin{split}
				\mathcal{I}^{\text{(1p)}\psi}_{s,\text{cpl}} &=\frac{1}{2\hbar^2}\sum_{\mathbf{q}_1,\mathbf{q}_2}\int_{-\infty}^{\infty} d\tau\cos (\omega_s \tau)  C_{s,\text{cpl}}(\mathbf{q}_1)C_{s,\text{cpl}}(\mathbf{q}_2)\widetilde{\mathcal{I}}^{\text{(1p)}\psi}_{s,\text{cpl}}.
			\end{split}\\
			\begin{split}
				\mathcal{I}^{\text{(1p)}\psi} &=\sum_{s,\text{cpl}} \mathcal{I}^{\text{(1p)}\psi}_{s,\text{cpl}},
			\end{split}
		\end{align}
	\end{subequations}
	Here we show explicitly the $\mathcal{I}^{\text{(1p)}\psi}$ terms presented in Eq.~\eqref{eq:Rel1p}
	\begin{subequations}
		\begin{align}
			\begin{split}
				&\mathcal{I}^{\text{(1p)}\eta1,\eta1,\eta1,\eta1}\\
				&=\frac{4\pi^2}{\nu_s} q^2 
				\Big( 2 n_s (q)+ 1 \Big)
				\int_{0}^{\pi} d\theta_{\mathbf{q}_2}
				\left(\prod_{j\in \{1,3\}} \frac{\pi \mathcal{G}_{\eta(\psi j)}^2}{\alpha_{\eta(\psi j)}}\right) \mathcal{Q}_z(\mathbf{q})
				\exp\left[- \frac{q^2 \sin^2 \theta_{\mathbf{q}_2}l_\eta^2}{2}\right],
			\end{split}\\
			\begin{split}
				&\mathcal{I}^{\text{(1p)}\eta1,\eta1,\eta'1,\eta'1}\\
				&=\frac{4\pi^2}{\nu_s} q^2 
				\Big( 2 n_s (q)+ 1 \Big)
				\int_{0}^{\pi} d\theta_{\mathbf{q}_2}
				\left(\prod_{j\in \{1,3\}} \frac{\pi \mathcal{G}_{\eta(\psi j)}^2}{\alpha_{\eta(\psi j)}}\right) \mathcal{Q}_z(\mathbf{q})
				\exp\left[- \frac{q^2 \sin^2 \theta_{\mathbf{q}_2}l_\eta^2}{2}\right]
				\\
				&\quad\,\times\text{BesselJ}\left[0,(\Delta h_{13}) x_0 q \sin \theta_{\mathbf{q}_2}\right],
			\end{split}\\
			\begin{split}
				&\mathcal{I}^{\text{(1p)}\eta1,\eta1,\eta'3,\eta'3}\\
				&=\frac{4\pi^2}{\nu_s} q^2 
				\Big( 2 n_s (q)+ 1 \Big)
				\int_{0}^{\pi} d\theta_{\mathbf{q}_2}
				\left(\prod_{j\in \{1,3\}} \frac{\pi \mathcal{G}_{\eta(\psi j)}^2}{\alpha_{\eta(\psi j)}}\right) \mathcal{Q}_z(\mathbf{q})
				\exp\left[- \frac{q^2 \sin^2 \theta_{\mathbf{q}_2}l_\eta^2}{2}\right]
				\\
				&\quad \times\text{BesselJ}\left[0,(\Delta h_{13}) x_0 q \sin \theta_{\mathbf{q}_2}\right] \left(1-\frac{l_{\eta'}^2q^2 \sin^2 \theta_{\mathbf{q}_2}}{4}\right),
			\end{split}
		\end{align}
	\end{subequations}
	where $\eta'$ denotes the opposite QD parity of $\eta$, e.g. if $\eta = L$, then $\eta'=R$ and vise versa.
	
	\begin{table}[t]
		\begin{tabular}{p{0.5cm}|p{2cm}|p{2.7cm}|p{1.8cm}|p{0.8cm}|p{2.7cm}|p{1.8cm}|p{0.8cm}|}
			\cline{2-8} 
			& &\multicolumn{3}{c|}{} & \multicolumn{3}{c|}{} \\[-10pt]
			& \hfil \multirow{3}{*}{$\rho^\psi$ (unit: 1)}&\multicolumn{3}{c|}{ 1$^{\text{st}}$ $\mathcal{I}^{\text{(1p)}\psi}$ in $\langle \cdots \rangle_\beta$}  & \multicolumn{3}{c|}{ 2$^{\text{nd}}$ $\mathcal{I}^{\text{(1p)}\psi}$ in $\langle \cdots \rangle_\beta$} \\
			\cline{3-8}
			& &\hfil  & \hfil  & \hfil  & \hfil   & \hfil & \\[-10pt]
			& &\hfil \multirow{2}{*}{$\mathcal{I}^{\text{(1p)}\psi}$} & \hfil numeric &\hfil \multirow{2}{*}{$\mathcal{D}$} & \hfil \multirow{2}{*}{$\mathcal{I}^{\text{(1p)}\psi}$} &  \hfil numeric & \hfil \multirow{2}{*}{$\mathcal{D}$} \\
			& & & \hfil ($10^{9}$Hz)  &  &  &  \hfil ($10^{9}$Hz) & \\
			\cline{2-8}
			& &\hfil  & \hfil  & \hfil   & \hfil   & \hfil   & \\[-10pt]
			& \hfil \multirow{2}{*}{$1.458\times10^{-6}$} &\hfil $ P_{L1,L1}P_{L1,L1}$ & \hfil $1.3794$&\hfil 0 &  \hfil $P_{L1,L1}P_{R1,R1}$ & \hfil  $1.3728$ & \hfil 1\\
			\cline{3-8}
			& &\hfil  & \hfil  & \hfil   & \hfil & \hfil & \\[-10pt]
			(a)& &\hfil $ P_{R1,R1} P_{R1,R1} $ & \hfil $1.3794$ & \hfil 0 & \hfil  $ P_{R1,R1} P_{L1,L1} $ & \hfil $1.3728$ &\hfil 1\\
			\cline{2-8}
			\multicolumn{8}{c}{} \\[5pt]
			\cline{2-8} 
			& &\multicolumn{3}{c|}{} & \multicolumn{3}{c|}{} \\[-10pt]
			& \hfil \multirow{3}{*}{$\rho^\psi$ (unit: 1)}&\multicolumn{3}{c|}{ 1$^{\text{st}}$ $\mathcal{I}^{\text{(1p)}\psi}$ in $\langle \cdots \rangle_\beta$}  & \multicolumn{3}{c|}{ 2$^{\text{nd}}$ $\mathcal{I}^{\text{(1p)}\psi}$ in $\langle \cdots \rangle_\beta$} \\
			\cline{3-8}
			& &\hfil  & \hfil  & \hfil  & \hfil   & \hfil & \\[-10pt]
			& &\hfil \multirow{2}{*}{$\mathcal{I}^{\text{(1p)}\psi}$} & \hfil numeric &\hfil \multirow{2}{*}{$\mathcal{D}$} & \hfil \multirow{2}{*}{$\mathcal{I}^{\text{(1p)}\psi}$} &  \hfil numeric & \hfil \multirow{2}{*}{$\mathcal{D}$} \\
			& & & \hfil ($10^{9}$Hz)  &  &  &  \hfil ($10^{9}$Hz) & \\
			\cline{2-8}
			& &\hfil  & \hfil  & \hfil   & \hfil   & \hfil   & \\[-10pt]
			& \hfil \multirow{2}{*}{$\approx 4\times10^{-8}$} &\hfil $ P_{L1,L1}P_{L1,L1}$ & \hfil $1.4850$&\hfil 0 &  \hfil $P_{L1,L1}P_{R2,R2}$ & \hfil  $1.4823$ & \hfil 1\\
			\cline{3-8}
			& &\hfil  & \hfil  & \hfil   & \hfil & \hfil & \\[-10pt]
			(b)& &\hfil $ P_{R2,R2} P_{R2,R2} $ & \hfil $1.4850$ & \hfil 0 & \hfil  $ P_{R2,R2} P_{L1,L1} $ & \hfil $1.4823$ &\hfil 1\\
			\cline{2-8}
		\end{tabular}
		\caption{(a) One-phonon relaxation rate, $\Gamma_1^{\text{1p}}$, of two electrons occupying a DQD device. The parameters are \cite{Kornich.14}: $\hbar \omega_L = \hbar \omega_R = 0.124$meV, $x_0=200$nm, $B_{||}=0.7$ T, $B_{\perp}=0$, $T = 50$mK, $l_\text{D} = 1\mu$m, $l_\text{R}=2\mu$m, $h=0.14\mu$eV, $a_z=2$nm,$U_{R1}=1$meV, $t_{L1,R1}=7.25\mu$eV, $2\Delta\sim 0.9$meV such that the exchange splitting $\hbar \omega_z =1.43\mu$eV. (b) One-phonon relaxation rate, $\Gamma_1^{\text{1p}}$, of four electrons occupying a DQD device. The parameters are : $\hbar \omega_L =7.5$meV, $\hbar \omega_R = 4$meV, $x_0=40$nm, $B_{||}=0$ T, $B_{\perp}=0.319$T, $T = 50$mK, $l_\text{D} = 1.26\mu$m, $l_\text{R}=1.72\mu$m, $h=0.14\mu$eV, $a_z=2$nm, $U_{R2}=50.349$meV, $U_{L1,R2}=41.790$meV, $U_{R2,R3}=50.209$meV, $\xi=0.485$meV, $t_{L1,R2}=27.700\mu$eV, $t_{L1,R3}=12.000\mu$eV, $\Delta= 4.154$meV such that the exchange splitting $\hbar \omega_z =4.414\mu$eV. For the four-electron case of interest here, different degree of admixture within both singlets and triplets subspace resulting in slightly different numerical values of $\rho^\psi$ for different terms, which is denoted as the approximation symbol, $\approx$ in (b).}
		\label{tab:2es4esDQDRel1p}
	\end{table}
	Table.~\ref{tab:2es4esDQDRel1p} shows that one-phonon relaxation rate mainly contributed by the difference between $1^\text{st}$ and $2^\text{nd}$ $\mathcal{I}^{\text{(1p)}\psi}$. In contrast to the case for two-phonon pure dephasing, $\Gamma_\varphi^{\text{(2p)}}$ (cf. Table.~\ref{tab:2esDQDPuredephase}), the numerical value of Bessel functions in the integrand, $\mathcal{I}^{\text{(1p)}\psi}$, does not differ much for different inter-dot distances, $x_0$, of interest here. This is due to the fact that $\mathcal{I}^{\text{(1p)}\psi}$ is evaluated at the wave vector, $q=\omega_z/\nu_s$, whose value is too small to induce considerable modification by Bessel function. For both cases in Table.~\ref{tab:2es4esDQDRel1p}, $q=4.20\times10^{5}$m$^{-1}$ ($J = 1.43\mu$eV) and $q = 1.32\times10^{6}$m$^{-1}$ ($J=4.414\mu$eV) for $(s,\text{cpl})=(l,\text{dp})$, resulting in Bessel function $\approx $ 1, cf.~Fig.~\ref{fig:compare2es4esCase}(b) and (d). Hence, only slightly larger difference between $1^\text{st}$ $\mathcal{I}^{(\text{1p})\psi}$ and $2^\text{nd}$ $\mathcal{I}^{(\text{1p})\psi}$ is observed for two-electron case as compared to four-electron case, cf.~Table.~\ref{tab:2es4esDQDRel1p}. Regardless, $\Gamma_1^{\text{(1p)}}$ is substantially suppressed for four electron case of interest here, as $\rho^\psi$ is about an order smaller compared to two electron case, due to the larger $J$ in consideration, resulting in smaller $\zeta$.
	\subsubsection{$\Gamma_1^\mathrm{1p}$ by spin-orbit interaction \normalfont{(SOI)}}\label{subsubsec:SOIFstOrdRel}
	For two-electron system, in the small detuning regime, i.e. $\Delta \simeq 0$, the lowest energy subspace can be obtained by projecting the two-electron Hamiltonian, $H_\text{2e}$, using the projector, cf.~Eq.~\eqref{eq:2eBases},
	\begin{equation}
		\mathcal{P}^{(S3)}_{2e} = |S(1,1)\rangle\langle S(1,1)|+|S(0,2)\rangle\langle S(2,0)|+|S(0,2)\rangle\langle S(0,2)|+|T(1,1)\rangle\langle T(1,1)|,
	\end{equation}
	such that we have
	\begin{equation}
		H_\text{2e}^{(S3)}=\mathcal{P}_{2e}^{(S3)}H_\text{2e}\mathcal{P}_{2e}^{(S3)},
	\end{equation}
	where the subscript 2e denotes a two-electron system. Diagonalizing $H_\text{2e}^{(S3)}$ gives the lowest singlet state, $|S'''(1,1)\rangle$. As shown in Ref.~\onlinecite{Kornich.14}, the phonon decoherence time, $T_2$, in small detuning regime, $\Delta \simeq 0$, is dominated by one-phonon relaxation, $\Gamma_1^\text{(1p)}$, mediated by SOI terms. Here, to gain an insight of $\Gamma_1^\text{(1p)}$ by SOI, we take the projection of $H_\text{2e}$ onto $\mathcal{P}^{(S4)}_{2e}$, which 
	\begin{equation}
		\begin{split}
			\mathcal{P}^{(S4)} = |S'''(1,1)\rangle \langle S'''(1,1)| +|T(1,1)\rangle\langle T(1,1)|+|T_+(1,1)\rangle\langle T_+(1,1)|+|T_+(1,1^\star)\rangle\langle T_+(1,1^\star)|.
		\end{split}
	\end{equation}
	Also, we split $\mathcal{P}^{(S4)}_{2e}$ into two subspaces,
	\begin{equation}
		\mathcal{P}^{(S4)}_{2e} = \mathcal{P}_{2e,1}^{(S4)} + \mathcal{P}^{(S4)}_{2e,2},
	\end{equation}
	where
	\begin{equation}
		\begin{split}
			\mathcal{P}^{(S4)}_{2e,1} &= |S'''(1,1)\rangle \langle S'''(1,1)| +|T_+(1,1)\rangle\langle T_+(1,1)|,\\
			\mathcal{P}^{(S4)}_{2e,2} &= |T(1,1)\rangle\langle T(1,1)|+|T_+(1,1^\star)\rangle\langle T_+(1,1^\star)|.
		\end{split}
	\end{equation}
	
	The resulting effective Hamiltonian,
	\begin{equation}
		\begin{split}
			H_{2e}^{(S4)}&=\mathcal{P}_{2e}^{(S4)}H_\text{2e}\mathcal{P}_{2e}^{(S4)}\\
			&=\left(
			\begin{array}{cccc}
				-J_S & 0 & \frac{\Omega}{\sqrt{2}} & \frac{\Omega_1}{\sqrt{2}} \\
				0 & 0 & 0& -\frac{\Omega_1}{\sqrt{2}}\\
				\frac{\Omega}{\sqrt{2}}&0 &E_\text{Z} & 0\\
				\frac{\Omega_1}{\sqrt{2}}& -\frac{\Omega_1}{\sqrt{2}} & 0 &\Delta E_{2e} + E_\text{Z}
			\end{array}
			\right)+E_\text{shift},
		\end{split}
	\end{equation}
	where $\Delta E_{2e}=\varepsilon_{R2}-\varepsilon_{R1}$, $E_\text{shift} = U_{L1,R1}$ and $J_S$ is the exchange splitting. Diagonalizing $H_{2e}^{(S4)}$ written in $\mathcal{P}^{(S4)}_{2e,1}$ and $\mathcal{P}^{(S4)}_{2e,2}$ individually, the eigenbases of the subspace spanned by $\mathcal{P}^{(S4)}_{2e,1}$ are
	\begin{equation}\label{eq:P1}
		\begin{split}
			|S'^{+}(1,1)\rangle &= -\cos \frac{\vartheta_S}{2} |S'''(1,1)\rangle + \sin \frac{\vartheta_S}{2} |T_+(1,1)\rangle,\\
			|T'_{+}(1,1)\rangle &= \sin \frac{\vartheta_S}{2} |S'''(1,1)\rangle + \cos \frac{\vartheta_S}{2} |T_+(1,1)\rangle,\\
		\end{split}
	\end{equation}
	while the eigenbases by $\mathcal{P}^{(S4)}_{2e,2}$ are
	\begin{equation}\label{eq:P2}
		\begin{split}
			|T'^{+}(1,1)\rangle &= \cos \frac{\vartheta_T}{2} |T(1,1)\rangle + \sin \frac{\vartheta_T}{2} |T_+(1,1^\star)\rangle,\\
			|T'_{+}(1,1^\star)\rangle &= -\sin \frac{\vartheta_T}{2} |T(1,1)\rangle + \cos \frac{\vartheta_T}{2} |T_+(1,1^\star)\rangle,\\
		\end{split}
	\end{equation}
	where $|S'^{+}(1,1)\rangle$ and $|T'^{+}(1,1)\rangle$ are the logical bases with minute admixture with $|T_+(1,1)\rangle$ and $|T_+(1,1^\star)\rangle$ respectively while $\tan \vartheta_S=\sqrt{2} \Omega/(E_\text{Z}+J_S)$, $\tan \vartheta_T=\sqrt{2} \Omega_1/(E_\text{Z}+\Delta E_{2e})$. $H_{2e}^{(S4)}$ is then transformed into the bases formed by the eigenstates of $\mathcal{P}^{(S4)}_{2e,1}$ and $\mathcal{P}^{(S4)}_{2e,2}$, written in the bases of $|S'^{+}(1,1)\rangle$, $|T'^{+}(1,1)\rangle$, $|T'_+(1,1)\rangle$ and $|T'_-(1,1)\rangle$,
	\begin{equation}
		\begin{split}
			&\left(U^{(S4)}_{2e}\right)^\dagger H_{2e}^{(S4)} U_{2e}^{(S4)} \\
			&=\left(
			\begin{array}{cc}
				-J_S \cos^2 \frac{\vartheta_S}{2}+E_\text{Z} \sin^2 \frac{\vartheta_S}{2}-\frac{\Omega}{\sqrt{2}} \sin \vartheta_S
				& -\frac{\Omega_1}{\sqrt{2}} \cos \frac{\theta_S}{2}\sin \frac{\theta_T}{2}
				\\
				-\frac{\Omega_1}{\sqrt{2}} \cos \frac{\theta_S}{2}\sin \frac{\theta_T}{2}
				& E_\text{Z} \sin^2 \frac{\vartheta_S}{2}+\Delta E_{2e} \sin^2 \frac{\vartheta_S}{2}-\frac{\Omega}{\sqrt{2}} \sin \vartheta_S
				\\
				0
				& \frac{\Omega_1}{\sqrt{2}} \cos \frac{\theta_S}{2}\sin \frac{\theta_T}{2}   
				\\
				-\frac{\Omega_1}{\sqrt{2}} \cos \frac{\theta_S}{2}\sin \frac{\theta_T}{2} 
				& 0
				\\
			\end{array}
			\right.
			\\
			&\quad\left.
			\begin{array}{cc}
				0
				& -\frac{\Omega_1}{\sqrt{2}} \cos \frac{\theta_S}{2}\sin \frac{\theta_T}{2} 
				\\
				\frac{\Omega_1}{\sqrt{2}} \cos \frac{\theta_S}{2}\sin \frac{\theta_T}{2} 
				&0
				\\
				-J_S \sin^2 \frac{\vartheta_S}{2}+E_\text{Z} \cos^2 \frac{\vartheta_S}{2}+\frac{\Omega}{\sqrt{2}} \sin \vartheta_S
				& \frac{\Omega_1}{\sqrt{2}} \cos \frac{\theta_S}{2}\sin \frac{\theta_T}{2} 
				\\
				\frac{\Omega_1}{\sqrt{2}} \cos \frac{\theta_S}{2}\sin \frac{\theta_T}{2}  
				&E_\text{Z} \cos^2 \frac{\vartheta_S}{2}+\Delta E_{2e} \cos^2 \frac{\vartheta_S}{2}+\frac{\Omega}{\sqrt{2}} \sin \vartheta_S \\
			\end{array}
			\right)
			,
		\end{split}
	\end{equation}
	where $U^{(S4)}_{2e}$ is the unitary transformation that takes $H_{2e}^{(S4)}$ to the subspace spanned by the eigenbases of the subspace spanned by $\mathcal{P}^{(S4)}_{2e,1}$ and $\mathcal{P}^{(S4)}_{2e,2}$. 
	
	The electron-phonon interaction $H_\text{e-ph}$, being projected onto the subspace spanned by $\mathcal{P}^{(S4)}$, is
	\begin{equation}
		\begin{split}
			&H^{(S4)}_{2e,\text{e-ph}}\\
			&=\mathcal{P}^{(S4)}_{2e} H_\text{e-ph} \mathcal{P}^{(S4)}_{2e}\\
			&=\left(
			\begin{array}{cccc}
				P_{L1,L1}+P_{R1,R1}+\sin \varsigma
				\left(P_{L1,R1}+P_{R1,L1}\right) & 0 & 0 & 0 \\
				0 & P_{L1,L1}+P_{R1,R1} & 0 & 0 \\
				0 & 0 & P_{L1,L1}+P_{R1,R1} & P_{R1,R2} \\
				0 & 0 & P_{R1,R2} & P_{L1,L1}+P_{R2,R2} \\
			\end{array}
			\right),
		\end{split}
	\end{equation}
	where $\tan \varsigma=4 t_{L1,R1}/(U-V_+)$ \cite{Kornich.14}. Transforming into the eigenbases of the subspace spanned by $\mathcal{P}^{(S4)}_{2e,1}$ and $\mathcal{P}^{(S4)}_{2e,2}$, written in the bases of $|S'^{+}(1,1)\rangle$, $|T'^{+}(1,1)\rangle$, $|T'_+(1,1)\rangle$ and $|T'_-(1,1)\rangle$ (cf.~Eq.~\eqref{eq:P1} and \eqref{eq:P2}), we have
	\begin{equation}\label{eq:ePhSOIRaw}
		\begin{split}
			& \widetilde{H}_{2e,\text{e-ph}}\\
			&=\left(U^{(S4)}_{2e}\right)^\dagger H_{2e,\text{e-ph}}^{(S4)} U_{2e}^{(S4)} \\
			&=\left(
			\begin{array}{cccc}
				P_{L1,L1}+P_{R1,R1}+ \sin \varsigma \left(P_{L1,R1}+P_{R1,L1}\right) \cos ^2\frac{\vartheta_S}{2} &
				\sin \frac{\vartheta_S}{2} \sin \frac{\vartheta_T}{2} P_{R1,R2} 
				\\
				\sin \frac{\vartheta_S}{2} \sin \frac{\vartheta_T}{2} P_{R1,R2} &
				P_{L1,L1}+\cos^2\vartheta_T P_{R1,R1} +\sin ^2\frac{\vartheta_T}{2} P_{R2,R2} 
				\\
				-\frac{1}{2} \sin \vartheta_S \sin \varsigma \left(P_{L1,R1}+P_{R1,L1}\right) &
				\cos \frac{\vartheta_S}{2} \sin\frac{\vartheta_T}{2} P_{R1,R2} 
				\\
				\cos \frac{\vartheta_T}{2} \sin \frac{\vartheta_S}{2} P_{R1,R2} &
				\frac{1}{2} \sin \vartheta_T\left(P_{R2,R2}-P_{R1,R1}\right)
				\\
			\end{array}
			\right.\\
			&\quad\left.
			\begin{array}{cccc} 
				-\frac{1}{2} \sin \vartheta_S \sin \varsigma\left(P_{L1,R1}+P_{R1,L1}\right) &
				\cos \frac{\vartheta_T}{2} \sin \frac{\vartheta_S}{2} P_{R1,R2}
				\\
				\cos \frac{\vartheta_S}{2} \sin \frac{\vartheta_T}{2} P_{R1,R2} &
				\frac{1}{2} \sin \vartheta_T \left(P_{R2,R2}-P_{R1,R1}\right) 
				\\
				P_{L1,L1}+P_{R1,R1}+\sin \varsigma \left(P_{L1,R1}+P_{R1,L1}\right) \sin^2\frac{\vartheta_S}{2} &
				\cos \frac{\vartheta_S}{2} \cos \frac{\vartheta_T}{2}P_{R1,R2} 
				\\
				\cos \frac{\vartheta_S}{2} \cos \frac{\vartheta_T}{2} P_{R1,R2}&
				P_{L1,L1}+\cos ^2\vartheta_T P_{R2,R2} +\sin ^2\frac{\vartheta_T}{2} P_{R1,R1}. \\
			\end{array}
			\right)\\
		\end{split}
	\end{equation}
	Dropping the terms $P_{\psi j,\psi k }$ with $\eta (\psi j)\neq \eta (\psi k)$ in Eq.~\eqref{eq:ePhSOIRaw}, we get
	\begin{equation}\label{eq:ePhSOI}
		\begin{split}
			& \widetilde{H}_{2e,\text{e-ph}}\\
			&=\left(U^{(S4)}_{2e}\right)^\dagger H_{2e,\text{e-ph}}^{(S4)} U_{2e}^{(S4)} \\
			&=\left(
			\begin{array}{cccc}
				P_{L1,L1}+P_{R1,R1} &
				\sin \frac{\vartheta_S}{2} \sin \frac{\vartheta_T}{2} P_{R1,R2} &
				0
				\\
				\sin \frac{\vartheta_S}{2} \sin \frac{\vartheta_T}{2} P_{R1,R2} &
				P_{L1,L1}+\cos^2\vartheta_T P_{R1,R1} +\sin ^2\frac{\vartheta_T}{2} P_{R2,R2} &
				\cos \frac{\vartheta_S}{2} \sin \frac{\vartheta_T}{2} P_{R1,R2}
				\\
				0&
				\cos \frac{\vartheta_S}{2} \sin\frac{\vartheta_T}{2} P_{R1,R2} &
				P_{L1,L1}+P_{R1,R1}
				\\
				\cos \frac{\vartheta_T}{2} \sin \frac{\vartheta_S}{2} P_{R1,R2} &
				\frac{1}{2} \sin \vartheta_T\left(P_{R2,R2}-P_{R1,R1}\right)&
				\cos \frac{\vartheta_S}{2} \cos \frac{\vartheta_T}{2} P_{R1,R2}
				\\
			\end{array}
			\right.\\
			&\quad\left.
			\begin{array}{cccc} 
				\cos \frac{\vartheta_T}{2} \sin \frac{\vartheta_S}{2} P_{R1,R2}
				\\
				\frac{1}{2} \sin \vartheta_T \left(P_{R2,R2}-P_{R1,R1}\right) 
				\\
				\cos \frac{\vartheta_S}{2} \cos \frac{\vartheta_T}{2}P_{R1,R2} 
				\\
				P_{L1,L1}+\cos ^2\vartheta_T P_{R2,R2} +\sin ^2\frac{\vartheta_T}{2} P_{R1,R1} \\
			\end{array}
			\right).
		\end{split}
	\end{equation}
	From Eq.~\eqref{eq:ePhSOI}, we obtain
	\begin{equation}\label{eq:BxSOI}
		\begin{split}
			\delta B_x &= \langle S'^+(1,1)|\widetilde{H}_{2e,\text{e-ph}}|T'^+(1,1)\rangle\\
			&=\sin \frac{\vartheta_S}{2} \sin \frac{\vartheta_T}{2} P_{R1,R2}, \\
			\delta B_y &= 0.
		\end{split}
	\end{equation}
	$\delta B_x$ gives rise to $\Gamma_1^\text{1p}$
	\begin{equation}\label{eq:Rel1pSOI}
		\begin{split}
			&\hbar^2 \Gamma_1^\text{1p} = 2\hbar^2 \left(J^+_{xx} (\omega_z)+J^+_{yy} (\omega_z)\right)^\text{1p}\\
			&=\sin^2 \frac{\vartheta_S}{2} \sin^2 \frac{\vartheta_T}{2} \int_{-\infty}^{\infty} d\tau\cos (\omega_z \tau)
			\langle P_{R1,R2}(\mathbf{q}_1)P_{R1,R2}(\mathbf{q}_2)\rangle_\beta.
		\end{split}
	\end{equation}
	
	Similarly, for four-electron system, we define the projector
	\begin{equation}
		\begin{split}
			\mathcal{P}^{(S5)} = |S(1,3)\rangle \langle S(1,3)| +|T(1,3)\rangle\langle T(1,3)|+|T_+(1,3)\rangle\langle T_+(1,3)|+|T_+(1,3^\star)\rangle\langle T_+(1,3^\star)|.
		\end{split}
	\end{equation}
	Also, we split $\mathcal{P}^{(S5)}$ into two subspaces,
	\begin{equation}
		\mathcal{P}^{(S5)} = \mathcal{P}_{1}^{(S5)} + \mathcal{P}^{(S5)}_{2},
	\end{equation}
	where
	\begin{equation}
		\begin{split}
			\mathcal{P}^{(S5)}_{1} &= |S(1,3)\rangle \langle S(1,3)| +|T_+(1,3)\rangle\langle T_+(1,3)|,\\
			\mathcal{P}^{(S5)}_{2} &= |T(1,3)\rangle\langle T(1,3)|+|T_+(1,3^\star)\rangle\langle T_+(1,3^\star)|.
		\end{split}
	\end{equation}
	The resulting four-electron effective Hamiltonian is
	\begin{equation}
		\begin{split}
			H_{e}^{(S5)}&=\mathcal{P}^{(S5)}H_\text{2e}\mathcal{P}^{(S5)}\\
			&=\left(
			\begin{array}{cccc}
				-J^{(1,3)} &
				0 &
				-\frac{\widetilde{\Omega}^-}{\sqrt{2}} &
				\frac{\widetilde{\Omega}^-_{1^*}}{\sqrt{2}}
				\\
				0 &
				0 &
				0 &
				\frac{\widetilde{\Omega}^-_{1^*}}{\sqrt{2}}
				\\
				-\frac{\widetilde{\Omega}^+}{\sqrt{2}}&
				0 &
				\Delta E+E_\text{Z} 
				& 0
				\\
				\frac{\widetilde{\Omega}^+_{1}}{\sqrt{2}}&
				\frac{\widetilde{\Omega}^+_{1}}{\sqrt{2}} &
				0 &
				\Delta E + E_\text{Z}
			\end{array}
			\right)+E'_\text{shift},
		\end{split}
	\end{equation}
	where $E'_\text{shift}=E_\text{shift}+J^{(1,3)}$ and $\Delta E = \varepsilon_{R3}-\varepsilon_{R2}$. The derivation of effective electron-phonon interaction for the four-electron case (4e) can be replicated from the two-electron (2e) case (Eq.~\eqref{eq:P1}-\eqref{eq:ePhSOI}), by replacing the relevant parameters as shown in Table.~\ref{tab:2e4e}.
	
	\begin{table}[t]
		\begin{tabular}{|p{3.2cm}|p{3cm}|p{8cm}|}
			\cline{1-3} 
			\hfil number of electron&\hfil2e & \hfil 4e\\
			\cline{1-3}
			&&  \\[-8pt]
			\hfil \multirow{9}{*}{parameters}&\hfil $\vartheta_S$& \hfil $\Theta_S$ \\
			\cline{2-3}
			&&  \\[-8pt]
			&\hfil $\vartheta_T$& \hfil $\Theta_T$ \\
			\cline{2-3}
			&&  \\[-8pt]
			&\hfil $J_S$& \hfil $J^{(1,3)}$ \\
			\cline{2-3}
			&&  \\[-8pt]
			&\hfil $\Delta E_{2e}$& \hfil $\Delta E$ \\
			\cline{2-3}
			&&  \\[-8pt]
			&\hfil $\Omega$& \hfil $-\widetilde{\Omega}^-$ \\
			\cline{2-3}
			&&  \\[-8pt]
			&\hfil $\Omega_1$& \hfil $-\widetilde{\Omega}^-_{1^*}$ \\
			\cline{2-3}
			&&  \\[-8pt]
			&\hfil $\varsigma$& \hfil 0 \\
			\cline{2-3}
			&&  \\[-8pt]
			&\hfil $\tan \vartheta_S= \frac{\sqrt{2}\Omega}{E_\text{Z}+J_S}$& \hfil $\tan \Theta_S= \frac{\sqrt{2}\widetilde{\Omega}}{E_\text{Z}+J^{(1,3)}}$\\[4pt]
			\cline{2-3}
			&&  \\[-8pt]
			&\hfil $\tan \vartheta_T=\frac{\sqrt{2}\Omega_1}{E_\text{Z}+\Delta E}$ & \hfil $\tan \Theta_T=\frac{\sqrt{2}\widetilde{\Omega}_1}{E_\text{Z}+\Delta E}$ \\[4pt]
			\cline{1-3}
			&&  \\[-8pt]
			\hfil \multirow{5}{*}{Slater determinants}&\hfil $|S'^+(1,1)\rangle$& \hfil $|S'^+(1,3)\rangle=-\cos \frac{\Theta_S}{2}|S(1,3)\rangle+\sin \frac{\Theta_S}{2}|T^+(1,3)\rangle$ \\
			\cline{2-3}
			&&  \\[-8pt]
			&\hfil $|T'_+(1,1)\rangle$& \hfil $|T'_+(1,3)\rangle=\sin \frac{\Theta_S}{2}|S(1,3)\rangle+\cos \frac{\Theta_S}{2}|T^+(1,3)\rangle$ \\
			\cline{2-3}
			&&  \\[-8pt]
			&\hfil $|T'^+(1,1)\rangle$& \hfil $|T'^+(1,3)\rangle=\cos \frac{\Theta_T}{2}|T(1,3)\rangle+\sin \frac{\Theta_S}{2}|T_+(1,3^\star)\rangle$ \\
			\cline{2-3}
			&&  \\[-8pt]
			&\hfil $|T'_+(1,1^\star)\rangle$& \hfil $|T'_+(1,3)\rangle=-\sin \frac{\Theta_T}{2}|T(1,3)\rangle+\cos \frac{\Theta_S}{2}|T_+(1,3^\star)\rangle$ \\
			\cline{1-3}
		\end{tabular}
		\caption{Correspondence between the parameters and Slater determinants adopted in the case of two electrons (2e) and four electrons (4e).}
		\label{tab:2e4e}
	\end{table}
	
	Denoting the unitary transformation that brings $H_{\text{e-ph}}^{(S5)} = \mathcal{P}^{(S5)}H_{\text{e-ph}}\mathcal{P}^{(S5)}$ to the subspace spanned by the eigenbases of  $\mathcal{P}_{1}^{(S5)}$ and $\mathcal{P}^{(S5)}_{2}$ as $U^{(S5)}$. The resulting electron-phonon interaction is
	\begin{equation}\label{eq:PhEl4esSOI}
		\begin{split}
			& \widetilde{H}_{\text{e-ph}}\\
			&=\left(U^{(S5)}\right)^\dagger H_{\text{e-ph}}^{(S4)} U^{(S5)} \\
			&=\left(
			\begin{array}{cccc}
				P_{L1,L1}+P_{R2,R2} &
				\sin \frac{\Theta_S}{2} \sin \frac{\Theta_T}{2} P_{R2,R3} &
				0 
				\\
				\sin \frac{\Theta_S}{2} \sin \frac{\Theta_T}{2} P_{R2,R3} &
				P_{L1,L1}+\cos^2\Theta_T P_{R2,R2} +\sin ^2\frac{\Theta_T}{2} P_{R3,R3} &
				\cos \frac{\Theta_S}{2} \sin \frac{\Theta_T}{2} P_{R2,R3}
				\\
				0&
				\cos \frac{\Theta_S}{2} \sin\frac{\Theta_T}{2} P_{R2,R3} &
				P_{L1,L1}+P_{R2,R2} 
				\\
				\sin \frac{\Theta_S}{2} \cos \frac{\Theta_T}{2} P_{R2,R3} &
				\frac{1}{2} \sin \Theta_T\left(P_{R3,R3}-P_{R2,R2}\right) &
				\cos \frac{\Theta_S}{2} \cos \frac{\Theta_T}{2} P_{R2,R3}
				\\
			\end{array}
			\right.\\
			&\quad\left.
			\begin{array}{c} 
				\cos \frac{\Theta_T}{2} \sin \frac{\Theta_S}{2} P_{R2,R3}
				\\
				\frac{1}{2} \sin \Theta_T \left(P_{R3,R3}-P_{R2,R2}\right) 
				\\
				\cos \frac{\Theta_S}{2} \cos \frac{\Theta_T}{2}P_{R2,R3} 
				\\
				P_{L1,L1}+\cos ^2\Theta_T P_{R3,R3} +\sin ^2\frac{\Theta_T}{2} P_{R2,R2} \\
			\end{array}
			\right),
		\end{split}
	\end{equation}
	In contrast to the two-electron case, $\widetilde{\Omega}_1= \langle \Phi_{R2}|\widetilde{p}_{[jkm]}|\Phi_{R3}\rangle$ is negligible as $\langle \phi_{R2}| \widetilde{p}_{[100]}| \phi_{R3}\rangle = \langle \phi_{R2}|  \widetilde{p}_{[010]}| \phi_{R3}\rangle = 0$, giving
	\begin{equation}
		\begin{split}
			\langle \Phi_{R2}|\widetilde{p}_{[jkm]}|\Phi_{R3}\rangle &= S_{R2,R3} \Big[\langle \psi_{R2} | \widetilde{p}_{[jkm]} | \psi_{R3} \rangle -g^*\langle \psi_{L1} | \widetilde{p}_{[jkm]} | \psi_{R3} \rangle - g^{**} \langle \psi_{R2} | \widetilde{p}_{[jkm]} | \psi_{L1} \rangle\\
			&\quad +g^{*}g^{**} \langle \psi_{L1} | \widetilde{p}_{[jkm]} | \psi_{L1} \rangle\Big]\\
			&=S_{R2,R3} \Big[-g^*\langle \psi_{L1} | \widetilde{p}_{[jkm]} | \psi_{R3} \rangle - g^{**} \langle \psi_{R2} | \widetilde{p}_{[jkm]} | \psi_{L1} \rangle\\
			&\quad +g^{*}g^{**} \langle \psi_{L1} | \widetilde{p}_{[jkm]} | \psi_{L1} \rangle\Big]\\
		\end{split}
	\end{equation}
	where $S_{R2,R3}= 1/\sqrt{\left(1-2s^*g^*+(g^*)^2\right)\left(1-2s^{**}g^{**}+(g^{**})^2\right)}$ and $g^{*}\ll1, g^{**}\ll 1$ as they encode the overlap between orbitals in the left and right dot. Hence, as apposed to the two-electron case, SOI does not lead to considerable relaxation in four-electron case of interest.
	
	\subsection{Phonon induced decoherence time}
	\begin{figure}[t]
		\centering{
			\includegraphics[width=0.7\columnwidth]{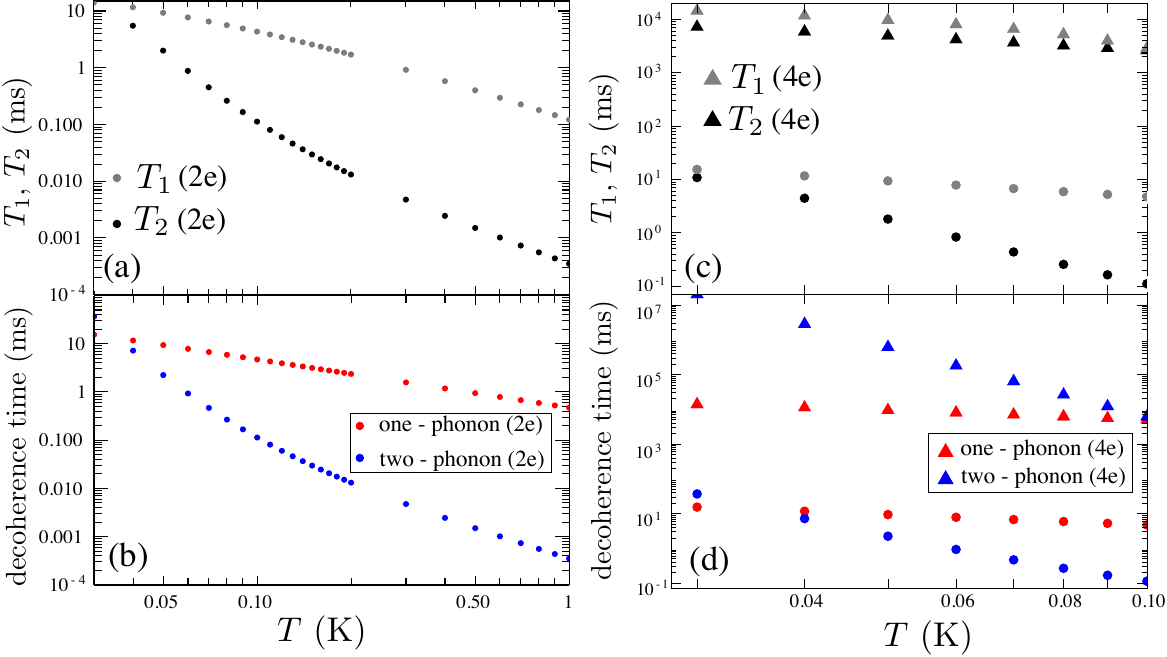}
		}
		\caption{(a), (c) The relaxation time, $T_1=1$ (gray circles), and decoherence time, $T_2$ (black circles), as function of temperature, $T$, for two-electron (2e) system \cite{Kornich.14} and four-electron (4e) system of interest. (b), (d) Decoherence time as function of temperature, $T$, arisen from one-phonon process, $1/\Gamma_1^{\text{(1p)}}$ (red circles), and two-phonon process (blue circles), $1/(\Gamma_\varphi^{\text{(2p)}}+\Gamma_1^{\text{(2p)}}/2)$, where the superscripts (1p) and (2p) denote one and two phonon process respectively.}
		\label{fig:2es4esT2}
	\end{figure}
	In this section, we show the calculated decoherence time, $T_2$, for the four-electron system (4e). We also show the result for two-electron (2e) system with dot parameters outlined in Ref.~\onlinecite{Kornich.14} for comparison. The results are shown in Fig.~\ref{fig:2es4esT2}. Fig.~\ref{fig:2es4esT2}(a) shows the decoherence time, $T_2$, and relaxation time, $T_1$, calculated up to second-order phonon process, as function of temperature while Fig.~\ref{fig:2es4esT2}(b) shows the decoherence time by one-phonon process, $1/\Gamma_1^{\text{(1p)}}$, and two-phonon process, $1/(\Gamma_\varphi^{\text{(2p)}}+\Gamma_1^{\text{(2p)}}/2)$, explicitly. Two-electron results in Fig.~\ref{fig:2es4esT2}(a) and (b) are identical with Ref.~\onlinecite{Kornich.14}. In Fig.~\ref{fig:2es4esT2}(c) and (d), we show the decoherence time for 0.03K$<T<0.1$K, within the range of temperature in which experiments operate GaAs spin qubit \cite{Dial.13,Nichol.17}. Fig.~\ref{fig:2es4esT2}(c) shows $T_2$ and $T_1$ for 4e system with parameters of interest in this work. The results in (a) are replicated here for comparison. It is observed that the $T_2$ and $T_1$ for 4e case are at least three orders longer as compare to 2e case. As $T_1$ is dominated by one-phonon relaxation process, we expect a suppressed $T_1$ in 4e case due to limited admixture with states of quadruple dot occupation and exponentially suppressed SOI terms, as discussed in Sec.~\ref{subsubsec:dBFstOrdRel} and Sec.~\ref{subsubsec:SOIFstOrdRel}. Fig.~\ref{fig:2es4esT2}(d) shows one-phonon and two-phonon process induced decoherence time for 4e case. Results in Fig.~\ref{fig:2es4esT2}(a) are also replicated here for comparison. Fig.~\ref{fig:2es4esT2}(d) shows that decoherence time in 4e case is dominated by one-phonon process, concurring with Fig.~\ref{fig:2es4esT2}(c), which shows $T_2 \approx 2T_1$. As discussed in Sec.~\ref{subsec:pureDep} and \ref{subsec:1pRel}, the difference between 2e and 4e cases, for the dot parameters of concern, can be attributed to the differences in electron Hamiltoinan, $H_e$, and the integral of electron wavefunction with electron density operator, $e^{i \mathbf{q}_m\cdot\mathbf{r}}$, leaving the temperature dependent part, the Bose-Einstein distribution, $n_s(q_m)$, unaffected. This results in both one and two-phonon decoherence time share the same temperature dependence for 2e and 4e cases, as observed in Fig.~\ref{fig:2es4esT2}(d).
	
	\section{Correspondence between Full CI results and Hubbard Model}\label{sec:CoressHubbardCI}
	\begin{figure}[t]
		\includegraphics[width=0.7\columnwidth]{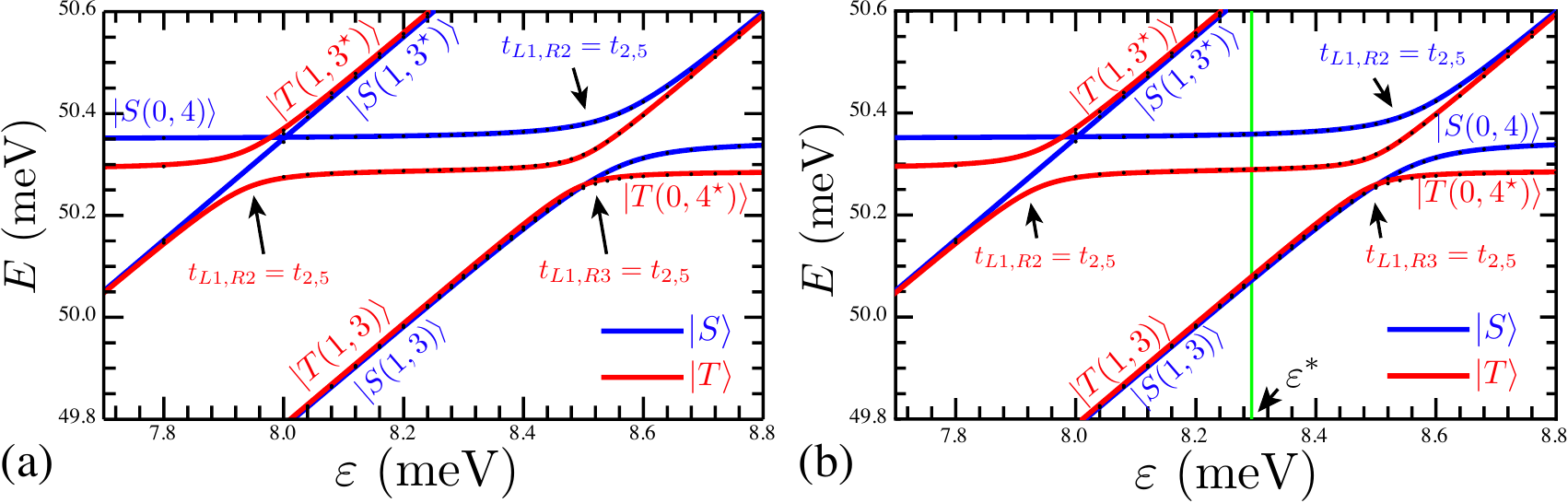}
		\caption{Eigenvalues of the four-electron states in a DQD near the transition point between $|T(1,3)\rangle$ and $|T(0,4^\star\rangle$. The eigenvalues obtained using Hubbard model are shown as blue and red solid line for singlet and triplet respectively. The eigenvalues obtained using full CI calculation are shown as black circles. (a) Eigenvalues of the system at $B=0.174$ T. The Hubbard parameters are: $U_{2,5}=U_{L1,R2} = 41.867$meV, $U_5=U_{R2}=50.532$meV, $U_{5,9}=U_{R2,R3}=50.393$meV, $\Delta E = 0.314$meV, $\xi=0.485$meV, $t_{2,5}=t_{L1,R2}=37\mu$eV, $t_{2,9}=t_{L1,R3}=29\mu$eV (b) Eigenvalues of the system at $B=0.319$ T. The Hubbard parameters are: $U_{2,5}=U_{L1,R2} = 41.790$meV, $U_5=U_{R2}=50.349$meV, $U_{5,9}=U_{R2,R3}=50.209$meV, $\Delta E = 0.564$meV, $\xi=0.485$meV, $t_{2,5}=t_{L1,R2}=37\mu$eV, $t_{2,9}=t_{L1,R3}=29\mu$eV. The green line indicates the effective exchange energy sweet spot, $\varepsilon^*$.
			\label{fig:HubbardModel0p687}
		}
	\end{figure}
	Figure \ref{fig:HubbardModel0p687} shows the lowest eigenvalues of four-electron states in a DQD device, where the color solid lines are the results obtained from Hubbard model while black circles are the results obtained using full CI calculation. The effective Hamiltonian obtained from Hubbard model can refer to Eq.~\eqref{eq:Ham4es} while excluding the polarized states, i.e. $S_z = \pm 1$, e.g. $|T_\pm(1,3)\rangle$. It can be observed that Hubbard model is able to recover the eigenvalues of full CI calculation with high accuracy.
	
	\section{Capacitive shift, $\beta_j$, and coupling, $\alpha$}
	Considering symmetric case as illustrated in Fig.~1 in the main text, i.e. two out-most QDs yield the same confinement strength $\hbar \omega_\text{out}$, two inner QDs yield the same confinement strength $\hbar \omega_\text{in}$, inter-dot distances ($2x_0$) in both DQD are the same, and detunings of $\mathbb{L}$ and $\mathbb{R}$ DQD are the same ($\varepsilon = \varepsilon_\mathbb{L} = \varepsilon_\mathbb{R}$), we have, from Eq.~(10), (11) and (13) in the main text,
	\begin{subequations}\label{eq:beta}
		\begin{align}
			\begin{split}
				&\beta \left(\varepsilon_\mathbb{L} = \varepsilon,\varepsilon_\mathbb{R} = \varepsilon\right)
				= \beta_\mathbb{L} \left(\varepsilon_\mathbb{L} = \varepsilon,\varepsilon_\mathbb{R} = \varepsilon\right) 
				= \beta_\mathbb{R} \left(\varepsilon_\mathbb{L} = \varepsilon,\varepsilon_\mathbb{R} = \varepsilon\right)
			\end{split}\\
			\begin{split}
				&=\left(\cos ^4\theta_T-\cos ^4\theta_S\right)U_{2,3} + \left(\cos ^2\theta_T-\cos  ^2\theta_S\right)\left(U_{2,4}+U_{2,8}\right)-2  \left(\cos ^4\theta_S-\cos ^2\theta_T\right)U_{2,8}-4 \sin ^2\theta_S  U_{1,8}\\
				&\quad +4  \sin ^2\theta_T U_{4,9}+\sin ^4\theta_T U_{9,12} +U_{5,8} \sin ^2\theta_S
				\left(\cos ^2\theta_S-3\right)+2 U_{8,9} \sin ^2\theta_T+2 U_{3,9}
				\sin ^2\theta_T \cos ^2\theta_T
			\end{split}\\
			\begin{split}\label{eq:betaApprox}
				&\approx \left(\cos ^4\theta_T-\cos ^4\theta_S\right)U_{2,3} + \left(\cos ^2\theta_T-\cos  ^2\theta_S\right)\left(U_{2,4}+U_{2,8}\right)+2  \left(\cos ^2\theta_T-\cos ^4\theta_S\right)U_{2,8}
			\end{split}\\
			\begin{split}\label{eq:betaApprox2}
				&\approx \left(\theta_S^2-\theta_T^2\right) \left(2U_{2,3}+U_{2,4}\right)+\left(5\theta_S^2-3\theta_T^2\right)U_{2,8}
			\end{split}
		\end{align}
	\end{subequations}
	\begin{subequations}
		\begin{align}\label{eq:alpha}
			\begin{split}
				&\alpha \left(\varepsilon_\mathbb{L} = \varepsilon,\varepsilon_\mathbb{R} = \varepsilon\right)
			\end{split}\\
			\begin{split}
				&= \left(\cos ^2\theta_S-\cos ^2\theta_T\right)^2U_{2,3}
				+ \sin ^2\theta_S \left(\cos 2 \theta_S-\cos 2 \theta_T\right)U_{2,8}
				+2 \sin ^2\theta_T \left(\cos ^2\theta_T-\cos ^2\theta_S\right) U_{3,9}\\
				&\quad+ \sin ^4 \theta_SU_{5,8}-2  \sin ^2\theta_S \sin ^2\theta_TU_{8,9}+\sin ^4\theta_TU_{9,12} 
			\end{split}\\
			\begin{split}\label{eq:alphaApprox}
				&\approx \left(\cos ^2\theta_S-\cos ^2\theta_T\right)^2 U_{2,3}
			\end{split}\\
			\begin{split}\label{eq:alphaApprox2}
				&=\frac{1}{16} \sin \left(\theta_S + \theta_T \right)^2 \sin \left(\theta_S - \theta_T \right)^2 U_{2,3}
			\end{split}\\
			\begin{split}\label{eq:alphaApprox3}
				&\approx\frac{1}{16} \left(\theta_S^2 - \theta_T^2 \right)^2  U_{2,3}
			\end{split}
		\end{align}
	\end{subequations}
	In the symmetric case, $V_{|ST\rangle}$ and $V_{|TS\rangle}$ are the same, resulting in $\beta_\mathbb{L} = \beta_\mathbb{R}$, cf.~Eq.~(10) in the main text. Eq.~\eqref{eq:betaApprox} and \eqref{eq:alphaApprox} are obtained by recognizing that $|\theta_S| \ll 1$ and $|\theta_T| \ll 1$ in the detuning range of interest (cf.~Fig.~2 in the main text), resulting in $\sin\theta_S\ll \cos \theta_S$ and $\sin\theta_T\ll \cos \theta_T$. In addition, Eq.~\eqref{eq:betaApprox2} is obtained by performing Taylor expansion on $\theta_S$ and $\theta_T$ in Eq.~\eqref{eq:betaApprox} to the lowest non-zero order while Eq.~\eqref{eq:alphaApprox3} is obtained by performing Taylor expansion on $\theta_S\pm \theta_T$ in Eq.~\eqref{eq:alphaApprox2}.
	
	\section{Two-qubit Gate Fidelity}\label{sec:twoQGateFidelity}
	In the main text, we evaluate the average gate fidelity  $F$ as \cite{Nielsen.02,Horodecki.99}:
	\begin{equation}
		F = \frac{d F_e+1}{d+1},
	\end{equation}
	where $d$ is the dimension of the system ($d=4$ for a two-qubit system). The entanglement fidelity, $F_e$, for a noisy two-qubit gate is defined by setting the initial state as a maximally entangled state $|\Psi_0\rangle$ of four qubits, two of which is applied upon by the gate. To calculate $F$ for two-qubit gates on singlet-triplet qubits, the initial state is $|\Psi_0\rangle = \frac{1}{2} \sum_{j,k=S,T}{|jk,jk\rangle}$, with the initial density matrix $\rho_{\Psi_0} =|\Psi_0\rangle\langle\Psi_0|=\frac{1}{4}\sum_{j,k,m,n=S,T}|jk,jk\rangle\langle mn,mn|$. The resulting density matrix after evolution in the noisy environment is then $\rho = (\mathcal{N}_\chi \otimes I)[\rho_{\Psi_0}]$, where $\mathcal{N}_\chi[\rho] = \widetilde{U}_\chi \rho \widetilde{U}_\chi^\dagger$ and $\widetilde{U}_\chi$ encapsulates the noisy effects. The entanglement fidelity is then  $F_e = \langle \Psi | (\mathcal{N}_\chi\otimes I)[\rho_{\Psi_0}] |\Psi\rangle$, where $|\Psi\rangle$ is the resulting state after an ideal evolution.
	
	\section{Echo pulse}
	The echo pulse is employed to cancel the quasistatic fluctuations of control parameters. To obtain an analytical result, considering quasistatic charge noise fluctuations,
	\begin{equation}\label{eq:noiseHcharge}
		\widetilde{U}_\chi = 
		\exp\left[-i \left\{\widetilde{H}\left(\delta \varepsilon^\text{QS} \right)\otimes I\right\} \frac{t_{2\pi}}{2} \right]
		\exp\left[- i \left\{\frac{\pi}{2}\left(\sigma_x^{(\mathbb{L})} + \sigma_x^{(\mathbb{R})}\right) \otimes I \right\} \right] 
		\exp\left[-i \left\{\widetilde{H}\left(\delta \varepsilon^\text{QS} \right)\otimes I\right\} \frac{t_{2\pi}}{2} \right],
	\end{equation}
	where $\widetilde{H}\left(\delta \varepsilon^\text{QS} \right)$ is the noisy Hamiltonian involving only the quasistatic charge noise fluctuations. 
	From Eq.~\eqref{eq:noiseHcharge} and Sec.~\ref{sec:twoQGateFidelity}, the gate fidelity is
	\begin{equation}\label{eq:alphaDeltaFid}
		F_e = \cos^2\left(2 \times \delta \alpha \times t_{2\pi}\right),
	\end{equation}
	where $\alpha\left(\varepsilon_\mathbb{L}+\delta \varepsilon_\mathbb{L}^\text{QS},\varepsilon_\mathbb{R}+\delta \varepsilon_\mathbb{R}^\text{QS}\right) = \alpha \left(\varepsilon_\mathbb{L},\varepsilon_\mathbb{R}\right)+\delta \alpha^\text{QS}$. Eq.~\eqref{eq:alphaDeltaFid} shows that a simple echo pulse is not able to cancel quasistatic fluctuations on $\alpha$.
	

\begin{thebibliography}{106}%
		\makeatletter
		\providecommand \@ifxundefined [1]{%
			\@ifx{#1\undefined}
		}%
		\providecommand \@ifnum [1]{%
			\ifnum #1\expandafter \@firstoftwo
			\else \expandafter \@secondoftwo
			\fi
		}%
		\providecommand \@ifx [1]{%
			\ifx #1\expandafter \@firstoftwo
			\else \expandafter \@secondoftwo
			\fi
		}%
		\providecommand \natexlab [1]{#1}%
		\providecommand \enquote  [1]{``#1''}%
		\providecommand \bibnamefont  [1]{#1}%
		\providecommand \bibfnamefont [1]{#1}%
		\providecommand \citenamefont [1]{#1}%
		\providecommand \href@noop [0]{\@secondoftwo}%
		\providecommand \href [0]{\begingroup \@sanitize@url \@href}%
		\providecommand \@href[1]{\@@startlink{#1}\@@href}%
		\providecommand \@@href[1]{\endgroup#1\@@endlink}%
		\providecommand \@sanitize@url [0]{\catcode `\\12\catcode `\$12\catcode
			`\&12\catcode `\#12\catcode `\^12\catcode `\_12\catcode `\%12\relax}%
		\providecommand \@@startlink[1]{}%
		\providecommand \@@endlink[0]{}%
		\providecommand \url  [0]{\begingroup\@sanitize@url \@url }%
		\providecommand \@url [1]{\endgroup\@href {#1}{\urlprefix }}%
		\providecommand \urlprefix  [0]{URL }%
		\providecommand \Eprint [0]{\href }%
		\providecommand \doibase [0]{http://dx.doi.org/}%
		\providecommand \selectlanguage [0]{\@gobble}%
		\providecommand \bibinfo  [0]{\@secondoftwo}%
		\providecommand \bibfield  [0]{\@secondoftwo}%
		\providecommand \translation [1]{[#1]}%
		\providecommand \BibitemOpen [0]{}%
		\providecommand \bibitemStop [0]{}%
		\providecommand \bibitemNoStop [0]{.\EOS\space}%
		\providecommand \EOS [0]{\spacefactor3000\relax}%
		\providecommand \BibitemShut  [1]{\csname bibitem#1\endcsname}%
		\let\auto@bib@innerbib\@empty
		\bibitem [{\citenamefont {Petta}\ \emph {et~al.}(2005)\citenamefont {Petta},
			\citenamefont {Johnson}, \citenamefont {Taylor}, \citenamefont {Laird},
			\citenamefont {Yacoby}, \citenamefont {Lukin}, \citenamefont {Marcus},
			\citenamefont {Hanson},\ and\ \citenamefont {Gossard}}]{Petta.05}%
		\BibitemOpen
		\bibfield  {author} {\bibinfo {author} {\bibfnamefont {J.~R.}\ \bibnamefont
				{Petta}}, \bibinfo {author} {\bibfnamefont {A.~C.}\ \bibnamefont {Johnson}},
			\bibinfo {author} {\bibfnamefont {J.~M.}\ \bibnamefont {Taylor}}, \bibinfo
			{author} {\bibfnamefont {E.~A.}\ \bibnamefont {Laird}}, \bibinfo {author}
			{\bibfnamefont {A.}~\bibnamefont {Yacoby}}, \bibinfo {author} {\bibfnamefont
				{M.~D.}\ \bibnamefont {Lukin}}, \bibinfo {author} {\bibfnamefont {C.~M.}\
				\bibnamefont {Marcus}}, \bibinfo {author} {\bibfnamefont {M.~P.}\
				\bibnamefont {Hanson}}, \ and\ \bibinfo {author} {\bibfnamefont {A.~C.}\
				\bibnamefont {Gossard}},\ }\href
		{https://www.science.org/doi/10.1126/science.1116955} {\bibfield  {journal}
			{\bibinfo  {journal} {Science}\ }\textbf {\bibinfo {volume} {309}},\ \bibinfo
			{pages} {2180} (\bibinfo {year} {2005})}\BibitemShut {NoStop}%
		\bibitem [{\citenamefont {Shulman}\ \emph {et~al.}(2012)\citenamefont
			{Shulman}, \citenamefont {Dial}, \citenamefont {Harvey}, \citenamefont
			{Bluhm}, \citenamefont {Umansky},\ and\ \citenamefont {Yacoby}}]{Shulman.12}%
		\BibitemOpen
		\bibfield  {author} {\bibinfo {author} {\bibfnamefont {M.~D.}\ \bibnamefont
				{Shulman}}, \bibinfo {author} {\bibfnamefont {O.~E.}\ \bibnamefont {Dial}},
			\bibinfo {author} {\bibfnamefont {S.~P.}\ \bibnamefont {Harvey}}, \bibinfo
			{author} {\bibfnamefont {H.}~\bibnamefont {Bluhm}}, \bibinfo {author}
			{\bibfnamefont {V.}~\bibnamefont {Umansky}}, \ and\ \bibinfo {author}
			{\bibfnamefont {A.}~\bibnamefont {Yacoby}},\ }\href
		{https://www.science.org/doi/10.1126/science.1217692} {\bibfield  {journal}
			{\bibinfo  {journal} {Science}\ }\textbf {\bibinfo {volume} {336}},\ \bibinfo
			{pages} {202} (\bibinfo {year} {2012})}\BibitemShut {NoStop}%
		\bibitem [{\citenamefont {Levy}(2002)}]{Levy.02}%
		\BibitemOpen
		\bibfield  {author} {\bibinfo {author} {\bibfnamefont {J.}~\bibnamefont
				{Levy}},\ }\href
		{https://journals.aps.org/prl/abstract/10.1103/PhysRevLett.89.147902}
		{\bibfield  {journal} {\bibinfo  {journal} {Phys. Rev. Lett.}\ }\textbf
			{\bibinfo {volume} {89}},\ \bibinfo {pages} {147902} (\bibinfo {year}
			{2002})}\BibitemShut {NoStop}%
		\bibitem [{\citenamefont {Wu}\ \emph {et~al.}(2014)\citenamefont {Wu},
			\citenamefont {Ward}, \citenamefont {Prance}, \citenamefont {Kim},
			\citenamefont {Gamble}, \citenamefont {Mohr}, \citenamefont {Shi},
			\citenamefont {Savage}, \citenamefont {Lagally}, \citenamefont {Friesen},
			\citenamefont {Coppersmith},\ and\ \citenamefont {Eriksson}}]{Wu.14}%
		\BibitemOpen
		\bibfield  {author} {\bibinfo {author} {\bibfnamefont {X.}~\bibnamefont
				{Wu}}, \bibinfo {author} {\bibfnamefont {D.~R.}\ \bibnamefont {Ward}},
			\bibinfo {author} {\bibfnamefont {J.~R.}\ \bibnamefont {Prance}}, \bibinfo
			{author} {\bibfnamefont {D.}~\bibnamefont {Kim}}, \bibinfo {author}
			{\bibfnamefont {J.~K.}\ \bibnamefont {Gamble}}, \bibinfo {author}
			{\bibfnamefont {R.~T.}\ \bibnamefont {Mohr}}, \bibinfo {author}
			{\bibfnamefont {Z.}~\bibnamefont {Shi}}, \bibinfo {author} {\bibfnamefont
				{D.~E.}\ \bibnamefont {Savage}}, \bibinfo {author} {\bibfnamefont {M.~G.}\
				\bibnamefont {Lagally}}, \bibinfo {author} {\bibfnamefont {M.}~\bibnamefont
				{Friesen}}, \bibinfo {author} {\bibfnamefont {S.~N.}\ \bibnamefont
				{Coppersmith}}, \ and\ \bibinfo {author} {\bibfnamefont {M.~A.}\ \bibnamefont
				{Eriksson}},\ }\href {https://www.pnas.org/doi/abs/10.1073/pnas.1412230111}
		{\bibfield  {journal} {\bibinfo  {journal} {Proc. Natl. Acad. Sci. U.S.A.}\
			}\textbf {\bibinfo {volume} {111}},\ \bibinfo {pages} {11938} (\bibinfo
			{year} {2014})}\BibitemShut {NoStop}%
		\bibitem [{\citenamefont {Maune}\ \emph {et~al.}(2012)\citenamefont {Maune},
			\citenamefont {Borselli}, \citenamefont {Huang}, \citenamefont {Ladd},
			\citenamefont {Deelman}, \citenamefont {Holabird}, \citenamefont {Kiselev},
			\citenamefont {Alvarado-Rodriguez}, \citenamefont {Ross}, \citenamefont
			{Schmitz}, \citenamefont {Sokolich}, \citenamefont {Watson}, \citenamefont
			{Gyure},\ and\ \citenamefont {Hunter}}]{Maune.12}%
		\BibitemOpen
		\bibfield  {author} {\bibinfo {author} {\bibfnamefont {B.~M.}\ \bibnamefont
				{Maune}}, \bibinfo {author} {\bibfnamefont {M.~G.}\ \bibnamefont {Borselli}},
			\bibinfo {author} {\bibfnamefont {B.}~\bibnamefont {Huang}}, \bibinfo
			{author} {\bibfnamefont {T.~D.}\ \bibnamefont {Ladd}}, \bibinfo {author}
			{\bibfnamefont {P.~W.}\ \bibnamefont {Deelman}}, \bibinfo {author}
			{\bibfnamefont {K.~S.}\ \bibnamefont {Holabird}}, \bibinfo {author}
			{\bibfnamefont {A.~A.}\ \bibnamefont {Kiselev}}, \bibinfo {author}
			{\bibfnamefont {I.}~\bibnamefont {Alvarado-Rodriguez}}, \bibinfo {author}
			{\bibfnamefont {R.~S.}\ \bibnamefont {Ross}}, \bibinfo {author}
			{\bibfnamefont {A.~E.}\ \bibnamefont {Schmitz}}, \bibinfo {author}
			{\bibfnamefont {M.}~\bibnamefont {Sokolich}}, \bibinfo {author}
			{\bibfnamefont {C.~A.}\ \bibnamefont {Watson}}, \bibinfo {author}
			{\bibfnamefont {M.~F.}\ \bibnamefont {Gyure}}, \ and\ \bibinfo {author}
			{\bibfnamefont {A.~T.}\ \bibnamefont {Hunter}},\ }\href
		{https://www.nature.com/articles/nature10707} {\bibfield  {journal} {\bibinfo
				{journal} {Nature (London)}\ }\textbf {\bibinfo {volume} {481}},\ \bibinfo
			{pages} {344} (\bibinfo {year} {2012})}\BibitemShut {NoStop}%
		\bibitem [{\citenamefont {Barthel}\ \emph {et~al.}(2010)\citenamefont
			{Barthel}, \citenamefont {Medford}, \citenamefont {Marcus}, \citenamefont
			{Hanson},\ and\ \citenamefont {Gossard}}]{Barthel.10}%
		\BibitemOpen
		\bibfield  {author} {\bibinfo {author} {\bibfnamefont {C.}~\bibnamefont
				{Barthel}}, \bibinfo {author} {\bibfnamefont {J.}~\bibnamefont {Medford}},
			\bibinfo {author} {\bibfnamefont {C.~M.}\ \bibnamefont {Marcus}}, \bibinfo
			{author} {\bibfnamefont {M.~P.}\ \bibnamefont {Hanson}}, \ and\ \bibinfo
			{author} {\bibfnamefont {A.~C.}\ \bibnamefont {Gossard}},\ }\href
		{https://link.aps.org/doi/10.1103/PhysRevLett.105.266808} {\bibfield
			{journal} {\bibinfo  {journal} {Phys. Rev. Lett.}\ }\textbf {\bibinfo
				{volume} {105}},\ \bibinfo {pages} {266808} (\bibinfo {year}
			{2010})}\BibitemShut {NoStop}%
		\bibitem [{\citenamefont {Shi}\ \emph {et~al.}(2011)\citenamefont {Shi},
			\citenamefont {Simmons}, \citenamefont {Prance}, \citenamefont {King~Gamble},
			\citenamefont {Friesen}, \citenamefont {Savage}, \citenamefont {Lagally},
			\citenamefont {Coppersmith},\ and\ \citenamefont {Eriksson}}]{Shi.11}%
		\BibitemOpen
		\bibfield  {author} {\bibinfo {author} {\bibfnamefont {Z.}~\bibnamefont
				{Shi}}, \bibinfo {author} {\bibfnamefont {C.~B.}\ \bibnamefont {Simmons}},
			\bibinfo {author} {\bibfnamefont {J.~R.}\ \bibnamefont {Prance}}, \bibinfo
			{author} {\bibfnamefont {J.}~\bibnamefont {King~Gamble}}, \bibinfo {author}
			{\bibfnamefont {M.}~\bibnamefont {Friesen}}, \bibinfo {author} {\bibfnamefont
				{D.~E.}\ \bibnamefont {Savage}}, \bibinfo {author} {\bibfnamefont {M.~G.}\
				\bibnamefont {Lagally}}, \bibinfo {author} {\bibfnamefont {S.~N.}\
				\bibnamefont {Coppersmith}}, \ and\ \bibinfo {author} {\bibfnamefont {M.~A.}\
				\bibnamefont {Eriksson}},\ }\href
		{https://aip.scitation.org/doi/10.1063/1.3666232} {\bibfield  {journal}
			{\bibinfo  {journal} {Appl. Phys. Lett.}\ }\textbf {\bibinfo {volume} {99}},\
			\bibinfo {pages} {233108} (\bibinfo {year} {2011})}\BibitemShut {NoStop}%
		\bibitem [{\citenamefont {Takeda}\ \emph {et~al.}(2020)\citenamefont {Takeda},
			\citenamefont {Noiri}, \citenamefont {Yoneda}, \citenamefont {Nakajima},\
			and\ \citenamefont {Tarucha}}]{Takeda.20}%
		\BibitemOpen
		\bibfield  {author} {\bibinfo {author} {\bibfnamefont {K.}~\bibnamefont
				{Takeda}}, \bibinfo {author} {\bibfnamefont {A.}~\bibnamefont {Noiri}},
			\bibinfo {author} {\bibfnamefont {J.}~\bibnamefont {Yoneda}}, \bibinfo
			{author} {\bibfnamefont {T.}~\bibnamefont {Nakajima}}, \ and\ \bibinfo
			{author} {\bibfnamefont {S.}~\bibnamefont {Tarucha}},\ }\href
		{https://journals.aps.org/prl/abstract/10.1103/PhysRevLett.124.117701}
		{\bibfield  {journal} {\bibinfo  {journal} {Phys. Rev. Lett.}\ }\textbf
			{\bibinfo {volume} {124}},\ \bibinfo {pages} {117701} (\bibinfo {year}
			{2020})}\BibitemShut {NoStop}%
		\bibitem [{\citenamefont {Cerfontaine}\ \emph
			{et~al.}(2020{\natexlab{a}})\citenamefont {Cerfontaine}, \citenamefont
			{Botzem}, \citenamefont {Ritzmann}, \citenamefont {Humpohl}, \citenamefont
			{Ludwig}, \citenamefont {Schuh}, \citenamefont {Bougeard}, \citenamefont
			{Wieck},\ and\ \citenamefont {Bluhm}}]{Cerfontaine.20}%
		\BibitemOpen
		\bibfield  {author} {\bibinfo {author} {\bibfnamefont {P.}~\bibnamefont
				{Cerfontaine}}, \bibinfo {author} {\bibfnamefont {T.}~\bibnamefont {Botzem}},
			\bibinfo {author} {\bibfnamefont {J.}~\bibnamefont {Ritzmann}}, \bibinfo
			{author} {\bibfnamefont {S.~S.}\ \bibnamefont {Humpohl}}, \bibinfo {author}
			{\bibfnamefont {A.}~\bibnamefont {Ludwig}}, \bibinfo {author} {\bibfnamefont
				{D.}~\bibnamefont {Schuh}}, \bibinfo {author} {\bibfnamefont
				{D.}~\bibnamefont {Bougeard}}, \bibinfo {author} {\bibfnamefont {A.~D.}\
				\bibnamefont {Wieck}}, \ and\ \bibinfo {author} {\bibfnamefont
				{H.}~\bibnamefont {Bluhm}},\ }\href
		{https://www.nature.com/articles/s41467-020-17865-3} {\bibfield  {journal}
			{\bibinfo  {journal} {Nat. Commun.}\ }\textbf {\bibinfo {volume} {11}},\
			\bibinfo {pages} {4144} (\bibinfo {year} {2020}{\natexlab{a}})}\BibitemShut
		{NoStop}%
		\bibitem [{\citenamefont {Eng}\ \emph {et~al.}(2015)\citenamefont {Eng},
			\citenamefont {Ladd}, \citenamefont {Smith}, \citenamefont {Borselli},
			\citenamefont {Kiselev}, \citenamefont {Fong}, \citenamefont {Holabird},
			\citenamefont {Hazard}, \citenamefont {Huang}, \citenamefont {Deelman},
			\citenamefont {Milosavljevic}, \citenamefont {Schmitz}, \citenamefont {Ross},
			\citenamefont {Gyure},\ and\ \citenamefont {Hunter}}]{Eng.15}%
		\BibitemOpen
		\bibfield  {author} {\bibinfo {author} {\bibfnamefont {K.}~\bibnamefont
				{Eng}}, \bibinfo {author} {\bibfnamefont {T.~D.}\ \bibnamefont {Ladd}},
			\bibinfo {author} {\bibfnamefont {A.}~\bibnamefont {Smith}}, \bibinfo
			{author} {\bibfnamefont {M.~G.}\ \bibnamefont {Borselli}}, \bibinfo {author}
			{\bibfnamefont {A.~A.}\ \bibnamefont {Kiselev}}, \bibinfo {author}
			{\bibfnamefont {B.~H.}\ \bibnamefont {Fong}}, \bibinfo {author}
			{\bibfnamefont {K.~S.}\ \bibnamefont {Holabird}}, \bibinfo {author}
			{\bibfnamefont {T.~M.}\ \bibnamefont {Hazard}}, \bibinfo {author}
			{\bibfnamefont {B.}~\bibnamefont {Huang}}, \bibinfo {author} {\bibfnamefont
				{P.~W.}\ \bibnamefont {Deelman}}, \bibinfo {author} {\bibfnamefont
				{I.}~\bibnamefont {Milosavljevic}}, \bibinfo {author} {\bibfnamefont {A.~E.}\
				\bibnamefont {Schmitz}}, \bibinfo {author} {\bibfnamefont {R.~S.}\
				\bibnamefont {Ross}}, \bibinfo {author} {\bibfnamefont {M.~F.}\ \bibnamefont
				{Gyure}}, \ and\ \bibinfo {author} {\bibfnamefont {A.~T.}\ \bibnamefont
				{Hunter}},\ }\href {https://www.science.org/doi/10.1126/sciadv.1500214}
		{\bibfield  {journal} {\bibinfo  {journal} {Sci. Adv.}\ }\textbf {\bibinfo
				{volume} {1}},\ \bibinfo {pages} {e1500214} (\bibinfo {year}
			{2015})}\BibitemShut {NoStop}%
		\bibitem [{\citenamefont {Noiri}\ \emph {et~al.}(2018)\citenamefont {Noiri},
			\citenamefont {Nakajima}, \citenamefont {Yoneda}, \citenamefont {Delbecq},
			\citenamefont {Stano}, \citenamefont {Otsuka}, \citenamefont {Takeda},
			\citenamefont {Amaha}, \citenamefont {Allison}, \citenamefont {Kawasaki},
			\citenamefont {Kojima}, \citenamefont {Ludwig}, \citenamefont {Wieck},
			\citenamefont {Loss},\ and\ \citenamefont {Tarucha}}]{Noiri.18}%
		\BibitemOpen
		\bibfield  {author} {\bibinfo {author} {\bibfnamefont {A.}~\bibnamefont
				{Noiri}}, \bibinfo {author} {\bibfnamefont {T.}~\bibnamefont {Nakajima}},
			\bibinfo {author} {\bibfnamefont {J.}~\bibnamefont {Yoneda}}, \bibinfo
			{author} {\bibfnamefont {M.~R.}\ \bibnamefont {Delbecq}}, \bibinfo {author}
			{\bibfnamefont {P.}~\bibnamefont {Stano}}, \bibinfo {author} {\bibfnamefont
				{T.}~\bibnamefont {Otsuka}}, \bibinfo {author} {\bibfnamefont
				{K.}~\bibnamefont {Takeda}}, \bibinfo {author} {\bibfnamefont
				{S.}~\bibnamefont {Amaha}}, \bibinfo {author} {\bibfnamefont
				{G.}~\bibnamefont {Allison}}, \bibinfo {author} {\bibfnamefont
				{K.}~\bibnamefont {Kawasaki}}, \bibinfo {author} {\bibfnamefont
				{Y.}~\bibnamefont {Kojima}}, \bibinfo {author} {\bibfnamefont
				{A.}~\bibnamefont {Ludwig}}, \bibinfo {author} {\bibfnamefont {A.~D.}\
				\bibnamefont {Wieck}}, \bibinfo {author} {\bibfnamefont {D.}~\bibnamefont
				{Loss}}, \ and\ \bibinfo {author} {\bibfnamefont {S.}~\bibnamefont
				{Tarucha}},\ }\href {https://www.nature.com/articles/s41467-018-07522-1}
		{\bibfield  {journal} {\bibinfo  {journal} {Nat. Commun.}\ }\textbf {\bibinfo
				{volume} {9}},\ \bibinfo {pages} {5066} (\bibinfo {year} {2018})}\BibitemShut
		{NoStop}%
		\bibitem [{\citenamefont {{Harvey-Collard}}\ \emph {et~al.}(2017)\citenamefont
			{{Harvey-Collard}}, \citenamefont {{Jock}}, \citenamefont {{Jacobson}},
			\citenamefont {{Baczewski}}, \citenamefont {{Mounce}}, \citenamefont
			{{Curry}}, \citenamefont {{Ward}}, \citenamefont {{Anderson}}, \citenamefont
			{{Manginell}}, \citenamefont {{Wendt}}, \citenamefont {{Rudolph}},
			\citenamefont {{Pluym}}, \citenamefont {{Lilly}}, \citenamefont
			{{Pioro-Ladrière}},\ and\ \citenamefont {{Carroll}}}]{Harvey.17}%
		\BibitemOpen
		\bibfield  {author} {\bibinfo {author} {\bibfnamefont {P.}~\bibnamefont
				{{Harvey-Collard}}}, \bibinfo {author} {\bibfnamefont {R.~M.}\ \bibnamefont
				{{Jock}}}, \bibinfo {author} {\bibfnamefont {N.~T.}\ \bibnamefont
				{{Jacobson}}}, \bibinfo {author} {\bibfnamefont {A.~D.}\ \bibnamefont
				{{Baczewski}}}, \bibinfo {author} {\bibfnamefont {A.~M.}\ \bibnamefont
				{{Mounce}}}, \bibinfo {author} {\bibfnamefont {M.~J.}\ \bibnamefont
				{{Curry}}}, \bibinfo {author} {\bibfnamefont {D.~R.}\ \bibnamefont {{Ward}}},
			\bibinfo {author} {\bibfnamefont {J.~M.}\ \bibnamefont {{Anderson}}},
			\bibinfo {author} {\bibfnamefont {R.~P.}\ \bibnamefont {{Manginell}}},
			\bibinfo {author} {\bibfnamefont {J.~R.}\ \bibnamefont {{Wendt}}}, \bibinfo
			{author} {\bibfnamefont {M.}~\bibnamefont {{Rudolph}}}, \bibinfo {author}
			{\bibfnamefont {T.}~\bibnamefont {{Pluym}}}, \bibinfo {author} {\bibfnamefont
				{M.~P.}\ \bibnamefont {{Lilly}}}, \bibinfo {author} {\bibfnamefont
				{M.}~\bibnamefont {{Pioro-Ladrière}}}, \ and\ \bibinfo {author}
			{\bibfnamefont {M.~S.}\ \bibnamefont {{Carroll}}},\ }in\ \href
		{https://ieeexplore.ieee.org/document/8268507} {\emph {\bibinfo {booktitle}
				{2017 IEEE International Electron Devices Meeting (IEDM)}}}\ (\bibinfo {year}
		{IEEE, New York, 2017})\ pp.\ \bibinfo {pages} {36.5.1--36.5.4}\BibitemShut
		{NoStop}%
		\bibitem [{\citenamefont {Koh}\ \emph {et~al.}(2012)\citenamefont {Koh},
			\citenamefont {Gamble}, \citenamefont {Friesen}, \citenamefont {Eriksson},\
			and\ \citenamefont {Coppersmith}}]{Koh.12}%
		\BibitemOpen
		\bibfield  {author} {\bibinfo {author} {\bibfnamefont {T.~S.}\ \bibnamefont
				{Koh}}, \bibinfo {author} {\bibfnamefont {J.~K.}\ \bibnamefont {Gamble}},
			\bibinfo {author} {\bibfnamefont {M.}~\bibnamefont {Friesen}}, \bibinfo
			{author} {\bibfnamefont {M.~A.}\ \bibnamefont {Eriksson}}, \ and\ \bibinfo
			{author} {\bibfnamefont {S.~N.}\ \bibnamefont {Coppersmith}},\ }\href
		{https://link.aps.org/doi/10.1103/PhysRevLett.109.250503} {\bibfield
			{journal} {\bibinfo  {journal} {Phys. Rev. Lett.}\ }\textbf {\bibinfo
				{volume} {109}},\ \bibinfo {pages} {250503} (\bibinfo {year}
			{2012})}\BibitemShut {NoStop}%
		\bibitem [{\citenamefont {Shi}\ \emph {et~al.}(2014)\citenamefont {Shi},
			\citenamefont {Simmons}, \citenamefont {Ward}, \citenamefont {Prance},
			\citenamefont {Wu}, \citenamefont {Koh}, \citenamefont {Gamble},
			\citenamefont {Savage}, \citenamefont {Lagally}, \citenamefont {Friesen},
			\citenamefont {Coppersmith},\ and\ \citenamefont {Eriksson}}]{Shi.14}%
		\BibitemOpen
		\bibfield  {author} {\bibinfo {author} {\bibfnamefont {Z.}~\bibnamefont
				{Shi}}, \bibinfo {author} {\bibfnamefont {C.~B.}\ \bibnamefont {Simmons}},
			\bibinfo {author} {\bibfnamefont {D.~R.}\ \bibnamefont {Ward}}, \bibinfo
			{author} {\bibfnamefont {J.~R.}\ \bibnamefont {Prance}}, \bibinfo {author}
			{\bibfnamefont {X.}~\bibnamefont {Wu}}, \bibinfo {author} {\bibfnamefont
				{T.~S.}\ \bibnamefont {Koh}}, \bibinfo {author} {\bibfnamefont {J.~K.}\
				\bibnamefont {Gamble}}, \bibinfo {author} {\bibfnamefont {D.~E.}\
				\bibnamefont {Savage}}, \bibinfo {author} {\bibfnamefont {M.~G.}\
				\bibnamefont {Lagally}}, \bibinfo {author} {\bibfnamefont {M.}~\bibnamefont
				{Friesen}}, \bibinfo {author} {\bibfnamefont {S.~N.}\ \bibnamefont
				{Coppersmith}}, \ and\ \bibinfo {author} {\bibfnamefont {M.~A.}\ \bibnamefont
				{Eriksson}},\ }\href {\doibase 10.1038/ncomms4020} {\bibfield  {journal}
			{\bibinfo  {journal} {Nat. Commun.}\ }\textbf {\bibinfo {volume} {5}},\
			\bibinfo {pages} {3020} (\bibinfo {year} {2014})}\BibitemShut {NoStop}%
		\bibitem [{\citenamefont {Shi}\ \emph {et~al.}(2012)\citenamefont {Shi},
			\citenamefont {Simmons}, \citenamefont {Prance}, \citenamefont {Gamble},
			\citenamefont {Koh}, \citenamefont {Shim}, \citenamefont {Hu}, \citenamefont
			{Savage}, \citenamefont {Lagally}, \citenamefont {Eriksson}, \citenamefont
			{Friesen},\ and\ \citenamefont {Coppersmith}}]{Shi.12}%
		\BibitemOpen
		\bibfield  {author} {\bibinfo {author} {\bibfnamefont {Z.}~\bibnamefont
				{Shi}}, \bibinfo {author} {\bibfnamefont {C.~B.}\ \bibnamefont {Simmons}},
			\bibinfo {author} {\bibfnamefont {J.~R.}\ \bibnamefont {Prance}}, \bibinfo
			{author} {\bibfnamefont {J.~K.}\ \bibnamefont {Gamble}}, \bibinfo {author}
			{\bibfnamefont {T.~S.}\ \bibnamefont {Koh}}, \bibinfo {author} {\bibfnamefont
				{Y.-P.}\ \bibnamefont {Shim}}, \bibinfo {author} {\bibfnamefont
				{X.}~\bibnamefont {Hu}}, \bibinfo {author} {\bibfnamefont {D.~E.}\
				\bibnamefont {Savage}}, \bibinfo {author} {\bibfnamefont {M.~G.}\
				\bibnamefont {Lagally}}, \bibinfo {author} {\bibfnamefont {M.~A.}\
				\bibnamefont {Eriksson}}, \bibinfo {author} {\bibfnamefont {M.}~\bibnamefont
				{Friesen}}, \ and\ \bibinfo {author} {\bibfnamefont {S.~N.}\ \bibnamefont
				{Coppersmith}},\ }\href
		{https://link.aps.org/doi/10.1103/PhysRevLett.108.140503} {\bibfield
			{journal} {\bibinfo  {journal} {Phys. Rev. Lett.}\ }\textbf {\bibinfo
				{volume} {108}},\ \bibinfo {pages} {140503} (\bibinfo {year}
			{2012})}\BibitemShut {NoStop}%
		\bibitem [{\citenamefont {Thorgrimsson}\ \emph {et~al.}(2017)\citenamefont
			{Thorgrimsson}, \citenamefont {Kim}, \citenamefont {Yang}, \citenamefont
			{Smith}, \citenamefont {Simmons}, \citenamefont {Ward}, \citenamefont
			{Foote}, \citenamefont {Corrigan}, \citenamefont {Savage}, \citenamefont
			{Lagally}, \citenamefont {Friesen}, \citenamefont {Coppersmith},\ and\
			\citenamefont {Eriksson}}]{Thorgrimsson.17}%
		\BibitemOpen
		\bibfield  {author} {\bibinfo {author} {\bibfnamefont {B.}~\bibnamefont
				{Thorgrimsson}}, \bibinfo {author} {\bibfnamefont {D.}~\bibnamefont {Kim}},
			\bibinfo {author} {\bibfnamefont {Y.-C.}\ \bibnamefont {Yang}}, \bibinfo
			{author} {\bibfnamefont {L.~W.}\ \bibnamefont {Smith}}, \bibinfo {author}
			{\bibfnamefont {C.~B.}\ \bibnamefont {Simmons}}, \bibinfo {author}
			{\bibfnamefont {D.~R.}\ \bibnamefont {Ward}}, \bibinfo {author}
			{\bibfnamefont {R.~H.}\ \bibnamefont {Foote}}, \bibinfo {author}
			{\bibfnamefont {J.}~\bibnamefont {Corrigan}}, \bibinfo {author}
			{\bibfnamefont {D.~E.}\ \bibnamefont {Savage}}, \bibinfo {author}
			{\bibfnamefont {M.~G.}\ \bibnamefont {Lagally}}, \bibinfo {author}
			{\bibfnamefont {M.}~\bibnamefont {Friesen}}, \bibinfo {author} {\bibfnamefont
				{S.~N.}\ \bibnamefont {Coppersmith}}, \ and\ \bibinfo {author} {\bibfnamefont
				{M.~A.}\ \bibnamefont {Eriksson}},\ }\href {\doibase
			10.1038/s41534-017-0034-2} {\bibfield  {journal} {\bibinfo  {journal} {npj
					Quantum Inf.}\ }\textbf {\bibinfo {volume} {3}},\ \bibinfo {pages} {32}
			(\bibinfo {year} {2017})}\BibitemShut {NoStop}%
		\bibitem [{\citenamefont {Koh}\ \emph {et~al.}(2013)\citenamefont {Koh},
			\citenamefont {Coppersmith},\ and\ \citenamefont {Friesen}}]{Koh.13}%
		\BibitemOpen
		\bibfield  {author} {\bibinfo {author} {\bibfnamefont {T.~S.}\ \bibnamefont
				{Koh}}, \bibinfo {author} {\bibfnamefont {S.~N.}\ \bibnamefont
				{Coppersmith}}, \ and\ \bibinfo {author} {\bibfnamefont {M.}~\bibnamefont
				{Friesen}},\ }\href {https://www.pnas.org/doi/abs/10.1073/pnas.1319875110}
		{\bibfield  {journal} {\bibinfo  {journal} {Proc. Natl. Acad. Sci. U.S.A.}\
			}\textbf {\bibinfo {volume} {110}},\ \bibinfo {pages} {19695} (\bibinfo
			{year} {2013})}\BibitemShut {NoStop}%
		\bibitem [{\citenamefont {Cao}\ \emph {et~al.}(2013)\citenamefont {Cao},
			\citenamefont {Li}, \citenamefont {Tu}, \citenamefont {Wang}, \citenamefont
			{Zhou}, \citenamefont {Xiao}, \citenamefont {Guo}, \citenamefont {Jiang},\
			and\ \citenamefont {Guo}}]{Cao.13}%
		\BibitemOpen
		\bibfield  {author} {\bibinfo {author} {\bibfnamefont {G.}~\bibnamefont
				{Cao}}, \bibinfo {author} {\bibfnamefont {H.-O.}\ \bibnamefont {Li}},
			\bibinfo {author} {\bibfnamefont {T.}~\bibnamefont {Tu}}, \bibinfo {author}
			{\bibfnamefont {L.}~\bibnamefont {Wang}}, \bibinfo {author} {\bibfnamefont
				{C.}~\bibnamefont {Zhou}}, \bibinfo {author} {\bibfnamefont {M.}~\bibnamefont
				{Xiao}}, \bibinfo {author} {\bibfnamefont {G.-C.}\ \bibnamefont {Guo}},
			\bibinfo {author} {\bibfnamefont {H.-W.}\ \bibnamefont {Jiang}}, \ and\
			\bibinfo {author} {\bibfnamefont {G.-P.}\ \bibnamefont {Guo}},\ }\href
		{https://www.nature.com/articles/ncomms2412} {\bibfield  {journal} {\bibinfo
				{journal} {Nat. Commun.}\ }\textbf {\bibinfo {volume} {4}},\ \bibinfo {pages}
			{1401} (\bibinfo {year} {2013})}\BibitemShut {NoStop}%
		\bibitem [{\citenamefont {Shinkai}\ \emph {et~al.}(2009)\citenamefont
			{Shinkai}, \citenamefont {Hayashi}, \citenamefont {Ota},\ and\ \citenamefont
			{Fujisawa}}]{Shinkai.09}%
		\BibitemOpen
		\bibfield  {author} {\bibinfo {author} {\bibfnamefont {G.}~\bibnamefont
				{Shinkai}}, \bibinfo {author} {\bibfnamefont {T.}~\bibnamefont {Hayashi}},
			\bibinfo {author} {\bibfnamefont {T.}~\bibnamefont {Ota}}, \ and\ \bibinfo
			{author} {\bibfnamefont {T.}~\bibnamefont {Fujisawa}},\ }\href
		{https://journals.aps.org/prl/abstract/10.1103/PhysRevLett.103.056802}
		{\bibfield  {journal} {\bibinfo  {journal} {Phys. Rev. Lett.}\ }\textbf
			{\bibinfo {volume} {103}},\ \bibinfo {pages} {056802} (\bibinfo {year}
			{2009})}\BibitemShut {NoStop}%
		\bibitem [{\citenamefont {Hayashi}\ \emph {et~al.}(2003)\citenamefont
			{Hayashi}, \citenamefont {Fujisawa}, \citenamefont {Cheong}, \citenamefont
			{Jeong},\ and\ \citenamefont {Hirayama}}]{Hayashi.03}%
		\BibitemOpen
		\bibfield  {author} {\bibinfo {author} {\bibfnamefont {T.}~\bibnamefont
				{Hayashi}}, \bibinfo {author} {\bibfnamefont {T.}~\bibnamefont {Fujisawa}},
			\bibinfo {author} {\bibfnamefont {H.~D.}\ \bibnamefont {Cheong}}, \bibinfo
			{author} {\bibfnamefont {Y.~H.}\ \bibnamefont {Jeong}}, \ and\ \bibinfo
			{author} {\bibfnamefont {Y.}~\bibnamefont {Hirayama}},\ }\href
		{https://journals.aps.org/prl/abstract/10.1103/PhysRevLett.91.226804}
		{\bibfield  {journal} {\bibinfo  {journal} {Phys. Rev. Lett.}\ }\textbf
			{\bibinfo {volume} {91}},\ \bibinfo {pages} {226804} (\bibinfo {year}
			{2003})}\BibitemShut {NoStop}%
		\bibitem [{\citenamefont {Petersson}\ \emph {et~al.}(2010)\citenamefont
			{Petersson}, \citenamefont {Petta}, \citenamefont {Lu},\ and\ \citenamefont
			{Gossard}}]{Petersson.10}%
		\BibitemOpen
		\bibfield  {author} {\bibinfo {author} {\bibfnamefont {K.~D.}\ \bibnamefont
				{Petersson}}, \bibinfo {author} {\bibfnamefont {J.~R.}\ \bibnamefont
				{Petta}}, \bibinfo {author} {\bibfnamefont {H.}~\bibnamefont {Lu}}, \ and\
			\bibinfo {author} {\bibfnamefont {A.~C.}\ \bibnamefont {Gossard}},\ }\href
		{https://journals.aps.org/prl/abstract/10.1103/PhysRevLett.105.246804}
		{\bibfield  {journal} {\bibinfo  {journal} {Phys. Rev. Lett.}\ }\textbf
			{\bibinfo {volume} {105}},\ \bibinfo {pages} {246804} (\bibinfo {year}
			{2010})}\BibitemShut {NoStop}%
		\bibitem [{\citenamefont {Dovzhenko}\ \emph {et~al.}(2011)\citenamefont
			{Dovzhenko}, \citenamefont {Stehlik}, \citenamefont {Petersson},
			\citenamefont {Petta}, \citenamefont {Lu},\ and\ \citenamefont
			{Gossard}}]{Dovzhenko.11}%
		\BibitemOpen
		\bibfield  {author} {\bibinfo {author} {\bibfnamefont {Y.}~\bibnamefont
				{Dovzhenko}}, \bibinfo {author} {\bibfnamefont {J.}~\bibnamefont {Stehlik}},
			\bibinfo {author} {\bibfnamefont {K.~D.}\ \bibnamefont {Petersson}}, \bibinfo
			{author} {\bibfnamefont {J.~R.}\ \bibnamefont {Petta}}, \bibinfo {author}
			{\bibfnamefont {H.}~\bibnamefont {Lu}}, \ and\ \bibinfo {author}
			{\bibfnamefont {A.~C.}\ \bibnamefont {Gossard}},\ }\href
		{https://journals.aps.org/prb/abstract/10.1103/PhysRevB.84.161302} {\bibfield
			{journal} {\bibinfo  {journal} {Phys. Rev. B}\ }\textbf {\bibinfo {volume}
				{84}},\ \bibinfo {pages} {161302} (\bibinfo {year} {2011})}\BibitemShut
		{NoStop}%
		\bibitem [{\citenamefont {Gorman}\ \emph {et~al.}(2005)\citenamefont {Gorman},
			\citenamefont {Hasko},\ and\ \citenamefont {Williams}}]{Gorman.05}%
		\BibitemOpen
		\bibfield  {author} {\bibinfo {author} {\bibfnamefont {J.}~\bibnamefont
				{Gorman}}, \bibinfo {author} {\bibfnamefont {D.~G.}\ \bibnamefont {Hasko}}, \
			and\ \bibinfo {author} {\bibfnamefont {D.~A.}\ \bibnamefont {Williams}},\
		}\href {https://journals.aps.org/prl/abstract/10.1103/PhysRevLett.95.090502}
		{\bibfield  {journal} {\bibinfo  {journal} {Phys. Rev. Lett.}\ }\textbf
			{\bibinfo {volume} {95}},\ \bibinfo {pages} {090502} (\bibinfo {year}
			{2005})}\BibitemShut {NoStop}%
		\bibitem [{\citenamefont {Shi}\ \emph {et~al.}(2013)\citenamefont {Shi},
			\citenamefont {Simmons}, \citenamefont {Ward}, \citenamefont {Prance},
			\citenamefont {Mohr}, \citenamefont {Koh}, \citenamefont {Gamble},
			\citenamefont {Wu}, \citenamefont {Savage}, \citenamefont {Lagally},
			\citenamefont {Friesen}, \citenamefont {Coppersmith},\ and\ \citenamefont
			{Eriksson}}]{Shi.13}%
		\BibitemOpen
		\bibfield  {author} {\bibinfo {author} {\bibfnamefont {Z.}~\bibnamefont
				{Shi}}, \bibinfo {author} {\bibfnamefont {C.~B.}\ \bibnamefont {Simmons}},
			\bibinfo {author} {\bibfnamefont {D.~R.}\ \bibnamefont {Ward}}, \bibinfo
			{author} {\bibfnamefont {J.~R.}\ \bibnamefont {Prance}}, \bibinfo {author}
			{\bibfnamefont {R.~T.}\ \bibnamefont {Mohr}}, \bibinfo {author}
			{\bibfnamefont {T.~S.}\ \bibnamefont {Koh}}, \bibinfo {author} {\bibfnamefont
				{J.~K.}\ \bibnamefont {Gamble}}, \bibinfo {author} {\bibfnamefont
				{X.}~\bibnamefont {Wu}}, \bibinfo {author} {\bibfnamefont {D.~E.}\
				\bibnamefont {Savage}}, \bibinfo {author} {\bibfnamefont {M.~G.}\
				\bibnamefont {Lagally}}, \bibinfo {author} {\bibfnamefont {M.}~\bibnamefont
				{Friesen}}, \bibinfo {author} {\bibfnamefont {S.~N.}\ \bibnamefont
				{Coppersmith}}, \ and\ \bibinfo {author} {\bibfnamefont {M.~A.}\ \bibnamefont
				{Eriksson}},\ }\href
		{https://journals.aps.org/prb/abstract/10.1103/PhysRevB.88.075416} {\bibfield
			{journal} {\bibinfo  {journal} {Phys. Rev. B}\ }\textbf {\bibinfo {volume}
				{88}},\ \bibinfo {pages} {075416} (\bibinfo {year} {2013})}\BibitemShut
		{NoStop}%
		\bibitem [{\citenamefont {Nichol}\ \emph {et~al.}(2017)\citenamefont {Nichol},
			\citenamefont {Orona}, \citenamefont {Harvey}, \citenamefont {Fallahi},
			\citenamefont {Gardner}, \citenamefont {Manfra},\ and\ \citenamefont
			{Yacoby}}]{Nichol.17}%
		\BibitemOpen
		\bibfield  {author} {\bibinfo {author} {\bibfnamefont {J.~M.}\ \bibnamefont
				{Nichol}}, \bibinfo {author} {\bibfnamefont {L.~A.}\ \bibnamefont {Orona}},
			\bibinfo {author} {\bibfnamefont {S.~P.}\ \bibnamefont {Harvey}}, \bibinfo
			{author} {\bibfnamefont {S.}~\bibnamefont {Fallahi}}, \bibinfo {author}
			{\bibfnamefont {G.~C.}\ \bibnamefont {Gardner}}, \bibinfo {author}
			{\bibfnamefont {M.~J.}\ \bibnamefont {Manfra}}, \ and\ \bibinfo {author}
			{\bibfnamefont {A.}~\bibnamefont {Yacoby}},\ }\href
		{https://www.nature.com/articles/s41534-016-0003-1} {\bibfield  {journal}
			{\bibinfo  {journal} {npj Quantum Inf.}\ }\textbf {\bibinfo {volume} {3}},\
			\bibinfo {pages} {3} (\bibinfo {year} {2017})}\BibitemShut {NoStop}%
		\bibitem [{\citenamefont {Taylor}\ \emph {et~al.}(2005)\citenamefont {Taylor},
			\citenamefont {Engel}, \citenamefont {D{\"u}r}, \citenamefont {Yacoby},
			\citenamefont {Marcus}, \citenamefont {Zoller},\ and\ \citenamefont
			{Lukin}}]{Taylor.05}%
		\BibitemOpen
		\bibfield  {author} {\bibinfo {author} {\bibfnamefont {J.~M.}\ \bibnamefont
				{Taylor}}, \bibinfo {author} {\bibfnamefont {H.~A.}\ \bibnamefont {Engel}},
			\bibinfo {author} {\bibfnamefont {W.}~\bibnamefont {D{\"u}r}}, \bibinfo
			{author} {\bibfnamefont {A.}~\bibnamefont {Yacoby}}, \bibinfo {author}
			{\bibfnamefont {C.~M.}\ \bibnamefont {Marcus}}, \bibinfo {author}
			{\bibfnamefont {P.}~\bibnamefont {Zoller}}, \ and\ \bibinfo {author}
			{\bibfnamefont {M.~D.}\ \bibnamefont {Lukin}},\ }\href
		{https://www.nature.com/articles/nphys174} {\bibfield  {journal} {\bibinfo
				{journal} {Nat. Phys.}\ }\textbf {\bibinfo {volume} {1}},\ \bibinfo {pages}
			{177} (\bibinfo {year} {2005})}\BibitemShut {NoStop}%
		\bibitem [{\citenamefont {Nielsen}\ \emph {et~al.}(2012)\citenamefont
			{Nielsen}, \citenamefont {Muller},\ and\ \citenamefont
			{Carroll}}]{Nielsen.12}%
		\BibitemOpen
		\bibfield  {author} {\bibinfo {author} {\bibfnamefont {E.}~\bibnamefont
				{Nielsen}}, \bibinfo {author} {\bibfnamefont {R.~P.}\ \bibnamefont {Muller}},
			\ and\ \bibinfo {author} {\bibfnamefont {M.~S.}\ \bibnamefont {Carroll}},\
		}\href {https://link.aps.org/doi/10.1103/PhysRevB.85.035319} {\bibfield
			{journal} {\bibinfo  {journal} {Phys. Rev. B}\ }\textbf {\bibinfo {volume}
				{85}},\ \bibinfo {pages} {035319} (\bibinfo {year} {2012})}\BibitemShut
		{NoStop}%
		\bibitem [{\citenamefont {Hiltunen}\ and\ \citenamefont
			{Harju}(2014)}]{Hiltunen.14}%
		\BibitemOpen
		\bibfield  {author} {\bibinfo {author} {\bibfnamefont {T.}~\bibnamefont
				{Hiltunen}}\ and\ \bibinfo {author} {\bibfnamefont {A.}~\bibnamefont
				{Harju}},\ }\href
		{https://journals.aps.org/prb/abstract/10.1103/PhysRevB.90.125303} {\bibfield
			{journal} {\bibinfo  {journal} {Phys. Rev. B}\ }\textbf {\bibinfo {volume}
				{90}},\ \bibinfo {pages} {125303} (\bibinfo {year} {2014})}\BibitemShut
		{NoStop}%
		\bibitem [{\citenamefont {Buterakos}\ \emph {et~al.}(2019)\citenamefont
			{Buterakos}, \citenamefont {Throckmorton},\ and\ \citenamefont
			{Das~Sarma}}]{Buterakos.19}%
		\BibitemOpen
		\bibfield  {author} {\bibinfo {author} {\bibfnamefont {D.}~\bibnamefont
				{Buterakos}}, \bibinfo {author} {\bibfnamefont {R.~E.}\ \bibnamefont
				{Throckmorton}}, \ and\ \bibinfo {author} {\bibfnamefont {S.}~\bibnamefont
				{Das~Sarma}},\ }\href
		{https://journals.aps.org/prb/abstract/10.1103/PhysRevB.100.075411}
		{\bibfield  {journal} {\bibinfo  {journal} {Phys. Rev. B}\ }\textbf {\bibinfo
				{volume} {100}},\ \bibinfo {pages} {075411} (\bibinfo {year}
			{2019})}\BibitemShut {NoStop}%
		\bibitem [{\citenamefont {Ramon}(2011)}]{Ramon.11}%
		\BibitemOpen
		\bibfield  {author} {\bibinfo {author} {\bibfnamefont {G.}~\bibnamefont
				{Ramon}},\ }\href
		{https://journals.aps.org/prb/abstract/10.1103/PhysRevB.84.155329} {\bibfield
			{journal} {\bibinfo  {journal} {Phys. Rev. B}\ }\textbf {\bibinfo {volume}
				{84}},\ \bibinfo {pages} {155329} (\bibinfo {year} {2011})}\BibitemShut
		{NoStop}%
		\bibitem [{\citenamefont {Calderon-Vargas}\ and\ \citenamefont
			{Kestner}(2015)}]{Calderon.15}%
		\BibitemOpen
		\bibfield  {author} {\bibinfo {author} {\bibfnamefont {F.~A.}\ \bibnamefont
				{Calderon-Vargas}}\ and\ \bibinfo {author} {\bibfnamefont {J.~P.}\
				\bibnamefont {Kestner}},\ }\href
		{https://link.aps.org/doi/10.1103/PhysRevB.91.035301} {\bibfield  {journal}
			{\bibinfo  {journal} {Phys. Rev. B}\ }\textbf {\bibinfo {volume} {91}},\
			\bibinfo {pages} {035301} (\bibinfo {year} {2015})}\BibitemShut {NoStop}%
		\bibitem [{\citenamefont {Wolfe}\ \emph {et~al.}(2017)\citenamefont {Wolfe},
			\citenamefont {Calderon-Vargas},\ and\ \citenamefont {Kestner}}]{Wolfe.17}%
		\BibitemOpen
		\bibfield  {author} {\bibinfo {author} {\bibfnamefont {M.~A.}\ \bibnamefont
				{Wolfe}}, \bibinfo {author} {\bibfnamefont {F.~A.}\ \bibnamefont
				{Calderon-Vargas}}, \ and\ \bibinfo {author} {\bibfnamefont {J.~P.}\
				\bibnamefont {Kestner}},\ }\href
		{https://link.aps.org/doi/10.1103/PhysRevB.96.201307} {\bibfield  {journal}
			{\bibinfo  {journal} {Phys. Rev. B}\ }\textbf {\bibinfo {volume} {96}},\
			\bibinfo {pages} {201307} (\bibinfo {year} {2017})}\BibitemShut {NoStop}%
		\bibitem [{\citenamefont {Stepanenko}\ and\ \citenamefont
			{Burkard}(2007)}]{Stepanenko.07}%
		\BibitemOpen
		\bibfield  {author} {\bibinfo {author} {\bibfnamefont {D.}~\bibnamefont
				{Stepanenko}}\ and\ \bibinfo {author} {\bibfnamefont {G.}~\bibnamefont
				{Burkard}},\ }\href {https://link.aps.org/doi/10.1103/PhysRevB.75.085324}
		{\bibfield  {journal} {\bibinfo  {journal} {Phys. Rev. B}\ }\textbf {\bibinfo
				{volume} {75}},\ \bibinfo {pages} {085324} (\bibinfo {year}
			{2007})}\BibitemShut {NoStop}%
		\bibitem [{\citenamefont {Yang}\ and\ \citenamefont
			{Das~Sarma}(2011)}]{Yang.11}%
		\BibitemOpen
		\bibfield  {author} {\bibinfo {author} {\bibfnamefont {S.}~\bibnamefont
				{Yang}}\ and\ \bibinfo {author} {\bibfnamefont {S.}~\bibnamefont
				{Das~Sarma}},\ }\href
		{https://journals.aps.org/prb/abstract/10.1103/PhysRevB.84.121306} {\bibfield
			{journal} {\bibinfo  {journal} {Phys. Rev. B}\ }\textbf {\bibinfo {volume}
				{84}},\ \bibinfo {pages} {121306} (\bibinfo {year} {2011})}\BibitemShut
		{NoStop}%
		\bibitem [{\citenamefont {Srinivasa}\ and\ \citenamefont
			{Taylor}(2015)}]{Srinivasa.15}%
		\BibitemOpen
		\bibfield  {author} {\bibinfo {author} {\bibfnamefont {V.}~\bibnamefont
				{Srinivasa}}\ and\ \bibinfo {author} {\bibfnamefont {J.~M.}\ \bibnamefont
				{Taylor}},\ }\href
		{https://journals.aps.org/prb/abstract/10.1103/PhysRevB.92.235301} {\bibfield
			{journal} {\bibinfo  {journal} {Phys. Rev. B}\ }\textbf {\bibinfo {volume}
				{92}},\ \bibinfo {pages} {235301} (\bibinfo {year} {2015})}\BibitemShut
		{NoStop}%
		\bibitem [{\citenamefont {Setser}\ and\ \citenamefont
			{Kestner}(2019)}]{Setser.19}%
		\BibitemOpen
		\bibfield  {author} {\bibinfo {author} {\bibfnamefont {A.~A.}\ \bibnamefont
				{Setser}}\ and\ \bibinfo {author} {\bibfnamefont {J.~P.}\ \bibnamefont
				{Kestner}},\ }\href {https://link.aps.org/doi/10.1103/PhysRevB.99.195403}
		{\bibfield  {journal} {\bibinfo  {journal} {Phys. Rev. B}\ }\textbf {\bibinfo
				{volume} {99}},\ \bibinfo {pages} {195403} (\bibinfo {year}
			{2019})}\BibitemShut {NoStop}%
		\bibitem [{\citenamefont {Frees}\ \emph {et~al.}(2019)\citenamefont {Frees},
			\citenamefont {Mehl}, \citenamefont {Gamble}, \citenamefont {Friesen},\ and\
			\citenamefont {Coppersmith}}]{Frees.19}%
		\BibitemOpen
		\bibfield  {author} {\bibinfo {author} {\bibfnamefont {A.}~\bibnamefont
				{Frees}}, \bibinfo {author} {\bibfnamefont {S.}~\bibnamefont {Mehl}},
			\bibinfo {author} {\bibfnamefont {J.~K.}\ \bibnamefont {Gamble}}, \bibinfo
			{author} {\bibfnamefont {M.}~\bibnamefont {Friesen}}, \ and\ \bibinfo
			{author} {\bibfnamefont {S.~N.}\ \bibnamefont {Coppersmith}},\ }\href
		{https://www.nature.com/articles/s41534-019-0190-7} {\bibfield  {journal}
			{\bibinfo  {journal} {npj Quantum Inf.}\ }\textbf {\bibinfo {volume} {5}},\
			\bibinfo {pages} {73} (\bibinfo {year} {2019})}\BibitemShut {NoStop}%
		\bibitem [{\citenamefont {Buterakos}\ \emph
			{et~al.}(2018{\natexlab{a}})\citenamefont {Buterakos}, \citenamefont
			{Throckmorton},\ and\ \citenamefont {Das~Sarma}}]{Buterakos.18.2}%
		\BibitemOpen
		\bibfield  {author} {\bibinfo {author} {\bibfnamefont {D.}~\bibnamefont
				{Buterakos}}, \bibinfo {author} {\bibfnamefont {R.~E.}\ \bibnamefont
				{Throckmorton}}, \ and\ \bibinfo {author} {\bibfnamefont {S.}~\bibnamefont
				{Das~Sarma}},\ }\href {https://link.aps.org/doi/10.1103/PhysRevB.97.045431}
		{\bibfield  {journal} {\bibinfo  {journal} {Phys. Rev. B}\ }\textbf {\bibinfo
				{volume} {97}},\ \bibinfo {pages} {045431} (\bibinfo {year}
			{2018}{\natexlab{a}})}\BibitemShut {NoStop}%
		\bibitem [{\citenamefont {Li}\ \emph {et~al.}(2012)\citenamefont {Li},
			\citenamefont {Hu},\ and\ \citenamefont {You}}]{Li.12}%
		\BibitemOpen
		\bibfield  {author} {\bibinfo {author} {\bibfnamefont {R.}~\bibnamefont
				{Li}}, \bibinfo {author} {\bibfnamefont {X.}~\bibnamefont {Hu}}, \ and\
			\bibinfo {author} {\bibfnamefont {J.~Q.}\ \bibnamefont {You}},\ }\href
		{https://journals.aps.org/prb/abstract/10.1103/PhysRevB.86.205306} {\bibfield
			{journal} {\bibinfo  {journal} {Phys. Rev. B}\ }\textbf {\bibinfo {volume}
				{86}},\ \bibinfo {pages} {205306} (\bibinfo {year} {2012})}\BibitemShut
		{NoStop}%
		\bibitem [{\citenamefont {Klinovaja}\ \emph {et~al.}(2012)\citenamefont
			{Klinovaja}, \citenamefont {Stepanenko}, \citenamefont {Halperin},\ and\
			\citenamefont {Loss}}]{Klinovaja.12}%
		\BibitemOpen
		\bibfield  {author} {\bibinfo {author} {\bibfnamefont {J.}~\bibnamefont
				{Klinovaja}}, \bibinfo {author} {\bibfnamefont {D.}~\bibnamefont
				{Stepanenko}}, \bibinfo {author} {\bibfnamefont {B.~I.}\ \bibnamefont
				{Halperin}}, \ and\ \bibinfo {author} {\bibfnamefont {D.}~\bibnamefont
				{Loss}},\ }\href
		{https://journals.aps.org/prb/abstract/10.1103/PhysRevB.86.085423} {\bibfield
			{journal} {\bibinfo  {journal} {Phys. Rev. B}\ }\textbf {\bibinfo {volume}
				{86}},\ \bibinfo {pages} {085423} (\bibinfo {year} {2012})}\BibitemShut
		{NoStop}%
		\bibitem [{\citenamefont {Mehl}\ \emph {et~al.}(2014)\citenamefont {Mehl},
			\citenamefont {Bluhm},\ and\ \citenamefont {DiVincenzo}}]{Mehl.14}%
		\BibitemOpen
		\bibfield  {author} {\bibinfo {author} {\bibfnamefont {S.}~\bibnamefont
				{Mehl}}, \bibinfo {author} {\bibfnamefont {H.}~\bibnamefont {Bluhm}}, \ and\
			\bibinfo {author} {\bibfnamefont {D.~P.}\ \bibnamefont {DiVincenzo}},\ }\href
		{https://link.aps.org/doi/10.1103/PhysRevB.90.045404} {\bibfield  {journal}
			{\bibinfo  {journal} {Phys. Rev. B}\ }\textbf {\bibinfo {volume} {90}},\
			\bibinfo {pages} {045404} (\bibinfo {year} {2014})}\BibitemShut {NoStop}%
		\bibitem [{\citenamefont {Wardrop}\ and\ \citenamefont
			{Doherty}(2014)}]{Wardrop.14}%
		\BibitemOpen
		\bibfield  {author} {\bibinfo {author} {\bibfnamefont {M.~P.}\ \bibnamefont
				{Wardrop}}\ and\ \bibinfo {author} {\bibfnamefont {A.~C.}\ \bibnamefont
				{Doherty}},\ }\href {https://link.aps.org/doi/10.1103/PhysRevB.90.045418}
		{\bibfield  {journal} {\bibinfo  {journal} {Phys. Rev. B}\ }\textbf {\bibinfo
				{volume} {90}},\ \bibinfo {pages} {045418} (\bibinfo {year}
			{2014})}\BibitemShut {NoStop}%
		\bibitem [{\citenamefont {Buterakos}\ \emph
			{et~al.}(2018{\natexlab{b}})\citenamefont {Buterakos}, \citenamefont
			{Throckmorton},\ and\ \citenamefont {Das~Sarma}}]{Buterakos.18}%
		\BibitemOpen
		\bibfield  {author} {\bibinfo {author} {\bibfnamefont {D.}~\bibnamefont
				{Buterakos}}, \bibinfo {author} {\bibfnamefont {R.~E.}\ \bibnamefont
				{Throckmorton}}, \ and\ \bibinfo {author} {\bibfnamefont {S.}~\bibnamefont
				{Das~Sarma}},\ }\href {https://link.aps.org/doi/10.1103/PhysRevB.98.035406}
		{\bibfield  {journal} {\bibinfo  {journal} {Phys. Rev. B}\ }\textbf {\bibinfo
				{volume} {98}},\ \bibinfo {pages} {035406} (\bibinfo {year}
			{2018}{\natexlab{b}})}\BibitemShut {NoStop}%
		\bibitem [{\citenamefont {Cerfontaine}\ \emph
			{et~al.}(2020{\natexlab{b}})\citenamefont {Cerfontaine}, \citenamefont
			{Otten}, \citenamefont {Wolfe}, \citenamefont {Bethke},\ and\ \citenamefont
			{Bluhm}}]{Cerfontaine.20.2}%
		\BibitemOpen
		\bibfield  {author} {\bibinfo {author} {\bibfnamefont {P.}~\bibnamefont
				{Cerfontaine}}, \bibinfo {author} {\bibfnamefont {R.}~\bibnamefont {Otten}},
			\bibinfo {author} {\bibfnamefont {M.~A.}\ \bibnamefont {Wolfe}}, \bibinfo
			{author} {\bibfnamefont {P.}~\bibnamefont {Bethke}}, \ and\ \bibinfo {author}
			{\bibfnamefont {H.}~\bibnamefont {Bluhm}},\ }\href
		{https://link.aps.org/doi/10.1103/PhysRevB.101.155311} {\bibfield  {journal}
			{\bibinfo  {journal} {Phys. Rev. B}\ }\textbf {\bibinfo {volume} {101}},\
			\bibinfo {pages} {155311} (\bibinfo {year} {2020}{\natexlab{b}})}\BibitemShut
		{NoStop}%
		\bibitem [{\citenamefont {Yang}\ \emph {et~al.}(2020)\citenamefont {Yang},
			\citenamefont {Coppersmith},\ and\ \citenamefont {Friesen}}]{Yang.20}%
		\BibitemOpen
		\bibfield  {author} {\bibinfo {author} {\bibfnamefont {Y.-C.}\ \bibnamefont
				{Yang}}, \bibinfo {author} {\bibfnamefont {S.~N.}\ \bibnamefont
				{Coppersmith}}, \ and\ \bibinfo {author} {\bibfnamefont {M.}~\bibnamefont
				{Friesen}},\ }\href {https://link.aps.org/doi/10.1103/PhysRevA.101.012338}
		{\bibfield  {journal} {\bibinfo  {journal} {Phys. Rev. A}\ }\textbf {\bibinfo
				{volume} {101}},\ \bibinfo {pages} {012338} (\bibinfo {year}
			{2020})}\BibitemShut {NoStop}%
		\bibitem [{\citenamefont {Abadillo-Uriel}\ \emph {et~al.}(2019)\citenamefont
			{Abadillo-Uriel}, \citenamefont {Eriksson}, \citenamefont {Coppersmith},\
			and\ \citenamefont {Friesen}}]{Abadillo.19}%
		\BibitemOpen
		\bibfield  {author} {\bibinfo {author} {\bibfnamefont {J.~C.}\ \bibnamefont
				{Abadillo-Uriel}}, \bibinfo {author} {\bibfnamefont {M.~A.}\ \bibnamefont
				{Eriksson}}, \bibinfo {author} {\bibfnamefont {S.~N.}\ \bibnamefont
				{Coppersmith}}, \ and\ \bibinfo {author} {\bibfnamefont {M.}~\bibnamefont
				{Friesen}},\ }\href {\doibase 10.1038/s41467-019-13548-w} {\bibfield
			{journal} {\bibinfo  {journal} {Nat. Commun.}\ }\textbf {\bibinfo {volume}
				{10}},\ \bibinfo {pages} {5641} (\bibinfo {year} {2019})}\BibitemShut
		{NoStop}%
		\bibitem [{\citenamefont {Abadillo-Uriel}\ \emph {et~al.}(2021)\citenamefont
			{Abadillo-Uriel}, \citenamefont {King}, \citenamefont {Coppersmith},\ and\
			\citenamefont {Friesen}}]{Abadillo.21}%
		\BibitemOpen
		\bibfield  {author} {\bibinfo {author} {\bibfnamefont {J.~C.}\ \bibnamefont
				{Abadillo-Uriel}}, \bibinfo {author} {\bibfnamefont {C.}~\bibnamefont
				{King}}, \bibinfo {author} {\bibfnamefont {S.~N.}\ \bibnamefont
				{Coppersmith}}, \ and\ \bibinfo {author} {\bibfnamefont {M.}~\bibnamefont
				{Friesen}},\ }\href {https://link.aps.org/doi/10.1103/PhysRevA.104.032612}
		{\bibfield  {journal} {\bibinfo  {journal} {Phys. Rev. A}\ }\textbf {\bibinfo
				{volume} {104}},\ \bibinfo {pages} {032612} (\bibinfo {year}
			{2021})}\BibitemShut {NoStop}%
		\bibitem [{\citenamefont {Chan}\ \emph {et~al.}(2021)\citenamefont {Chan},
			\citenamefont {Kestner},\ and\ \citenamefont {Wang}}]{Chan.21}%
		\BibitemOpen
		\bibfield  {author} {\bibinfo {author} {\bibfnamefont {G.~X.}\ \bibnamefont
				{Chan}}, \bibinfo {author} {\bibfnamefont {J.~P.}\ \bibnamefont {Kestner}}, \
			and\ \bibinfo {author} {\bibfnamefont {X.}~\bibnamefont {Wang}},\ }\href
		{https://link.aps.org/doi/10.1103/PhysRevB.103.L161409} {\bibfield  {journal}
			{\bibinfo  {journal} {Phys. Rev. B}\ }\textbf {\bibinfo {volume} {103}},\
			\bibinfo {pages} {L161409} (\bibinfo {year} {2021})}\BibitemShut {NoStop}%
		\bibitem [{\citenamefont {Barnes}\ \emph {et~al.}(2011)\citenamefont {Barnes},
			\citenamefont {Kestner}, \citenamefont {Nguyen},\ and\ \citenamefont
			{Das~Sarma}}]{Barnes.11}%
		\BibitemOpen
		\bibfield  {author} {\bibinfo {author} {\bibfnamefont {E.}~\bibnamefont
				{Barnes}}, \bibinfo {author} {\bibfnamefont {J.~P.}\ \bibnamefont {Kestner}},
			\bibinfo {author} {\bibfnamefont {N.~T.~T.}\ \bibnamefont {Nguyen}}, \ and\
			\bibinfo {author} {\bibfnamefont {S.}~\bibnamefont {Das~Sarma}},\ }\href
		{https://journals.aps.org/prb/abstract/10.1103/PhysRevB.84.235309} {\bibfield
			{journal} {\bibinfo  {journal} {Phys. Rev. B}\ }\textbf {\bibinfo {volume}
				{84}},\ \bibinfo {pages} {235309} (\bibinfo {year} {2011})}\BibitemShut
		{NoStop}%
		\bibitem [{\citenamefont {Leon}\ \emph {et~al.}(2021)\citenamefont {Leon},
			\citenamefont {Yang}, \citenamefont {Hwang}, \citenamefont {Camirand~Lemyre},
			\citenamefont {Tanttu}, \citenamefont {Huang}, \citenamefont {Huang},
			\citenamefont {Hudson}, \citenamefont {Itoh}, \citenamefont {Laucht},
			\citenamefont {Pioro-Ladri{\`e}re}, \citenamefont {Saraiva},\ and\
			\citenamefont {Dzurak}}]{Leon.21}%
		\BibitemOpen
		\bibfield  {author} {\bibinfo {author} {\bibfnamefont {R.~C.~C.}\
				\bibnamefont {Leon}}, \bibinfo {author} {\bibfnamefont {C.~H.}\ \bibnamefont
				{Yang}}, \bibinfo {author} {\bibfnamefont {J.~C.~C.}\ \bibnamefont {Hwang}},
			\bibinfo {author} {\bibfnamefont {J.}~\bibnamefont {Camirand~Lemyre}},
			\bibinfo {author} {\bibfnamefont {T.}~\bibnamefont {Tanttu}}, \bibinfo
			{author} {\bibfnamefont {W.}~\bibnamefont {Huang}}, \bibinfo {author}
			{\bibfnamefont {J.~Y.}\ \bibnamefont {Huang}}, \bibinfo {author}
			{\bibfnamefont {F.~E.}\ \bibnamefont {Hudson}}, \bibinfo {author}
			{\bibfnamefont {K.~M.}\ \bibnamefont {Itoh}}, \bibinfo {author}
			{\bibfnamefont {A.}~\bibnamefont {Laucht}}, \bibinfo {author} {\bibfnamefont
				{M.}~\bibnamefont {Pioro-Ladri{\`e}re}}, \bibinfo {author} {\bibfnamefont
				{A.}~\bibnamefont {Saraiva}}, \ and\ \bibinfo {author} {\bibfnamefont
				{A.~S.}\ \bibnamefont {Dzurak}},\ }\href
		{https://doi.org/10.1038/s41467-021-23437-w} {\bibfield  {journal} {\bibinfo
				{journal} {Nat. Commun.}\ }\textbf {\bibinfo {volume} {12}},\ \bibinfo
			{pages} {3228} (\bibinfo {year} {2021})}\BibitemShut {NoStop}%
		\bibitem [{\citenamefont {Mehl}\ and\ \citenamefont
			{DiVincenzo}(2013)}]{Mehl.13}%
		\BibitemOpen
		\bibfield  {author} {\bibinfo {author} {\bibfnamefont {S.}~\bibnamefont
				{Mehl}}\ and\ \bibinfo {author} {\bibfnamefont {D.~P.}\ \bibnamefont
				{DiVincenzo}},\ }\href {\doibase 10.1103/PhysRevB.88.161408} {\bibfield
			{journal} {\bibinfo  {journal} {Phys. Rev. B}\ }\textbf {\bibinfo {volume}
				{88}},\ \bibinfo {pages} {161408} (\bibinfo {year} {2013})}\BibitemShut
		{NoStop}%
		\bibitem [{\citenamefont {Deng}\ \emph {et~al.}(2018)\citenamefont {Deng},
			\citenamefont {Calderon-Vargas}, \citenamefont {Mayhall},\ and\ \citenamefont
			{Barnes}}]{Deng.18}%
		\BibitemOpen
		\bibfield  {author} {\bibinfo {author} {\bibfnamefont {K.}~\bibnamefont
				{Deng}}, \bibinfo {author} {\bibfnamefont {F.~A.}\ \bibnamefont
				{Calderon-Vargas}}, \bibinfo {author} {\bibfnamefont {N.~J.}\ \bibnamefont
				{Mayhall}}, \ and\ \bibinfo {author} {\bibfnamefont {E.}~\bibnamefont
				{Barnes}},\ }\href {https://link.aps.org/doi/10.1103/PhysRevB.97.245301}
		{\bibfield  {journal} {\bibinfo  {journal} {Phys. Rev. B}\ }\textbf {\bibinfo
				{volume} {97}},\ \bibinfo {pages} {245301} (\bibinfo {year}
			{2018})}\BibitemShut {NoStop}%
		\bibitem [{\citenamefont {Malinowski}\ \emph {et~al.}(2018)\citenamefont
			{Malinowski}, \citenamefont {Martins}, \citenamefont {Smith}, \citenamefont
			{Bartlett}, \citenamefont {Doherty}, \citenamefont {Nissen}, \citenamefont
			{Fallahi}, \citenamefont {Gardner}, \citenamefont {Manfra}, \citenamefont
			{Marcus},\ and\ \citenamefont {Kuemmeth}}]{Malinowski.18}%
		\BibitemOpen
		\bibfield  {author} {\bibinfo {author} {\bibfnamefont {F.~K.}\ \bibnamefont
				{Malinowski}}, \bibinfo {author} {\bibfnamefont {F.}~\bibnamefont {Martins}},
			\bibinfo {author} {\bibfnamefont {T.~B.}\ \bibnamefont {Smith}}, \bibinfo
			{author} {\bibfnamefont {S.~D.}\ \bibnamefont {Bartlett}}, \bibinfo {author}
			{\bibfnamefont {A.~C.}\ \bibnamefont {Doherty}}, \bibinfo {author}
			{\bibfnamefont {P.~D.}\ \bibnamefont {Nissen}}, \bibinfo {author}
			{\bibfnamefont {S.}~\bibnamefont {Fallahi}}, \bibinfo {author} {\bibfnamefont
				{G.~C.}\ \bibnamefont {Gardner}}, \bibinfo {author} {\bibfnamefont {M.~J.}\
				\bibnamefont {Manfra}}, \bibinfo {author} {\bibfnamefont {C.~M.}\
				\bibnamefont {Marcus}}, \ and\ \bibinfo {author} {\bibfnamefont
				{F.}~\bibnamefont {Kuemmeth}},\ }\href
		{https://link.aps.org/doi/10.1103/PhysRevX.8.011045} {\bibfield  {journal}
			{\bibinfo  {journal} {Phys. Rev. X}\ }\textbf {\bibinfo {volume} {8}},\
			\bibinfo {pages} {011045} (\bibinfo {year} {2018})}\BibitemShut {NoStop}%
		\bibitem [{\citenamefont {Martins}\ \emph {et~al.}(2017)\citenamefont
			{Martins}, \citenamefont {Malinowski}, \citenamefont {Nissen}, \citenamefont
			{Fallahi}, \citenamefont {Gardner}, \citenamefont {Manfra}, \citenamefont
			{Marcus},\ and\ \citenamefont {Kuemmeth}}]{Martins.17}%
		\BibitemOpen
		\bibfield  {author} {\bibinfo {author} {\bibfnamefont {F.}~\bibnamefont
				{Martins}}, \bibinfo {author} {\bibfnamefont {F.~K.}\ \bibnamefont
				{Malinowski}}, \bibinfo {author} {\bibfnamefont {P.~D.}\ \bibnamefont
				{Nissen}}, \bibinfo {author} {\bibfnamefont {S.}~\bibnamefont {Fallahi}},
			\bibinfo {author} {\bibfnamefont {G.~C.}\ \bibnamefont {Gardner}}, \bibinfo
			{author} {\bibfnamefont {M.~J.}\ \bibnamefont {Manfra}}, \bibinfo {author}
			{\bibfnamefont {C.~M.}\ \bibnamefont {Marcus}}, \ and\ \bibinfo {author}
			{\bibfnamefont {F.}~\bibnamefont {Kuemmeth}},\ }\href
		{https://link.aps.org/doi/10.1103/PhysRevLett.119.227701} {\bibfield
			{journal} {\bibinfo  {journal} {Phys. Rev. Lett.}\ }\textbf {\bibinfo
				{volume} {119}},\ \bibinfo {pages} {227701} (\bibinfo {year}
			{2017})}\BibitemShut {NoStop}%
		\bibitem [{Bar()}]{Barrier}%
		\BibitemOpen
		\href@noop {} {}\bibinfo {note} {We consider a piecewise potential function
			to facilitate the full CI calculations. Some works that employ analytical
			solutions of a double well potential may serve as alternatives
			\cite{Xie.15,Sitnitsky.18}, however, those only focus on one-dimensional
			problem. An accurate description of the electron system in a DQD device
			requires a two-dimensional treatment \cite{Li.10,Burkard.99,Barnes.11}.
			Although analytical solutions on a two-dimensional double well potential may
			serve as an improvement for the numerical simulations, those works are still
			lacking in literature.}\BibitemShut {Stop}%
		\bibitem [{\citenamefont {Chan}\ and\ \citenamefont {Wang}(2022)}]{ChanGX.22}%
		\BibitemOpen
		\bibfield  {author} {\bibinfo {author} {\bibfnamefont {G.~X.}\ \bibnamefont
				{Chan}}\ and\ \bibinfo {author} {\bibfnamefont {X.}~\bibnamefont {Wang}},\
		}\href {https://link.aps.org/doi/10.1103/PhysRevB.105.245409} {\bibfield
			{journal} {\bibinfo  {journal} {Phys. Rev. B}\ }\textbf {\bibinfo {volume}
				{105}},\ \bibinfo {pages} {245409} (\bibinfo {year} {2022})}\BibitemShut
		{NoStop}%
		\bibitem [{sm()}]{sm}%
		\BibitemOpen
		\href@noop {} {}\bibinfo {note} {See Supplemental Material at [URL will be
			inserted by publisher] for more numerical results, theoretical analyses, and
			discussions, which includes
			Refs.~\cite{Stein.00,Castin.04,Higginbotham.14,Stano.05,Stano.06,Golovach.08,Raith.12,Nielsen.02,Horodecki.99,Borhani.06,Golovach.04,Veldhorst.14,Yang.13,Friesen.07,Boykin.04,Yoneda.18,Chan.18,Mi.18,Connors.22,Zhao.19,Veldhorst.15,Watson.18,Huang.19,Leon.20,Ercan.21,Hendrickx.20,Hendrickx.18,Hardy.19,Sammak.19,Terrazos.21,Liles.18,Mutter.21,Mutter.20,Hu.07,Sigillito.15}}\BibitemShut
		{NoStop}%
		\bibitem [{SSC()}]{SSConditions}%
		\BibitemOpen
		\href@noop {} {}\bibinfo {note} {For the competition to occur between the hybridization in the singlet states and the hybridization in the triplet states, there are
			three conditions to be met for the single-qubit case: (1) The local exchange
			energy is negative $\left(J<0\right)$ in the large detuning regime where
			$\left(n_{\mathbb{R}\mathrm{L}},n_{\mathbb{R}\mathrm{R}}\right)=\left(0,4\right)$,
			cf.~the right end of Fig. 2(i). (2) The tunnelling energy from the
			singly-occupied dot to the lowest valence orbital in the few-electron dot is
			larger or equal to the tunnelling energy to the first excited valence
			orbital, i.e., $t_{\mathbb{R}\mathrm{L}1,\mathbb{R}\mathrm{R}2} \geq
			t_{\mathbb{R}\mathrm{L}1,\mathbb{R}\mathrm{R}3}$ . (3) The energy difference
			between the fully-occupied singlet and the fully occupied triplet, i.e.,
			$\left| E_{\vert S\rangle} -E_{\vert T\rangle} \right|$ is not too large
			(cf.~Eq.~\eqref{eq:Heff} for the definitions of $E_{\vert S\rangle}$ and
			$E_{\vert T\rangle}$). A direct consequence when the above three conditions
			are met is the non-monotonic behavior of the local exchange energy $J$ as a
			function of detuning $\varepsilon$ in the single-qubit case, cf.~Fig. 2(i) in
			the main text \cite{ChanGX.22}.}\BibitemShut {Stop}%
		\bibitem [{\citenamefont {Boter}\ \emph {et~al.}(2020)\citenamefont {Boter},
			\citenamefont {Xue}, \citenamefont {Kr\"ahenmann}, \citenamefont {Watson},
			\citenamefont {Premakumar}, \citenamefont {Ward}, \citenamefont {Savage},
			\citenamefont {Lagally}, \citenamefont {Friesen}, \citenamefont
			{Coppersmith}, \citenamefont {Eriksson}, \citenamefont {Joynt},\ and\
			\citenamefont {Vandersypen}}]{Boter.20}%
		\BibitemOpen
		\bibfield  {author} {\bibinfo {author} {\bibfnamefont {J.~M.}\ \bibnamefont
				{Boter}}, \bibinfo {author} {\bibfnamefont {X.}~\bibnamefont {Xue}}, \bibinfo
			{author} {\bibfnamefont {T.}~\bibnamefont {Kr\"ahenmann}}, \bibinfo {author}
			{\bibfnamefont {T.~F.}\ \bibnamefont {Watson}}, \bibinfo {author}
			{\bibfnamefont {V.~N.}\ \bibnamefont {Premakumar}}, \bibinfo {author}
			{\bibfnamefont {D.~R.}\ \bibnamefont {Ward}}, \bibinfo {author}
			{\bibfnamefont {D.~E.}\ \bibnamefont {Savage}}, \bibinfo {author}
			{\bibfnamefont {M.~G.}\ \bibnamefont {Lagally}}, \bibinfo {author}
			{\bibfnamefont {M.}~\bibnamefont {Friesen}}, \bibinfo {author} {\bibfnamefont
				{S.~N.}\ \bibnamefont {Coppersmith}}, \bibinfo {author} {\bibfnamefont
				{M.~A.}\ \bibnamefont {Eriksson}}, \bibinfo {author} {\bibfnamefont
				{R.}~\bibnamefont {Joynt}}, \ and\ \bibinfo {author} {\bibfnamefont
				{L.~M.~K.}\ \bibnamefont {Vandersypen}},\ }\href
		{https://link.aps.org/doi/10.1103/PhysRevB.101.235133} {\bibfield  {journal}
			{\bibinfo  {journal} {Phys. Rev. B}\ }\textbf {\bibinfo {volume} {101}},\
			\bibinfo {pages} {235133} (\bibinfo {year} {2020})}\BibitemShut {NoStop}%
		\bibitem [{\citenamefont {Reed}\ \emph {et~al.}(2016)\citenamefont {Reed},
			\citenamefont {Maune}, \citenamefont {Andrews}, \citenamefont {Borselli},
			\citenamefont {Eng}, \citenamefont {Jura}, \citenamefont {Kiselev},
			\citenamefont {Ladd}, \citenamefont {Merkel}, \citenamefont {Milosavljevic},
			\citenamefont {Pritchett}, \citenamefont {Rakher}, \citenamefont {Ross},
			\citenamefont {Schmitz}, \citenamefont {Smith}, \citenamefont {Wright},
			\citenamefont {Gyure},\ and\ \citenamefont {Hunter}}]{Reed.16}%
		\BibitemOpen
		\bibfield  {author} {\bibinfo {author} {\bibfnamefont {M.~D.}\ \bibnamefont
				{Reed}}, \bibinfo {author} {\bibfnamefont {B.~M.}\ \bibnamefont {Maune}},
			\bibinfo {author} {\bibfnamefont {R.~W.}\ \bibnamefont {Andrews}}, \bibinfo
			{author} {\bibfnamefont {M.~G.}\ \bibnamefont {Borselli}}, \bibinfo {author}
			{\bibfnamefont {K.}~\bibnamefont {Eng}}, \bibinfo {author} {\bibfnamefont
				{M.~P.}\ \bibnamefont {Jura}}, \bibinfo {author} {\bibfnamefont {A.~A.}\
				\bibnamefont {Kiselev}}, \bibinfo {author} {\bibfnamefont {T.~D.}\
				\bibnamefont {Ladd}}, \bibinfo {author} {\bibfnamefont {S.~T.}\ \bibnamefont
				{Merkel}}, \bibinfo {author} {\bibfnamefont {I.}~\bibnamefont
				{Milosavljevic}}, \bibinfo {author} {\bibfnamefont {E.~J.}\ \bibnamefont
				{Pritchett}}, \bibinfo {author} {\bibfnamefont {M.~T.}\ \bibnamefont
				{Rakher}}, \bibinfo {author} {\bibfnamefont {R.~S.}\ \bibnamefont {Ross}},
			\bibinfo {author} {\bibfnamefont {A.~E.}\ \bibnamefont {Schmitz}}, \bibinfo
			{author} {\bibfnamefont {A.}~\bibnamefont {Smith}}, \bibinfo {author}
			{\bibfnamefont {J.~A.}\ \bibnamefont {Wright}}, \bibinfo {author}
			{\bibfnamefont {M.~F.}\ \bibnamefont {Gyure}}, \ and\ \bibinfo {author}
			{\bibfnamefont {A.~T.}\ \bibnamefont {Hunter}},\ }\href
		{https://link.aps.org/doi/10.1103/PhysRevLett.116.110402} {\bibfield
			{journal} {\bibinfo  {journal} {Phys. Rev. Lett.}\ }\textbf {\bibinfo
				{volume} {116}},\ \bibinfo {pages} {110402} (\bibinfo {year}
			{2016})}\BibitemShut {NoStop}%
		\bibitem [{\citenamefont {Kornich}\ \emph {et~al.}(2014)\citenamefont
			{Kornich}, \citenamefont {Kloeffel},\ and\ \citenamefont
			{Loss}}]{Kornich.14}%
		\BibitemOpen
		\bibfield  {author} {\bibinfo {author} {\bibfnamefont {V.}~\bibnamefont
				{Kornich}}, \bibinfo {author} {\bibfnamefont {C.}~\bibnamefont {Kloeffel}}, \
			and\ \bibinfo {author} {\bibfnamefont {D.}~\bibnamefont {Loss}},\ }\href
		{https://link.aps.org/doi/10.1103/PhysRevB.89.085410} {\bibfield  {journal}
			{\bibinfo  {journal} {Phys. Rev. B}\ }\textbf {\bibinfo {volume} {89}},\
			\bibinfo {pages} {085410} (\bibinfo {year} {2014})}\BibitemShut {NoStop}%
		\bibitem [{SOI()}]{SOINote}%
		\BibitemOpen
		\href@noop {} {}\bibinfo {note} {The SOI integral is zero for bare
			(non-orthonormalized) F-D states. Therefore, the SOI integral for
			orthonormalized F-D states ($\Phi_j$) consists of the linear combination of
			SOI integrals of the bare F-D states, of which the coefficients being the
			overlap terms between the orbitals in the left and right dot. This results in
			a negligible magnitude for the SOI integral given by $\langle
			\Phi_{\mathbb{R}\mathrm{R}2}\vert
			H_\text{SOI}\vert\Phi_{\mathbb{R}\mathrm{R}3}\rangle$, where $H_\text{SOI}$
			is the spin-orbit interaction. (See Eq.~S-116 and the following discussion in
			the Supplemental Material for details.)}\BibitemShut {NoStop}%
		\bibitem [{\citenamefont {Medford}\ \emph {et~al.}(2012)\citenamefont
			{Medford}, \citenamefont {Cywi\ifmmode~\acute{n}\else \'{n}\fi{}ski},
			\citenamefont {Barthel}, \citenamefont {Marcus}, \citenamefont {Hanson},\
			and\ \citenamefont {Gossard}}]{Medford.12}%
		\BibitemOpen
		\bibfield  {author} {\bibinfo {author} {\bibfnamefont {J.}~\bibnamefont
				{Medford}}, \bibinfo {author} {\bibfnamefont {L.}~\bibnamefont
				{Cywi\ifmmode~\acute{n}\else \'{n}\fi{}ski}}, \bibinfo {author}
			{\bibfnamefont {C.}~\bibnamefont {Barthel}}, \bibinfo {author} {\bibfnamefont
				{C.~M.}\ \bibnamefont {Marcus}}, \bibinfo {author} {\bibfnamefont {M.~P.}\
				\bibnamefont {Hanson}}, \ and\ \bibinfo {author} {\bibfnamefont {A.~C.}\
				\bibnamefont {Gossard}},\ }\href
		{https://link.aps.org/doi/10.1103/PhysRevLett.108.086802} {\bibfield
			{journal} {\bibinfo  {journal} {Phys. Rev. Lett.}\ }\textbf {\bibinfo
				{volume} {108}},\ \bibinfo {pages} {086802} (\bibinfo {year}
			{2012})}\BibitemShut {NoStop}%
		\bibitem [{\citenamefont {Rudner}\ \emph {et~al.}(2011)\citenamefont {Rudner},
			\citenamefont {Koppens}, \citenamefont {Folk}, \citenamefont {Vandersypen},\
			and\ \citenamefont {Levitov}}]{Rudner.11}%
		\BibitemOpen
		\bibfield  {author} {\bibinfo {author} {\bibfnamefont {M.~S.}\ \bibnamefont
				{Rudner}}, \bibinfo {author} {\bibfnamefont {F.~H.~L.}\ \bibnamefont
				{Koppens}}, \bibinfo {author} {\bibfnamefont {J.~A.}\ \bibnamefont {Folk}},
			\bibinfo {author} {\bibfnamefont {L.~M.~K.}\ \bibnamefont {Vandersypen}}, \
			and\ \bibinfo {author} {\bibfnamefont {L.~S.}\ \bibnamefont {Levitov}},\
		}\href {https://link.aps.org/doi/10.1103/PhysRevB.84.075339} {\bibfield
			{journal} {\bibinfo  {journal} {Phys. Rev. B}\ }\textbf {\bibinfo {volume}
				{84}},\ \bibinfo {pages} {075339} (\bibinfo {year} {2011})}\BibitemShut
		{NoStop}%
		\bibitem [{\citenamefont {Dial}\ \emph {et~al.}(2013)\citenamefont {Dial},
			\citenamefont {Shulman}, \citenamefont {Harvey}, \citenamefont {Bluhm},
			\citenamefont {Umansky},\ and\ \citenamefont {Yacoby}}]{Dial.13}%
		\BibitemOpen
		\bibfield  {author} {\bibinfo {author} {\bibfnamefont {O.~E.}\ \bibnamefont
				{Dial}}, \bibinfo {author} {\bibfnamefont {M.~D.}\ \bibnamefont {Shulman}},
			\bibinfo {author} {\bibfnamefont {S.~P.}\ \bibnamefont {Harvey}}, \bibinfo
			{author} {\bibfnamefont {H.}~\bibnamefont {Bluhm}}, \bibinfo {author}
			{\bibfnamefont {V.}~\bibnamefont {Umansky}}, \ and\ \bibinfo {author}
			{\bibfnamefont {A.}~\bibnamefont {Yacoby}},\ }\href
		{https://link.aps.org/doi/10.1103/PhysRevLett.110.146804} {\bibfield
			{journal} {\bibinfo  {journal} {Phys. Rev. Lett.}\ }\textbf {\bibinfo
				{volume} {110}},\ \bibinfo {pages} {146804} (\bibinfo {year}
			{2013})}\BibitemShut {NoStop}%
		\bibitem [{\citenamefont {Yang}\ \emph {et~al.}(2019)\citenamefont {Yang},
			\citenamefont {Coppersmith},\ and\ \citenamefont {Friesen}}]{Yang.19}%
		\BibitemOpen
		\bibfield  {author} {\bibinfo {author} {\bibfnamefont {Y.-C.}\ \bibnamefont
				{Yang}}, \bibinfo {author} {\bibfnamefont {S.~N.}\ \bibnamefont
				{Coppersmith}}, \ and\ \bibinfo {author} {\bibfnamefont {M.}~\bibnamefont
				{Friesen}},\ }\href {\doibase 10.1038/s41534-019-0127-1} {\bibfield
			{journal} {\bibinfo  {journal} {npj Quantum Inf.}\ }\textbf {\bibinfo
				{volume} {5}},\ \bibinfo {pages} {12} (\bibinfo {year} {2019})}\BibitemShut
		{NoStop}%
		\bibitem [{Fid()}]{FidelityNote}%
		\BibitemOpen
		\href@noop {} {}\bibinfo {note} {The definition of gate fidelity $F$ and the
			initial input state are explained in Sec.~VIII in the Supplemental
			Material.}\BibitemShut {Stop}%
		\bibitem [{\citenamefont {Xie}\ \emph {et~al.}(2015)\citenamefont {Xie},
			\citenamefont {Wang},\ and\ \citenamefont {Fu}}]{Xie.15}%
		\BibitemOpen
		\bibfield  {author} {\bibinfo {author} {\bibfnamefont {Q.}~\bibnamefont
				{Xie}}, \bibinfo {author} {\bibfnamefont {L.}~\bibnamefont {Wang}}, \ and\
			\bibinfo {author} {\bibfnamefont {J.}~\bibnamefont {Fu}},\ }\href
		{https://doi.org/10.1088/0031-8949/90/4/045204} {\bibfield  {journal}
			{\bibinfo  {journal} {Phys. Scr.}\ }\textbf {\bibinfo {volume} {90}},\
			\bibinfo {pages} {045204} (\bibinfo {year} {2015})}\BibitemShut {NoStop}%
		\bibitem [{\citenamefont {Sitnitsky}(2018)}]{Sitnitsky.18}%
		\BibitemOpen
		\bibfield  {author} {\bibinfo {author} {\bibfnamefont {A.~E.}\ \bibnamefont
				{Sitnitsky}},\ }\href
		{https://www.sciencedirect.com/science/article/pii/S2210271X18302093}
		{\bibfield  {journal} {\bibinfo  {journal} {Comput. Theor. Chem.}\ }\textbf
			{\bibinfo {volume} {1138}},\ \bibinfo {pages} {15} (\bibinfo {year}
			{2018})}\BibitemShut {NoStop}%
		\bibitem [{\citenamefont {Li}\ \emph {et~al.}(2010)\citenamefont {Li},
			\citenamefont {Cywi\ifmmode~\acute{n}\else \'{n}\fi{}ski}, \citenamefont
			{Culcer}, \citenamefont {Hu},\ and\ \citenamefont {Das~Sarma}}]{Li.10}%
		\BibitemOpen
		\bibfield  {author} {\bibinfo {author} {\bibfnamefont {Q.}~\bibnamefont
				{Li}}, \bibinfo {author} {\bibfnamefont {L.}~\bibnamefont
				{Cywi\ifmmode~\acute{n}\else \'{n}\fi{}ski}}, \bibinfo {author}
			{\bibfnamefont {D.}~\bibnamefont {Culcer}}, \bibinfo {author} {\bibfnamefont
				{X.}~\bibnamefont {Hu}}, \ and\ \bibinfo {author} {\bibfnamefont
				{S.}~\bibnamefont {Das~Sarma}},\ }\href {\doibase 10.1103/PhysRevB.81.085313}
		{\bibfield  {journal} {\bibinfo  {journal} {Phys. Rev. B}\ }\textbf {\bibinfo
				{volume} {81}},\ \bibinfo {pages} {085313} (\bibinfo {year}
			{2010})}\BibitemShut {NoStop}%
		\bibitem [{\citenamefont {Burkard}\ \emph {et~al.}(1999)\citenamefont
			{Burkard}, \citenamefont {Loss},\ and\ \citenamefont
			{DiVincenzo}}]{Burkard.99}%
		\BibitemOpen
		\bibfield  {author} {\bibinfo {author} {\bibfnamefont {G.}~\bibnamefont
				{Burkard}}, \bibinfo {author} {\bibfnamefont {D.}~\bibnamefont {Loss}}, \
			and\ \bibinfo {author} {\bibfnamefont {D.~P.}\ \bibnamefont {DiVincenzo}},\
		}\href {\doibase 10.1103/PhysRevB.59.2070} {\bibfield  {journal} {\bibinfo
				{journal} {Phys. Rev. B}\ }\textbf {\bibinfo {volume} {59}},\ \bibinfo
			{pages} {2070} (\bibinfo {year} {1999})}\BibitemShut {NoStop}%
		\bibitem [{\citenamefont {Stein}(2000)}]{Stein.00}%
		\BibitemOpen
		\bibfield  {author} {\bibinfo {author} {\bibfnamefont {J.~Y.}\ \bibnamefont
				{Stein}},\ }\href
		{https://onlinelibrary.wiley.com/doi/book/10.1002/047120059X} {\emph
			{\bibinfo {title} {Digital signal processing : a computer science
					perspective}}}\ (\bibinfo  {publisher} {Wiley},\ \bibinfo {address} {New
			York},\ \bibinfo {year} {2000})\BibitemShut {NoStop}%
		\bibitem [{\citenamefont {Castin}(2004)}]{Castin.04}%
		\BibitemOpen
		\bibfield  {author} {\bibinfo {author} {\bibfnamefont {Y.}~\bibnamefont
				{Castin}},\ }\href
		{https://jp4.journaldephysique.org/en/articles/jp4/abs/2004/04/Course4/Course4.html}
		{\bibfield  {journal} {\bibinfo  {journal} {J. Phys. IV France}\ }\textbf
			{\bibinfo {volume} {116}},\ \bibinfo {pages} {89} (\bibinfo {year}
			{2004})}\BibitemShut {NoStop}%
		\bibitem [{\citenamefont {Higginbotham}\ \emph {et~al.}(2014)\citenamefont
			{Higginbotham}, \citenamefont {Kuemmeth}, \citenamefont {Hanson},
			\citenamefont {Gossard},\ and\ \citenamefont {Marcus}}]{Higginbotham.14}%
		\BibitemOpen
		\bibfield  {author} {\bibinfo {author} {\bibfnamefont {A.~P.}\ \bibnamefont
				{Higginbotham}}, \bibinfo {author} {\bibfnamefont {F.}~\bibnamefont
				{Kuemmeth}}, \bibinfo {author} {\bibfnamefont {M.~P.}\ \bibnamefont
				{Hanson}}, \bibinfo {author} {\bibfnamefont {A.~C.}\ \bibnamefont {Gossard}},
			\ and\ \bibinfo {author} {\bibfnamefont {C.~M.}\ \bibnamefont {Marcus}},\
		}\href {https://link.aps.org/doi/10.1103/PhysRevLett.112.026801} {\bibfield
			{journal} {\bibinfo  {journal} {Phys. Rev. Lett.}\ }\textbf {\bibinfo
				{volume} {112}},\ \bibinfo {pages} {026801} (\bibinfo {year}
			{2014})}\BibitemShut {NoStop}%
		\bibitem [{\citenamefont {Stano}\ and\ \citenamefont
			{Fabian}(2005)}]{Stano.05}%
		\BibitemOpen
		\bibfield  {author} {\bibinfo {author} {\bibfnamefont {P.}~\bibnamefont
				{Stano}}\ and\ \bibinfo {author} {\bibfnamefont {J.}~\bibnamefont {Fabian}},\
		}\href {https://journals.aps.org/prb/abstract/10.1103/PhysRevB.72.155410}
		{\bibfield  {journal} {\bibinfo  {journal} {Phys. Rev. B}\ }\textbf {\bibinfo
				{volume} {72}},\ \bibinfo {pages} {155410} (\bibinfo {year}
			{2005})}\BibitemShut {NoStop}%
		\bibitem [{\citenamefont {Stano}\ and\ \citenamefont
			{Fabian}(2006)}]{Stano.06}%
		\BibitemOpen
		\bibfield  {author} {\bibinfo {author} {\bibfnamefont {P.}~\bibnamefont
				{Stano}}\ and\ \bibinfo {author} {\bibfnamefont {J.}~\bibnamefont {Fabian}},\
		}\href {https://journals.aps.org/prl/abstract/10.1103/PhysRevLett.96.186602}
		{\bibfield  {journal} {\bibinfo  {journal} {Phys. Rev. Lett.}\ }\textbf
			{\bibinfo {volume} {96}},\ \bibinfo {pages} {186602} (\bibinfo {year}
			{2006})}\BibitemShut {NoStop}%
		\bibitem [{\citenamefont {Golovach}\ \emph {et~al.}(2008)\citenamefont
			{Golovach}, \citenamefont {Khaetskii},\ and\ \citenamefont
			{Loss}}]{Golovach.08}%
		\BibitemOpen
		\bibfield  {author} {\bibinfo {author} {\bibfnamefont {V.~N.}\ \bibnamefont
				{Golovach}}, \bibinfo {author} {\bibfnamefont {A.}~\bibnamefont {Khaetskii}},
			\ and\ \bibinfo {author} {\bibfnamefont {D.}~\bibnamefont {Loss}},\ }\href
		{https://link.aps.org/doi/10.1103/PhysRevB.77.045328} {\bibfield  {journal}
			{\bibinfo  {journal} {Phys. Rev. B}\ }\textbf {\bibinfo {volume} {77}},\
			\bibinfo {pages} {045328} (\bibinfo {year} {2008})}\BibitemShut {NoStop}%
		\bibitem [{\citenamefont {Raith}\ \emph {et~al.}(2012)\citenamefont {Raith},
			\citenamefont {Stano}, \citenamefont {Baruffa},\ and\ \citenamefont
			{Fabian}}]{Raith.12}%
		\BibitemOpen
		\bibfield  {author} {\bibinfo {author} {\bibfnamefont {M.}~\bibnamefont
				{Raith}}, \bibinfo {author} {\bibfnamefont {P.}~\bibnamefont {Stano}},
			\bibinfo {author} {\bibfnamefont {F.}~\bibnamefont {Baruffa}}, \ and\
			\bibinfo {author} {\bibfnamefont {J.}~\bibnamefont {Fabian}},\ }\href
		{https://link.aps.org/doi/10.1103/PhysRevLett.108.246602} {\bibfield
			{journal} {\bibinfo  {journal} {Phys. Rev. Lett.}\ }\textbf {\bibinfo
				{volume} {108}},\ \bibinfo {pages} {246602} (\bibinfo {year}
			{2012})}\BibitemShut {NoStop}%
		\bibitem [{\citenamefont {Nielsen}(2002)}]{Nielsen.02}%
		\BibitemOpen
		\bibfield  {author} {\bibinfo {author} {\bibfnamefont {M.~A.}\ \bibnamefont
				{Nielsen}},\ }\href
		{https://www.sciencedirect.com/science/article/pii/S0375960102012720}
		{\bibfield  {journal} {\bibinfo  {journal} {Phys. Lett. A}\ }\textbf
			{\bibinfo {volume} {303}},\ \bibinfo {pages} {249 } (\bibinfo {year}
			{2002})}\BibitemShut {NoStop}%
		\bibitem [{\citenamefont {Horodecki}\ \emph {et~al.}(1999)\citenamefont
			{Horodecki}, \citenamefont {Horodecki},\ and\ \citenamefont
			{Horodecki}}]{Horodecki.99}%
		\BibitemOpen
		\bibfield  {author} {\bibinfo {author} {\bibfnamefont {M.}~\bibnamefont
				{Horodecki}}, \bibinfo {author} {\bibfnamefont {P.}~\bibnamefont
				{Horodecki}}, \ and\ \bibinfo {author} {\bibfnamefont {R.}~\bibnamefont
				{Horodecki}},\ }\href
		{https://journals.aps.org/pra/abstract/10.1103/PhysRevA.60.1888} {\bibfield
			{journal} {\bibinfo  {journal} {Phys. Rev. A}\ }\textbf {\bibinfo {volume}
				{60}},\ \bibinfo {pages} {1888} (\bibinfo {year} {1999})}\BibitemShut
		{NoStop}%
		\bibitem [{\citenamefont {Borhani}\ \emph {et~al.}(2006)\citenamefont
			{Borhani}, \citenamefont {Golovach},\ and\ \citenamefont
			{Loss}}]{Borhani.06}%
		\BibitemOpen
		\bibfield  {author} {\bibinfo {author} {\bibfnamefont {M.}~\bibnamefont
				{Borhani}}, \bibinfo {author} {\bibfnamefont {V.~N.}\ \bibnamefont
				{Golovach}}, \ and\ \bibinfo {author} {\bibfnamefont {D.}~\bibnamefont
				{Loss}},\ }\href {\doibase 10.1103/PhysRevB.73.155311} {\bibfield  {journal}
			{\bibinfo  {journal} {Phys. Rev. B}\ }\textbf {\bibinfo {volume} {73}},\
			\bibinfo {pages} {155311} (\bibinfo {year} {2006})}\BibitemShut {NoStop}%
		\bibitem [{\citenamefont {Golovach}\ \emph {et~al.}(2004)\citenamefont
			{Golovach}, \citenamefont {Khaetskii},\ and\ \citenamefont
			{Loss}}]{Golovach.04}%
		\BibitemOpen
		\bibfield  {author} {\bibinfo {author} {\bibfnamefont {V.~N.}\ \bibnamefont
				{Golovach}}, \bibinfo {author} {\bibfnamefont {A.}~\bibnamefont {Khaetskii}},
			\ and\ \bibinfo {author} {\bibfnamefont {D.}~\bibnamefont {Loss}},\ }\href
		{\doibase 10.1103/PhysRevLett.93.016601} {\bibfield  {journal} {\bibinfo
				{journal} {Phys. Rev. Lett.}\ }\textbf {\bibinfo {volume} {93}},\ \bibinfo
			{pages} {016601} (\bibinfo {year} {2004})}\BibitemShut {NoStop}%
		\bibitem [{\citenamefont {Veldhorst}\ \emph {et~al.}(2014)\citenamefont
			{Veldhorst}, \citenamefont {Hwang}, \citenamefont {Yang}, \citenamefont
			{Leenstra}, \citenamefont {de~Ronde}, \citenamefont {Dehollain},
			\citenamefont {Muhonen}, \citenamefont {Hudson}, \citenamefont {Itoh},
			\citenamefont {Morello},\ and\ \citenamefont {Dzurak}}]{Veldhorst.14}%
		\BibitemOpen
		\bibfield  {author} {\bibinfo {author} {\bibfnamefont {M.}~\bibnamefont
				{Veldhorst}}, \bibinfo {author} {\bibfnamefont {J.~C.~C.}\ \bibnamefont
				{Hwang}}, \bibinfo {author} {\bibfnamefont {C.~H.}\ \bibnamefont {Yang}},
			\bibinfo {author} {\bibfnamefont {A.~W.}\ \bibnamefont {Leenstra}}, \bibinfo
			{author} {\bibfnamefont {B.}~\bibnamefont {de~Ronde}}, \bibinfo {author}
			{\bibfnamefont {J.~P.}\ \bibnamefont {Dehollain}}, \bibinfo {author}
			{\bibfnamefont {J.~T.}\ \bibnamefont {Muhonen}}, \bibinfo {author}
			{\bibfnamefont {F.~E.}\ \bibnamefont {Hudson}}, \bibinfo {author}
			{\bibfnamefont {K.~M.}\ \bibnamefont {Itoh}}, \bibinfo {author}
			{\bibfnamefont {A.}~\bibnamefont {Morello}}, \ and\ \bibinfo {author}
			{\bibfnamefont {A.~S.}\ \bibnamefont {Dzurak}},\ }\href
		{https://doi.org/10.1038/nnano.2014.216} {\bibfield  {journal} {\bibinfo
				{journal} {Nat. Nanotechnol.}\ }\textbf {\bibinfo {volume} {9}},\ \bibinfo
			{pages} {981} (\bibinfo {year} {2014})}\BibitemShut {NoStop}%
		\bibitem [{\citenamefont {Yang}\ \emph {et~al.}(2013)\citenamefont {Yang},
			\citenamefont {Rossi}, \citenamefont {Ruskov}, \citenamefont {Lai},
			\citenamefont {Mohiyaddin}, \citenamefont {Lee}, \citenamefont {Tahan},
			\citenamefont {Klimeck}, \citenamefont {Morello},\ and\ \citenamefont
			{Dzurak}}]{Yang.13}%
		\BibitemOpen
		\bibfield  {author} {\bibinfo {author} {\bibfnamefont {C.~H.}\ \bibnamefont
				{Yang}}, \bibinfo {author} {\bibfnamefont {A.}~\bibnamefont {Rossi}},
			\bibinfo {author} {\bibfnamefont {R.}~\bibnamefont {Ruskov}}, \bibinfo
			{author} {\bibfnamefont {N.~S.}\ \bibnamefont {Lai}}, \bibinfo {author}
			{\bibfnamefont {F.~A.}\ \bibnamefont {Mohiyaddin}}, \bibinfo {author}
			{\bibfnamefont {S.}~\bibnamefont {Lee}}, \bibinfo {author} {\bibfnamefont
				{C.}~\bibnamefont {Tahan}}, \bibinfo {author} {\bibfnamefont
				{G.}~\bibnamefont {Klimeck}}, \bibinfo {author} {\bibfnamefont
				{A.}~\bibnamefont {Morello}}, \ and\ \bibinfo {author} {\bibfnamefont
				{A.~S.}\ \bibnamefont {Dzurak}},\ }\href {https://doi.org/10.1038/ncomms3069}
		{\bibfield  {journal} {\bibinfo  {journal} {Nat. Commun.}\ }\textbf {\bibinfo
				{volume} {4}},\ \bibinfo {pages} {2069} (\bibinfo {year} {2013})}\BibitemShut
		{NoStop}%
		\bibitem [{\citenamefont {Friesen}\ \emph {et~al.}(2007)\citenamefont
			{Friesen}, \citenamefont {Chutia}, \citenamefont {Tahan},\ and\ \citenamefont
			{Coppersmith}}]{Friesen.07}%
		\BibitemOpen
		\bibfield  {author} {\bibinfo {author} {\bibfnamefont {M.}~\bibnamefont
				{Friesen}}, \bibinfo {author} {\bibfnamefont {S.}~\bibnamefont {Chutia}},
			\bibinfo {author} {\bibfnamefont {C.}~\bibnamefont {Tahan}}, \ and\ \bibinfo
			{author} {\bibfnamefont {S.~N.}\ \bibnamefont {Coppersmith}},\ }\href
		{\doibase 10.1103/PhysRevB.75.115318} {\bibfield  {journal} {\bibinfo
				{journal} {Phys. Rev. B}\ }\textbf {\bibinfo {volume} {75}},\ \bibinfo
			{pages} {115318} (\bibinfo {year} {2007})}\BibitemShut {NoStop}%
		\bibitem [{\citenamefont {Boykin}\ \emph {et~al.}(2004)\citenamefont {Boykin},
			\citenamefont {Klimeck}, \citenamefont {Eriksson}, \citenamefont {Friesen},
			\citenamefont {Coppersmith}, \citenamefont {von Allmen}, \citenamefont
			{Oyafuso},\ and\ \citenamefont {Lee}}]{Boykin.04}%
		\BibitemOpen
		\bibfield  {author} {\bibinfo {author} {\bibfnamefont {T.~B.}\ \bibnamefont
				{Boykin}}, \bibinfo {author} {\bibfnamefont {G.}~\bibnamefont {Klimeck}},
			\bibinfo {author} {\bibfnamefont {M.~A.}\ \bibnamefont {Eriksson}}, \bibinfo
			{author} {\bibfnamefont {M.}~\bibnamefont {Friesen}}, \bibinfo {author}
			{\bibfnamefont {S.~N.}\ \bibnamefont {Coppersmith}}, \bibinfo {author}
			{\bibfnamefont {P.}~\bibnamefont {von Allmen}}, \bibinfo {author}
			{\bibfnamefont {F.}~\bibnamefont {Oyafuso}}, \ and\ \bibinfo {author}
			{\bibfnamefont {S.}~\bibnamefont {Lee}},\ }\href
		{https://doi.org/10.1063/1.1637718} {\bibfield  {journal} {\bibinfo
				{journal} {Appl. Phys. Lett.}\ }\textbf {\bibinfo {volume} {84}},\ \bibinfo
			{pages} {115} (\bibinfo {year} {2004})}\BibitemShut {NoStop}%
		\bibitem [{\citenamefont {Yoneda}\ \emph {et~al.}(2018)\citenamefont {Yoneda},
			\citenamefont {Takeda}, \citenamefont {Otsuka}, \citenamefont {Nakajima},
			\citenamefont {Delbecq}, \citenamefont {Allison}, \citenamefont {Honda},
			\citenamefont {Kodera}, \citenamefont {Oda}, \citenamefont {Hoshi},
			\citenamefont {Usami}, \citenamefont {Itoh},\ and\ \citenamefont
			{Tarucha}}]{Yoneda.18}%
		\BibitemOpen
		\bibfield  {author} {\bibinfo {author} {\bibfnamefont {J.}~\bibnamefont
				{Yoneda}}, \bibinfo {author} {\bibfnamefont {K.}~\bibnamefont {Takeda}},
			\bibinfo {author} {\bibfnamefont {T.}~\bibnamefont {Otsuka}}, \bibinfo
			{author} {\bibfnamefont {T.}~\bibnamefont {Nakajima}}, \bibinfo {author}
			{\bibfnamefont {M.~R.}\ \bibnamefont {Delbecq}}, \bibinfo {author}
			{\bibfnamefont {G.}~\bibnamefont {Allison}}, \bibinfo {author} {\bibfnamefont
				{T.}~\bibnamefont {Honda}}, \bibinfo {author} {\bibfnamefont
				{T.}~\bibnamefont {Kodera}}, \bibinfo {author} {\bibfnamefont
				{S.}~\bibnamefont {Oda}}, \bibinfo {author} {\bibfnamefont {Y.}~\bibnamefont
				{Hoshi}}, \bibinfo {author} {\bibfnamefont {N.}~\bibnamefont {Usami}},
			\bibinfo {author} {\bibfnamefont {K.~M.}\ \bibnamefont {Itoh}}, \ and\
			\bibinfo {author} {\bibfnamefont {S.}~\bibnamefont {Tarucha}},\ }\href
		{https://doi.org/10.1038/s41565-017-0014-x} {\bibfield  {journal} {\bibinfo
				{journal} {Nat. Nanotechnol.}\ }\textbf {\bibinfo {volume} {13}},\ \bibinfo
			{pages} {102} (\bibinfo {year} {2018})}\BibitemShut {NoStop}%
		\bibitem [{\citenamefont {Chan}\ \emph {et~al.}(2018)\citenamefont {Chan},
			\citenamefont {Huang}, \citenamefont {Yang}, \citenamefont {Hwang},
			\citenamefont {Hensen}, \citenamefont {Tanttu}, \citenamefont {Hudson},
			\citenamefont {Itoh}, \citenamefont {Laucht}, \citenamefont {Morello},\ and\
			\citenamefont {Dzurak}}]{Chan.18}%
		\BibitemOpen
		\bibfield  {author} {\bibinfo {author} {\bibfnamefont {K.~W.}\ \bibnamefont
				{Chan}}, \bibinfo {author} {\bibfnamefont {W.}~\bibnamefont {Huang}},
			\bibinfo {author} {\bibfnamefont {C.~H.}\ \bibnamefont {Yang}}, \bibinfo
			{author} {\bibfnamefont {J.~C.~C.}\ \bibnamefont {Hwang}}, \bibinfo {author}
			{\bibfnamefont {B.}~\bibnamefont {Hensen}}, \bibinfo {author} {\bibfnamefont
				{T.}~\bibnamefont {Tanttu}}, \bibinfo {author} {\bibfnamefont {F.~E.}\
				\bibnamefont {Hudson}}, \bibinfo {author} {\bibfnamefont {K.~M.}\
				\bibnamefont {Itoh}}, \bibinfo {author} {\bibfnamefont {A.}~\bibnamefont
				{Laucht}}, \bibinfo {author} {\bibfnamefont {A.}~\bibnamefont {Morello}}, \
			and\ \bibinfo {author} {\bibfnamefont {A.~S.}\ \bibnamefont {Dzurak}},\
		}\href {https://link.aps.org/doi/10.1103/PhysRevApplied.10.044017} {\bibfield
			{journal} {\bibinfo  {journal} {Phys. Rev. Applied}\ }\textbf {\bibinfo
				{volume} {10}},\ \bibinfo {pages} {044017} (\bibinfo {year}
			{2018})}\BibitemShut {NoStop}%
		\bibitem [{\citenamefont {Mi}\ \emph {et~al.}(2018)\citenamefont {Mi},
			\citenamefont {Kohler},\ and\ \citenamefont {Petta}}]{Mi.18}%
		\BibitemOpen
		\bibfield  {author} {\bibinfo {author} {\bibfnamefont {X.}~\bibnamefont
				{Mi}}, \bibinfo {author} {\bibfnamefont {S.}~\bibnamefont {Kohler}}, \ and\
			\bibinfo {author} {\bibfnamefont {J.~R.}\ \bibnamefont {Petta}},\ }\href
		{\doibase 10.1103/PhysRevB.98.161404} {\bibfield  {journal} {\bibinfo
				{journal} {Phys. Rev. B}\ }\textbf {\bibinfo {volume} {98}},\ \bibinfo
			{pages} {161404} (\bibinfo {year} {2018})}\BibitemShut {NoStop}%
		\bibitem [{\citenamefont {Connors}\ \emph {et~al.}(2022)\citenamefont
			{Connors}, \citenamefont {Nelson}, \citenamefont {Edge},\ and\ \citenamefont
			{Nichol}}]{Connors.22}%
		\BibitemOpen
		\bibfield  {author} {\bibinfo {author} {\bibfnamefont {E.~J.}\ \bibnamefont
				{Connors}}, \bibinfo {author} {\bibfnamefont {J.}~\bibnamefont {Nelson}},
			\bibinfo {author} {\bibfnamefont {L.~F.}\ \bibnamefont {Edge}}, \ and\
			\bibinfo {author} {\bibfnamefont {J.~M.}\ \bibnamefont {Nichol}},\ }\href
		{https://doi.org/10.1038/s41467-022-28519-x} {\bibfield  {journal} {\bibinfo
				{journal} {Nat. Commun.}\ }\textbf {\bibinfo {volume} {13}},\ \bibinfo
			{pages} {940} (\bibinfo {year} {2022})}\BibitemShut {NoStop}%
		\bibitem [{\citenamefont {Zhao}\ \emph {et~al.}(2019)\citenamefont {Zhao},
			\citenamefont {Tanttu}, \citenamefont {Tan}, \citenamefont {Hensen},
			\citenamefont {Chan}, \citenamefont {Hwang}, \citenamefont {Leon},
			\citenamefont {Yang}, \citenamefont {Gilbert}, \citenamefont {Hudson},
			\citenamefont {Itoh}, \citenamefont {Kiselev}, \citenamefont {Ladd},
			\citenamefont {Morello}, \citenamefont {Laucht},\ and\ \citenamefont
			{Dzurak}}]{Zhao.19}%
		\BibitemOpen
		\bibfield  {author} {\bibinfo {author} {\bibfnamefont {R.}~\bibnamefont
				{Zhao}}, \bibinfo {author} {\bibfnamefont {T.}~\bibnamefont {Tanttu}},
			\bibinfo {author} {\bibfnamefont {K.~Y.}\ \bibnamefont {Tan}}, \bibinfo
			{author} {\bibfnamefont {B.}~\bibnamefont {Hensen}}, \bibinfo {author}
			{\bibfnamefont {K.~W.}\ \bibnamefont {Chan}}, \bibinfo {author}
			{\bibfnamefont {J.~C.~C.}\ \bibnamefont {Hwang}}, \bibinfo {author}
			{\bibfnamefont {R.~C.~C.}\ \bibnamefont {Leon}}, \bibinfo {author}
			{\bibfnamefont {C.~H.}\ \bibnamefont {Yang}}, \bibinfo {author}
			{\bibfnamefont {W.}~\bibnamefont {Gilbert}}, \bibinfo {author} {\bibfnamefont
				{F.~E.}\ \bibnamefont {Hudson}}, \bibinfo {author} {\bibfnamefont {K.~M.}\
				\bibnamefont {Itoh}}, \bibinfo {author} {\bibfnamefont {A.~A.}\ \bibnamefont
				{Kiselev}}, \bibinfo {author} {\bibfnamefont {T.~D.}\ \bibnamefont {Ladd}},
			\bibinfo {author} {\bibfnamefont {A.}~\bibnamefont {Morello}}, \bibinfo
			{author} {\bibfnamefont {A.}~\bibnamefont {Laucht}}, \ and\ \bibinfo {author}
			{\bibfnamefont {A.~S.}\ \bibnamefont {Dzurak}},\ }\href
		{https://doi.org/10.1038/s41467-019-13416-7} {\bibfield  {journal} {\bibinfo
				{journal} {Nat. Commun.}\ }\textbf {\bibinfo {volume} {10}},\ \bibinfo
			{pages} {5500} (\bibinfo {year} {2019})}\BibitemShut {NoStop}%
		\bibitem [{\citenamefont {Veldhorst}\ \emph {et~al.}(2015)\citenamefont
			{Veldhorst}, \citenamefont {Yang}, \citenamefont {Hwang}, \citenamefont
			{Huang}, \citenamefont {Dehollain}, \citenamefont {Muhonen}, \citenamefont
			{Simmons}, \citenamefont {Laucht}, \citenamefont {Hudson}, \citenamefont
			{Itoh}, \citenamefont {Morello},\ and\ \citenamefont
			{Dzurak}}]{Veldhorst.15}%
		\BibitemOpen
		\bibfield  {author} {\bibinfo {author} {\bibfnamefont {M.}~\bibnamefont
				{Veldhorst}}, \bibinfo {author} {\bibfnamefont {C.~H.}\ \bibnamefont {Yang}},
			\bibinfo {author} {\bibfnamefont {J.~C.~C.}\ \bibnamefont {Hwang}}, \bibinfo
			{author} {\bibfnamefont {W.}~\bibnamefont {Huang}}, \bibinfo {author}
			{\bibfnamefont {J.~P.}\ \bibnamefont {Dehollain}}, \bibinfo {author}
			{\bibfnamefont {J.~T.}\ \bibnamefont {Muhonen}}, \bibinfo {author}
			{\bibfnamefont {S.}~\bibnamefont {Simmons}}, \bibinfo {author} {\bibfnamefont
				{A.}~\bibnamefont {Laucht}}, \bibinfo {author} {\bibfnamefont {F.~E.}\
				\bibnamefont {Hudson}}, \bibinfo {author} {\bibfnamefont {K.~M.}\
				\bibnamefont {Itoh}}, \bibinfo {author} {\bibfnamefont {A.}~\bibnamefont
				{Morello}}, \ and\ \bibinfo {author} {\bibfnamefont {A.~S.}\ \bibnamefont
				{Dzurak}},\ }\href {https://doi.org/10.1038/nature15263} {\bibfield
			{journal} {\bibinfo  {journal} {Nature}\ }\textbf {\bibinfo {volume} {526}},\
			\bibinfo {pages} {410} (\bibinfo {year} {2015})}\BibitemShut {NoStop}%
		\bibitem [{\citenamefont {Watson}\ \emph {et~al.}(2018)\citenamefont {Watson},
			\citenamefont {Philips}, \citenamefont {Kawakami}, \citenamefont {Ward},
			\citenamefont {Scarlino}, \citenamefont {Veldhorst}, \citenamefont {Savage},
			\citenamefont {Lagally}, \citenamefont {Friesen}, \citenamefont
			{Coppersmith}, \citenamefont {Eriksson},\ and\ \citenamefont
			{Vandersypen}}]{Watson.18}%
		\BibitemOpen
		\bibfield  {author} {\bibinfo {author} {\bibfnamefont {T.~F.}\ \bibnamefont
				{Watson}}, \bibinfo {author} {\bibfnamefont {S.~G.~J.}\ \bibnamefont
				{Philips}}, \bibinfo {author} {\bibfnamefont {E.}~\bibnamefont {Kawakami}},
			\bibinfo {author} {\bibfnamefont {D.~R.}\ \bibnamefont {Ward}}, \bibinfo
			{author} {\bibfnamefont {P.}~\bibnamefont {Scarlino}}, \bibinfo {author}
			{\bibfnamefont {M.}~\bibnamefont {Veldhorst}}, \bibinfo {author}
			{\bibfnamefont {D.~E.}\ \bibnamefont {Savage}}, \bibinfo {author}
			{\bibfnamefont {M.~G.}\ \bibnamefont {Lagally}}, \bibinfo {author}
			{\bibfnamefont {M.}~\bibnamefont {Friesen}}, \bibinfo {author} {\bibfnamefont
				{S.~N.}\ \bibnamefont {Coppersmith}}, \bibinfo {author} {\bibfnamefont
				{M.~A.}\ \bibnamefont {Eriksson}}, \ and\ \bibinfo {author} {\bibfnamefont
				{L.~M.~K.}\ \bibnamefont {Vandersypen}},\ }\href
		{https://doi.org/10.1038/nature25766} {\bibfield  {journal} {\bibinfo
				{journal} {Nature}\ }\textbf {\bibinfo {volume} {555}},\ \bibinfo {pages}
			{633} (\bibinfo {year} {2018})}\BibitemShut {NoStop}%
		\bibitem [{\citenamefont {Huang}\ \emph {et~al.}(2019)\citenamefont {Huang},
			\citenamefont {Yang}, \citenamefont {Chan}, \citenamefont {Tanttu},
			\citenamefont {Hensen}, \citenamefont {Leon}, \citenamefont {Fogarty},
			\citenamefont {Hwang}, \citenamefont {Hudson}, \citenamefont {Itoh},
			\citenamefont {Morello}, \citenamefont {Laucht},\ and\ \citenamefont
			{Dzurak}}]{Huang.19}%
		\BibitemOpen
		\bibfield  {author} {\bibinfo {author} {\bibfnamefont {W.}~\bibnamefont
				{Huang}}, \bibinfo {author} {\bibfnamefont {C.~H.}\ \bibnamefont {Yang}},
			\bibinfo {author} {\bibfnamefont {K.~W.}\ \bibnamefont {Chan}}, \bibinfo
			{author} {\bibfnamefont {T.}~\bibnamefont {Tanttu}}, \bibinfo {author}
			{\bibfnamefont {B.}~\bibnamefont {Hensen}}, \bibinfo {author} {\bibfnamefont
				{R.~C.~C.}\ \bibnamefont {Leon}}, \bibinfo {author} {\bibfnamefont {M.~A.}\
				\bibnamefont {Fogarty}}, \bibinfo {author} {\bibfnamefont {J.~C.~C.}\
				\bibnamefont {Hwang}}, \bibinfo {author} {\bibfnamefont {F.~E.}\ \bibnamefont
				{Hudson}}, \bibinfo {author} {\bibfnamefont {K.~M.}\ \bibnamefont {Itoh}},
			\bibinfo {author} {\bibfnamefont {A.}~\bibnamefont {Morello}}, \bibinfo
			{author} {\bibfnamefont {A.}~\bibnamefont {Laucht}}, \ and\ \bibinfo {author}
			{\bibfnamefont {A.~S.}\ \bibnamefont {Dzurak}},\ }\href
		{https://doi.org/10.1038/s41586-019-1197-0} {\bibfield  {journal} {\bibinfo
				{journal} {Nature}\ }\textbf {\bibinfo {volume} {569}},\ \bibinfo {pages}
			{532} (\bibinfo {year} {2019})}\BibitemShut {NoStop}%
		\bibitem [{\citenamefont {Leon}\ \emph {et~al.}(2020)\citenamefont {Leon},
			\citenamefont {Yang}, \citenamefont {Hwang}, \citenamefont {Lemyre},
			\citenamefont {Tanttu}, \citenamefont {Huang}, \citenamefont {Chan},
			\citenamefont {Tan}, \citenamefont {Hudson}, \citenamefont {Itoh},
			\citenamefont {Morello}, \citenamefont {Laucht}, \citenamefont
			{Pioro-Ladri{\`e}re}, \citenamefont {Saraiva},\ and\ \citenamefont
			{Dzurak}}]{Leon.20}%
		\BibitemOpen
		\bibfield  {author} {\bibinfo {author} {\bibfnamefont {R.~C.~C.}\
				\bibnamefont {Leon}}, \bibinfo {author} {\bibfnamefont {C.~H.}\ \bibnamefont
				{Yang}}, \bibinfo {author} {\bibfnamefont {J.~C.~C.}\ \bibnamefont {Hwang}},
			\bibinfo {author} {\bibfnamefont {J.~C.}\ \bibnamefont {Lemyre}}, \bibinfo
			{author} {\bibfnamefont {T.}~\bibnamefont {Tanttu}}, \bibinfo {author}
			{\bibfnamefont {W.}~\bibnamefont {Huang}}, \bibinfo {author} {\bibfnamefont
				{K.~W.}\ \bibnamefont {Chan}}, \bibinfo {author} {\bibfnamefont {K.~Y.}\
				\bibnamefont {Tan}}, \bibinfo {author} {\bibfnamefont {F.~E.}\ \bibnamefont
				{Hudson}}, \bibinfo {author} {\bibfnamefont {K.~M.}\ \bibnamefont {Itoh}},
			\bibinfo {author} {\bibfnamefont {A.}~\bibnamefont {Morello}}, \bibinfo
			{author} {\bibfnamefont {A.}~\bibnamefont {Laucht}}, \bibinfo {author}
			{\bibfnamefont {M.}~\bibnamefont {Pioro-Ladri{\`e}re}}, \bibinfo {author}
			{\bibfnamefont {A.}~\bibnamefont {Saraiva}}, \ and\ \bibinfo {author}
			{\bibfnamefont {A.~S.}\ \bibnamefont {Dzurak}},\ }\href
		{https://doi.org/10.1038/s41467-019-14053-w} {\bibfield  {journal} {\bibinfo
				{journal} {Nat. Commun.}\ }\textbf {\bibinfo {volume} {11}},\ \bibinfo
			{pages} {797} (\bibinfo {year} {2020})}\BibitemShut {NoStop}%
		\bibitem [{\citenamefont {Ercan}\ \emph {et~al.}(2021)\citenamefont {Ercan},
			\citenamefont {Coppersmith},\ and\ \citenamefont {Friesen}}]{Ercan.21}%
		\BibitemOpen
		\bibfield  {author} {\bibinfo {author} {\bibfnamefont {H.~E.}\ \bibnamefont
				{Ercan}}, \bibinfo {author} {\bibfnamefont {S.~N.}\ \bibnamefont
				{Coppersmith}}, \ and\ \bibinfo {author} {\bibfnamefont {M.}~\bibnamefont
				{Friesen}},\ }\href {\doibase 10.1103/PhysRevB.104.235302} {\bibfield
			{journal} {\bibinfo  {journal} {Phys. Rev. B}\ }\textbf {\bibinfo {volume}
				{104}},\ \bibinfo {pages} {235302} (\bibinfo {year} {2021})}\BibitemShut
		{NoStop}%
		\bibitem [{\citenamefont {Hendrickx}\ \emph {et~al.}(2020)\citenamefont
			{Hendrickx}, \citenamefont {Franke}, \citenamefont {Sammak}, \citenamefont
			{Scappucci},\ and\ \citenamefont {Veldhorst}}]{Hendrickx.20}%
		\BibitemOpen
		\bibfield  {author} {\bibinfo {author} {\bibfnamefont {N.~W.}\ \bibnamefont
				{Hendrickx}}, \bibinfo {author} {\bibfnamefont {D.~P.}\ \bibnamefont
				{Franke}}, \bibinfo {author} {\bibfnamefont {A.}~\bibnamefont {Sammak}},
			\bibinfo {author} {\bibfnamefont {G.}~\bibnamefont {Scappucci}}, \ and\
			\bibinfo {author} {\bibfnamefont {M.}~\bibnamefont {Veldhorst}},\ }\href
		{https://doi.org/10.1038/s41586-019-1919-3} {\bibfield  {journal} {\bibinfo
				{journal} {Nature}\ }\textbf {\bibinfo {volume} {577}},\ \bibinfo {pages}
			{487} (\bibinfo {year} {2020})}\BibitemShut {NoStop}%
		\bibitem [{\citenamefont {Hendrickx}\ \emph {et~al.}(2018)\citenamefont
			{Hendrickx}, \citenamefont {Franke}, \citenamefont {Sammak}, \citenamefont
			{Kouwenhoven}, \citenamefont {Sabbagh}, \citenamefont {Yeoh}, \citenamefont
			{Li}, \citenamefont {Tagliaferri}, \citenamefont {Virgilio}, \citenamefont
			{Capellini}, \citenamefont {Scappucci},\ and\ \citenamefont
			{Veldhorst}}]{Hendrickx.18}%
		\BibitemOpen
		\bibfield  {author} {\bibinfo {author} {\bibfnamefont {N.~W.}\ \bibnamefont
				{Hendrickx}}, \bibinfo {author} {\bibfnamefont {D.~P.}\ \bibnamefont
				{Franke}}, \bibinfo {author} {\bibfnamefont {A.}~\bibnamefont {Sammak}},
			\bibinfo {author} {\bibfnamefont {M.}~\bibnamefont {Kouwenhoven}}, \bibinfo
			{author} {\bibfnamefont {D.}~\bibnamefont {Sabbagh}}, \bibinfo {author}
			{\bibfnamefont {L.}~\bibnamefont {Yeoh}}, \bibinfo {author} {\bibfnamefont
				{R.}~\bibnamefont {Li}}, \bibinfo {author} {\bibfnamefont {M.~L.~V.}\
				\bibnamefont {Tagliaferri}}, \bibinfo {author} {\bibfnamefont
				{M.}~\bibnamefont {Virgilio}}, \bibinfo {author} {\bibfnamefont
				{G.}~\bibnamefont {Capellini}}, \bibinfo {author} {\bibfnamefont
				{G.}~\bibnamefont {Scappucci}}, \ and\ \bibinfo {author} {\bibfnamefont
				{M.}~\bibnamefont {Veldhorst}},\ }\href
		{https://doi.org/10.1038/s41467-018-05299-x} {\bibfield  {journal} {\bibinfo
				{journal} {Nat. Commun.}\ }\textbf {\bibinfo {volume} {9}},\ \bibinfo {pages}
			{2835} (\bibinfo {year} {2018})}\BibitemShut {NoStop}%
		\bibitem [{\citenamefont {Hardy}\ \emph {et~al.}(2019)\citenamefont {Hardy},
			\citenamefont {Harris}, \citenamefont {Su}, \citenamefont {Chuang},
			\citenamefont {Moussa}, \citenamefont {Maurer}, \citenamefont {Li},
			\citenamefont {Lu},\ and\ \citenamefont {Luhman}}]{Hardy.19}%
		\BibitemOpen
		\bibfield  {author} {\bibinfo {author} {\bibfnamefont {W.~J.}\ \bibnamefont
				{Hardy}}, \bibinfo {author} {\bibfnamefont {C.~T.}\ \bibnamefont {Harris}},
			\bibinfo {author} {\bibfnamefont {Y.-H.}\ \bibnamefont {Su}}, \bibinfo
			{author} {\bibfnamefont {Y.}~\bibnamefont {Chuang}}, \bibinfo {author}
			{\bibfnamefont {J.}~\bibnamefont {Moussa}}, \bibinfo {author} {\bibfnamefont
				{L.~N.}\ \bibnamefont {Maurer}}, \bibinfo {author} {\bibfnamefont {J.-Y.}\
				\bibnamefont {Li}}, \bibinfo {author} {\bibfnamefont {T.-M.}\ \bibnamefont
				{Lu}}, \ and\ \bibinfo {author} {\bibfnamefont {D.~R.}\ \bibnamefont
				{Luhman}},\ }\href {\doibase 10.1088/1361-6528/ab061e} {\bibfield  {journal}
			{\bibinfo  {journal} {Nanotechnology}\ }\textbf {\bibinfo {volume} {30}},\
			\bibinfo {pages} {215202} (\bibinfo {year} {2019})}\BibitemShut {NoStop}%
		\bibitem [{\citenamefont {Sammak}\ \emph {et~al.}(2019)\citenamefont {Sammak},
			\citenamefont {Sabbagh}, \citenamefont {Hendrickx}, \citenamefont {Lodari},
			\citenamefont {Paquelet~Wuetz}, \citenamefont {Tosato}, \citenamefont {Yeoh},
			\citenamefont {Bollani}, \citenamefont {Virgilio}, \citenamefont {Schubert},
			\citenamefont {Zaumseil}, \citenamefont {Capellini}, \citenamefont
			{Veldhorst},\ and\ \citenamefont {Scappucci}}]{Sammak.19}%
		\BibitemOpen
		\bibfield  {author} {\bibinfo {author} {\bibfnamefont {A.}~\bibnamefont
				{Sammak}}, \bibinfo {author} {\bibfnamefont {D.}~\bibnamefont {Sabbagh}},
			\bibinfo {author} {\bibfnamefont {N.~W.}\ \bibnamefont {Hendrickx}}, \bibinfo
			{author} {\bibfnamefont {M.}~\bibnamefont {Lodari}}, \bibinfo {author}
			{\bibfnamefont {B.}~\bibnamefont {Paquelet~Wuetz}}, \bibinfo {author}
			{\bibfnamefont {A.}~\bibnamefont {Tosato}}, \bibinfo {author} {\bibfnamefont
				{L.}~\bibnamefont {Yeoh}}, \bibinfo {author} {\bibfnamefont {M.}~\bibnamefont
				{Bollani}}, \bibinfo {author} {\bibfnamefont {M.}~\bibnamefont {Virgilio}},
			\bibinfo {author} {\bibfnamefont {M.~A.}\ \bibnamefont {Schubert}}, \bibinfo
			{author} {\bibfnamefont {P.}~\bibnamefont {Zaumseil}}, \bibinfo {author}
			{\bibfnamefont {G.}~\bibnamefont {Capellini}}, \bibinfo {author}
			{\bibfnamefont {M.}~\bibnamefont {Veldhorst}}, \ and\ \bibinfo {author}
			{\bibfnamefont {G.}~\bibnamefont {Scappucci}},\ }\href
		{https://onlinelibrary.wiley.com/doi/abs/10.1002/adfm.201807613} {\bibfield
			{journal} {\bibinfo  {journal} {Adv. Funct. Mater.}\ }\textbf {\bibinfo
				{volume} {29}},\ \bibinfo {pages} {1807613} (\bibinfo {year}
			{2019})}\BibitemShut {NoStop}%
		\bibitem [{\citenamefont {Terrazos}\ \emph {et~al.}(2021)\citenamefont
			{Terrazos}, \citenamefont {Marcellina}, \citenamefont {Wang}, \citenamefont
			{Coppersmith}, \citenamefont {Friesen}, \citenamefont {Hamilton},
			\citenamefont {Hu}, \citenamefont {Koiller}, \citenamefont {Saraiva},
			\citenamefont {Culcer},\ and\ \citenamefont {Capaz}}]{Terrazos.21}%
		\BibitemOpen
		\bibfield  {author} {\bibinfo {author} {\bibfnamefont {L.~A.}\ \bibnamefont
				{Terrazos}}, \bibinfo {author} {\bibfnamefont {E.}~\bibnamefont
				{Marcellina}}, \bibinfo {author} {\bibfnamefont {Z.}~\bibnamefont {Wang}},
			\bibinfo {author} {\bibfnamefont {S.~N.}\ \bibnamefont {Coppersmith}},
			\bibinfo {author} {\bibfnamefont {M.}~\bibnamefont {Friesen}}, \bibinfo
			{author} {\bibfnamefont {A.~R.}\ \bibnamefont {Hamilton}}, \bibinfo {author}
			{\bibfnamefont {X.}~\bibnamefont {Hu}}, \bibinfo {author} {\bibfnamefont
				{B.}~\bibnamefont {Koiller}}, \bibinfo {author} {\bibfnamefont {A.~L.}\
				\bibnamefont {Saraiva}}, \bibinfo {author} {\bibfnamefont {D.}~\bibnamefont
				{Culcer}}, \ and\ \bibinfo {author} {\bibfnamefont {R.~B.}\ \bibnamefont
				{Capaz}},\ }\href {https://link.aps.org/doi/10.1103/PhysRevB.103.125201}
		{\bibfield  {journal} {\bibinfo  {journal} {Phys. Rev. B}\ }\textbf {\bibinfo
				{volume} {103}},\ \bibinfo {pages} {125201} (\bibinfo {year}
			{2021})}\BibitemShut {NoStop}%
		\bibitem [{\citenamefont {Liles}\ \emph {et~al.}(2018)\citenamefont {Liles},
			\citenamefont {Li}, \citenamefont {Yang}, \citenamefont {Hudson},
			\citenamefont {Veldhorst}, \citenamefont {Dzurak},\ and\ \citenamefont
			{Hamilton}}]{Liles.18}%
		\BibitemOpen
		\bibfield  {author} {\bibinfo {author} {\bibfnamefont {S.~D.}\ \bibnamefont
				{Liles}}, \bibinfo {author} {\bibfnamefont {R.}~\bibnamefont {Li}}, \bibinfo
			{author} {\bibfnamefont {C.~H.}\ \bibnamefont {Yang}}, \bibinfo {author}
			{\bibfnamefont {F.~E.}\ \bibnamefont {Hudson}}, \bibinfo {author}
			{\bibfnamefont {M.}~\bibnamefont {Veldhorst}}, \bibinfo {author}
			{\bibfnamefont {A.~S.}\ \bibnamefont {Dzurak}}, \ and\ \bibinfo {author}
			{\bibfnamefont {A.~R.}\ \bibnamefont {Hamilton}},\ }\href
		{https://doi.org/10.1038/s41467-018-05700-9} {\bibfield  {journal} {\bibinfo
				{journal} {Nat. Commun.}\ }\textbf {\bibinfo {volume} {9}},\ \bibinfo {pages}
			{3255} (\bibinfo {year} {2018})}\BibitemShut {NoStop}%
		\bibitem [{\citenamefont {Mutter}\ and\ \citenamefont
			{Burkard}(2021)}]{Mutter.21}%
		\BibitemOpen
		\bibfield  {author} {\bibinfo {author} {\bibfnamefont {P.~M.}\ \bibnamefont
				{Mutter}}\ and\ \bibinfo {author} {\bibfnamefont {G.}~\bibnamefont
				{Burkard}},\ }\href
		{https://link.aps.org/doi/10.1103/PhysRevResearch.3.013194} {\bibfield
			{journal} {\bibinfo  {journal} {Phys. Rev. Research}\ }\textbf {\bibinfo
				{volume} {3}},\ \bibinfo {pages} {013194} (\bibinfo {year}
			{2021})}\BibitemShut {NoStop}%
		\bibitem [{\citenamefont {Mutter}\ and\ \citenamefont
			{Burkard}(2020)}]{Mutter.20}%
		\BibitemOpen
		\bibfield  {author} {\bibinfo {author} {\bibfnamefont {P.~M.}\ \bibnamefont
				{Mutter}}\ and\ \bibinfo {author} {\bibfnamefont {G.}~\bibnamefont
				{Burkard}},\ }\href {https://link.aps.org/doi/10.1103/PhysRevB.102.205412}
		{\bibfield  {journal} {\bibinfo  {journal} {Phys. Rev. B}\ }\textbf {\bibinfo
				{volume} {102}},\ \bibinfo {pages} {205412} (\bibinfo {year}
			{2020})}\BibitemShut {NoStop}%
		\bibitem [{\citenamefont {Hu}\ \emph {et~al.}(2007)\citenamefont {Hu},
			\citenamefont {Churchill}, \citenamefont {Reilly}, \citenamefont {Xiang},
			\citenamefont {Lieber},\ and\ \citenamefont {Marcus}}]{Hu.07}%
		\BibitemOpen
		\bibfield  {author} {\bibinfo {author} {\bibfnamefont {Y.}~\bibnamefont
				{Hu}}, \bibinfo {author} {\bibfnamefont {H.~O.~H.}\ \bibnamefont
				{Churchill}}, \bibinfo {author} {\bibfnamefont {D.~J.}\ \bibnamefont
				{Reilly}}, \bibinfo {author} {\bibfnamefont {J.}~\bibnamefont {Xiang}},
			\bibinfo {author} {\bibfnamefont {C.~M.}\ \bibnamefont {Lieber}}, \ and\
			\bibinfo {author} {\bibfnamefont {C.~M.}\ \bibnamefont {Marcus}},\ }\href
		{https://doi.org/10.1038/nnano.2007.302} {\bibfield  {journal} {\bibinfo
				{journal} {Nat. Nanotechnol.}\ }\textbf {\bibinfo {volume} {2}},\ \bibinfo
			{pages} {622} (\bibinfo {year} {2007})}\BibitemShut {NoStop}%
		\bibitem [{\citenamefont {Sigillito}\ \emph {et~al.}(2015)\citenamefont
			{Sigillito}, \citenamefont {Jock}, \citenamefont {Tyryshkin}, \citenamefont
			{Beeman}, \citenamefont {Haller}, \citenamefont {Itoh},\ and\ \citenamefont
			{Lyon}}]{Sigillito.15}%
		\BibitemOpen
		\bibfield  {author} {\bibinfo {author} {\bibfnamefont {A.~J.}\ \bibnamefont
				{Sigillito}}, \bibinfo {author} {\bibfnamefont {R.~M.}\ \bibnamefont {Jock}},
			\bibinfo {author} {\bibfnamefont {A.~M.}\ \bibnamefont {Tyryshkin}}, \bibinfo
			{author} {\bibfnamefont {J.~W.}\ \bibnamefont {Beeman}}, \bibinfo {author}
			{\bibfnamefont {E.~E.}\ \bibnamefont {Haller}}, \bibinfo {author}
			{\bibfnamefont {K.~M.}\ \bibnamefont {Itoh}}, \ and\ \bibinfo {author}
			{\bibfnamefont {S.~A.}\ \bibnamefont {Lyon}},\ }\href {\doibase
			10.1103/PhysRevLett.115.247601} {\bibfield  {journal} {\bibinfo  {journal}
				{Phys. Rev. Lett.}\ }\textbf {\bibinfo {volume} {115}},\ \bibinfo {pages}
			{247601} (\bibinfo {year} {2015})}\BibitemShut {NoStop}%
	\end{thebibliography}
\end{document}